\documentclass[a4paper, 12pt]{article}

\usepackage{jheppub}

\usepackage{mssjo}

\usepackage{cleveref}
\usepackage{subcaption}
\usepackage{multicol}   
\usepackage{enumitem}   
\usepackage{longtable}  
\usepackage{placeins}   

\usepackage{pgf,tikz}
\usetikzlibrary{patterns, patterns.meta, calc, snakes, decorations, shapes, external, fadings}
\newcommand{\hatchborderampl}{4pt}
\newcommand{\weakhatchborderampl}{3pt}

\tikzset{
    dot/.style={circle, draw, fill=#1, inner sep=1.4pt},
    dot/.default={none},
    pdot/.style={regular polygon, regular polygon sides=3, draw, fill=#1, inner sep=1pt},
    pdot/.default={none},
    zdot/.style={regular polygon, regular polygon sides=4, draw, fill=#1, inner sep=1.2pt},
    zdot/.default={none},
    ndot/.style={regular polygon, regular polygon sides=3, shape border rotate=180, draw, fill=#1, inner sep=1pt},
    ndot/.default={none},
    hatchborder/.style={
        postaction={decorate, draw, thin}, decoration={border, segment length=2pt, amplitude=\hatchborderampl, angle=225}
    },
    weakhatchborder/.style={
        postaction={decorate, draw, thin}, decoration={border, segment length=3pt, amplitude=\weakhatchborderampl, angle=225}
    },
    invhatchborder/.style={
        postaction={decorate, draw, thin}, decoration={border, segment length=2pt, amplitude=\hatchborderampl, angle=45}
    },
    weakinvhatchborder/.style={
        postaction={decorate, draw, thin}, decoration={border, segment length=3pt, amplitude=\weakhatchborderampl, angle=135}
    },
    relevant/.style={very thick, hatchborder},
    irrelevant/.style={thin, opacity=0.5, weakhatchborder},
    facet/.style={color=., fill=., fill opacity=#1, join=round}, facet/.default={.8},
    axes to fid/.style={thick, black, densely dotted},
    surface to fid/.style={thick, black, stealth-}
}

\newcommand{\constrline}[3][]{%
    \draw[#1] %
    let \p1 = (#2) in%
    ($ #3*1cm*1cm/(veclen(\p{1})*veclen(\p{1}))*(\p{1}) $)  +($ (0,0)!10!90:(\p{1}) $) -- +($ (0,0)!10!270:(\p{1}) $)%
} 

\newcommand{\xconstrline}[5]{%
    \draw[#3, #1, name path=#4path#2] %
    let \p1 = (#4#2) in%
    ($ #5*1cm*1cm/(veclen(\p{1})*veclen(\p{1}))*(\p{1}) $)  +($ (0,0)!10!90:(\p{1}) $) -- +($ (0,0)!10!270:(\p{1}) $)%
}
\newcommand{\pconstrline}[2][]{\xconstrline{#1}{#2}{pcolour}{p}{1}}
\newcommand{\nconstrline}[2][]{\xconstrline{#1}{#2}{ncolour}{n}{-1}}
\newcommand{\zconstrline}[2][]{\xconstrline{#1}{#2}{zcolour}{z}{0}}

\usepackage{pgfplots}
\pgfplotsset{compat=1.16}
\usepgfplotslibrary{fillbetween}
\usepgfplotslibrary{colorbrewer}
\usepgfplotslibrary{colormaps}
\usepgfplotslibrary{groupplots}
\usepgfplotslibrary{patchplots}

\newcommand{\relmark}{pentagon*}
\pgfplotsset{
    pin near coord/.style args={#1/#2}{
        scatter/@pre marker code/.append code={
            \ifnum\coordindex=#1 \node[pin=#2]{};\fi
        }
    },
    pins near some coords/.style={ 
        scatter,
        scatter/@pre marker code/.code={},
        scatter/@post marker code/.code={},%
        pin near coord/.list={#1} 
    },
    discard if/.style 2 args={
        x filter/.code={
            \edef\tempa{\thisrow{#1}}
            \edef\tempb{#2}
            \ifx\tempa\tempb
                
            \fi
        }
    },
    discard if not/.style 2 args={
        x filter/.code={
            \edef\tempa{\thisrow{#1}}
            \edef\tempb{#2}
            \ifx\tempa\tempb
            \else
                
            \fi
        }
    },
    colormap={invhot}{
        [1cm]rgb255(0cm)=(255,255,255) rgb255(6cm)=(255,255,0) rgb255(8cm)=(255,0,0)
    }
}

\definecolour{plotI}{HTML}{0077BB}
\definecolour{plotII}{HTML}{33BBEE}
\definecolour{plotIII}{HTML}{009988}
\definecolour{plotIV}{HTML}{EE7733}
\definecolour{plotV}{HTML}{CC3311}
\definecolour{plotVI}{HTML}{EE3377}
\definecolour{plotgrey}{HTML}{BBBBBB}

\newcommand{\Mphys}{M_\text{phys}}
\newcommand{\pix}{\pi_{16}}

\newcommand{\tbe}{\tilde\beta}
\newcommand{\phantomplus}{\phantom{{}+{}}}
\newcommand{\phantomeq}{\phantom{{}={}}}

\newcommand{\Ze}[3][]{    Z^{#2,#3}_{#1}}
\newcommand{\ze}[3][]{\zeta^{#2,#3}_{#1}}
\newcommand{\Ph}[2]{\Phi^{(#1)}_{#2}}
\newcommand{\Ps}[2][]{\Psi^{#1}_{#2}}

\definecolour{SU2colour}{HTML}{228833}
\newcommand{\SUIIcolcol}{\textcolour{SU2colour}{\bfseries green}}
\definecolour{SU3colour}{HTML}{66CCEE}
\newcommand{\SUIIIcolcol}{\textcolour{SU3colour}{\bfseries cyan}}
\definecolour{SU4colour}{HTML}{AA3377}
\newcommand{\SUIVcolcol}{\textcolour{SU4colour}{\bfseries purple}}

\newcommand{\intfile}[3]{../code/constrscan/scans/#1/#2/#3/integrals/}
\newcommand{\plotcolour}[2]{black!#2!#1}
\newcommand{\plotname}[3]{#1_#2_#3}

\newcommand{\plotint}[6]{%
    \addplot [name path=\plotname{NLO}{#2}{#3}, thick, \plotcolour{#4}{#5}, no marks] table [x={lambda}, y expr=\thisrow{#2}+#6]{\intfile{NLO}{#1}{D2}integral_s#3.dat};%
    \addplot [name path=\plotname{NNLO}{#2}{#3}, dashed, \plotcolour{#4}{#5}, no marks] table [x={lambda}, y expr=\thisrow{#2}+#6]{\intfile{NNLO}{#1}{D2}integral_s#3.dat};%
    \addplot[color=\plotcolour{#4}{#5}, opacity=.5] fill between[of=\plotname{NLO}{#2}{#3} and \plotname{NNLO}{#2}{#3}];%
}

\newcommand{\plotseries}[3]{%
        \plotint{#1}{#2}{0_2}{#1colour}{0}{#3}%
        \plotint{#1}{#2}{p1_2}{#1colour}{25}{#3}%
        \plotint{#1}{#2}{p2_2}{#1colour}{50}{#3}%
        \plotint{#1}{#2}{p3_2}{#1colour}{75}{#3}%
}

\usepackage{array}
\newcolumntype{C}{>{$}c<{$}} 
\newcolumntype{R}{>{$}r<{$}} 
\newcolumntype{L}{>{$}l<{$}} 

\usepackage{amssymb}
\usepackage{amsthm}

\newcommand{\constr}[2]{\left\langle #1,#2\right\rangle} 

\newcommand{\ind}[2][]{\left\lfloor#2\right\rfloor_{#1}} 
\newcommand{\cind}[2][]{\left\lceil#2\right\rceil^{#1}}   

\DeclareRobustCommand{\mark}[2]{\tikzexternaldisable\raisebox{.2ex}{\tikz \draw[#1] (0,0) node[#2] {};}\tikzexternalenable}
\DeclareRobustCommand{\Mark}[2]{\mark{#1}{#2=#1}} 

\crefname{condition}{condition}{conditions}
\crefname{property}{property}{properties}
\crefname{definition}{definition}{definitions}
\crefname{case}{case}{cases}
\crefname{category}{category}{categories}
\crefname{correspondence}{correspondence}{correspondences}
\crefname{improvement}{}{}

\newtheorem{proposition}{Proposition}[section]
\newtheorem{corollary}{Corollary}[section]
\newtheorem{lemma}{Lemma}[section]

\renewcommand{\proof}{\noindent\textsc{Proof. }}
\renewcommand{\qed}{\hfill$\square$}
\newcommand{\QED}{\hfill$\blacksquare$}

\makeatletter
\newenvironment{mcases}[1][ll]{
    \let\@ifnextchar\new@ifnextchar
    \left\lbrace
    \def\arraystretch{1.2}%
    \array{@{}l@{\quad}#1@{}}
}{
    \endarray\right.
}
\makeatother

\DeclareMathOperator{\hull}{Hull}

\DeclareMathOperator{\cl}{cl}
\DeclareMathOperator{\Int}{int}
\DeclareMathOperator{\Span}{span}
\newcommand{\mathand}{\mathbin{\text{ and }}}

\newcommand{\sumI}[1]{\sum_{i\in I_{#1}}}
\newcommand{\pml}{{\pm1}}
\newcommand{\mpl}{{\mp1}}

\newcommand{\reg}{{\s A}}   
\newcommand{\sat}{{\s B}}   
\newcommand{\rel}{{\s R}}   
\renewcommand{\deg}[1]{d_{#1}}  

\newcommand{\lift}[2][]{\ell_{#1}(#2)}

\newcommand{\lOmega}{\omega^\ell_0}

\newcommand{\psP}[1]{\s P_{#1}}
\newcommand{\psZ}[2]{\s Z^{(#1)}_{#2}}
\newcommand{\psN}[2]{\s N^{(#1)}_{#2}}
\newcommand{\psH}[2]{\s H^{(#1)}_{#2}}
\newcommand{\psG}[2]{\s G^{(#1)}_{#2}}
\newcommand{\psJ}[2]{\s J_{#1}(#2)}
\newcommand{\isL}[2]{L^{(#1)}_{#2}}
\newcommand{\isM}[2]{M^{(#1)}_{#2}}
\newcommand{\isN}[2]{N^{(#1)}_{#2}}
\newcommand{\isJ}[2]{J^{#1}_{#2}}

\definecolour{pcolour}{HTML}{DDAA33}
\newcommand{\pcolour}{yellow}
\newcommand{\pmark}{\mark{pcolour}{pdot}}
\newcommand{\pMark}{\Mark{pcolour}{pdot}}
\newcommand{\pcolcol}{\textcolour{pcolour}{\bfseries\pcolour}}

\definecolour{zcolour}{HTML}{004488}
\newcommand{\zcolour}{blue}
\newcommand{\zmark}{\mark{zcolour}{zdot}}
\newcommand{\zMark}{\Mark{zcolour}{zdot}}
\newcommand{\zcolcol}{\textcolour{zcolour}{\bfseries\zcolour}}

\definecolour{ncolour}{HTML}{BB5566}
\newcommand{\ncolour}{red}
\newcommand{\nmark}{\mark{ncolour}{ndot}}
\newcommand{\nMark}{\Mark{ncolour}{ndot}}
\newcommand{\ncolcol}{\textcolour{ncolour}{\bfseries\ncolour}}

\newcommand{\nangle}{0}
\newcommand{\zangle}{0}
\newcommand{\pangle}{0}
\newcommand{\rotatecoordinate}[1]{%
    \coordinate (n#1) at ($ (0,0)!1! \nangle:(r#1) $);%
    \coordinate (z#1) at ($ (0,0)!1! \zangle:(r#1) $);%
    \coordinate (p#1) at ($ (0,0)!1! \pangle:(r#1) $);}
\newcommand{\regioncoords}[1][]{%
    \coordinate (orig) at (0,0 #1);
    \coordinate (r1)  at ( 0  ,3.5 #1);%
    \coordinate (r2)  at ( 2  ,2.4 #1);%
    \coordinate (r3)  at ( 2  ,1.2 #1);%
    \coordinate (r4)  at (  .4, .7 #1);%
    \coordinate (r5)  at (- .5, .8 #1);%
    \coordinate (r6)  at (-1.5,1.2 #1);%
    \coordinate (r7)  at (-1.4,2.7 #1);%
    \coordinate (r8)  at ( 1.4,1.5 #1);%
    \coordinate (r9)  at (  .6,1.3 #1);%
    \coordinate (r10) at (  .8,2.3 #1);%
    \coordinate (r11) at (- .9,2.5 #1);%
    \coordinate (r12) at (- .4,1.7 #1);%
    \rotatecoordinate{1};%
    \rotatecoordinate{2};%
    \rotatecoordinate{3};%
    \rotatecoordinate{4};%
    \rotatecoordinate{5};%
    \rotatecoordinate{6};%
    \rotatecoordinate{7};%
    \rotatecoordinate{8};%
    \rotatecoordinate{9};%
    \rotatecoordinate{10};%
    \rotatecoordinate{11};%
    \rotatecoordinate{12};%
}
\newcommand{\coordaxes}{%
    \draw[thin] (-4,0) -- (4,0);%
    \draw[thin] (0,-4) -- (0,4);%
    \clip (0,0) circle[radius=4];%
}

\title{NNLO Positivity Bounds\\on Chiral Perturbation Theory\\for a General Number of Flavours}
\author[a]{Benjamin Alvarez,}
\author[b]{Johan Bijnens,}
\author[b]{Mattias Sjö}

\affiliation[a]{Aix Marseille Univ, Univ Toulon,\\ 
    CNRS, CPT, Marseille, France}
\affiliation[b]{Department of Astronomy and Theoretical Physics, Lund University,\\
    Box 43, SE 22100 Lund, Sweden}

\emailAdd{benjamin-alvarez@univ-tln.fr}
\emailAdd{bijnens@thep.lu.se}
\emailAdd{mattias.sjo@thep.lu.se}

\abstract{
    We present positivity bounds, derived from the principles of analyticity, unitarity and crossing symmetry, that constrain the low-energy constants of chiral perturbation theory.
    Bounds are produced for 2, 3 or more flavours in meson-meson scattering with equal meson masses, up to and including next-to-next-to-leading order (NNLO), using the second and higher derivatives of the amplitude.
    We enhance the bounds by using the most general isospin combinations posible (or higher-flavour counterparts thereof) and by analytically integrating the low-energy range of the discontinuities.
    In addition, we present a powerful and general mathematical framework for efficiently managing large numbers of positivity bounds.
}

\preprint{\begin{minipage}{3cm} LU TP 21-50 \\ December 2021 \end{minipage}}

\begin{document}

\maketitle


\section{Introduction}
\label{sec:intro}
Chiral perturbation theory (\chpt) is the most widespread theory for low-energy quantum chromodynamics (QCD). 
It is an effective field theory (EFT) which reformulates the non-perturbative behaviour of low-energy QCD as a perturbative theory of new degrees of freedom, physically interpreted as bound states of quarks. 
When constructed using $n$ light quark flavours, the degrees of freedom are the $n^2-1$ light pseudoscalar mesons: the pions for $n=2$, with the kaons and eta added for $n=3$. 
\chpt\ was developed by Gasser \& Leutwyler \cite{Gasser:1983yg,Gasser:1984gg} based on earlier work by Weinberg \cite{Weinberg:1978kz}; see \cite{Scherer:2012xha,Pich:2018ltt} for modern introductions with further references. 

At leading order in the low-energy expansion, the only parameters of \chpt\ are the meson mass and decay constant, but higher orders introduce a rapidly increasing number of Wilson coefficients or low-energy constants (LECs) which, while in principle derivable from the underlying QCD dynamics, must in practice be seen as unknowns. 
At next-to-leading order (NLO), the LECs can be measured reasonably well with experimental or lattice methods, although the precision is typically only one or two significant digits. 
At next-to-next-to-leading order (NNLO), only tentative results are presently available. 
For a review of LEC measurements, see \cite{Bijnens:2014lea}.

All quantum field theories must obey the axioms of unitarity, analyticity and crossing symmetry, and normally do so by construction. 
However, it turns out that these axioms are not automatically satisfied by EFTs such as \chpt\ when perturbativity is assumed at a fixed order in the expansion. 
Therefore, imposing the axioms actually adds new information, typically by placing bounds on the scattering amplitudes. 
Pioneering work was done by Martin \cite{Martin:1969ina} before the development of \chpt\ as such.
Bounds on NLO two-flavour \chpt\ amplitudes, which in turn translate to bounds on the LECs, were first obtained in \cite{Pham:1985cr,Ananthanarayan:1994hf,Pennington:1994kc} and extended in \cite{Dita:1998mh,Distler:2006if}. 
Further improvements were made in \cite{Manohar:2008tc} and extended to three-flavour \chpt\ in \cite{Mateu:2008gv}.
There is ongoing research in extending these methods, both specific to \chpt\ and with broader scope; recent examples include \cite{Bellazzini:2020cot,Caron-Huot:2020cmc,Sinha:2020win,Zahed:2021fkp}.

The method of \cite{Manohar:2008tc,Mateu:2008gv}, which serves as the basis of our method, is to apply dispersion relations (a consequence of analyticity) to a meson-meson scattering amplitude decomposed into isospin components (for higher flavours, the Clebsch-Gordan decomposition is used).
Then, crossing symmetry and the optical theorem (a consequence of unitarity) are applied to give a positivity condition on the decomposed amplitude.
With the amplitude calculated in terms of the LECs to some order, this results in bounds on linear combinations of LECs.
More recently, stronger bounds have been obtained in \cite{Wang:2020jxr,Tolley:2020gtv} by improving this method; put extremely simply, this was done with more sophisticated use of dispersion relations and crossing symmetry, respectively.
Put similarly simply, our work instead improves the handling of the isospin decompositions and the LEC bounds themselves, although some improvements similar to \cite{Wang:2020jxr} are also made.
Perhaps more importantly, we perform the first extension to NNLO \chpt\ with any number of flavours (two flavours was treated in \cite{Wang:2020jxr}), albeit with the simplification that all mesons have the same mass.
The LECs are independent of the chosen masses, although the bounds do depend on the mass. 
At NLO they depend only on the ratio of the meson mass and the subtraction scale $\mu$, at NNLO also on the ratio of the meson mass and decay constant.

Preliminary results of this work are presented in the Lund University master thesis \cite{AlvarezThesis}.
Our work is structured as follows: \cref{sec:chPT} introduces \chpt\ and its LECs; \cref{sec:ampl} (backed by \cref{sec:LEC-details}) presents the $2\to2$ meson scattering amplitude used to obtain the bounds; \cref{sec:constr} (backed by \cref{sec:constraints}) introduces the mathematical framework used to manage them; \cref{sec:bounds} (backed by \cref{sec:Jbar}) presents the method of \cite{Manohar:2008tc,Mateu:2008gv} and the improvements made to it; and \cref{sec:results} displays the most interesting bounds we obtain, with final remarks given in \cref{sec:final}.

\section{Chiral perturbation theory}
\label{sec:chPT}
$n$-flavour \chpt\ is based around a non-linear sigma model (NLSM), whose degrees of freedom are the $n^2-1$ Nambu-Goldstone bosons that arise when the chiral symmetry $G=\SU(n)_L\times\SU(n)_R$  of $n$-flavour massless QCD is spontaneously broken into its diagonal subgroup $H=\SU(n)_V$.
The Goldstone bosons live in the coset space $G/H$, which is isomorphic to $\SU(n)$. 

The presence of quark masses, electroweak interactions, etc.\ can be accounted for by including four external $n\times n$ flavour-space matrix fields --- $s$ (scalar), $p$ (pseudo\-scalar), $v_\mu$ (vector) and $a_\mu$ (axial vector)%
\footnote{
    One can add more types of externals fields to \chpt. 
    Examples are symmetric or antisymmetric tensors \cite{Cata:2007ns,Donoghue:1991qv}. 
    These extensions are not relevant for this work.}
--- into the massless QCD Lagrangian.
These additions were introduced in \cite{Gasser:1983yg,Gasser:1984gg}, and endow the Nambu-Goldstone bosons with masses and interactions that allow them to accurately model the light pseudoscalar mesons, turning the $\SU(n)$ NLSM into \chpt\ proper.

The Nambu-Goldstone boson fields can be organised into a $n\times n$ flavour-space matrix field $u(\phi)$ \cite{Coleman:1969sm,Callan:1969sn}.
Under the chiral transformation $(g_L,g_R)\in G$, $u(\phi)$ transforms as
\begin{equation}
    u(\phi) \transf{} g_R\, u(\phi)\, h\big[g_L,g_R,u(\phi)\big] = h\big[g_L,g_R,u(\phi)\big]\, u(\phi)\,g_L^\dag,
\end{equation}
where $h\in H$ is defined by this transformation.
By requiring that $G$ can be made local while leaving the extended QCD Lagrangian invariant, it can be shown that
\begin{equation}
    \label{eq:def-chi-lr}
    \begin{aligned}
        \chi \equiv 2B(s + ip)      &\transf{} g_R\chi g_L^\dag,\\
        \ell_\mu \equiv v_\mu-a_\mu &\transf{} g_L\ell_\mu g_L^\dag - i\p_\mu g_Lg_L^\dag,\\
        r_\mu \equiv v_\mu+a_\mu    &\transf{} g_R r_\mu g_R^\dag - i\p_\mu g_Rg_R^\dag,
    \end{aligned}
\end{equation}
where $B$ is a constant related to the leading-order (LO) meson decay constant and the $\langle \bar q q\rangle$ condensate.

It is possible to rewrite $u(\phi),\chi,\ell_\mu,r_\mu$ in a basis of fields that transform entirely in terms of $g_L$ and $g_R$, as is done in \cite{Gasser:1984gg} to derive the NLO \chpt\ Lagrangian.
We instead choose to follow \cite{Ecker:1988te,Bijnens:1999sh,Bijnens:2018lez} and rewrite them in a basis of fields that all transform as $X\to hXh^\dag$:
\begin{equation}
    \begin{aligned}
        u_\mu           &\equiv i\left[u^\dag(\p_\mu - ir_\mu)u - u(\p_\mu - i\ell_\mu)u^\dag\right],\\
        \chi_\pm        &\equiv u^\dag\chi u^\dag \pm u\chi^\dag u,\\
        f^{\mu\nu}_\pm  &\equiv u F^{\mu\nu}_L u^\dag \pm u^\dag F^{\mu\nu}_R u,
    \end{aligned}
\end{equation}
where $F_L^{\mu\nu} \equiv \p^\mu\ell^\nu - \p^\nu\ell^\mu - i\comm{\ell^\mu}{\ell^\nu}$ and similarly for $F_R^{\mu\nu}$ and $r^\mu$.
These transformation properties are conserved under the covariant derivative $\nabla_\mu$ defined as
\begin{equation}
    \nabla_\mu X = \p_\mu X + \comm{\Gamma_\mu}{X},\qquad 
    \Gamma_\mu \equiv \frac12\left[u^\dag(\p_\mu - ir_\mu)u + u(\p_\mu - i\ell_\mu)u^\dag\right].
\end{equation}

\subsection{The \chpt\ Lagrangian}\label{sec:lagr}
There exists an infinite number of possible Lagrangian terms consistent with the symmetries of \chpt.
They can be organised into a power-counting hierarchy in the small energy-momentum scale $p$, where $u_\mu,\nabla_\mu=\O(p)$ and $\chi_\pm,f^{\mu\nu}_\pm=\O(p^2)$.
Thus,
\begin{equation}
    \lagr_\text{\chpt} = \lagr_2 + \lagr_4 + \lagr_6 + \ldots,
\end{equation}
where $\lagr_{2n}$ is $\O(p^{2n})$; odd powers are forbidden by parity.
The coefficient of each term in $\lagr_{2n}$ is a separate LEC.%
\footnote{
    Some ``terms'', like the one associated with $\hat L_{10}$ in \cref{eq:NLO-lagr} below, actually consist of several terms. 
    These transform into each other under the discrete symmetries of the Lagrangian, and must therefore appear with the same LEC.}

The LO Lagrangian is
\begin{equation}\label{eq:LO-lagr}
    \lagr_2 = \frac{F^2}{4}\tr{u_\mu u^\mu + \chi_+},
\end{equation}
where $F$ is a LEC related to the LO meson decay constant, and $\tr{\ldots}$ indicates a trace over flavour-space indices.
The LEC of the $\chi_+$ term is $\frac{BF^2}{4}$ as defined in \cref{eq:def-chi-lr}.
By requiring that the kinetic term is canonically normalised, one can fix $u(\phi) = 1 + \frac{it^a\phi^a}{F\sqrt 2} + \ldots$, where $t^a$ are the generators of $SU(n)$ and Einstein's summation convention is used.
The higher-order terms depend on the choice of parametrisation, which influences the computation of amplitudes but not the amplitudes themselves.

The next-to-leading-order (NLO) Lagrangian, which was first determined in \cite{Gasser:1984gg}, is in terms of our basis%
\footnote{
    There are two additional \emph{contact terms} proportional to $\tr{\chi_+^2 - \chi_-^2}$ and $\tr{f_+^{\mu\nu}f_{\mu\nu}^+ + f_-^{\mu\nu}f^-_{\mu\nu}}$.
    They are needed for renormalisation but make no physical contributions to the amplitudes considered here.}
\begin{equation}\label{eq:NLO-lagr}
    \begin{aligned}
        \lagr_4 = 
            &\phantomplus \hat L_0\tr{u_\mu u_\nu u^\mu u^\nu} + \hat L_1\tr{u_\mu u^\mu}^2 
                + \hat L_2\tr{u_\mu u_\nu}\tr{u^\mu u^\nu} + \hat L_3\tr{(u_\mu u^\mu)^2}\\
            &+ \hat L_4\tr{u_\mu u^\mu}\tr{\chi_+} + \hat L_5\tr{u_\mu u^\mu \chi_+} + \hat L_6\tr{\chi_+}^2 + \hat L_7\tr{\chi_-}^2 + \hat L_8\tr{\chi_+^2 + \chi_-^2}\\
            &- i\hat L_9\tr{f_+^{\mu\nu}u_\mu u_\nu} + \hat L_{10}\tr{f_+^{\mu\nu}f_{\mu\nu}^+ - f_-^{\mu\nu}f^-_{\mu\nu}},
    \end{aligned}
\end{equation}
where the LECs are $\hat L_i$.
The analogous NNLO Lagrangian with 112 LECs $K_i$ was determined in \cite{Bijnens:1999sh}.
The 1862-LEC NNNLO Lagrangian, which we do not use here, was determined in \cite{Bijnens:2018lez}.

For small $n$, the Cayley-Hamilton identity reduces the number of independent terms, and consequently the number of LECs.
At $n=3$, it is standard to eliminate $\hat L_0$; the remaining LECs are conventionally labelled $L_i$ with $i$ preserved.
At $n=2$, it is customary to also redefine the LECs slightly, resulting in the $l_i$ of the original Gasser-Leutwyler convention \cite{Gasser:1983yg}.
At NNLO, the 112+3 $K_i$ (ordinary+contact terms) are reduced to 90+4 $C_i$ at $n=3$ and 52+4 $c_i$ at $n=2$ as detailed in \cite{Bijnens:1999sh}.
For more details on the Lagrangians for different $n$, see \cite{Bijnens:2014lea,Bijnens:1999hw}.

The NLO renormalisation was first carried out in \cite{Gasser:1983yg,Gasser:1984gg}, and the extension to NNLO in \cite{Bijnens:1999hw}; for more information on \chpt\ renormalisation, see \cite{Bijnens:1997vq}. 
A slightly altered $\overline{\text{MS}}$ scheme is conventionally used, with renormalisation scale $\mu=0.77$~GeV.
The renormalised LECs are denoted $X_i^r$ where $X=\ell,L,\hat L,$ etc.
At $n=2$ flavours it is conventional to use $\bar\ell_i$ instead, related to $\ell_i^r$ through
\begin{equation}\label{eq:renorm}
    \ell^r_i = \frac{\gamma_i}{32\pi^2}\left[\bar\ell_i + \ln\left(\frac{\Mphys^2}{\mu^2}\right)\right]
\end{equation}
where $\Mphys$ is the chosen meson mass and $\gamma_i$ are coefficients found in \cite{Gasser:1983yg}.
Effectively, \cref{eq:renorm} sets the renormalisation scale to $\Mphys$ for $\bar\ell_i$.

\section{Scattering amplitudes}
\label{sec:ampl}

In this section, and in the remainder of the paper, we will restrict ourselves to a simplified version of \chpt.
Firstly, we will not include the external (axial) vector fields $a_\mu,v_\mu$ in the Lagrangian, which essentially amounts to ignoring electroweak corrections to the amplitude.
Secondly, we will assume that all mesons have the same mass $\Mphys$, as mentioned in the introduction.
While this limits the phenomenological applicability of three-flavour \chpt, it is a reasonable approximation that simplifies the procedure for obtaining bounds (see \cref{sec:bounds}).
More importantly, the full NNLO amplitude is currently not available in the general-mass case; available results only cover $\pi\pi$ scattering in two- \cite{Bijnens:1995yn,Bijnens:1997vq} and three-flavour \cite{Bijnens:2004eu} \chpt, as well as $\pi K$ scattering \cite{Bijnens:2004bu}, and are not expressed in terms of elementary functions.
With equal masses, we normalise all Mandelstam variables so that $s+t+u=4$. 

For the general equal-mass $n$-flavour scattering process $a+b\to c+d$, there are nine independent flavour structures possible: the six distinct index permutations on $\tr{t^at^bt^ct^d}$ and the three on $\tr{t^at^b}\tr{t^ct^d}$.
Due to charge conjugation symmetry, a permutation is not independent of its reverse.
Thus, the scattering amplitude $M$ may be decomposed as
\begin{equation}
    \label{eq:ampl-general}
    \begin{aligned}
        M(s,t,u) 
            =&\phantomplus \big[\tr{t^at^bt^ct^d} + \tr{t^dt^ct^bt^a}\big]B(s,t,u) \\
            &+ \big[\tr{t^at^ct^dt^b} + \tr{t^bt^dt^ct^a}\big]B(t,u,s) \\
            &+ \big[\tr{t^at^dt^bt^c} + \tr{t^ct^bt^dt^a}\big]B(u,s,t) \\
            &+ \kdu ab\kdu cd C(s,t,u) + \kdu ac\kdu bd C(t,u,s) + \kdu ad\kdu bc C(u,s,t),
    \end{aligned}
\end{equation}
where $s,t,u$ are the normalised Mandelstam variables, and crossing symmetry imposes that only two distinct functions $B,C$ are used.%
\footnote{
    These functions have the symmetries $B(s,t,u)=B(u,t,s)$ and $C(s,t,u)=C(s,u,t)$, which is consistent with the symmetries of the respective flavour structures.
    Likewise, $A(s,t,u)=A(s,u,t)$ holds in \cref{eq:ampl-su2}.}
This is the form used in \cite{Bijnens:2011fm}, where the functions $B,C$ are given to NNLO for $\SU(n)$ equal-mass \chpt.
The NLO results were first obtained in \cite{Gasser:1983yg,Chivukula:1992gi}.

\subsection{Other forms of the amplitude}
With two flavours, the traces can be evaluated in terms of Kronecker $\delta$'s, giving%
\footnote{
    This form can be traced back to the original current-algebra calculation \cite{Weinberg:1966kf} of the $\pi\pi$ amplitude.}
\begin{equation}
    \label{eq:ampl-su2}
    M(s,t,u) = \kdu ab\kdu cd A(s,t,u) + \kdu ac\kdu bd A(t,u,s) + \kdu ad\kdu bc A(u,s,t),
\end{equation}
which is the form used in \cite{Manohar:2008tc} (up to reordering the arguments as permitted by the symmetries of $A$).
In terms of the functions above,
\begin{equation}    
    \label{eq:A}
    A(s,t,u) = C(s,t,u) + B(s,t,u) + B(t,u,s) - B(u,s,t).
\end{equation}
the function $A$ was first determined to NLO in \cite{Gasser:1983yg}.

With $n$ flavours, the traces can be evaluated using the anticommutation relation $\acomm{t^a}{t^b} = \tfrac2n\kdu ab + d^{abc}t^c$ to give%
\footnote{
    The relevant identity is 
    \begin{equation*} 
        \tr{t^at^bt^ct^d} + \tr{t^at^dt^ct^b} 
            = \tfrac12\left(d^{abe}d^{cde} + d^{ade}d^{cbe} - d^{ace}d^{bde}\right)
            + \tfrac2n\left(\kdu ab\kdu cd + \kdu ad\kdu cb - \kdu ac \kdu bd\right).
    \end{equation*}
    It is most easily derived by first using $t^at^b = \tfrac1n\kdu ab + \tfrac12(d^{abc}+if^{abc})t^c$ repeatedly, and then removing all occurrences of $f$ with the Jacobi-like identity 
    \begin{equation*}
        f^{abe}f^{cde} = d^{ace}d^{bde} - d^{bce}d^{ade} + \tfrac4n\left(\kdu ac\kdu bd - \kdu ad\kdu bc\right),
    \end{equation*}
    which is derived from the observation that $\comm{\comm{t^a}{t^b}}{t^c} = \acomm{\acomm{t^b}{t^c}}{t^a} - \acomm{\acomm{t^c}{t^a}}{t^b}$.
}
\begin{equation}
    \label{eq:ampl-d}
    \begin{aligned}
        M(s,t,u) 
            =&\phantomplus d^{abe}d^{cde}B'(s,t,u) + d^{ace}d^{bde}B'(t,u,s) + d^{ade}d^{bce}B'(u,s,t) \\
            &+ \kdu ab\kdu cd C'(s,t,u) + \kdu ac\kdu bd C'(t,u,s) + \kdu ad\kdu bc C'(u,s,t),
    \end{aligned}
\end{equation}
where
\begin{equation}
    \begin{aligned}
        B'(s,t,u) &= \tfrac12\big[B(s,t,u) + B(t,u,s) - B(u,s,t)\big],\\
        C'(s,t,u) &= C(s,t,u) + \tfrac4n B'(s,t,u).
    \end{aligned}
\end{equation}
With three flavours, the Cayley-Hamilton theorem%
    \footnote{
        More specifically the $n=3$ Cayley-Hamilton theorem, recast as the $\SU(3)$-specific identity\\[-1em]
        \begin{equation*}    
            3\left(d^{abe}d^{cde} + d^{bce}d^{ade} + d^{cae}d^{bde}\right)
            = 2\left(\kdu ab\kdu cd + \kdu bc\kdu ad + \kdu ca\kdu bd\right).
        \end{equation*}
        \\[-2.5em] 
    }
allows for the removal of one term at the expense of symmetry, leaving
\begin{equation}
    \label{eq:ampl-su3}
    \begin{aligned}
        M(s,t,u) 
            =&\phantomplus \kdu ab\kdu cd A_1(s,t,u) + \kdu ac\kdu bd A_2(s,t,u) + \kdu ad\kdu bc A_3(s,t,u)\\            
             &+ d^{abe}d^{cde}B_1(s,t,u) + d^{ace}d^{bde}B_2(s,t,u) \\
    \end{aligned}
\end{equation}
where 
\begin{equation}
    \label{eq:AB-su3}
    \begin{gathered}
        B_1(s,t,u) = B(t,u,s)-B(u,s,t),\qquad
        B_2(s,t,u) = B(t,u,s)-B(s,t,u),\\
        \begin{aligned}
            A_1(s,t,u) &= C(s,t,u) + B(s,t,u) + \tfrac13 B_1(s,t,u),\\
            A_2(s,t,u) &= C(t,u,s) + B(u,s,t) + \tfrac13 B_2(s,t,u),\\
            A_3(s,t,u) &= C(u,s,t) + B(s,t,u) + B(u,s,t) - B(t,u,s).
        \end{aligned}\\
    \end{gathered}
\end{equation}
This is the form used in \cite{Mateu:2008gv}.

\subsection{Structure of the amplitude}\label{sec:structure}
The functions $B(s,t,u)$ and $C(s,t,u)$ consist of one part that is polynomial in the Mandelstam variables and contains the LECs, plus the so-called \emph{unitarity correction} that is non-polynomial in the Mandelstam variables.%
\footnote{
    This split is not uniquely defined, but we adhere to the conventions of \cite{Bijnens:2011fm}.}
The polynomial parts are quadratic at NLO and cubic at NNLO.
At NLO, the unitarity correction does not contain any LECs; at NNLO, the unitarity correction depends on the NLO LECs.

The unitarity correction at NLO depends on the function $\bar J$, which originates in the loop integral as shown in \cite{Gasser:1983yg}.
The NNLO unitarity correction introduces four analogous functions $k_i, i=1,\ldots 4$ \cite{Gasser:1998qt,Bijnens:1995yn,Bijnens:2004eu,Bijnens:2011fm}.
More details about these functions can be found in \cref{sec:Jbar}.

The LEC content of the amplitude considered here is more limited than that of the full \chpt\ Lagrangian.
About half of the Lagrangian terms are dropped by not including the external (axial) vector fields, and a significant part of the NNLO Lagrangian cannot appear in a 4-particle process below NNNLO.
Also, the number of LECs is reduced by the Cayley-Hamilton theorem in the 2- and 3-flavour case as described in \cref{sec:lagr}.
Lastly, $\hat L_7,K_{12},K_{24},K_{30},K_{34},K_{36},K_{41}$ and $K_{42}$, i.e.\ those whose Lagrangian terms contain $\tr{\chi_-}$, disappear in the equal-mass limit.%
\footnote{
    This can be understood by noting that $\chi_-$ has odd parity, so all terms in its expansion contain an odd number of pseudoscalar fields.
    If the even-parity Lagrangian term contains two traces of odd-parity objects, it can therefore only result in six-point vertices or larger, since the trace of a single field vanishes.
    Therefore, $K_{12},K_{24}$ etc.\ do not appear in the NNLO four-point amplitude, whereas $\hat L_7$ only appears in $s,t,u$-independent tadpole diagrams.
    As will be shown in \cref{sec:bounds}, we only consider $s$-derivatives of the amplitude, so also $\hat L_7$ disappears for our purposes.}
Even with these reductions, there are still 35 (27 at $n=3$, 18 at $n=2$) NNLO LECs that are involved in the amplitude at hand, in addition to 8 (7, 4) NLO LECs.

\subsection{Irreducible amplitudes}\label{sec:irred}
The scattered particles are in the adjoint representation of $\SU(n)$.
The Clebsch-Gordan decomposition of the initial and final states is therefore%
\footnote{
    \cite{Chivukula:1992gi} contains an intuitive description of how the decomposition is performed.}
\begin{equation}
    \label{eq:irreps}
    \text{Adj}\otimes\text{Adj} = R_I + R_S + R_A + R_S^A + R_A^S + R_S^S + R_A^A,
\end{equation}
where $R_I$ is the singlet representation, and the sub(super)scripts on the other representations indicate lower (upper) index pairs that are symmetric ($S$) or antisymmetric ($A$).
Details on the representations and their dimensions can be found in \cite{Neville:1963zz,Bijnens:2011fm}.
From this, it follows that the scattering amplitude can be decomposed in terms of seven corresponding irreducible amplitudes $T_J$.
In terms of \cref{eq:ampl-general}, these are
\begin{equation}
    \label{eq:ampl-irred}
    \begin{alignedat}{4}
        R_I:&&           T_I &= 2\frac{n^2-1}{n}\big[B(s,t,u)+B(t,u,s)\big] - \frac2n B(u,s,t) \\
                            &&&\phantomeq + (n^2-1)C(s,t,u) + C(t,u,s) + C(u,s,t),\\
        R_S:&&           T_S &= \frac{n^2-4}{n}\big[B(s,t,u)+B(t,u,s)\big] - \frac4n B(u,s,t) \\
                            &&&\phantomeq + C(t,u,s) + C(u,s,t),\\
        R_A:&&           T_A &= n\big[B(t,u,s)-B(s,t,u)\big] + C(t,u,s) - C(u,s,t),\\
        R^A_S,R^S_A:&\qquad&   T_{AS} = T_{SA} 
                            &= C(t,u,s) - C(u,s,t),\\
        R^S_S:&&         T_{SS} 
                            &= 2B(u,s,t)+C(t,u,s)+C(u,s,t),\\
        R^A_A:&&         T_{AA} 
                            &= -2B(u,s,t)+C(t,u,s)+C(u,s,t).
    \end{alignedat}
\end{equation}
Only six amplitudes are needed, since $T_{SA}$ and $T_{AS}$ are identical due to crossing symmetry etc., as mentioned in \cite{Mateu:2008gv}.

In $\SU(3)$, the $R_A^A$ representation vanishes, so only five amplitudes are needed.
In \cite{Mateu:2008gv}, the representations are labelled by their dimensions, which are $1,8,8,10$ and $27$ in the order they appear in \cref{eq:irreps,eq:ampl-irred}.

In $\SU(2)$, only $R_I$, $R_A$ and $R^S_S$ remain and have dimension $1,3$ and $5$, respectively.
The corresponding amplitudes can be identified with the isospin components $T^0, T^1$ and $T^2$, respectively.
In terms of \cref{eq:ampl-su2}, they are
\begin{equation}
    \label{eq:ampl-isospin}
    \begin{aligned}
        T^0 &= 3A(s,t,u) + A(t,u,s) + A(u,s,t),\\ 
        T^1 &= A(t,u,s)-A(u,s,t),\\ 
        T^2 &= A(t,u,s)+A(u,s,t).
    \end{aligned}
\end{equation}
This well-known relation can be derived from \cref{eq:A,eq:ampl-irred}.

\subsection{Eigenstate amplitudes}\label{sec:eigenstates}
A general amplitude can be expressed as $a_JT^J$, where the index $J$ runs over the representations in the order they appear in \cref{eq:irreps}.
For a physically applicable scattering process, however, the initial and final states 
should typically be taken as a product of mass eigenstates such as $\pi,K$ and $\eta$.
This corresponds to fixing $a_J$ to a small selection of values so that $T(ab\to cd) = a_J(ab\to cd) T^J$. 
Here, as in \cite{Manohar:2008tc,Mateu:2008gv}, we consider only elastic scattering of eigenstates, with $a_J(ab\to ab) \equiv a_J(ab)$.

With two flavours, where $J$ runs over $I,A,SS$ (alternatively, isospin $0,1,2$), the eigenstates are%
\footnote{
    $a_J$ is invariant under particle/antiparticle exchange, so $a_J(\pi^+\pi^+)=a_J(\pi^-\pi^-)\equiv a_J(\pi^\pm\pi^\pm)$.
    Note, however, that $a_J(\pi^\pm\pi^\mp)\neq a_J(\pi^\pm\pi^\pm)$ --- they are instead related by crossing; see \cref{eq:crossing-su2-su3}.}
\begin{equation}\label{eq:2flav-eigen}
    a_J(\pi^0\pi^0)     = \begin{pmatrix}   \tfrac13    &   0           &   \tfrac23    \end{pmatrix},\qquad
    a_J(\pi^0\pi^\pm)   = \begin{pmatrix}   0           &   \tfrac12    &   \tfrac12    \end{pmatrix},\qquad
    a_J(\pi^\pm\pi^\pm) = \begin{pmatrix}   0           &   0           &   1           \end{pmatrix},
\end{equation}
and with three flavours, where $J$ runs over $I,S,A,AS,SS$, they are%
\footnote{
    Here, $\pi$ without superscript stands for any of $\pi^\pm$ or $\pi^0$ (and similarly for $K$) whenever $a_J$ is agnostic about the particular choice.
    We use $a_J(ab,cd,\ldots)$ for $a_J(ab)=a_J(cd)=\ldots$.}
\begin{equation}\label{eq:3flav-eigen}
    \begin{alignedat}{6}
        a_J(\pi^0\pi^0)     &= \begin{pmatrix}   \tfrac18    &   \tfrac15    &   0           &   0               &   \tfrac{27}{40}  \end{pmatrix},&\qquad
        a_J(\pi^\pm\pi^\pm,K^\pm K^\pm,K^0K^0) 
                            &= \begin{pmatrix}   0           &   0           &   0           &   0               &   1               \end{pmatrix},\\
        a_J(\pi^0\pi^\pm)   &= \begin{pmatrix}   0           &   0           &   \tfrac13    &   \tfrac16        &   \tfrac12        \end{pmatrix},&\qquad
        a_J(K^\pm\pi^\pm,K^\pm\pi^\mp,K^0K^\pm)   
                            &= \begin{pmatrix}   0           &   0           &   0           &   \tfrac12        &   \tfrac12        \end{pmatrix},\\
        a_J(K^0\pi^\pm)     &= \begin{pmatrix}   0           &   \tfrac3{10} &   \tfrac16    &   \tfrac13        &   \tfrac15        \end{pmatrix},&\qquad
        a_J(K\pi^0)         &= \begin{pmatrix}   0           &   \tfrac3{20} &   \tfrac1{12} &   \tfrac5{12}     &   \tfrac7{20}     \end{pmatrix},\\
        a_J(\pi\eta)        &= \begin{pmatrix}   0           &   \tfrac15    &   0           &   \tfrac12        &   \tfrac3{10}     \end{pmatrix},&\qquad
        a_J(K\eta)          &= \begin{pmatrix}   0           &   \tfrac1{20} &   \tfrac14    &   \tfrac14        &   \tfrac9{20}     \end{pmatrix};
    \end{alignedat}
\end{equation}
see e.g.\ \cite{Manohar:2008tc,Mateu:2008gv}, respectively.%
\footnote{
    \Cref{eq:3flav-eigen} differs from the values given in \cite{Mateu:2008gv}: there was an error or misprint in $a_J(\pi^0\pi^\pm)$, and all eigenstates were not included, with $a_J(K^\pm\pi^\pm)$ given as $a_J(K\pi)$.}
With four or more flavours, \chpt\ loses its applicability as low-energy QCD since there are only three light quarks in the Standard Model.
Therefore, there is little sense in considering eigenstates for $n$ flavours, although we can note that $T(\pi^\pm\pi^\pm\to\pi^\pm\pi^\pm) = T_{SS}$ regardless of $n$.

One of our extensions over previous work is that we use all possible values for the $a_J$, rather than restricting them to eigenstates (see \cref{sec:bounds} for what constitutes ``possible'').
This can be done without complications, since the mass eigenstates are completely degenerate in the equal-mass limit.
However, it is still useful to view those states that remain mass eigenstates in the unequal-mass case as special.
Below, by ``eigenstate'' we will specifically mean scattering between these states.
Note that by treating general $a_J$, we effectively include inelastic scattering such as \mbox{$a_J(\pi^0\pi^0\to\pi^+\pi^-) = \begin{pmatrix} \tfrac13 & 0 & -\tfrac13 \end{pmatrix}$}.
However, it turns out that inelastic scattering is useless for our purposes by invariably failing to satisfy \cref{eq:main:cond}.
This (in addition to \cite{Manohar:2008tc,Mateu:2008gv}) is why this section has focused mainly on elastic scattering.

\subsection{Crossing symmetry}\label{sec:crossing}
Since all amplitudes can be expressed as $a_JT^J$, crossing symmetry implies that channel crossing must take the form of a linear transformation of $a_J$.
For $s\exch u$ crossing, the transformation $T^I(u,t,s) = C_u^{IJ}T^J(s,t,u)$ is given by \cite{AlvarezThesis,Neville:1963zz}
\begin{equation}
    \label{eq:crossing}
    C_u^{IJ} = \begin{pmatrix}
        \frac{1}{n^2-1} &   1                       &   -1          &   \frac{4-n^2}{2} &   \frac{n^2(n+3)}{4(n+1)}         &   \frac{n^2(n-3)}{4(n-1)}         \\
        \frac{1}{n^2-1} &   \frac{n^2-12}{2(n^2-4)} &   -\tfrac12   &   1               &   \frac{n^2(n+3)}{4(n+1)(n+2)}    &   \frac{n^2(3-n)}{4(n-1)(n-2)}    \\
        \frac{1}{1-n^2} &   -\tfrac12               &   \tfrac12    &   0               &   \frac{n(n+3)}{4(n+1)}           &   \frac{n(3-n)}{4(n-1)}           \\
        \frac{1}{1-n^2} &   \frac{2}{n^2-1}         &   0           &   \tfrac12        &   \frac{n(n+3)}{4(n+1)(n+2)}      &   \frac{n(n-3)}{4(n-1)(n-2)}      \\
        \frac{1}{n^2-1} &   \frac{1}{2+n}           &   \frac1n     &   \frac{n-2}{2n}  &   \frac{n^2+n+2}{4(n+1)(n+2)}     &   \frac{n-3}{4(n-1)}              \\
        \frac{1}{n^2-1} &   \frac{1}{2-n}           &   -\frac1n    &   \frac{n+2}{2n}  &   \frac{n+3}{4(n+1)}              &   \frac{n^2-n+2}{4(n-1)(n-2)}     \\   
    \end{pmatrix}
\end{equation}
which also works at $n=2,3$ by removing appropriate rows and columns:
\begin{equation}
    \label{eq:crossing-su2-su3}
    \left.C_u^{IJ}\right|_{\SU(2)} = \frac{1}{6}\begin{pmatrix}
        2 & -6 & 10 \\
        -2 & 3 & 5 \\
        2 & 3 & 1 \\
    \end{pmatrix}
    \qquad
    \left.C_u^{IJ}\right|_{\SU(3)} = \begin{pmatrix}
        \frac18 & 1  & -1  & -\frac52 & \frac{27}8  \\
        \frac18 & -\frac3{10} & -\frac12 & 1  & \frac98  \\
        -\frac18 & -\frac12 & \frac12 & 0  & \frac{27}{40} \\
        -\frac18 & \frac25 & 0  & \frac12 & \frac9{40}  \\
        \frac18 & \frac15 & \frac13 & \frac16 & \frac7{40}  \\ 
    \end{pmatrix}.
\end{equation} 
These versions can be found in \cite{Manohar:2008tc,Mateu:2008gv} respectively.

\section{Linear constraints}\label{sec:constr}
In this section, we will introduce a mathematical language of \emph{linear constraints}. 
This formalism is introduced before positivity bounds (see \cref{sec:bounds}) so that they can be established in full generality.
In order to make the handling of the bounds as general and powerful as possible, we dedicate this section to developing some useful mathematical definitions and results.%
\footnote{
    In this section employ mathematical notation that, depending on the background of the reader, may not be entirely familiar.
    We also define new notation for our own purposes.
    A glossary covering all potentially unfamiliar notation is provided in \cref{sec:glossary}.}

\subsection{Definition and combination of constraints}\label{sec:constr-comb}
For a set of parameters $b_i$ (e.g. the LECs), a linear constraint takes the general form
\begin{equation}\label{eq:constraint}
    \alpha_1b_1 + \alpha_2 b_2 + \ldots + \alpha_n b_n - c \geq 0,
\end{equation}
where $c,\alpha_i$ are known coefficients.
By treating $\alpha_i,b_i$ as components of vectors, this is equivalent to
\begin{equation}
    \label{eq:constraint-as-product}
    \v\alpha\cdot\v b \geq c.
\end{equation}
We say that $\v b$ lives in the \emph{parameter space}, whereas $\v\alpha$ lives in the \emph{constraint space}.%
\footnote{
    We consistently use Roman letters for vectors in parameter space and Greek letters for vectors in constraint space.
    In general, parameter space may be any finite-dimensional real vector space, with constraint space considered as its dual.} 
Since $\v\alpha$ and $c$ can be rescaled by any positive scalar without changing the inequality, any linear constraint can be described by the pair $\constr{\v\alpha}{ c}$ with $c\in\{1,0,-1\}$. 

We say that a point $\v b$ \emph{satisfies} a constraint $\constr{\v\alpha}{c}$ if $\v\alpha\cdot\v b \geq c$.
We denote by $\sat\left(\constr{\v\alpha}{c}\right)$ the subset of parameter space that satisfies $\constr{\v\alpha}{c}$.  For any $\v\alpha$, it is clear that the origin $\v b=\v 0$ is contained in $\sat\left(\constr{\v\alpha}{-1}\right)$ but not in $\sat\left(\constr{\v\alpha}{1}\right)$, and lies on the boundary of $\sat\left(\constr{\v\alpha}{ 0}\right)$ (except when $\v\alpha=\v 0$).

The LECs will typically be subject to many linear constraints simultaneously.
We will normally use the letter $\Omega$ to denote a constraint, either a single one like $\constr{\v\alpha}{c}$ or a combination of several such constraints.
Given two constraints $\Omega,\Omega'$, we write the constraint that imposes both of them simultaneously as $\Omega+\Omega'$.
A point $\v b$ satisfies $\Omega+\Omega'$ if and only if it satisfies both $\Omega$ and $\Omega'$; thus, the $\sat$ notation naturally generalises through $\sat(\Omega+\Omega') \equiv \sat(\Omega) \cap \sat(\Omega')$.
For combinations of many constraints, we will generalise $+$ into e.g.\ $\Omega=\sum_{i}\constr{\v\alpha_i}{ c_i}$.

\subsection{Stronger and weaker constraints}\label{sec:stronger-weaker}
A hierarchy can be established among the constraints based on how strong (restrictive) they are.
For instance, $b_1\geq 1$ is stronger than $b_1\geq 0$.
We will write the stronger-than relation as $\Omega\geq\Omega'$, which holds if all points that satisfy $\Omega$ also satisfy $\Omega'$.  Thus, $\Omega\geq\Omega'$ is equivalent to $\sat(\Omega)\subseteq\sat(\Omega')$.
Naturally, we say $\Omega=\Omega'$ if $\sat(\Omega)=\sat(\Omega')$, and say $\Omega>\Omega'$ if $\Omega\geq\Omega'$ but $\Omega\neq\Omega'$.
Just like subset relations, our stronger-than relation is not a total ordering, as there exist many pairs of constraints $\Omega,\Omega'$ where neither is stronger than the other.
From our definitions, it trivially follows that
\begin{equation}\label{eq:constr-ids}
    \begin{gathered}
        (\Omega + \Omega') \geq \Omega, \qquad (\Omega + \Omega') \geq \Omega',\\
        (\Omega + \Omega') = \Omega \equ \quad \Omega \geq \Omega',
    \end{gathered}
\end{equation}
so that if $\Omega\not\geq\Omega'$ and $\Omega'\not\geq\Omega$, their combination $\Omega+\Omega'$ is indeed a new, strictly stronger constraint.
Furthermore, we see that, for all $\lambda > 0$, $\kappa>1$, and $\v\alpha\neq\v 0$,
\begin{gather}
    \label{eq:change-c}
    \constr{\lambda \v\alpha}{1} > \constr{\lambda\v\alpha}{0} > \constr{\lambda \v\alpha}{-1},\\
    \label{eq:rescale}
    \constr{\kappa\v\alpha}{1} < \constr{\v\alpha}{1}, \qquad \constr{\lambda\v\alpha}{ 0} = \constr{\v\alpha}{ 0},\qquad \constr{\kappa\v\alpha}{ -1} > \constr{\v\alpha}{ -1}.
\end{gather}
From the $c=0$ version of \cref{eq:rescale}, we see that $\constr{\v\alpha}{0}$ is not a unique representation of the constraint, since we can freely rescale $\v\alpha$ without changing it.
We may remove this ambiguity by constraining $\v\alpha$ to be a unit vector.

There exists a constraint $\Omega_\infty$, equivalent to $\constr{\v 0}{1}$ or e.g.\ $\constr{\v\alpha}{1} + \constr{-\v\alpha}{1}$, that is not satisfied by any point.
It follows that $\Omega_\infty\geq\Omega$ and $\Omega_\infty+\Omega=\Omega_\infty$ for any $\Omega$.
A constraint that is satisfied by all points, i.e.\ $\constr{\v 0}{-1}$ or $\constr{\v 0}{0}$, will be called a \emph{trivial constraint}.

\subsection{Determining the relationship between constraints}
We will now present a general result, which determines if a given linear constraint $\constr{\v\beta}{c}$ is weaker than an arbitrarily complicated constraint $\Omega$.
This will serve as the basis for all our uses of linear constraints.%
\footnote{
    \Cref{proposition 4.1}, along with a version of the notation we use here, was defined in \cite{AlvarezThesis}, although the proof was completely different.
    An incorrect version of \cref{proposition 4.2} was also presented without proof.
    To the best of our knowledge, these results are novel, although the relevant literature is vast and lies outside our area of expertise.
    The closest we have found is \cite{preparata1979finding}, although their algorithm requires knowing a point that satisfies $\Omega$, relies on more complicated mathematical machinery, and does not include all the extensions presented further below in \cref{sec:relevant,sec:constraints}.} 
For complete proofs, more details, and practical applications, see \cref{sec:constraints}.

Consider a set of linear constraints $\constr{\v\alpha_i}{c}$ for $i$ in some finite set $I\subset\N$.%
\footnote{
    It is crucial that only finite combinations of constraints are considered, and it will normally be tacitly assumed that all sets like $I$ are finite.
    A limited extension to infinite sets is covered in \cref{sec:infinite}.}
Note that $c$ is the same for all constraints.
Then let $\omega_c \equiv \sum_{i\in I}\constr{\v\alpha_i}{ c}$;%
\footnote{
    We use lowercase $\omega$ here to emphasise that it is not a general constraint.
    A similar treatment of general $\Omega$ is given below.} 
an example of such a constraint is given in \cref{fig:constr}.
Then define $\reg(\omega_c)$ as the set of all points that can be expressed as
\begin{equation}\label{eq:def-reg}
    \sum_{i\in I}\lambda_i\v\alpha_i, \qquad \lambda_i \geq 0, \quad \sum_{i\in I}\lambda_i \in \Lambda_c,
\end{equation}
where
\begin{equation}\label{eq:ranges-for-Lambda}
    \Lambda_1 = [1, \infty), \qquad \Lambda_0 = [0,\infty),\qquad \Lambda_\inv = [0,1].
\end{equation}
The shape of $\reg(\omega_c)$ is illustrated in \cref{fig:alvarez}.
With these definitions, the following holds:
\begin{proposition}[determining if constraint is weaker, special case]\label{proposition 4.1}
    Let $\constr{\v\beta}{c}$ be a single linear constraint, and let $\omega_c\neq\Omega_\infty$ be defined as above.
Then $\constr{\v\beta}{c} \leq \omega_c$ if and only if $\v\beta\in\reg(\omega_c)$.
\end{proposition}
\noindent This is proven in \cref{sec:proof}.
If $\omega_c$ is a single linear constraint, this result reduces down to \cref{eq:rescale}.
The condition $\omega_c\neq\Omega_\infty$ is necessary, since there exist corner cases where $\omega_c=\Omega_\infty$ but $\reg(\omega_c)$ fails to cover the entire constraint space.%
\footnote{
    A trivial example of this is $\omega_1=\constr{\v 0}{1}$, where $\reg(\omega_1)=\{\v 0\}$.} 
However, if $\reg(\omega_c)$ does cover the entire space, then it is certain that $\omega_c=\Omega_\infty$.

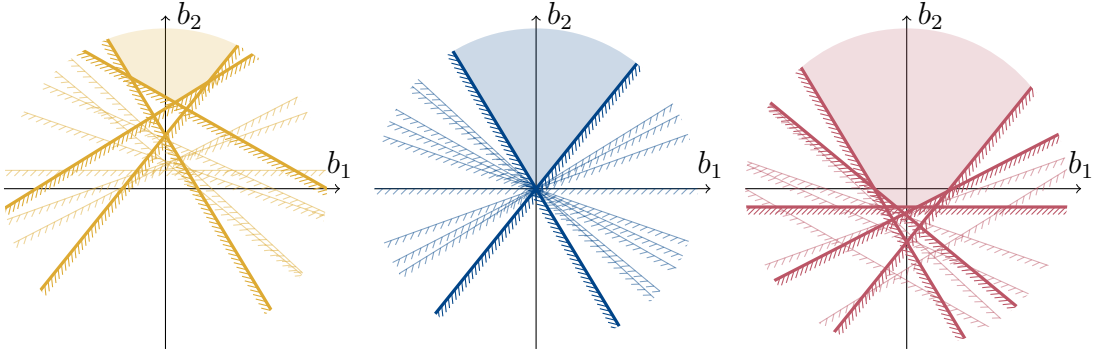
\begin{figure}[hbtp]
    \begin{tikzpicture}[scale=.85]
        \regioncoords
        \draw[thin, ->] (-2.5,0) -- (2.7,0) node [above] {$b_1$};
        \draw[thin, ->] (0,-2.5) -- (0,2.7) node [right] {$b_2$};
        \begin{scope}
            \clip (0,0) circle[radius=2.5];
            \draw[name path=circ] (0,0) circle[radius=4];

            \pconstrline[irrelevant]{1};
            \pconstrline[irrelevant]{2};
            \pconstrline[relevant]{3};
            \pconstrline[relevant]{4};
            \pconstrline[relevant]{5};
            \pconstrline[relevant]{6};
            \pconstrline[irrelevant]{7};
            \pconstrline[irrelevant]{8};
            \pconstrline[irrelevant]{9};
            \pconstrline[irrelevant]{10};
            \pconstrline[irrelevant]{11};
            
            \path[name intersections={of=ppath3 and circ}] coordinate (pi0) at (intersection-1); 
            \path[name intersections={of=ppath3 and ppath4}] coordinate (pi1) at (intersection-1); 
            \path[name intersections={of=ppath4 and ppath5}] coordinate (pi2) at (intersection-1); 
            \path[name intersections={of=ppath5 and ppath6}] coordinate (pi3) at (intersection-1);
            \path[name intersections={of=ppath6 and circ}] coordinate (pi4) at (intersection-1); 
            
            \path[fill=pcolour, opacity=0.2] (pi0) -- (pi1) -- (pi2) -- (pi3) -- (pi4) -- (0,5) -- cycle;
        \end{scope}
    \end{tikzpicture}
    \begin{tikzpicture}[scale=.85]
        \regioncoords
        \draw[thin, ->] (-2.5,0) -- (2.7,0) node [above] {$b_1$};
        \draw[thin, ->] (0,-2.5) -- (0,2.7) node [right] {$b_2$};
        \begin{scope}
            \clip (0,0) circle[radius=2.5];
            \draw[name path=circ] (0,0) circle[radius=4];

            \zconstrline[irrelevant]{1};
            \zconstrline[irrelevant]{2};
            \zconstrline[relevant]{3};
            \zconstrline[irrelevant]{4};
            \zconstrline[irrelevant]{5};
            \zconstrline[relevant]{6};
            \zconstrline[irrelevant]{7};
            \zconstrline[irrelevant]{8};
            \zconstrline[irrelevant]{9};
            \zconstrline[irrelevant]{10};
            \zconstrline[irrelevant]{11};
            
            \path[name intersections={of=zpath3 and circ}] coordinate (zi0) at (intersection-1); 
            \path[name intersections={of=zpath3 and zpath6}] coordinate (zi1) at (intersection-1);
            \path[name intersections={of=zpath6 and circ}] coordinate (zi2) at (intersection-1); 
            
            \path[fill=zcolour, opacity=0.2] (zi0) -- (zi1) -- (zi2)  -- (0,5) -- cycle;
            
        \end{scope}
    \end{tikzpicture}
    \begin{tikzpicture}[scale=.85]
        \regioncoords
        \draw[thin, ->] (-2.5,0) -- (2.7,0) node [above] {$b_1$};
        \draw[thin, ->] (0,-2.5) -- (0,2.7) node [right] {$b_2$};
        \begin{scope}
            \clip (0,0) circle[radius=2.5];
            \draw[name path=circ] (0,0) circle[radius=4];

            \nconstrline[relevant]{1};
            \nconstrline[relevant]{2};
            \nconstrline[relevant]{3};
            \nconstrline[irrelevant]{4};
            \nconstrline[irrelevant]{5};
            \nconstrline[relevant]{6};
            \nconstrline[relevant]{7};
            \nconstrline[irrelevant]{8};
            \nconstrline[irrelevant]{9};
            \nconstrline[irrelevant]{10};
            \nconstrline[irrelevant]{11};
            
            \path[name intersections={of=npath3 and circ}] coordinate (ni0) at (intersection-1); 
            \path[name intersections={of=npath3 and npath2}] coordinate (ni1) at (intersection-1); 
            \path[name intersections={of=npath2 and npath1}] coordinate (ni2) at (intersection-1); 
            \path[name intersections={of=npath1 and npath7}] coordinate (ni3) at (intersection-1); 
            \path[name intersections={of=npath7 and npath6}] coordinate (ni4) at (intersection-1);
            \path[name intersections={of=npath6 and circ}] coordinate (ni5) at (intersection-1); 

            \path[fill=ncolour, opacity=0.2] (ni0) -- (ni1) -- (ni2) -- (ni3) -- (ni4) -- (ni5) -- (0,5) -- cycle;
        \end{scope}
    \end{tikzpicture}
    \caption{
        A cropped depiction of twelve random two-dimensional constraints $\constr{\v\alpha_i}{c}$ for $c=+1$ (left, \pcolcol), $c=0$ (middle, \zcolcol) and $c=-1$ (right, \ncolcol) illustrated as the parameter-space lines $\v\alpha_i\cdot\v b=c$.
        The side of the line that is \emph{excluded} by the constraint is hatched.
        The region $\sat(\omega_c)$ for $\omega_c=\sum_i\constr{\v\alpha_i}{c}$, i.e.\ the set of points that satisfy all the constraints, is shaded.
        The lines corresponding to \emph{relevant} constraints (i.e.\ those that actually delimit $\sat(\omega_c)$; this is more closely defined in \cref{sec:relevant}) are drawn more strongly than the rest.}
    \label{fig:constr}
\end{figure}

\renewcommand{\nangle}{0}
\renewcommand{\zangle}{0}
\renewcommand{\pangle}{0}
\begin{figure}[hbtp]
    \begin{tikzpicture}[scale=.6]
        \regioncoords
        \draw[thin, ->] (-2.5,0) -- (2.5,0) node [above] {$\alpha_1$};
        \draw[thin, ->] (0,-.5) -- (0,4.3) node [right] {$\alpha_2$};
        \begin{scope}
            \clip (0,0) circle[radius=4];

            \draw[thick, pcolour, every node/.style={pdot=pcolour},
                    pattern = north east lines, pattern color=pcolour!50]
                (p3) node{} -- (p4) node{} -- (p5) node{} -- (p6) node{} 
                    -- ($ 5*(p6) $) -- ($ 5*(p3) $)  -- cycle;
            \draw[pcolour, every node/.style={pdot}]
                (p3)-- (p2) node{} -- (p1) node{} -- (p7) node{} -- (p6);
            \path[pcolour, every node/.style={pdot}]
                (p8) node{} (p9) node{} (p10) node{} (p11) node{} (p12) node{};
            \draw[pcolour, dashed]
                (p3) -- (0,0) -- (p6);
                
            \draw (0,0) node[dot=black] {};
        \end{scope}
    \end{tikzpicture}
    \begin{tikzpicture}[scale=.6]
        \regioncoords
        \draw[thin, ->] (-2.5,0) -- (2.5,0) node [above] {$\alpha_1$};
        \draw[thin, ->] (0,-.5) -- (0,4.3) node [right] {$\alpha_2$};
        \begin{scope}
            \clip (0,0) circle[radius=4];

            \draw[thick, zcolour, every node/.style={zdot=zcolour},
                    pattern = north east lines, pattern color=zcolour!50]
                (0,0) -- (z3) node{} -- ($ 5*(z3) $) -- ($ 5*(z6) $) -- (z6) node{} -- cycle;, 
            \draw[zcolour, every node/.style={zdot}]
                (z3)-- (z2) node{} -- (z1) node{} -- (z7) node{} -- (z6) -- (z5) node{} -- (z4)  node{} -- cycle;
            \path[zcolour, every node/.style={zdot}]
                (z8) node{} (z9) node{} (z10) node{} (z11) node{} (z12) node{};
                
            \draw (0,0) node[dot=black] {};
        \end{scope}
    \end{tikzpicture}
    \begin{tikzpicture}[scale=.6]
        \regioncoords
        \draw[thin, ->] (-2.5,0) -- (2.5,0) node [above] {$\alpha_1$};
        \draw[thin, ->] (0,-.5) -- (0,4.3) node [right] {$\alpha_2$};
        \begin{scope}
            \clip (0,0) circle[radius=4];

            \draw[thick, ncolour, every node/.style={ndot=ncolour},
                    pattern = north east lines, pattern color=ncolour!50]
                (n1) node{} -- (n2) node{} -- (n3) node{} -- (0,0) -- (n6) node{} -- (n7) node{} -- cycle;
            \draw[ncolour, every node/.style={ndot}]
                (n3)-- (n4) node{} -- (n5) node{} -- (n6);
            \path[ncolour, every node/.style={ndot}]
                (n8) node{} (n9) node{} (n10) node{} (n11) node{} (n12) node{};
            \draw[ncolour, dashed]
                (n3) -- ($ 5*(n3) $) (n6) -- ($ 5*(n6) $);
                
            \draw (0,0) node[dot=black] {};
        \end{scope}
    \end{tikzpicture}
    \vspace{-2cm}
    \caption[Examples of the regions $\reg(\omega_\inv)$, $\reg(\omega_0)$ and $\reg(\omega_1)$.]{Examples of the regions $\reg(\omega_1)$ (left, \pcolcol), $\reg(\omega_0)$ (middle, \zcolcol) and $\reg(\omega_\inv)$ (right, \ncolcol) in constraint space, using the same $\v\alpha_i$ as in \cref{fig:constr}.
    The $\v\alpha_i$ are represented as points (\pmark, \zmark, \nmark, respectively) in the space, and the relevant ones are filled (again, relevancy is defined in \cref{sec:relevant}).  The convex hulls (as defined in \cref{eq:hull}) of the $\v\alpha_i$ are outlined.  For comparison to \cref{fig:constr}, it is helpful to remember that $\v\alpha_i$ are normal vectors to the lines shown there, and that larger $|\v\alpha_i|$ correspond to lines passing \emph{closer} to the origin. 
    
    \qquad Note how, given identical $\v\alpha_i$, $\reg(\omega_0)$ is the union of $\reg(\omega_1)$ and $\reg(\omega_\inv)$ (this is easy to see from \cref{eq:ranges-for-Lambda}) whereas the set of relevant constraints is the intersection of the respective sets.}
    \label{fig:alvarez}
\end{figure}
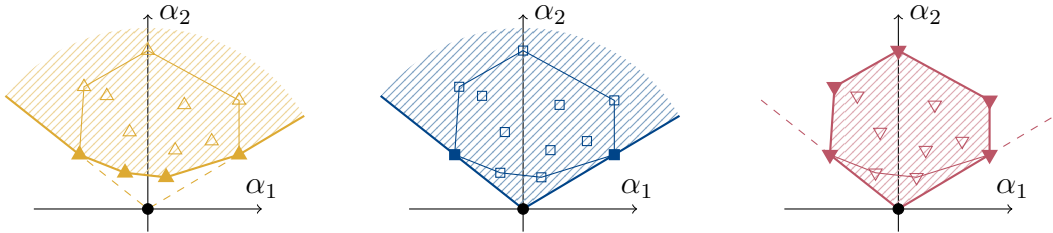

In \cref{fig:alvarez}, we may note that $\reg(\omega_c)$ is closely related to the \emph{convex hull} of the $\v\alpha_i$.
In essence, $\reg(\omega_c)$ is obtained by forming the hull, and then also including all points that give weaker constraints under \cref{eq:rescale}.
We may also note that the convex hull can be defined as
\begin{equation}\label{eq:hull}
    \hull\big(\{\v\alpha_i\}_{i\in I}\big) = \setbuild{ \sum_{i\in I}\lambda_i\v\alpha_i}{\lambda_i\geq 0,\;\sum_{i\in I}\lambda_i = 1},
\end{equation}
which is very similar to \cref{eq:def-reg}.

Now, let us handle the general case.
The most general combination of a finite number of linear constraints can be expressed as
\begin{equation}\label{eq:def-Omega}
    \Omega = \sumI{1}\constr{\v\alpha_i}{ 1}     + \sumI{0}\constr{\v\alpha_i}{ 0} + \sumI{-1}\constr{\v\alpha_i}{ -1},
\end{equation} 
where $I_c$ are some disjoint, finite, and possibly empty sets.
We may compactly write this as $\Omega=\sumI{}\constr{\v\alpha_i}{c_i}$ where $I\equiv I_{+1}\cup I_0\cup I_{-1}$ and $c_i=c$ if $i\in I_c$.

Similarly to $\reg(\omega_c)$, let $\reg_c(\Omega)$ be the set of all points that can be expressed as (recall that $c\in\{1,0,-1\}$)
\begin{equation}\label{eq:def-gen-reg}
    \sumI{-1}\lambda_i\v\alpha_i + \sumI{0}\lambda_i\v\alpha_i + \sumI{1}\lambda_i\v\alpha_i, 
    \qquad\lambda_i \geq 0,
\end{equation}
with $\lambda_i$ constrained by the condition
\begin{equation}\label{eq:gen-condition}
    \sumI{1}\lambda_i - \sumI{-1}\lambda_i \geq c,
\end{equation}
An illustration of $\reg_c(\Omega)$ can be found in \cref{sec:visual}.
With these definitions, the following holds:
\begin{proposition}[determining if constraint is weaker, general case]\label{proposition 4.2}
    Let $\constr{\v\beta}{c}$ be a linear constraint, and let $\Omega\neq\Omega_\infty$ be defined as above.
Then $\constr{\v\beta}{c} \leq \Omega$ if and only if $\v\beta\in\reg_c(\Omega)$.
\end{proposition}
\noindent This is proven in \cref{sec:proof-gen}.
If only one of the $I_c$ is nonempty, this reduces down to \cref{proposition 4.1}.

While it is not as useful for the purposes of \cref{proposition 4.2}, one may note that \cref{eq:def-gen-reg,eq:gen-condition} can be more succinctly stated as
\begin{equation}\label{eq:def-reg-alt}
    \reg_c(\Omega) = \setbuild{\sum_{i\in I}\lambda_i\v\alpha_i}{\lambda_i\geq 0, \sum_{i\in I}\lambda_ic_i\geq c}.
\end{equation}
This definition of $\reg_c(\Omega)$ works also if $c,c_i$ are not constrained to $\{-1,0,1\}$.

\subsection{Representations and degeneracy}\label{sec:relevant}
Checking if $\v b$ satisfies $\Omega$ becomes computationally expensive if $\Omega$ is the combination of many different linear constraints. 
However, $\Omega$ is usually not uniquely determined by how it is expressed as a sum of linear constraints, and it is possible to vastly reduce that redundancy.
To that end, we define a \emph{representation} of a constraint $\Omega$ as any finite set $\s S$ of linear constraints with the property%
\footnote{
    Clearly, all $\Omega$ also admit representation as a sum of an \emph{infinite} number of constraints. 
    However, we will not consider such representations, and \cref{proposition 4.3} below generally only holds if $\Omega$ can be expressed as a finite sum. 
    See \cref{sec:infinite} for a discussion about infinite sums of constraints.}
\begin{equation}
    \Omega = \sum_{\constr{\v\alpha}{c}\in \s S}\constr{\v\alpha}{c}.
\end{equation}
If it is implicit which representation is used for $\Omega$, we may call the $\constr{\v\alpha}{c}\in\s S$ the \emph{elements of $\Omega$}.

It is clear that there exist \emph{minimal representations}, i.e.\ representations with the smallest number of elements.
As we will see below, there is often a unique minimal representation, which we will label $\rel(\Omega)$.
However, there is an important exception to this: when $\sat(\Omega)$ is contained in a hyperplane.
This happens when there are some $\v\delta,d$ such that $\v\delta\cdot\v b=d$ for all $\v b\in\sat(\Omega)$, or equivalently $\constr{\v\delta}{d}+\constr{-\v\delta}{-d}\leq\Omega$.
We will call $\Omega$ \emph{degenerate} if so is the case, and \emph{non-degenerate} otherwise.%
\footnote{
    As defined here, $\Omega_\infty$ would be considered a special case of a degenerate constraint.
    In the closer study of degenerate constraints given in \cref{sec:degenerate}, it turns out to be more useful to consider $\Omega_\infty$ seperately, viewing it as neither degenerate nor non-degenerate.} 
With this in mind, we can state the following result:
\begin{proposition}[finding relevant constraints, non-degenerate case]\label{proposition 4.3}
    If $\Omega$ is a non-degenerate constraint, there exists a minimal representation $\rel(\Omega)$ that is unique up to the normalisation of its elements. 
    Furthermore, for any representation $\s S$ of $\Omega$, the relation $\rel(\Omega)\subseteq\s S$ is true up to normalisation.
    
    The elements of $\rel(\Omega)$ are exactly those $\constr{\v\alpha}{c}\leq\Omega$ for which there is some $\v b\in\sat(\Omega)$ such that $\v\alpha\cdot\v b = c$ and $\v\beta\cdot\v b > d$ for all $\constr{\v\beta}{d}\leq\Omega$ with  $\constr{\v\beta}{d}\neq\constr{\v\alpha}{c}$.
\end{proposition}
\noindent This is proven in \cref{sec:proof-relevant}.
Due to this uniqueness, and the fact that $\rel(\Omega)$ is a subset of any representation, we will call the elements of $\rel(\Omega)$ the \emph{relevant elements of $\Omega$}, and call all other elements of any representation \emph{irrelevant}, since they can be discarded without altering $\Omega$.
A more practical way of finding $\rel(\Omega)$, based on \cref{proposition 4.2}, is given in \cref{sec:practical-relevant}.

When $\Omega$ is degenerate, there is typically no unique minimal representation, although there is still a straightforward way to find \emph{some} minimal representation, which we will also label $\rel(\Omega)$.
This generalisation of \cref{proposition 4.3} is discussed in \cref{sec:degenerate}, along with a more general method of replacing any degenerate constraint with a non-degenerate analogue in a lower-dimensional space.
Note, however, that degenerate constraints are only a corner case with little practical relevance: a small perturbation, e.g.\ by numerical error, to the elements of a degenerate constraint will either render it non-degenerate, or render it equal to $\Omega_\infty$.

\section{Positivity bounds}
\label{sec:bounds}
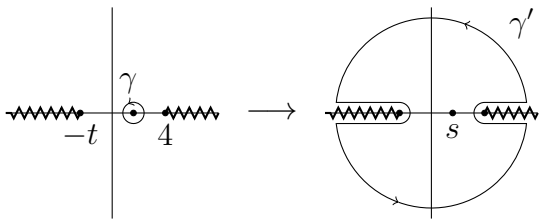
\begin{figure}[hbtp]
    \tikzset{
        branchcut/.style={thick, decorate, decoration={zigzag, segment length=1.4mm, amplitude=.7mm}}
    }
    \centering
    \begin{tikzpicture}[scale=.7]
        \begin{scope}
            \draw (-2,0) -- (2,0);
            \draw (0,-2) -- (0,2);
            
            \draw[fill=black] (-.6,0) circle[radius=1.5pt] node[below] {$-t$};
            \draw[branchcut] (-0.6,0) -- (-2,0);
            
            \draw[fill=black] (1,0) circle[radius=1.5pt] node[below] {$4$};
            \draw[branchcut] (1,0) -- (2,0);
            
            \draw[fill=black] (.4,0) circle[radius=1.5pt];
            \draw[->] (.4,0)  +(120:.2) arc (120:480:.2) node[above] {$\gamma$};
        \end{scope}
        \begin{scope}[shift={(3,0)}]
            \node (0,0) {$\quad\longrightarrow\quad$};
        \end{scope}
        \begin{scope}[shift={(6,0)}]
            \draw (-2,0) -- (2,0);
            \draw (0,-2) -- (0,2);
        
            \draw[fill=black] (-.6,0) circle[radius=1.5pt];
            \draw[branchcut] (-0.6,0) -- (-2,0);
        
            \draw[fill=black] (1,0) circle[radius=1.5pt];
            \draw[branchcut] (1,0) -- (2,0);
        
            \draw[fill=black] (.4,0) circle[radius=1.5pt] node[below] {$s$};
                
                \node at (1.7,1.7) {$\gamma'$};
            \begin{scope}
                \clip (0,0) circle[radius=1.8];
                \draw (2,-.2) -- (1,-.2) arc (270:90:.2) -- (2,.2);
                \draw (-2,-.2) -- (-.6,-.2) arc (-90:90:.2) -- (-2,.2);
            \end{scope}
            \begin{scope}
                \clip (-2,3) rectangle (2,.2);
                \draw[->] (1.8,0) arc (0:70:1.8);
                \draw     (70:1.8) arc (70:180:1.8);
            \end{scope}
            \begin{scope}
                \clip (-2,-2.2) rectangle (2,-.2);
                \draw[->] (-1.8,0) arc (180:250:1.8);
                \draw     (250:1.8) arc (250:360:1.8);
            \end{scope}
        \end{scope}
    \end{tikzpicture}
    \caption{The contour integral in the $z$-plane around $z=s$ used in the dispersion relation.}
        \label{fig:contour}
\end{figure}

Equipped with the notion of linear constraints, we are ready to move on to the main topic of this paper: positivity bounds.  
(For a more detailed version of this derivation, see \cite{Manohar:2008tc}; various generalisations can be found in e.g.\ \cite{Wang:2020jxr,Tolley:2020gtv}.)
We start by writing down the fixed-$t$ dispersion relation for the amplitude $a_JT^J$:
\begin{equation}
    a_JT^J(s,t) = \frac{1}{2\pi i}\oint_\gamma \d z \frac{a_JT^J(z,t)}{z-s}.
\end{equation}
The amplitude has two branch cuts along the real axis: a right-hand cut starting at $z=4$ corresponding to the $s$-channel, and a left-hand cut starting at $z=-t$ corresponding to the $u$-channel.
The discontinuity across these cuts is $T(z+i\e)-T(z-i\e) = 2i\Im T(z+i\e)$.
For real $s$ in the span $-t<s<4$, we deform the contour $\gamma$ as shown in \cref{fig:contour}.
We can then reexpress the integral in terms of the discontinuities, which may require derivatives (subtractions) to make the contour at infinity vanish.
Using the crossing relation derived in \cref{sec:crossing} to rewrite the $u$-channel cut in terms of $s$, the result is
\begin{equation}
    \label{eq:dispersion}
    a_J\frac{\d^k}{\d s^k} T^J(s,t) = \frac{k!}{\pi}\int_4^\infty \d z
        \left[
            \frac{a_J}{(z-s)^{k+1}} + \frac{(-1)^k a_I C_u^{IJ}}{(z-u)^{k+1}}
        \right] \Im T^J(z + i\e, t).
\end{equation}
The Froissart bound \cite{Froissart:1961ux} shows that the integral converges whenever $k\geq2$, since $T^J(z+i\e,t)=\O(s\ln^2 s)$.%
\footnote{
    Note that $T^J(z+i\e,t)$ on the right-hand side of \cref{eq:dispersion} is the exact, non-perturbative amplitude --- see e.g.\ \cite{Manohar:2008tc}. Indeed, the perturbative \chpt\ amplitude at any fixed order grows polynomially with $s$, so it violates the Froissart bound.
    We can insert the fixed-order perturbative amplitude at the left-hand side thanks to the smallness of $s$ (and $t$), which guarantees good agreement with the exact one.}
We will discuss specific values for $k$ in \cref{sec:nderiv}; here, we keep it general.

\begin{figure}
    \begin{minipage}[c]{0.6\textwidth}
        \centering
        \begin{tikzpicture}[scale=1.2,>=stealth] 
            \path[fill=red] (-1,1) -- (1,1) -- (1,-1) -- cycle;
            \draw[very thick] (0,0) -- (1,0) -- (1,1) -- (-1,1) -- cycle;
            
            \draw[thick, dotted] (1,-2) -- (1,3) node[very near end, sloped, below] {\footnotesize$s=4$};
            \draw[thick, dotted] (-2,1) -- (3,1) node[very near end, sloped, above] {\footnotesize$t=4$};
            \draw[thick, dotted] (-2,2) -- (2,-2) node[very near end, sloped, below,yshift=.5ex,xshift=.5ex] {\footnotesize$u=4$};
            \draw[thick, dotted] (-2,3) -- (3,-2) node[very near start, sloped, below,yshift=.5ex,xshift=.5ex] {\footnotesize$u=0$};
            
            \path[pattern=north east lines] (0,1) -- (0,3) -- (-2,3) -- cycle;
            \path[pattern=north east lines] (1,0) -- (3,0) -- (3,-2) -- cycle;
            \path[pattern=north east lines] (0,0) -- (0,-2) -- (-2,-2) -- (-2,0) -- cycle;
            
            \draw[->] (-2,0) -- (3.1,0) node[above] {$s$};
            \draw[->] (0,-2) -- (0,3.1) node[right] {$t$};
            \draw[->] (3,3) -- (-2.1,-2.1) node[above left] {$u$};
        \end{tikzpicture}
    \end{minipage}
    \begin{minipage}[c]{0.4\textwidth}
        \caption[The Mandelstam plane, with the part in which the positivity conditions are valid outlined.]{
            The plane of normalised Mandelstam variables.
            The \selfcolour{red} triangle is the region where the amplitude is real and free from singularities or branch cuts.
            The positivity conditions \cref{eq:main} are valid inside the outlined part.
            The hatched regions with $s,t$ or $u$ positive are the physical regions for the respective channels. 
        }
        \label{fig:mandelstam}
    \end{minipage}
\end{figure}
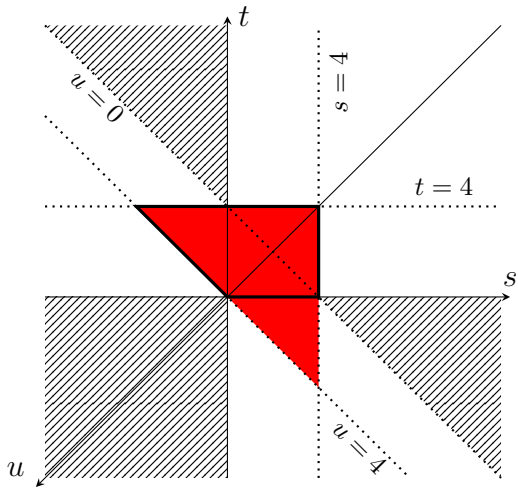

Above threshold, the partial-wave expansion of the amplitude takes the form\footnote{There is a limited domain of validity for this expansion, but it does not affect the range of $s,t$ used by us.
Again, see \cite{Manohar:2008tc} for details.}
\begin{equation}
    \label{eq:part-wave}
    T^J(s,t) = \sum_{\ell=0}^\infty (2\ell + 1)f^J_\ell(s) P_\ell\left(1 + \frac{2t}{s-4}\right),
\end{equation}
where $f^J_\ell$ are partial wave amplitudes, $P_\ell$ are Legendre polynomials, and the expression in parentheses is the cosine of the scattering angle.
The optical theorem then gives
\begin{equation}\label{eq:optical}
    \Im f^J_\ell(s) = s\beta(s)\sigma^J_\ell(s),\qquad \beta(s)\equiv\sqrt{1-\frac{4}{s}},
\end{equation}
which is positive above threshold since the partial-wave cross-sections $\sigma^J_\ell$ are always positive.
Therefore,
\begin{equation}
    \Im T^J(s,t) = \sum_{\ell=0}^\infty (2\ell + 1)s\beta(s)\sigma^I_\ell P_\ell\left(1 + \frac{2t}{s-4}\right)
\end{equation}
is positive above threshold as long as $P_\ell$ is.
Since $P_\ell(z)\geq 0$ when $z\geq 1$, \cref{eq:dispersion} therefore imposes the constraint that, for any $t\in[0,4], s\in[-t,4]$ and any representation index $J$,
\begin{subequations}\label{eq:main}
    \begin{align}
        a_J\frac{\d^k}{\d s^k}T^J(s,t) &\geq 0 \label{eq:main:constr}\\ 
        \text{if}\quad a_I\left\{\delta^{IJ}\left[\frac{z-u}{z-s}\right]^{k+1} + (-1)^k C_u^{IJ}\right\} &\geq 0\quad\text{for all $z\geq 4$}.\label{eq:main:cond}
    \end{align}
\end{subequations}
The region in the $s,t$ plane where this holds is shown in \cref{fig:mandelstam}.
Note that $u\in[-4,4]$, so the expression in square brackets above is always positive.

Up to and including NNLO, the second derivative of $T^J$ is linear in all LECs, so we obtain from \cref{eq:main:constr} an expression of the form
\begin{equation}
    \label{eq:constr-LEC}
    \sum_i \alpha_i \hat L_i^r + \sum_j \beta_j K_j^r + \gamma \geq 0,
\end{equation}
where the coefficients $\alpha_i,\beta_i,\gamma$ are functions of $s,t$ and $a_J$, but not of the LECs.
This constitutes a linear constraint, and each valid choice of $s,t$ and $a_J$ potentially yields a different constraint.
The result of combining these constraints will be that only a limited region in parameter space ($\sat(\Omega)$ in the notation of \cref{sec:constr}) satisfies the positivity bounds.
With some luck, the boundary of this region is close enough to the experimentally measured value to improve on its uncertainty (carefully considering also the uncertainty of the bounds).

\subsection{Conditions on $a_J$}\label{sec:aJ-cond}
If we demand that \cref{eq:main:cond} holds in the entire allowed $s,t$ region, we see that the factor in square brackets can be made arbitrarily large or small by varying $s,u,z$.
Therefore, we obtain the independent conditions $a_J\geq 0$ and $(-1)^k a_IC_u^{IJ}\geq 0$.
However, we may apply the dispersion relation independently to each fixed $s,t$.
Then, \cref{eq:main:cond} can be made less restrictive, and a wider range of constraints on the LECs can be generated.
This also includes permitting odd $k$ for some $s,t$.

While we may fix $s$ and $t$ (which in turn fixes $u$), we must still allow $z$ to cover its entire range.
Therefore, finding all valid $a_J$ for given $s,t$ presents some practical issues.
We solve this by using the technology of \cref{sec:constr}, since \cref{eq:main:cond} is a set of linear constraints on the vector $a_I$; we may write it compactly as $\v a\cdot\v\beta^J(z)\geq 0$.
Noting that $\left(\frac{z-u}{z-s}\right)^{k+1}$ is monotonic as a function of $z\in[4,\infty)$, we see that it is always possible to write $\v\beta^J(z) = \mu\v\beta^J(4) + (1-\mu)\v\beta^J(\infty)$ for $\mu\in[0,1]$.
By \cref{proposition 4.1,proposition 4.3} (see also \cref{proposition B.2-relevant}), it follows that only $\v\beta^J(4)$ and $\v\beta^J(\infty)$ are relevant constraints on $\v a$.
Thus, it is sufficient to evaluate \cref{eq:main:cond} at $z=4$ and $z=\infty$, rather than letting $z$ cover its entire range.

\begin{figure}[hbtp]
    \centering
    \begin{tikzpicture}[>=stealth, scale=2]
        \coordinate (Iss) at ( 0.555, 1.333);
        \coordinate (Ass) at ( 1.074, 1.111);
        \coordinate (Aa)  at ( 0.333, 0    );
        \coordinate (ia)  at (-0.296,-0.444);
        \coordinate (Ii)  at ( 0    , 0.5  );
        \coordinate (ass) at ( 1.592, 0.889);
        \coordinate (iss) at ( 0.296, 1.444);
    
        \fill[grey!80] (-1.666,-2) -- (1,2) -- (1.666,2) -- (-1,-2) -- cycle;
        \fill[path fading=north, white] (-1.666,-2) -- (-1.333,-1.5) -- (-.666,-1.5) -- (-1,-2) -- cycle;
        \fill[path fading=south, white] (1.666,2) -- (1.333,1.5) -- (.666,1.5) -- (1,2) -- cycle;
        \fill[orange] (0,0) rectangle (.333,.5);     
        
        \coordinate (pf-iss) at ( 0    , 1    );
        \coordinate (pf-ass) at ( 1    , 0    );
        \coordinate (pf-ia)  at ( 0    , 0    );
        
        \coordinate (ph-iss) at ( 0.296, 1.444);
        \coordinate (ph-ass) at ( 1.592, 0.899);
        \coordinate (ph-ia)  at (-0.296,-0.444);
        
        \coordinate (mh-iss) at (-0.630, 0.056);
        \coordinate (mh-ass) at (-0.259,-1.889);
        \coordinate (mh-ia)  at ( 0.629, 0.944);
        
        \coordinate (mf-iss) at (-0.333, 0.5  );
        \coordinate (mf-ass) at ( 0.333,-1    );
        \coordinate (mf-ia)  at ( 0.333, 0.5  );
        
        \draw[thick, pattern=vertical lines]   (pf-iss) -- (pf-ass) -- (pf-ia) -- cycle;
        \draw[thick, pattern=north east lines] (ph-iss) -- (ph-ass) -- (ph-ia) -- cycle;
        \draw[thick, pattern=north west lines] (mh-iss) -- (mh-ass) -- (mh-ia) -- cycle;
        \draw[thick, pattern=horizontal lines] (mf-iss) -- (mf-ass) -- (mf-ia) -- cycle;
        
        \draw[thin,->] (-2,0) -- (2,0) node[right] {$a_I$};
        \draw[thin,->] (0,-2) -- (0,2) node[right] {$a_A$};
        \draw[thin,->] (2,2) -- (-2,-2) node[above left] {$a_{SS}$};
        \foreach \i in {-1.5,-1,-0.5,0.5,1,1.5} {
            \draw[thin] (0,\i) -- (-.1,\i) node[left] {\tiny $\i$};
            \draw[thin] (\i,0) -- (\i,-.1) node[below] {\tiny $\i$};, 
        }
        \foreach \i in {-1,-2,2,3,4}
            \draw[thin] ($ (.5,.5) - .5*(\i,\i) $) -- +(135:.1) node[above] {\tiny $\i$};
        
        \fill[orange, opacity=.5] (0,0) rectangle (.333,.5);    
                    
        \draw[blue] (0,0)    node[circle,inner sep=1.5pt, fill, pin={[pin edge={thick,blue!50}, pin distance= 1cm]150:$\pi^\pm\pi^\pm$}] {};
        \draw[blue] (.333,0) node[circle,inner sep=1.5pt, fill, pin={[pin edge={thick,blue!50}, pin distance=.7cm]-60:$\pi^0\pi^0$}] {};
        \draw[blue] (0,.5)   node[circle,inner sep=1.5pt, fill, pin={[pin edge={thick,blue!50}, pin distance=.5cm]130:$\pi^0\pi^\pm$}] {};
        \draw[blue] (.333,.5)   node[circle,inner sep=1.5pt, fill, pin={[pin edge={thick,blue!50}, pin distance=1.5cm]-15:$\pi^\pm\pi^\mp$}] {};
        
        \draw (1.5,-1)    node  {$\tikzineq{\draw[pattern=vertical lines  ] (0,0) rectangle (1em,1em);} \: s=+4\phantom{.5}$};
        \draw (1.5,-1.25) node  {$\tikzineq{\draw[pattern=north east lines] (0,0) rectangle (1em,1em);} \: s=+0.5$};
        \draw (1.5,-1.5)  node  {$\tikzineq{\draw[pattern=north west lines] (0,0) rectangle (1em,1em);} \: s=-0.5$};
        \draw (1.5,-1.75) node  {$\tikzineq{\draw[pattern=horizontal lines] (0,0) rectangle (1em,1em);} \: s=-4\phantom{.5}$};
    \end{tikzpicture}
    \caption[Illustration of which $a_I,a_A,a_{SS}$, normalised so that $a_I+a_A+a_{SS}=1$, are permitted by \cref{eq:main:cond} for $n=2$, $k=2$ at $t=4$ and various fixed $s$]
        {
            Illustration of which $a_I,a_A,a_{SS}$, normalised so that $a_I+a_A+a_{SS}=1$, are permitted by \cref{eq:main:cond} for $n=2$, $k=2$ at $t=4$ and various fixed $s$.
            The shaded band is the region permitted by the $z=\infty$ bounds (see the discussion in \cref{sec:aJ-cond}).
            It is independent of $s$ and extends to infinity.
            The hatched triangles are the regions permitted by the $z=4$ bounds for various $s$ as indicated.
            Thus, the $a_J$ permitted at fixed $s,t$ is the overlap between the triangle and the band.
            The \selfcolour{orange} rectangle is the region permitted at all $s,t$.
            
            \qquad The \selfcolour{blue} points represent the eigenstates, including $a_J(\pi^\pm\pi^\mp) = C^{IJ}_u a_I(\pi^\pm\pi^\pm)$ in addition to those given in \cref{eq:2flav-eigen}.
            Not shown is the inelastic scattering \mbox{$a_J(\pi^0\pi^0\to\pi^+\pi^-)=\begin{pmatrix}\tfrac13&0&-\tfrac13 \end{pmatrix}$}, which never satisfies \cref{eq:main:cond}.
        
            \qquad For $n\geq2$, the permitted region has an analogous shape, albeit in 4 ($n=3$) and 5 ($n\geq4$) dimensions, respectively.
            Like for $n=2$, the eigenstate scattering amplitudes are mainly located in the corners of the always-permitted region.}
    \label{fig:aJ}
\end{figure}
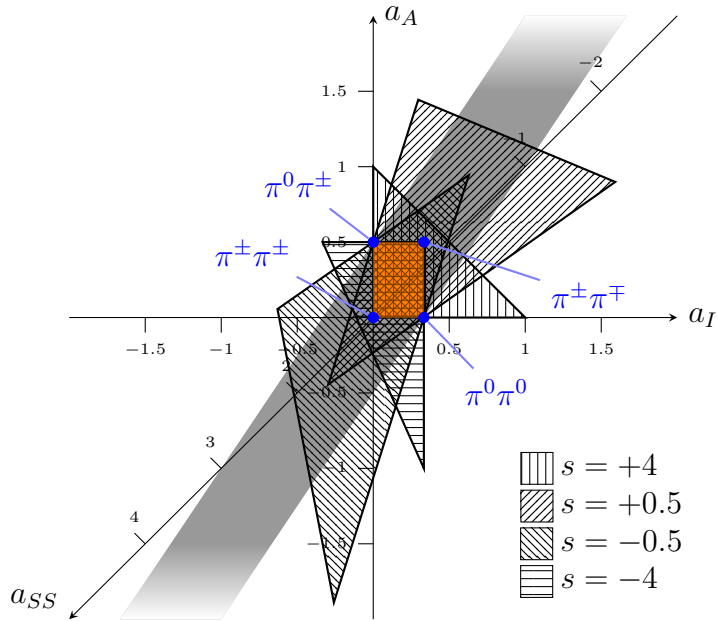

Another practical problem is that the set of allowed $a_J$ is typically unbounded.
However, \cref{eq:main:constr} is independent of the magnitude of $a_J$.
The obvious solution is to fix the normalisation of the vector $\v a$, but this is problematic since a linear constraint on $\v a$ is not necessarily a linear constraint on $\frac{1}{|\v a|}\v a$.
Instead, we may simply rescale $\v a$ so that $\sum_J a_J = 1$.
This does not cover all possible $\v a$ (for that, we must also look at $\sum_J a_J = 0$ and $\sum_J a_J = -1$), but it turns out that \cref{eq:main:cond} is only satisfied by $\v a$ for which this works.
Using this, constraints on $a_I$ are shown in \cref{fig:aJ}.

\subsection{The number of derivatives}\label{sec:nderiv}
As mentioned before, \cref{eq:main} requires $k\geq 2$ to be valid, and $k=2$ is sufficient; indeed, \cite{Manohar:1996cq} claims that this value produces the best bounds.
However, nothing prevents us from taking more derivatives, and with our generalised methods, we do find new relevant bounds from larger $k$; see e.g.\ \cref{fig:SU2-l12,fig:SU3-L123} below.
Also \cite{Wang:2020jxr} makes use of higher derivatives.

At NLO, the LECs only enter through the second-order polynomial part of the amplitude, so the third and higher derivatives are parameter-independent and do not generate any bounds.
This is not the case at NNLO, where the polynomial part is third-order, and where the non-polynomial unitarity correction contains NLO LECs.
Therefore, $k=3$ should yield another set of bounds on the NNLO LECs, and $k\geq 4$ should add bounds on the NLO LECs not obtainable from the NLO-only amplitude.

It also turns out that odd $k$ cannot be used at any order in the 2-flavour case.
To see this, look explicitly at \cref{eq:main:cond} at $z=\infty$:
\begin{equation}
    \delta^{IJ} - C_u^{IJ} = \tfrac16 c_1^Ic_2^J\qquad\text{where}\quad
    c_1 = \begin{pmatrix} 2 & 1 & -1 \end{pmatrix}, c_2 = \begin{pmatrix} 2 & 3 & -5 \end{pmatrix}
\end{equation}
Due to what seems to be a coincidence, the matrix factorises into a direct product, and since $c_2$ has different-sign elements, no nonzero $a_I$ satisfies $(a_I c_1^I)c_2^J\geq0$ for all $J$.
No such coincidences hinder odd $k$ at 3 or more flavours, and we have explicitly found $a_J$ that satisfy \cref{eq:main:cond} with odd $k$ at 3 and 4 flavours (these turn out to produce very weak bounds, though).
Even $k>2$ remain permitted also at 2 flavours.

\subsection{The value of $t$}\label{sec:t=4}
There is no immediately obvious reason to favour any specific part of the allowed $s,t$ region when producing bounds.
However, one may note that at NLO with $k\geq 2$, the only part of $\frac{\d^k}{\d s^k} T^J(s,t)$ that depends on $s,t$ is the LEC-independent unitarity correction, which manifests itself as the term labelled $\gamma$ in \cref{eq:constr-LEC}.
Therefore, at fixed $a_J$ the most restrictive bound is obtained by minimising $\gamma$.
It turns out that given $s$, the magnitude of $\frac{\d^k}{\d s^k} T^J(s,t)$ tends to increase with $t$, with minima and maxima always falling along the $t=4$ line.
Therefore, it is expected that all relevant constraints should be found with $t=4$.
While we see no clear \emph{a priori} reason for it to be so, we have verified it by scanning the entire $s,t$ range for bounds; all relevant ones were found at $t=4$, within numerical uncertainty.

At NNLO, also the $\alpha_i$ in \cref{eq:constr-LEC} may depend on $s,t$, so the simple argument above does not hold.
However, the NNLO corrections are far too small to affect the overall shape of the amplitude, so for NLO LECs, fixing $t=4$ should remain sufficient.
The situation is yet more complicated for the NNLO LECs, since it turns out that certain combinations of them only feature in the amplitude when $t\neq 4$ (see the next section).
Therefore, complete NNLO bounds require using the full $s,t$ range.

\subsection{Independently bounded parameters}\label{sec:param}
While the parameter space affected by our bounds is technically the full space of (N)NLO LECs, it is of course impractical to work in such a large and redundant space.
Many LECs do not receive any bounds at all by not appearing in the Lagrangian relevant for meson-meson scattering (see \cref{sec:structure}), and others only appear in fixed linear combinations.
Specifically, all NLO LECs that appear in the amplitude do so independently, but the NNLO LECs only appear in combinations; consequently, it is not possible to obtain bounds on the individual $K_i^r$.
For instance, we shall see below that the combination $K_4^r - 2K_2^r$ appears in the amplitude and therefore receives bounds,%
\footnote{
    This combination is, up to a scale factor, $\Delta_3$ as defined below and explicitly given in \cref{sec:LEC-details}.} but its complement $2K_4^r + K_2^r$, which does not appear, is free to assume any value.
Therefore, nothing can be said about the values of $K_2^r$ and $K_4^r$ themselves.

We will therefore reexpress our parameter space in terms of NLO LECs in addition to a new set of independently bounded parameters built from the NNLO LECs.
Their form can be deduced from the polynomial parts of the functions $B,C$ described in \cref{sec:ampl}, which, following \cite{Bijnens:2011fm}, are
\begin{subequations}\label{eq:polynomial}
    \begin{align}
        B_P(s,t,u) &= \gamma_1 + \gamma_2t + \gamma_3t^2 + \gamma_4(s-u)^2 + \gamma_5t^3 + \gamma_6t(s-u)^2 + \text{(NLO)},\\
        C_P(s,t,u) &= \delta_1 + \delta_2s + \delta_3s^2 + \delta_4(t-u)^2 + \delta_5s^3 + \delta_6s(t-u)^2 + \text{(NLO)}.
    \end{align}
\end{subequations}
Here, $\gamma_i,\delta_i$ are linear combinations of the NNLO LECs, and ``(NLO)'' contains all NLO LECs, constant terms, etc.%
\footnote{
    This differs from the convention in \cite{Bijnens:2011fm}, where the NLO terms are included in $\gamma_i,\delta_i$.
    The ``NNLO parts'' that we extract here are easily read off from the appendices to that paper.}%
$^,$%
\footnote{
    \cite{Wang:2020jxr} use a similar approach in their 2-flavour NNLO bounds, but do not separate NLO and NNLO parts.
    This results in a smaller parameter space (6 dimensions compared to our 8; see \cref{sec:paramspace}), but our approach has the benefit of separating the relatively well-determined $\bar l_i$ from the much more uncertain NNLO values, allowing for figures such as \cref{fig:SU2-l34}.
    The fact that our parametrisation remains partly redundant is not a major issue, since we always fix some parameters rather than working in the full space.}
Therefore, bounds on the NNLO LECs only come in the form of bounds on $\gamma_i,\delta_i$.
Of course, not all $\gamma_i,\delta_i$ are bounded either --- those with $i<3$ vanish in the second derivative of the amplitude and therefore receive no bounds, and those with $i<6$ vanish in the third.
The remaining combinations are also not necessarily independent, so we will proceed to remould them into a better set of parameters.

\subsubsection{General number of flavours}
As is discussed in \cref{sec:t=4}, all relevant NLO bounds appear at $t=4$, so we may expect that this particular $t$-value is special also at NNLO.
Therefore, we express the polynomial parts in terms of $s$ and $\bar t\equiv 4 - t$, using $u=\bar t - s$:
\begin{equation}\label{eq:Bdecomp}
    \begin{aligned}
        B_P(s,t,u)  &= s^2\Gamma_1 + \bar t s^2(\Gamma_4 - 3\Gamma_3) + \ldots,\\
        B_P(t,u,s)  &= s^2\Gamma_2 - s^3\Gamma_3 + \bar t s^2\Gamma_4 + \ldots,\\
        B_P(u,s,t)  &= s^2\Gamma_2 + s^3\Gamma_3 + \bar t s^2(\Gamma_4 - 3\Gamma_3) + \ldots,
    \end{aligned}
\end{equation}
where ``$\ldots$'' consists of terms that vanish in the second derivative of the amplitude.
Here, we have defined the parameters
\begin{equation}\label{eq:GammaLECs}
    \begin{alignedat}{4}
        \Gamma_1    &=  4\gamma_4 + 16\gamma_6,             &\qquad 
        \Gamma_2    &=  \gamma_3 + \gamma_4 + 8\gamma_6,    \\
        \Gamma_3    &=  \gamma_5 + \gamma_6,                &\qquad 
        \Gamma_4    &=  3\gamma_5 - \gamma_6,
    \end{alignedat}
\end{equation}
of which $\Gamma_4$ only receives bounds when $t\neq 4$ due to the presence of $\bar t$.
Similarly,
\begin{equation}\label{eq:Cdecomp}
    \begin{aligned}
        C_P(s,t,u)  &= s^2\Delta_2 + s^3\Delta_3 + \bar t s^2(\Delta_4 - 3\Delta_3) + \ldots,   \\
        C_P(t,u,s)  &= s^2\Delta_1 + \bar t s^2(\Delta_4 - 3\Delta_3) + \ldots,                 \\
        C_P(u,s,t)  &= s^2\Delta_2 - s^3\Delta_3 + \bar t s^2\Delta_4 + \ldots,
    \end{aligned}
\end{equation}
where $\Delta_i$ are defined in terms of $\delta_i$ identically to \cref{eq:GammaLECs}.

These 8 parameters $\Delta_i,\Gamma_i$ constitute a minimal set of parameters for NNLO bounds with a general number of flavours;
explicit expressions are given in \cref{sec:LEC-details}.%
\footnote{
    While they can technically be considered LECs, we will avoid confusion by referring to the $\Gamma_i,\Delta_i$ as \emph{NNLO parameters}, reserving ``LEC'' for the coefficients appearing in the standard form of the Lagrangian.}
At 2 and 3 flavours, the Cayley-Hamilton identity allows for further reduction of the number of parameters.

\subsubsection{Two flavours}
Decomposing the polynomial part of $A(s,t,u)$ using \cref{eq:A,eq:Bdecomp,eq:Cdecomp} reveals
\begin{equation}\label{eq:Adecomp}
    \begin{aligned}
        A_P(s,t,u)  &= s^2\Theta_2 - s^3\Theta_3 + \bar t s^2\Theta_4 + \ldots + \ldots,        \\
        A_P(t,u,s)  &= s^2\Theta_1 + \bar t s^2(\Theta_4 - 3\Theta_3),                          \\
        A_P(u,s,t)  &= s^2\Theta_2 + s^3\Theta_3 + \bar t s^2(\Theta_4 - 3\Theta_3) + \ldots,   
    \end{aligned}
\end{equation}
where
\begin{equation}\label{eq:ThetaLECs}
    \begin{alignedat}{6}
        \Theta_1    &= 2\gamma_3 - 2\gamma_4 + 4\delta_4 + 16\delta_6,              &\qquad     
        \Theta_2    &= 4\gamma_4 + 16\gamma_6 + \delta_3 + \delta_4 + 8\delta_6,    \\
        \Theta_3    &= 2\gamma_5 + 2\gamma_6 - \delta_5 - \delta_6,                 &\qquad
        \Theta_4    &= 3\gamma_5 - \gamma_6 - 4\delta_6.
    \end{alignedat}
\end{equation}
$\Theta_4$, like $\Gamma_4$ and $\Delta_4$, is only bounded when $t\neq 4$.
Explicit expressions and experiment-based reference values for $\Theta_i$ are given in \cref{sec:LEC-details}.

\subsubsection{Three flavours}
A similar but less elegant simplification is possible in the 3-flavour case, using \cref{eq:ampl-su3}:%
\footnote{
    A more symmetric result would have been obtained by eliminating the $d^{ace}d^{bde}$ term in \cref{eq:ampl-d} instead, but we choose to follow \cite{Mateu:2008gv}.}
\begin{equation}
    \begin{aligned}
        B_{1P}(s,t,u) &= -2s^3\Gamma_3 + 3\bar t s^2\Gamma_3,                          \\
        B_{2P}(s,t,u) &= s^2\Xi_1 - s^3\Gamma_3 + 3\bar t s^2\Xi_4,                    \\
        A_{1P}(s,t,u) &= s^2\Xi_2 + s^3(\Delta_3 - \tfrac23\Gamma_3) + \bar t s^2\Xi_4,\\
        A_{2P}(s,t,u) &= s^2\Xi_3 + \tfrac23 s^3\Gamma_3 + \bar t s^2\Xi_4,\\
        A_{3P}(s,t,u) &= s^2\Xi_2 + s^3(2\Gamma_3 - \Delta_3) + \bar t s^2(\Xi_4 + 3\Delta_3 - 4\Gamma_3),
    \end{aligned}
\end{equation}
where
\begin{equation}
    \begin{alignedat}{6}
        \Xi_1    &= \gamma_3 - 3\gamma_4 - 8\gamma_6,                                   &\qquad     
        \Xi_2    &= 5\gamma_4 + 16\gamma_6 + \delta_3 + \delta_4 + 8\delta_6,           \\
        \Xi_3    &= \tfrac43\gamma_3 + \tfrac{16}3\gamma_6 + 4\delta_4 + 16\delta_6,    &\qquad
        \Xi_4    &= \gamma_5 - 3\gamma_6 - 4\delta_6.
    \end{alignedat}
    \label{eq:XiLECs}
\end{equation}
Again, expressions and values are given in \cref{sec:LEC-details}.

\subsubsection{The full parameter space}\label{sec:paramspace}
\Cref{tab:LECs} summarises the parameters affected by our bounds at different orders and number of flavours.
Note how the dimension of the space ranges from 2 (NLO 2-flavour) to 16 (NNLO $n$-flavour, $n\geq4$).
If $t$ is fixed to 4 at NNLO, this is reduced by 1 if $n\leq 3$ and by 2 otherwise; as discussed in \cref{sec:t=4}, $t=4$ is the only relevant value at NLO.
The difficulties associated with large parameter spaces are discussed in \cref{sec:results}.
\begin{table}[hbtp]
    \centering
    {\renewcommand{\arraystretch}{1.4}
    \begin{tabular}{llll}
        \hline\hline
        Flavours    &   NLO                                 &   \multicolumn{2}{l}{NNLO}                           \\
        \hline
        2           &   $\bar l_1,\bar l_2$                 &   \underline{\underline{$\bar l_1,\ldots\bar l_4$}};                  &   $\Theta_1,\Theta_2,\underline{\Theta_3},(\Theta_4)$             \\
        3           &   $L_1^r,L_2^r,L_3^r$                 &   \underline{\underline{$L_1^r,\ldots,L_6^r,L_8^r$}};                 &   $\Xi_1,\Xi_2,\Xi_3,(\Xi_4),\underline{\Gamma_3, \Delta_3}$      \\
        $\geq 4$    &   $\hat L_0^r,\ldots,\hat L_3^r$      &   \underline{\underline{$\hat L_0^r,\ldots,\hat L_6^r,\hat L_8^r$}};  &   $\Gamma_1,\Gamma_2,\underline{\Gamma_3},(\Gamma_4)$, $\Delta_1,\Delta_2,\underline{\Delta_3},(\Delta_4)$
    \end{tabular}}
    \caption{Summary of the NLO LECs and NNLO parameters appearing in the second derivative of the amplitude.
        The parameters $\Gamma_i',\Delta_i',\Theta_i',\Xi_i'$ as defined in \cref{eq:GammaLECs,eq:ThetaLECs,eq:XiLECs}.
        The LECs and parameters that remain in the third derivative are underlined, and those that remain also in the fourth and above are doubly underlined.
        Parameters that only feature in the amplitude when $t\neq4$ are placed in parentheses.}
    \label{tab:LECs}
\end{table}

\subsection{The absence of catastrophic divergences}
At NLO and above, the coefficient $\gamma$ in \cref{eq:constr-LEC} diverges in the limit $s\to0$ or $u\to0$.
If the divergence is towards positive infinity, this is not a problem --- it simply means that the positivity bound becomes trivial in these limits.
However, divergence towards negative infinity would be catastrophic, since no finite LECs could satisfy the positivity condition.
If there were some value of $a_J$ for which the divergence is in this direction, the theory would be inconsistent.

The situation becomes more complicated at NNLO, where also $\alpha_i$ diverge.
If $\alpha_i$ diverge at the same rate or faster than $\gamma$, the positivity conditions remain sensible also in these limits, but if $\gamma$ were to diverge towards negative infinity faster than $\alpha_i$, we would again have inconsistencies.

As $s$ approaches 4 from below, the $k$th derivative of the amplitude diverges as odd powers (up to $2k-1$) of $1/\delta$, where $s=4(1-\delta^2)$; see \cref{sec:Jbar-s4} for details.
Let $q^J$ be the coefficient of the leading divergence $\d^kT^J/\d s^k$ that contributes to $\gamma$.
Then consistency requires $a_Jq^J\geq 0$ for all valid $a_J$.
Since \cref{eq:main:constr} requires $a_J\geq 0$ in the limit $s\to 4$, this is satisfied if $q^J\geq 0$.
At both NLO and NNLO for any number of flavours $n$, this turns out to be true for the $s\to 4$ divergence (this was already noted in \cite{Manohar:2008tc} for $n=2$).
Also, the divergences of $\alpha_i$ are of equal or lower powers than those of $\gamma$.

The same divergence structure appears in the $u\to4$ limit, but here the coefficients of the leading divergences are not necessarily positive.
However, we may use crossing symmetry to rewrite
\begin{equation}
    \frac{\d^k}{\d s^k}T^I(s,t) = (-1)^k\frac{\d^k}{\d u^k}C_u^{IJ}T^J(u,t).
\end{equation}
Here, we can simply relabel $u$ as $s$.
The coefficient of the leading divergence is here $(-1)^k C_u^{IJ} q^J$, and since \cref{eq:main:constr} requires $(-1)^ka_IC_u^{IJ}\geq 0$, in the limit $u\to4$, the fact that $q^J\geq 0$ in the $s\to4$ limit guarantees that there are no catastrophic divergences in the $u\to4$ limit either.

Since $\d\delta/\d s=-1/8\delta$, taking another derivative does not change the sign of $q^J$.
Therefore, if no catastrophic divergences appear at the first $k$ where $\gamma$ diverges, they will not appear at larger $k$ either.\footnote{The first divergence happens at $k=1$ for $J\in\{I,S,SS,AA\}$ and at $k=2$ for $J=A$, regardless of $n$.
There is no divergence for $J=AS$, since $T^{AS}(s,t,u)$ does not contain $\bar J(s)$ or $k_i(s)$.}

\subsection{Integrals above threshold}\label{sec:int}
The right-hand side of \cref{eq:dispersion} is, in its standard application, a non-perturbative quantity, about which the only knowledge we have is the fact that it is positive.
However, \chpt\ is a low-energy theory, so its amplitude at any order should be an excellent approximation of the true
amplitude for energies sufficiently close to threshold.
Taking inspiration from the approach used in \cite{Wang:2020jxr}, we may therefore explicitly evaluate the lowest part of the integral on the right-hand side of \cref{eq:dispersion} and subtract it from both sides.
Specifically, we define
\begin{equation}
    D_k^J(\lambda, v, t) = \frac{k!}{\pi}\int_4^\lambda \frac{\d z}{(z-v)^{k+1}} \Im T^J(z+i\e, t)
\end{equation}
and modify \cref{eq:dispersion} to
\begin{equation}\label{eq:mod-dispersion}
    a_J\left[\frac{\d^k}{\d s^k} T^J(s,t) - D_k^J(\lambda, s, t) - (-1)^k C^{JI}D_k^I(\lambda, u, t)\right]
    = \frac{k!}{\pi}\int_\lambda^\infty \d z
        \left[
            \ldots
        \right] \Im T^J(z + i\e, t).
\end{equation}
This has two benefits:
\begin{enumerate}[label=\roman*)]
    \item When $s,t,u$ and $a_J$ satisfy the conditions of \cref{eq:main}, both $D^J_k$ and the right-hand side are positive, so we obtain a stronger positivity bound.
    \item The right-hand side of \cref{eq:mod-dispersion} is positive under a wider range of conditions than that of \cref{eq:dispersion}, so we obtain more positivity bounds.
        (This is because the constraint $\v a\cdot\v\alpha^J(4)\geq 0$  is replaced by the weaker $\v a\cdot\v\alpha^J(\lambda)\geq0$, recalling the notation and discussion in \cref{sec:aJ-cond}).
        Some of the new bounds are weakened by $D_k^J$ being negative, but they may still contribute. 
\end{enumerate}
The size of $\lambda$ presents a tradeoff: larger values amplify the benefits of using it, but also decrease the accuracy of the relation as the fixed-order \chpt\ amplitude strays from the exact result.
The integral also requires some mathematical machinery; $D_k^J$ is by no means a simple function, but we determine it up to NNLO in \cref{sec:integrals}
(of course, it could also have been done numerically).
By evaluating the NNLO corrections, we obtain a good idea of the accuracy of the NLO result.

An upper bound on $\lambda$ is provided by \cite{Chivukula:1992gi}, which determines the breakdown scale of $n$-flavour \chpt\ to be $s\sim (4\pi F)^2/n$. 
Using the value $F=92.2(1)\:\text{MeV}$ adopted by \cite{Bijnens:2014lea} along with $\Mphys=M_\pi$, this places the breakdown at $\lambda\approx 35$ for $n=2$ and at $\lambda\approx 25$ for $n=3$. 
Thus, we cannot expect sensible results for $\lambda$ anywhere close to this, and certainly not above it.%
\footnote{
    Note that $(2M_K)^2\approx 14$ and $(2M_\eta)\approx16$ are already quite close to the breakdown scale.
    This offers some motivation as to why the equal-mass approximation is reasonable also at 3 flavours: with unequal, real-world masses, 3-flavour \chpt\ operates close to the limits of its range of validity even with nonrelativistic particles, which offsets the accuracy gained by increasing the realism of the model.}

\FloatBarrier
\pagebreak
\section{Results}\label{sec:results}
Here, we present the constraints obtained using the methods described in the preceding sections.

Following \cref{sec:constr}, we will use the letter $\Omega$ to denote each collection of constraints, and $\sat(\Omega)$ to denote the sets of parameter-space points that satisfy these.
We will compare each $\sat(\Omega)$ to a reference point, taken as the central value of the LEC estimates in \cite{Bijnens:2014lea}.
These values can also be found in \cref{tab:fiducial} in \cref{sec:LEC-details}.

In all but the simplest cases, parameter space has too many dimensions to be visualised as a whole.
Therefore, we will show lower-dimensional slices, with all omitted parameters set to their reference values.
We will primarily show two-dimensional slices, since they are the easiest to understand, although some three-dimensional slices will be needed as well.
It is not practical to show an exhaustive set of slices, so we will focus on grouping parameters that are, in some loose sense, related.

As a visual aid and a rough indicator of constraint strength, we define the quantity $\rho(\constr{\v\alpha}{c})$ to be the shortest distance between the reference point and the hyperplane $\v\alpha\cdot\v b = c$.%
\footnote{
    Note that this refers to distance in the full parameter space, which does not directly correspond to distance in the subspaces shown in the figures.}
Alternatively, we may use $\hat\rho(\constr{\v\alpha}{c})$, which is the analogous distance if the space where all parameters are rescaled so that their reference values are 1.

\subsection{Two flavours}\label{sec:results-2flav}
Two-flavour \chpt\ constraints are in many regards quite simple: there are only 2 parameters at NLO and 7 at NNLO; there is only one reasonable choice for $M_\text{phys}$, namely $M_\pi$; and as shown in \cref{sec:nderiv}, we do not have to consider odd numbers of derivatives.
The NLO constraints have been extensively studied in e.g.\ \cite{Manohar:2008tc,Wang:2020jxr,Tolley:2020gtv}, whereas the NNLO constraints are novel to this work. 

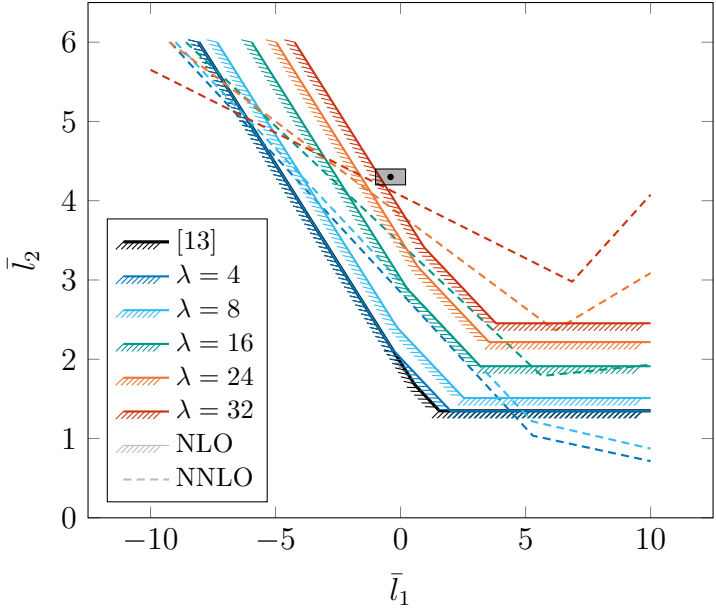
\begin{figure}[hbtp]
    \begin{minipage}[t]{0.55\textwidth}
        \raisebox{-\height}{
            \hspace{-1.7cm}
            \begin{tikzpicture}
                \begin{axis}[scale=1.2,
                    ylabel={$\bar l_2$}, xlabel={$\bar l_1$},
                    xmin=-12.5,xmax=12.5, ymin=0, ymax=6.5,
                    legend pos=south west, legend cell align=left,
                    legend image code/.code={
                        \draw (0cm,0cm) -- (0.6cm,0cm);
                    },
                    legend style={font=\footnotesize}]
                
                    \addlegendimage{black, very thick, hatchborder};
                    \addlegendimage{plotI,  thick, hatchborder};
                    \addlegendimage{plotII,  thick, hatchborder};
                    \addlegendimage{plotIII,  thick, hatchborder};
                    \addlegendimage{plotIV,  thick, hatchborder};
                    \addlegendimage{plotV,  thick, hatchborder};
                    \addlegendimage{plotgrey,  hatchborder};
                    \addlegendimage{plotgrey,  thick, densely dashed};

                    \draw[
                        black, join=round, very thick, hatchborder
                    ]  
                        (-8.075, 6)
                        -- (0.567,1.679)
                        -- (1.554,1.35)
                        -- (10,1.35);
                    
                    \draw[
                        plotI, join=round, thick, hatchborder
                    ]
                        (-8.07517, 5.99979)
                        -- (-0.324077, 2.12455)
                        -- (1.99952, 1.35006)
                        -- (9.9999, 1.35014);
                        \draw[
                        plotI, join=round, thick, densely dashed
                    ]
                        (-9.21773, 5.99961)
                        -- (5.2909, 1.03618)
                        -- (10.0005, 0.714619);
                        
                    \draw[
                        plotII, join=round, thick, hatchborder
                    ]
                        (-7.33653, 6.0002)
                        -- (-0.202206, 2.43291)
                        -- (-0.0944658, 2.38661)
                        -- (2.52571, 1.51313)
                        -- (2.54915, 1.51135)
                        -- (10.0003, 1.51097);
                              
                    \draw[
                        plotII, join=round, thick, densely dashed
                    ]  
                        (-9.00084, 6.0002)
                        -- (5.26049, 1.21766)
                        -- (9.99989, 0.872907);

                    \draw[
                        plotIII, join=round, thick, hatchborder
                    ]
                        (-5.95224, 5.99957)
                        -- (0.291645, 2.87854)
                        -- (0.275445, 2.88637)
                        -- (3.18997, 1.91242)
                        -- (3.19887, 1.91232)
                        -- (9.99989, 1.91201);
                              
                    \draw[
                        plotIII, join=round, thick, densely dashed
                    ]
                        (-8.58342, 5.9998)
                        -- (5.64901, 1.79017)
                        -- (5.6502, 1.78936)
                        -- (10.0005, 1.93105);
                    
                    \draw[
                        plotIV, join=round, thick, hatchborder
                    ]
                        (-4.975, 5.99993)
                        -- (0.668122, 3.17853)
                        -- (0.718448, 3.15709)
                        -- (3.53839, 2.21682)
                        -- (10, 2.21654);
                                  
                    \draw[
                        plotIV, join=round, thick, densely dashed
                    ]
                        (-9.25112, 6.00055)
                        -- (6.17655, 2.3673)
                        -- (6.1867, 2.36605)
                        -- (10, 3.08739);
                    
                    \draw[
                        plotV, join=round, thick, hatchborder
                    ]
                        (-4.23312, 5.99979)
                        -- (0.907885, 3.42967)
                        -- (0.97265, 3.40197)
                        -- (3.8153, 2.45432)
                        -- (3.83135, 2.45368)
                        -- (10.0001, 2.45369);
                             
                    \draw[
                        plotV, join=round, thick, densely dashed
                    ]  
                        (-10.0002, 5.65137)
                        -- (6.85547, 2.98035)
                        -- (6.86942, 2.9802)
                        -- (9.99945, 4.07537);

                    \filldraw [black] (-0.4, 4.3)circle[radius=1pt];
                    \draw[black, fill=black, fill opacity=0.3] (-1,4.2) rectangle (0.2,4.4);

                    \legend{$[13]$,$\lambda=4$,$\lambda=8$,$\lambda=16$,$\lambda=24$,$\lambda=32$,NLO,NNLO,,,,,,,};
                \end{axis}
            \end{tikzpicture}
        }
    \end{minipage}
    \begin{minipage}[t]{0.45\textwidth}
        \caption[Two-flavour two-derivative constraints on $\bar l_1,\bar l_2$.]{
            Two-flavour two-deriva\-tive constraints on $\bar l_1,\bar l_2$ for various $\lambda$, as indicated in the legend.
            The presentation is similar to \cref{fig:constr}: each version of $\sat(\Omega)$ is outlined, with the hatched side indicating the points excluded by the constraints.
            The bounds from \cite{Manohar:2008tc} are also drawn.
            The  reference point \mbox{$\bar l_1=-0.4(6)$}, \mbox{$\bar l_2=4.3(1)$} is drawn as a dot with an uncertainty region around it.
            For each set of constraints, the direct NNLO counterpart (i.e.\ using the same $\lambda,s,t$ and $a_J$) is drawn as a dashed outline.}
        \label{fig:SU2-l12}
    \end{minipage}
\end{figure}

\Cref{fig:SU2-l12} shows constraints obtained using 2 derivatives and various amounts of above-threshold integration.
The non-integrated ($\lambda=4$) constraints are slightly stronger than those in \cite{Manohar:2008tc}, which only considered eigenstate $a_J$ rather than the full space, but the constraints do not come close to the experimental uncertainty of the reference point without using $\lambda$ that are far too large for the results to be trusted (recall that perturbative breakdown is expected at $\lambda\approx 35$).
In \cite{Wang:2020jxr,Tolley:2020gtv}, comparable bounds are obtained with slightly less extreme $\lambda$, but in both cases, $\lambda$ needs to be rather large to start cutting into the experimental uncertainty.
The large discrepancy between the NLO and NNLO versions of the constraints indicate that the bounds are highly unreliable for all but the smallest $\lambda$ used. 
Even with $\lambda=4$, the difference is quite significant. 

\begin{figure}[hbtp]
    \begin{minipage}[t]{0.6\textwidth}
        \raisebox{-\height}{
            \hspace{-1.7cm}
            \begin{tikzpicture}
                \begin{axis}[scale=1.2,
                    ylabel={$\bar l_2$}, xlabel={$\bar l_1$},
                    xmin=-12.5,xmax=13.5, ymin=.5, ymax=9.5,
                    legend pos=north east, legend cell align=left,
                    legend image code/.code={
                        \draw (0cm,0cm) -- (0.6cm,0cm);
                    },
                    legend style={font=\footnotesize}]
                
                    \addlegendimage{plotI, thick, hatchborder};
                    \addlegendimage{plotIII, thick, hatchborder};
                    \addlegendimage{plotV, thick, hatchborder};
                    \addlegendimage{plotgrey, thick, hatchborder};
                    \addlegendimage{plotgrey, thick, densely dashed, hatchborder};
                    \addlegendimage{plotgrey, thick, densely dotted, hatchborder};
                    
                    \draw[
                        plotI, join=round, opacity=.5, thin, weakhatchborder
                    ]
                        (-12.0008,7.9626)
                        -- (-0.324077,2.12455)
                        -- (1.99952, 1.35006)
                        -- (11.9996,1.35006);
                    
                    \draw[
                        plotI, join=round, thick, 
                        postaction={decorate,draw,thin}, invhatchborder
                    ]
                        (11.9995, 1.52292)
                        -- (1.51172, 2.23894)
                        -- (-1.31549, 2.69241)
                        -- (-11.9993, 6.34788);

                    \draw[
                        plotI, join=round, thick, densely dashed, invhatchborder
                    ]
                        (-2.37543, 9.00002)
                        -- (5.38005, 5.32254)
                        -- (12.0004, 2.06887);

                    \draw[
                        plotV, join=round, thick, densely dotted, invhatchborder
                    ]
                        (-2.51767, 8.99968)
                        -- (11.9994, 1.79949);
                    
                    \draw[
                        plotIII, join=round, thin, weakhatchborder
                    ]
                        (-11.9996, 8.33135)
                        -- (-0.202206, 2.43291)
                        -- (-0.0944658, 2.38661)
                        -- (2.52571, 1.51313)
                        -- (2.54915, 1.51135)
                        -- (12.0005, 1.51186);
                    
                    \draw[
                        plotIII, join=round, thick, invhatchborder
                    ]
                        (11.9995, 1.77469)
                        -- (2.55997, 2.51404)
                        -- (1.18698, 2.66546)
                        -- (0.238266, 2.79402)
                        -- (-1.32693, 3.20141)
                        -- (-10.8474, 6.39585)
                        -- (-11.0738, 6.47263)
                        -- (-11.2926, 6.54763)
                        -- (-11.5254, 6.62962)
                        -- (-11.8298, 6.73578)
                        -- (-11.9995, 6.79618);

                    \draw[
                        plotIII, join=round, thick, densely dashed, invhatchborder
                    ]
                        (-11.3214, 9.00016)  
                        -- (-10.9932, 8.95568)  
                        -- (-9.91955, 8.77647)  
                        -- (-8.71913, 8.54577)  
                        -- (-7.37468, 8.25707)  
                        -- (3.94601, 5.62959)  
                        -- (12.0003, 1.4603)  
                        (-6.04257, 1.00015)  
                        -- (-7.76656, 2.83051)  
                        -- (-8.07238, 3.18279)  
                        --(-12.0005, 7.94612);  

                    \draw[
                        plotV, join=round, thin, weakhatchborder
                    ]
                        (-11.9999, 8.71883)
                        -- (0.0317066, 2.70276)
                        -- (0.0618442, 2.68807)
                        -- (2.94007, 1.72857)
                        -- (12.0002, 1.72886);
                        
                    \draw[
                        plotV, join=round, thick, hatchborder
                    ]
                        (-12.0002, 7.41195)  
                        -- (-6.3951, 5.328)  
                        -- (-6.23606, 5.26938)  
                        -- (-6.0821, 5.21449)  
                        -- (-5.95212, 5.16781)  
                        -- (-5.83237, 5.12652)  
                        -- (-5.70553, 5.08279)  
                        -- (-5.53063, 5.02404)  
                        -- (-5.36457, 4.96969)  
                        -- (-5.2976, 4.9482)  
                        -- (-5.25458, 4.93383)  
                        -- (-5.16758, 4.90607)  
                        -- (-1.2058, 3.64465)  
                        -- (0.524492, 3.24422)  
                        -- (12.0002, 2.15303);  

                    \draw[
                        plotV, join=round, thick, densely dashed, hatchborder
                    ]
                        (11.6441, 0.999967)  
                        -- (2.09598, 5.68563)  
                        -- (-6.23222, 6.54122)  
                        -- (-6.10807, 6.05474)  
                        -- (-3.79306, 3.04983)  
                        -- (-3.68259, 2.91218)  
                        -- (-3.37137, 2.54966)  
                        -- (-3.04907, 2.20042)  
                        -- (-2.65723, 1.80026)  
                        -- (-2.18061, 1.33768)  
                        -- (-1.81728, 1.00001);  

                    \filldraw [black] (-0.4, 4.3)circle[radius=1pt];
                    \draw[black, fill=black, fill opacity=0.3] (-1,4.2) rectangle (0.2,4.4);
                    
                    \legend{$\lambda=4$,$\lambda=8$,$\lambda=12$,$k=2$,$k=4$,$k=6$,,,,,,,,,,};
                        
                \end{axis}
            \end{tikzpicture}
        }
    \end{minipage}
    \begin{minipage}[t]{0.4\textwidth}
        \caption{NNLO bounds on $\bar l_1,\bar l_2$ with $k=2,4,6$ derivatives, displayed similarly to \cref{fig:SU2-l12} but over a slightly wider part of parameter space.
        For comparison to \cref{fig:SU2-l12}, the corresponding NLO bounds are drawn with weak solid lines.
        The six-derivative bounds are nearly independent of $\lambda$, so all versions are not drawn. }
        \label{fig:SU2-l12-deriv}
    \end{minipage}
\end{figure}
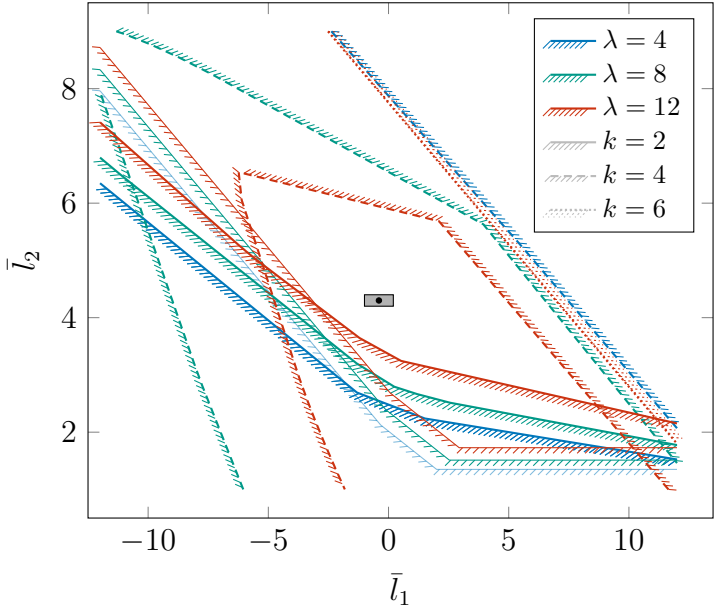

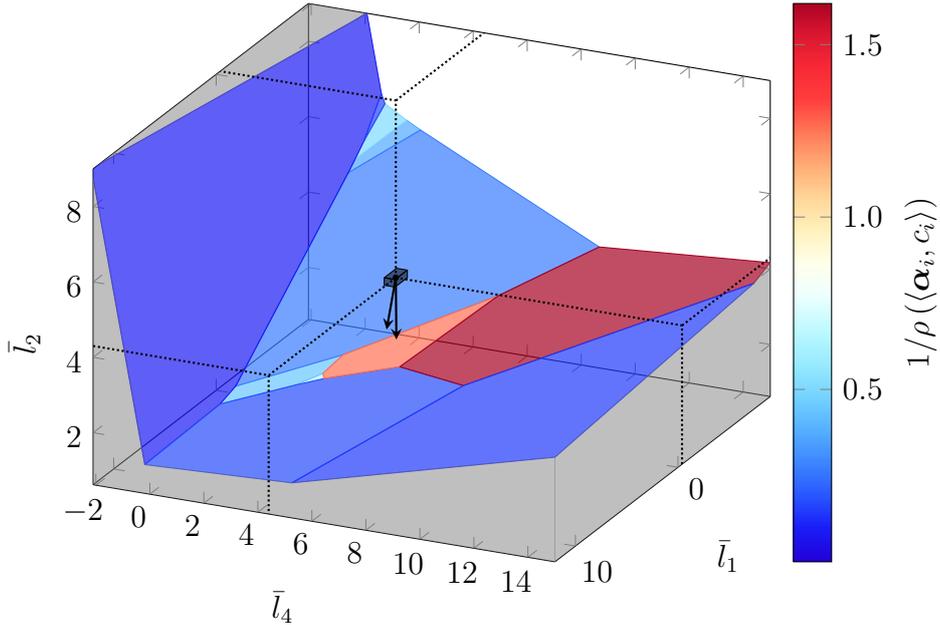
\begin{figure}[hbtp]
    \centering
    \begin{tikzpicture}
        \begin{axis}[
                scale=1.3,
                view={115}{30}, xlabel={$\bar l_1$}, ylabel={$\bar l_4$}, zlabel={$\bar l_2$},
                ymin=-2.1,zmax=9,
                point meta min=0,
                colormap/temp, colorbar, 
                colorbar style={
                    ymin=0, ytick={500,1000,1500}, yticklabels={0.5,1.0,1.5},ylabel={$1/\rho\left(\constr{\v\alpha_i}{c_i}\right)$}
                }
            ]

            \addplot3 [patch, patch table with point meta={\resultpath/NNLO/SU2/M0.135/D2/lam4/visualisation/l1l2l4/constr_0_table.dat}, opacity=0] 
                    table {\resultpath/NNLO/SU2/M0.135/D2/lam4/visualisation/l1l2l4/constr_0_coords.dat};
                    
            \coordinate (p0) at (2.45687, 2.78489, 2.16497);
            \coordinate (p1) at (3.62124, 2.19905, 2.02714);
            \coordinate (p2) at (2.76311, 2.93834, 2.10863);
            \coordinate (p3) at (6.03483, 0.333061, 1.80534);
            \coordinate (p4) at (4.73109, 0.417079, 2.01574);
            \coordinate (p5) at (1.02221, 2.98983, 2.38529);
            \coordinate (p6) at (-6.34339, 0.628077, 5.7395);
            \coordinate (p7) at (-8.99975, 1.54053, 6.33773);
            \coordinate (p8) at (-8.99959, 0.734727, 6.68118);
            \coordinate (p9) at (-8.99964, 0.734731, 6.68121);
            \coordinate (p10) at (-7.20673, 0.637736, 6.05656);
            \coordinate (p11) at (-8.99995, 2.05013, 6.14263);
            \coordinate (p12) at (-5.72345, 0.618611, 5.52164);
            \coordinate (p13) at (-4.82557, 6.51791, 3.15324);
            \coordinate (p14) at (-8.99997, 2.05014, 6.14265);
            \coordinate (p15) at (-9.00005, 8.66681, 3.83028);
            \coordinate (p16) at (-9.00011, 8.66687, 3.83031);
            \coordinate (p17) at (-9.00027, 0.599601, 6.93826);
            \coordinate (p18) at (-9.00046, 0.5995, 6.9384);
            \coordinate (p19) at (-9.00011, 0.0655247, 9.00011);
            \coordinate (p20) at (11.9999, -0.190636, 1.38205);
            \coordinate (p21) at (11.9999, -0.190636, 1.38205);
            \coordinate (p22) at (12.0001, -2.16421, 9.00087);
            \coordinate (p23) at (0.947462, 5.05858, 2.29736);
            \coordinate (p24) at (12.0002, 5.25865, 1.54884);
            \coordinate (p25) at (12.0002, 5.25865, 1.54883);
            \coordinate (p26) at (1.87995, 7.76865, 2.31686);
            \coordinate (p27) at (-9.00038, 14.9994, 4.17021);
            \coordinate (p28) at (-7.40094, 14.9995, 3.95485);
            \coordinate (p29) at (-8.99982, 1.54165, 6.33777);
            \coordinate (p30) at (-7.40099, 14.9996, 3.95487);
            \coordinate (p31) at (12.0006, 15.0007, 3.39418);
            
            \draw[color of colormap={655.888/1.61929}, facet]
                (p0) -- (p1) -- (p2) -- cycle;

            \draw[color of colormap={362.615/1.61929}, facet]
                (p29) -- (p11) -- (p12) -- (p6) -- (p7) -- cycle;
                
            \draw[color of colormap={442.001/1.61929}, facet]
                (p0) 
                -- (p1) 
                -- (p3) 
                -- (p4) 
                -- (p5) 
                -- cycle;
                
            \draw[color of colormap={505.925/1.61929}, facet]
                (p6) 
                -- (p7) 
                -- (p8) 
                -- (p9) 
                -- (p10) 
                -- cycle;
                
            \draw[color of colormap={264.256/1.61929}, facet]
                (p4) 
                -- (p5) 
                -- (p13) -- (p15)  
                -- (p16) 
                -- (p14) 
                -- (p11) 
                -- (p12) 
                -- cycle;
                
            \draw[color of colormap={81.0492/1.61929}, facet]
                (p3) 
                -- (p4) -- (p12)  
                -- (p6) 
                -- (p10) 
                -- (p17) 
                -- (p18) 
                -- (p19) 
                -- (p22)  
                -- (p21) 
                -- (p20) 
                -- cycle;
                
            \draw[color of colormap={159.898/1.61929}, facet]
                (p8) -- (p10) -- (p17) -- cycle;
                
            \draw[color of colormap={171.168/1.61929}, facet]
                (p1) -- (p2) -- (p23) -- (p26) -- (p25) -- (p20) -- (p3) -- cycle;
                
            \draw[color of colormap={1250.66/1.61929}, facet]
                (p0) -- (p2) -- (p23) -- (p13) -- (p5) -- cycle;
                
            \draw[color of colormap={1619.29/1.61929}, facet]
                (p27) -- (p28) -- (p26) -- (p23) -- (p13) -- (p15) -- cycle;
                
            \draw[color of colormap={139.375/1.61929}, facet]
                (p28) -- (p30) -- (p31) -- (p25) -- (p26) -- cycle;
                                
            \def\xfid{-0.4} \def\yfid{4.4}  \def\zfid{4.3}
            \def\xmin{-9}   \def\ymin{-2.1} \def\zmin{0}
            \def\xmax{12}   \def\ymax{15}   \def\zmax{9}
            
            \coordinate (fid) at (\xfid,\yfid,\zfid);
            
            \draw[grey, fill, opacity=.5]
                (p22) -- (p21) -- (p25) -- (p31) -- (\xmax,\ymax,\zmin) -- (\xmax,\ymin,\zmin) -- cycle;
            \draw[grey, fill, opacity=.5]
                (p31) -- (p28) -- (p27) -- (\xmin,\ymax,\zmin) -- (\xmax,\ymax,\zmin) -- cycle;
            \draw[grey, fill, opacity=.5]
                (p22) -- (p19) -- (\xmin,\ymin,\zmax) -- cycle;
            
            \draw[surface to fid]
                (-0.930352, 3.85828, 2.74993) -- (fid);
            \draw[surface to fid]
                (-0.673776, 4.31767, 2.5934)  -- (fid);
                
            \draw[axes to fid]
                (fid) -- (\xfid,\yfid,\zmax) -- (\xfid,\ymin,\zmax) (\xfid,\yfid,\zmax) -- (\xmin,\yfid,\zmax)
                (fid) -- (\xfid,\ymax,\zfid) -- (\xfid,\ymax,\zmin) (\xfid,\ymax,\zfid) -- (\xmin,\ymax,\zfid)
                (fid) -- (\xmax,\yfid,\zfid) -- (\xmax,\ymin,\zfid) (\xmax,\yfid,\zfid) -- (\xmax,\yfid,\zmin);
                
            \filldraw [black] (\xfid,\yfid,\zfid) circle[radius=1pt] {};

            \draw[black, fill=black, fill opacity=0.5, join=round] 
                (0.2,4.6,4.4) -- (0.2,4.6,4.2) -- (-1,4.6,4.2) -- (-1,4.6,4.4) -- cycle;
            \draw[black, fill=black, fill opacity=0.5, join=round] 
                (0.2,4.6,4.4) -- (0.2,4.2,4.4) -- (0.2,4.2,4.2) -- (0.2,4.6,4.2)  -- cycle;
            \draw[black, fill=black, fill opacity=0.5, join=round] 
                (0.2,4.6,4.4) -- (0.2,4.2,4.4) -- (-1,4.2,4.4) -- (-1,4.6,4.4) -- cycle;
        \end{axis}
    \end{tikzpicture}
    \caption[NNLO two-derivative constraints on $\bar l_1,\bar l_2$ and $\bar l_4$ with $\lambda=4$]{
        NNLO two-derivative constraints on $\bar l_1,\bar l_2$ and $\bar l_4$ with $\lambda=4$; a cross-section at $\bar l_4=4.4$ would yield part of \cref{fig:SU2-l12}.
        The constraint surfaces are coloured according to their proximity $1/\rho\left(\constr{\v\alpha_i}{c_i}\right)$ to the reference point, which is drawn similar to the 2D plots. 
        The ``open space'' bounded by the constraint surfaces is part of $\sat(\Omega)$; the greyed-out region is excluded by the constraints.
    
        \qquad To clarify its spatial position, the reference point is connected to the boundaries of the plotted region with black dotted lines parallel to the coordinate axes, whenever doing so is possible without intersecting a constraint surface.
        It is similarly connected to neighbouring surfaces with black arrows.
        These are orthogonal to the respective surfaces, even though the different scales of the axes makes it not appear so.}
    \label{fig:SU2-l124}
\end{figure}

\Cref{fig:SU2-l12-deriv} shows similar NNLO constraints over a more conservative $\lambda$ range; however, here we display the effects of higher even derivatives (recall again that odd derivatives need not be considered with 2 flavours).
Unlike the ones shown in \cref{fig:SU2-l12}, these constraints impose \emph{upper} bounds on the LECs as well as lower bounds.
Note that in the upper-left part of the plot, the two-derivative bounds are less restrictive than their NLO counterpart.
This can partly be seen as an artefact of introducing multiple new parameters and fixing them to imprecise experimental values (for instance, \cref{fig:SU2-l124} shows that smaller values of $\bar l_4$ strengthen the bounds on $\bar l_1,\bar l_2$), but one must keep in mind that switching to a more refined theory can both strengthen and relax the predictions.

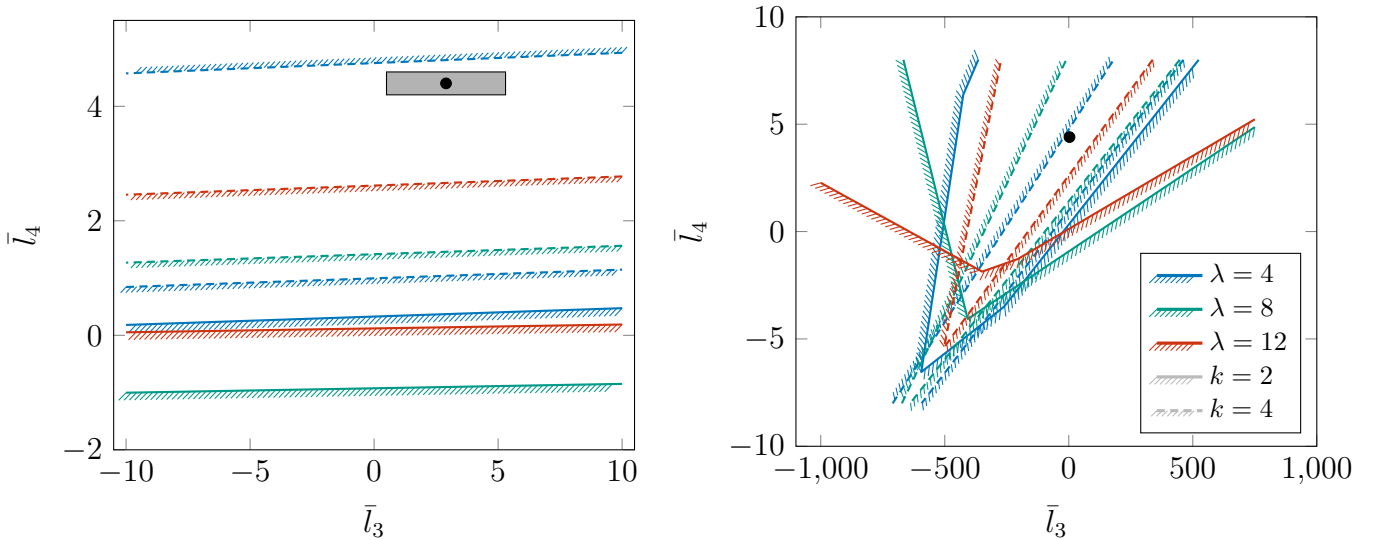
\begin{figure}[hbtp]
    \hspace{-1.5cm}
    \hbox{
        \begin{tikzpicture}
            \begin{axis}[ylabel={$\bar l_4$}, xlabel={$\bar l_3$},
                legend pos=south east, legend cell align=left,
                legend image code/.code={
                    \draw (0cm,0cm) -- (0.6cm,0cm);
                },
                legend style={font=\footnotesize},
                xmin=-10.5,xmax=10.5, ymin=-2,ymax=5.5]
            
                \addlegendimage{red, very thick, hatchborder};
                \addlegendimage{pcolour, very thick, hatchborder};
                \addlegendimage{blue, very thick, hatchborder};
                \addlegendimage{gray, very thick, hatchborder};
                \addlegendimage{gray, very thick, densely dashed, hatchborder};
                \addlegendimage{gray, very thick, densely dotted, hatchborder};
                
                \draw[
                    plotI, join=round, thick, invhatchborder
                ]
                    (10.0001, 0.474028) -- (-9.99962, 0.181029);

                \draw[
                    plotI, join=round, thick, densely dashed, hatchborder
                ]  
                    (-10.0004, 0.843889) -- (10.0001, 1.14747) 
                    (10.0002, 4.93429) -- (-10.0005, 4.57466);
                    
                \draw[
                    plotIII, join=round, thick, hatchborder
                ]  
                    (-9.99989, -1.0027) -- (9.99978, -0.848235);

                \draw[
                    plotIII, join=round, thick, densely dashed, invhatchborder
                ]  
                    (10, 1.56377) -- (-9.99955, 1.2691);

                \draw[
                    plotV, join=round, thick, invhatchborder
                ]  
                    (9.99987, 0.188635) -- (-10.0001, 0.0524908);
                    
                \draw[
                    plotV, join=round, thick, densely dashed, invhatchborder
                ]  
                    (10, 2.77502) -- (-9.99955, 2.45587);

                \filldraw [black] (2.9, 4.4)circle[radius=2pt];
                \draw[black, fill=black, fill opacity=0.3] (0.5,4.2) rectangle (5.3,4.6);
                    
            \end{axis}
        \end{tikzpicture}
        \begin{tikzpicture}
            \begin{axis}[
                xmin=-1100, xmax=1000, ymin=-10, ymax=10,
                xlabel={$\bar l_3$},ylabel={\raisebox{-1em}{$\bar l_4$}},
                legend pos=south east, legend cell align=left,
                legend image code/.code={
                    \draw (0cm,0cm) -- (0.6cm,0cm);
                },
                legend style={font=\footnotesize}]
            
                \addlegendimage{plotI, very thick, hatchborder};
                \addlegendimage{plotIII, very thick, hatchborder};
                \addlegendimage{plotV, very thick, hatchborder};
                \addlegendimage{plotgrey, very thick, hatchborder};
                \addlegendimage{plotgrey, very thick, densely dashed, hatchborder};
                
                \draw[
                    plotI, join=round, thick, hatchborder
                ]  
                    (-364.915, 8.00053)
                    -- (-425.543, 6.36676)
                    -- (-594.662, -6.54193)
                    -- (-262.132, -3.51291)
                    -- (-262.139, -3.51301)
                    -- (523.714, 7.99791);

                \draw[
                    plotI, join=round, thick, densely dashed, invhatchborder
                ]  
                    (461.999, 7.998)
                    -- (-593.287, -7.99831)
                    (-710.014, -8.00266)
                    -- (-492.186, -4.0817)
                    -- (180.777, 8.00014);

                \draw[
                    plotIII, join=round, thick, invhatchborder
                ]  
                    (750.018, 4.86686)
                    -- (-407.473, -4.07228)
                    -- (-666.692, 8.0011);

                \draw[
                    plotIII, join=round, thick, densely dashed, invhatchborder
                ]  
                    (446.971, 7.9979)
                    -- (-639.382, -7.99947)
                    (-672.821, -8.00065)
                    -- (-10.2216, 7.99947);

                \draw[
                    plotV, join=round, thick, hatchborder
                ] 
                    (-1000, 2.27101)
                    -- (-349.385, -1.86502)
                    -- (-205.514, -1.27891)
                    -- (-205.515, -1.27892)
                    -- (750.475, 5.23081);

                \draw[
                    plotV, join=round, thick, densely dashed, invhatchborder
                ]  
                    (337.902, 8.00054)
                    -- (-502.897, -5.39663)
                    -- (-502.896, -5.39661)
                    -- (-272.846, 8.00031);
                    
                \filldraw [black] (2.9, 4.4)circle[radius=2pt];
                \draw[black, fill=black, fill opacity=0.3] (0.5,4.2) rectangle (5.3,4.6);

                \legend{$\lambda=4$,$\lambda=8$,$\lambda=12$,$k=2$,$k=4$,,,,,,,,,,};
                    
            \end{axis}
        \end{tikzpicture}
    }
    \caption{
        NNLO $k$-derivative constraints on $\bar l_3,\bar l_4$, displayed like in \cref{fig:SU2-l12-deriv}, using similarly scaled axes (left) or an extreme scale on the $\bar l_3$ axis (right) to show the full constraint structure.
        In all cases, the bounds are satisfied by the reference point.
        The six-derivative bounds (which like in \cref{fig:SU2-l12} are nearly $\lambda$-independent) exactly overlap with the $\lambda=4$ two-derivative bounds and are not drawn. }
    \label{fig:SU2-l34}
\end{figure}

\Cref{fig:SU2-l34} shows similar bounds on $\bar l_3$ and $\bar l_4$, which are the only NLO LECs other than $\bar l_1,\bar l_2$ that appear in the NNLO amplitude.
The bounds on $\bar l_3$ are extremely weak, since $\bar l_3$ figures in the amplitude with much smaller prefactors than the other LECs.
It can be partly understood by noting that the $\bar l_3$ term in the Lagrangian, unlike the other terms, does not contain the field $u_\mu$.
Interestingly, the upper bounds on $\bar l_4$ become \emph{weaker} as $\lambda$ is increased. This does not necessarily contradict the arguments made in \cref{sec:int}, due to the complicated NNLO situation where both $\v\alpha$ and $c$ in a constraint may depend on $\lambda$. 
Nevertheless, it is surprising to see, and does not seem to appear in other bounds, such as those on $\bar l_{1,2}$.

\Cref{fig:SU2-l124} combines the $\bar l_4$ bounds with the $\bar l_1,\bar l_2$ bounds to form a summary of the effective bounds on the two-flavour NLO LECs obtained in this paper.
Unfortunately, only the $k=2$, $\lambda\approx4$ constraints have a shape that is sensible to show in three dimensions.
\begin{figure}[hbtp]
    \hspace{-1.7cm}
    \begin{tikzpicture}
        \begin{axis}[
                view={170}{50}, xlabel={$10^{3} \Theta_3$}, ylabel={$10^{3} \Theta_2$}, zlabel={$10^{3} \Theta_1$},
                colormap/temp, point meta min=0, point meta max = 3500,
                xmin=-4,xmax=5,
                ymin=-20,ymax=15,
                zmin=-5,zmax=50
            ]
                \addplot3 [patch, patch table with point meta={\tresultpath/onlyNNLO/SU2/M0.135/D2/lam4/visualisation/T123/constr_0_table.dat}, opacity=0] 
                            table {\tresultpath/onlyNNLO/SU2/M0.135/D2/lam4/visualisation/T123/constr_0_coords.dat};
                            
                \coordinate (p0) at (-0.13158, -10.9478, 18.0256);
                \coordinate (p1) at (-0.126158, -10.8942, 17.9463);
                \coordinate (p2) at (-0.136994, -10.9913, 18.0972);
                \coordinate (p3) at (-0.126111, -10.8938, 17.9457);
                \coordinate (p4) at (-0.136994, -10.9913, 18.0972);
                \coordinate (p5) at (-0.126142, 2.0662, 4.98467);
                \coordinate (p6) at (-0.137, 2.12166, 4.98469);
                \coordinate (p7) at (-0.149777, 2.21364, 4.98493);
                \coordinate (p8) at (-0.149759, -11.1104, 18.3081);
                \coordinate (p9) at (-0.153215, 2.24708, 4.98482);
                \coordinate (p10) at (-0.153206, -11.1481, 18.3796);
                \coordinate (p11) at (-0.153207, -11.1482, 18.3796);
                \coordinate (p12) at (-0.110531, 2.01009, 4.98464);
                \coordinate (p13) at (-0.087871, 1.96263, 4.98453);
                \coordinate (p14) at (-0.14452, 2.17188, 4.98504);
                \coordinate (p15) at (1.0277, 1.29342, 4.98461);
                \coordinate (p16) at (-1.30212, 15.0005, 4.98484);
                \coordinate (p17) at (2.87084, 15, 4.98501);
                \coordinate (p18) at (1.45594, 3.5424, 4.98451);
                \coordinate (p19) at (1.0277, 1.29342, 4.98462);
                \coordinate (p20) at (-4.00028, -0.0688701, 49.9998);
                \coordinate (p21) at (-2.20439, -20.0012, 50.0004);
                \coordinate (p22) at (-0.947675, -20.001, 36.0512);
                \coordinate (p23) at (-0.174679, -11.4179, 18.8894);
                \coordinate (p24) at (-1.30212, 15.0005, 4.98485);
                \coordinate (p25) at (-3.99843, 14.9998, 34.9308);
                \coordinate (p26) at (-0.0778782, -10.5671, 17.5083);
                \coordinate (p27) at (-0.0604053, -10.4523, 17.3833);
                \coordinate (p28) at (-0.0878734, -10.6183, 17.5665);
                \coordinate (p29) at (0.0341688, -9.68659, 16.5606);
                \coordinate (p30) at (0.0662538, -9.37752, 16.2335);
                \coordinate (p31) at (-0.0878948, -10.6184, 17.5666);
                \coordinate (p32) at (-0.11053, -10.7685, 17.7626);
                \coordinate (p33) at (-0.14455, -11.0602, 18.2163);
                \coordinate (p34) at (-0.141935, -11.0375, 18.1768);
                \coordinate (p35) at (1.45594, 3.5424, 4.98451);
                \coordinate (p36) at (3.25486, -2.39268, 35.7427);
                \coordinate (p37) at (0.0662549, -9.37768, 16.2338);
                \coordinate (p38) at (0.0662682, -19.9994, 37.4755);
                \coordinate (p39) at (3.80087, -6.65177, 50.0021);
                \coordinate (p40) at (1.25886, -20.0006, 49.9988);
                \coordinate (p41) at (1.0277, 1.29342, 4.98462);
                \coordinate (p42) at (-0.0604064, -10.4525, 17.3836);
                \coordinate (p43) at (-0.0779136, -19.9994, 36.3749);
                \coordinate (p44) at (-0.0778807, -10.5675, 17.5088);
                \coordinate (p45) at (-0.0604056, -20.0021, 36.4805);
                \coordinate (p46) at (-0.174684, -19.9999, 36.05);
                \coordinate (p47) at (-0.947668, -20.0008, 36.051);
                \coordinate (p48) at (-0.141955, -11.0374, 18.1767);
                \coordinate (p49) at (-0.144546, -11.0599, 18.2158);
                \coordinate (p50) at (-0.153288, -11.1494, 18.3816);
                \coordinate (p51) at (-0.141973, -20, 36.0998);
                \coordinate (p52) at (-0.149482, -11.1073, 18.3031);
                \coordinate (p53) at (-0.110328, -10.7669, 17.76);
                \coordinate (p54) at (-0.131583, -10.9481, 18.026);
                \coordinate (p55) at (-0.0883002, -10.6203, 17.5697);
                \coordinate (p56) at (-0.12588, -10.892, 17.9427);
                \coordinate (p57) at (-0.131535, -20.0008, 36.132);
                \coordinate (p58) at (-0.0603403, -10.4516, 17.3821);
                \coordinate (p59) at (0.0341688, -9.6866, 16.5607);
                \coordinate (p60) at (0.0341302, -9.68654, 16.5605);
                \coordinate (p61) at (0.0341982, -20.0021, 37.1877);
                \coordinate (p62) at (0.0662361, -9.37767, 16.2338);
                \coordinate (p63) at (4.99988, 4.856, 49.9988);
                \coordinate (p64) at (5.00155, 15.0008, 34.7854);
                \coordinate (p65) at (3.25468, 15.0005, 9.65344);
                \coordinate (p66) at (-0.141929, -11.037, 18.176);
                \coordinate (p67) at (-0.136987, -10.9907, 18.0963);
                \coordinate (p68) at (1.45593, 3.54302, 4.98449);
                \coordinate (p69) at (2.87084, 15, 4.985);
                \coordinate (p70) at (2.87081, 14.9999, 4.98496);
                
                \draw[plotIII, facet]
                        (p0)-- (p2)-- (p4)-- (p6)-- (p5)-- (p3)-- (p1)
                        -- cycle;
                \draw[plotIII, facet]
                        (p7)-- (p8)-- (p10)-- (p11)-- (p9)
                        -- cycle;
                \draw[plotIII, facet]
                        (p9)-- (p7)-- (p14)-- (p6)-- (p5)-- (p12)-- (p13)-- (p15)-- (p19)-- (p18)-- (p17)-- (p16)
                        -- cycle;
                \draw[plotIII, facet]
                        (p11)-- (p9)-- (p16)-- (p24)-- (p25)-- (p20)-- (p21)-- (p22)-- (p23)
                        -- cycle;
                \draw[plotIII, facet]
                        (p26)-- (p28)-- (p13)-- (p15)-- (p30)-- (p29)-- (p27)
                        -- cycle;
                \draw[plotIII, facet]
                        (p28)-- (p31)-- (p32)-- (p12)-- (p13)
                        -- cycle;
                \draw[plotIII, facet]
                        (p3)-- (p5)-- (p12)-- (p32)
                        -- cycle;
                \draw[plotIII, facet]
                        (p33)-- (p8)-- (p7)-- (p14)
                        -- cycle;
                \draw[plotIII, facet]
                        (p33)-- (p34)-- (p4)-- (p6)-- (p14)
                        -- cycle;
                \draw[color of colormap={2275.89/3.5}, facet]
                        (p35)-- (p36)-- (p39)-- (p40)-- (p38)-- (p51)-- (p56)-- (p50)-- (p37)-- (p41)
                        -- cycle;
                \draw[color of colormap={547.925/3.5}, facet]
                        (p48)-- (p49)-- (p52)-- (p50)-- (p22)-- (p46)-- (p51)
                        -- cycle;
                \draw[color of colormap={921.964/3.5}, facet]
                        (p63)-- (p64)-- (p65)-- (p36)-- (p39)
                        -- cycle;
                \draw[color of colormap={1409.79/3.5}, facet]
                        (p68)-- (p70)-- (p69)-- (p65)-- (p36)-- (p35)
                        -- cycle;

                \def\xmin{-4}\def\xmax{5}
                \def\ymin{-20}\def\ymax{15}
                \def\zmin{-5}\def\zmax{50}
                
                \draw[grey, fill, opacity=.5]
                    (p25) -- (p24) -- (p16) -- (p17) -- (p65) -- (p64) -- (\xmax,\ymax,\zmin) -- (\xmin,\ymax,\zmin) -- cycle;
                \draw[grey, fill, opacity=.5]
                    (p64) -- (p63) -- (\xmax,\ymin,\zmax) -- (\xmax,\ymin,\zmin) -- (\xmax,\ymax,\zmin) -- cycle;
                \draw[grey, fill, opacity=.5]
                    (p63) -- (p39) -- (p40) -- (\xmax,\ymin,\zmax) -- cycle;
                \draw[grey, fill, opacity=.5]
                    (p21) -- (p20) -- (\xmin,\ymin,\zmax) -- cycle;

                \def\xfid{-0.16} \def\yfid{0.6776} \def\zfid{0.344}
                    
                \draw[axes to fid] (\xfid,\ymax,\zfid) -- (\xfid,\yfid,\zfid);
                \draw[axes to fid] (\xfid,\ymax,\zmin) -- (\xfid,\ymax,\zfid) -- (\xmin,\ymax,\zfid);
                \draw[axes to fid] (\xfid,\yfid,\zmax) -- (\xfid,\yfid,\zfid);
                \draw[axes to fid] (\xfid,\ymin,\zmax) -- (\xfid,\yfid,\zmax) -- (\xmin,\yfid,\zmax);
                    
                \filldraw [black] (\xfid,\yfid,\zfid)circle[radius=1pt];
                    
        \end{axis}
    \end{tikzpicture}
    \begin{tikzpicture} 
        \begin{axis}[
                view={170}{50}, xlabel={$10^{3} \Theta_3$},
                colormap/temp, colorbar, point meta min=0, point meta max = 3500,
                colorbar style={
                    ymin=0, ytick={1000,2000,3000}, yticklabels={1,2,3},ylabel={$10^3/\rho\left(\constr{\v\alpha_i}{c_i}\right)$}
                },
                xmin=-4,xmax=5,
                ymin=-20,ymax=15,
                zmin=-5,zmax=50
            ]
            
                \addplot3 [patch, patch table with point meta={\tresultpath/onlyNNLO/SU2/M0.135/D2/lam8/visualisation/T123/constr_0_table.dat}, opacity=0] 
                            table {\tresultpath/onlyNNLO/SU2/M0.135/D2/lam8/visualisation/T123/constr_0_coords.dat};
                            
                \coordinate (p0) at (-0.153372, -0.0736698, 0.4821);
                \coordinate (p1) at (-0.1534, 1.07342, -0.664988);
                \coordinate (p2) at (-0.153398, 1.0734, -0.664979);
                \coordinate (p3) at (-0.100274, 1.04154, -0.665019);
                \coordinate (p4) at (-0.100274, 1.04154, -0.665019);
                \coordinate (p5) at (-0.167118, -0.452086, 0.91357);
                \coordinate (p6) at (0.699759, 6.32277, -0.665219);
                \coordinate (p7) at (0.795536, -19.9975, 39.7634);
                \coordinate (p8) at (-0.166423, -0.425889, 0.881467);
                \coordinate (p9) at (-0.158674, -0.200741, 0.620205);
                \coordinate (p10) at (-0.167171, -0.455127, 0.917414);
                \coordinate (p11) at (-0.266806, -4.63906, 6.20686);
                \coordinate (p12) at (-0.15869, -0.200762, 0.620267);
                \coordinate (p13) at (-0.162385, -0.301434, 0.73438);
                \coordinate (p14) at (0.509018, -19.9995, 36.929);
                \coordinate (p15) at (0.795653, 6.03605, 0.712287);
                \coordinate (p16) at (-0.164852, -0.376294, 0.821965);
                \coordinate (p17) at (-0.158711, 1.08444, -0.66497);
                \coordinate (p18) at (0.509041, -20.0005, 36.9307);
                \coordinate (p19) at (-1.6505, -20.0018, 36.9271);
                \coordinate (p20) at (4.56357, 15.0023, 50.0007);
                \coordinate (p21) at (1.41032, -20.002, 50.0022);
                \coordinate (p22) at (0.795701, 6.03642, 0.711718);
                \coordinate (p23) at (1.64971, 15.0005, 1.47989);
                \coordinate (p24) at (0.795571, -19.9983, 39.7652);
                \coordinate (p25) at (-0.167081, 1.12644, -0.665026);
                \coordinate (p26) at (-0.164948, 1.11066, -0.665013);
                \coordinate (p27) at (0.699776, 6.32293, -0.665235);
                \coordinate (p28) at (-0.16717, 1.12731, -0.664986);
                \coordinate (p29) at (-0.166442, 1.12069, -0.664981);
                \coordinate (p30) at (-0.162379, 1.09796, -0.66504);
                \coordinate (p31) at (1.46105, 14.9997, -0.664975);
                \coordinate (p32) at (-1.41705, 15, -0.664941);
                \coordinate (p33) at (-0.158713, 1.08445, -0.664977);
                \coordinate (p34) at (-0.1534, 1.07342, -0.664989);
                \coordinate (p35) at (-0.100274, 1.04154, -0.665019);
                \coordinate (p36) at (-0.167176, -0.455142, 0.917443);
                \coordinate (p37) at (-2.82841, -20.0011, 50.0001);
                \coordinate (p38) at (-3.99994, -6.99407, 50.0018);
                \coordinate (p39) at (-3.99944, 14.9987, 28.0057);
                \coordinate (p40) at (-0.167238, -0.455414, 0.917783);
                \coordinate (p41) at (-0.266796, -4.63888, 6.20662);
                \coordinate (p42) at (-0.166397, -0.426176, 0.881757);
                \coordinate (p43) at (-0.16637, -0.426107, 0.881713);
                \coordinate (p44) at (-0.164937, -0.376277, 0.821929);
                \coordinate (p45) at (-0.164854, -0.376298, 0.821974);
                \coordinate (p46) at (-0.162383, -0.30143, 0.734371);
                \coordinate (p47) at (-0.158693, -0.200765, 0.620278);
                \coordinate (p48) at (-0.167078, -0.451874, 0.913352);
                \coordinate (p49) at (0.795665, 6.03615, 0.711072);
                \coordinate (p50) at (0.699761, 6.3228, -0.665221);
                \coordinate (p51) at (-0.162396, -0.301534, 0.734427);
                \coordinate (p52) at (-0.167158, -0.451812, 0.913226);
                
                \draw[color of colormap={3500/3.5}, facet]
                        (p0)-- (p2)-- (p1)-- (p3)-- (p4)
                        -- cycle;
                \draw[color of colormap={3500/3.5}, facet]
                        (p5)-- (p10)-- (p11)-- (p14)-- (p7)-- (p15)-- (p6)-- (p4)-- (p0)-- (p9)-- (p12)-- (p13)-- (p16)-- (p8)
                        -- cycle;
                \draw[color of colormap={3500/3.5}, facet]
                        (p0)-- (p2)-- (p17)-- (p9)
                        -- cycle;
                \draw[color of colormap={468.756/3.5}, facet]
                        (p18)-- (p19)-- (p11)
                        -- cycle;
                \draw[color of colormap={2445.17/3.5}, facet]
                        (p20)-- (p21)-- (p24)-- (p22)-- (p23)
                        -- cycle;
                \draw[color of colormap={3500/3.5}, facet]
                        (p25)-- (p28)-- (p32)-- (p31)-- (p27)-- (p35)-- (p34)-- (p33)-- (p7)-- (p30)-- (p26)-- (p29)
                        -- cycle;
                \draw[color of colormap={3500/3.5}, facet]
                        (p36)-- (p28)-- (p32)-- (p39)-- (p38)-- (p37)-- (p19)-- (p11)-- (p41)-- (p40)
                        -- cycle;
                \draw[color of colormap={3500/3.5}, facet]
                        (p29)-- (p42)-- (p43)-- (p45)-- (p44)-- (p26)
                        -- cycle;
                \draw[color of colormap={3500/3.5}, facet]
                        (p30)-- (p46)-- (p47)-- (p7)
                        -- cycle;
                \draw[color of colormap={2603.39/3.5}, facet]
                        (p22)-- (p23)-- (p31)-- (p27)-- (p50)-- (p49)
                        -- cycle;
                \draw[color of colormap={3500/3.5}, facet]
                        (p26)-- (p44)-- (p51)-- (p46)-- (p30)
                        -- cycle;
                \draw[color of colormap={3500/3.5}, facet]
                        (p25)-- (p48)-- (p52)-- (p42)-- (p29)
                        -- cycle;

                \def\xmin{-4}\def\xmax{5}
                \def\ymin{-20}\def\ymax{15}
                \def\zmin{-5}\def\zmax{50}

                \draw[grey, fill, opacity=.5]
                    (p39) -- (p32) -- (p31) -- (p23) -- (p20) -- (\xmax,\ymax,\zmax) -- (\xmax,\ymax,\zmin) -- (\xmin,\ymax,\zmin) -- cycle;
                \draw[grey, fill, opacity=.5]
                    (p20) -- (p21) -- (\xmax,\ymin,\zmax) -- (\xmax,\ymax,\zmax) -- cycle;
                \draw[grey, fill, opacity=.5]
                    (p37) -- (p38) -- (\xmin,\ymin,\zmax) -- cycle;
                \draw[grey, fill, opacity=.5]
                    (\xmax,\ymax,\zmax) -- (\xmax,\ymin,\zmax) -- (\xmax,\ymin,\zmin) -- (\xmax,\ymax,\zmin) -- cycle;

                \def\xfid{-0.16} \def\yfid{0.6776} \def\zfid{0.344}
                    
                \draw[axes to fid] (\xfid,\ymax,\zfid) -- (\xfid,\yfid,\zfid);
                \draw[axes to fid] (\xfid,\ymax,\zmin) -- (\xfid,\ymax,\zfid) -- (\xmin,\ymax,\zfid);
                \draw[axes to fid] (\xfid,\yfid,\zmax) -- (\xfid,\yfid,\zfid);
                \draw[axes to fid] (\xfid,\ymin,\zmax) -- (\xfid,\yfid,\zmax) -- (\xmin,\yfid,\zmax);
                    
                \filldraw [black] (\xfid,\yfid,\zfid)circle[radius=1pt];

        \end{axis}
    \end{tikzpicture}
    
    \hspace{-1.7cm}
    \begin{tikzpicture} 
        \begin{axis}[
                view={170}{50}, xlabel={$10^{3} \Theta_4$}, ylabel={$10^{3} \Theta_2$}, zlabel={$10^{3} \Theta_1$},
                colormap/temp, point meta min=0, point meta max = 3500,
                xmin=-3,xmax=5,
                ymin=-7,ymax=10,
                zmin=-3,zmax=10
            ]
                \addplot3 [patch, patch table with point meta={\tresultpath/onlyNNLO/SU2/M0.135/D2/lam4/visualisation/T124/constr_0_table.dat}, opacity=0] 
                            table {\tresultpath/onlyNNLO/SU2/M0.135/D2/lam4/visualisation/T124/constr_0_coords.dat};
                            
                \coordinate (p0) at (-0.239174, 2.26235, 5.06613);
                \coordinate (p1) at (-0.250551, 2.2563, 5.10544);
                \coordinate (p2) at (-0.250561, -2.63803, 9.99969);
                \coordinate (p3) at (-0.250559, -2.63801, 9.99959);
                \coordinate (p4) at (-0.239168, -2.67226, 9.9995);
                \coordinate (p5) at (-0.23917, -2.67228, 9.99957);
                \coordinate (p6) at (-0.26678, 2.26427, 5.16234);
                \coordinate (p7) at (-0.266801, -2.57298, 9.99962);
                \coordinate (p8) at (-0.231377, 2.27394, 5.03862);
                \coordinate (p9) at (-0.231384, -2.68774, 10);
                \coordinate (p10) at (-0.231385, -2.68776, 10.0001);
                \coordinate (p11) at (-0.324139, 2.38431, 5.36352);
                \coordinate (p12) at (0.311221, 10.0004, 3.13947);
                \coordinate (p13) at (0.311221, 10.0004, 3.13947);
                \coordinate (p14) at (-0.226212, 2.28642, 5.02062);
                \coordinate (p15) at (0.311225, 4.16761, 3.13943);
                \coordinate (p16) at (-0.290254, 2.2999, 5.24487);
                \coordinate (p17) at (-1.64875, 9.99927, 9.99927);
                \coordinate (p18) at (-1.64844, 7.0216, 10.0004);
                \coordinate (p19) at (-0.226232, -2.6922, 10.0002);
                \coordinate (p20) at (-0.226234, -2.69326, 10.0003);
                \coordinate (p21) at (5.00021, 4.16768, 3.13931);
                \coordinate (p22) at (5.00022, -2.69331, 10.0004);
                \coordinate (p23) at (-0.324178, -2.25212, 10.0002);
                \coordinate (p24) at (-0.324179, -2.25213, 10.0002);
                \coordinate (p25) at (-1.64845, 7.02162, 10.0004);
                \coordinate (p26) at (-0.266803, -2.573, 9.99969);
                \coordinate (p27) at (-0.290184, -2.4558, 10.0004);
                \coordinate (p28) at (-0.290189, -2.45584, 10.0006);
                \coordinate (p29) at (5.00046, 9.99976, 3.1401);
                
                \draw[plotIII, facet]
                        (p0)-- (p4)-- (p5)-- (p2)-- (p3)-- (p1)
                        -- cycle;
                \draw[plotIII, facet]
                        (p1)-- (p3)-- (p7)-- (p6)
                        -- cycle;
                \draw[plotIII, facet]
                        (p0)-- (p4)-- (p10)-- (p9)-- (p8)
                        -- cycle;
                \draw[plotIII, facet]
                        (p1)-- (p0)-- (p8)-- (p14)-- (p15)-- (p12)-- (p13)-- (p17)-- (p18)-- (p11)-- (p16)-- (p6)
                        -- cycle;
                \draw[plotIII, facet]
                        (p19)-- (p14)-- (p15)-- (p21)-- (p22)-- (p20)
                        -- cycle;
                \draw[plotIII, facet]
                        (p23)-- (p11)-- (p18)-- (p25)-- (p24)
                        -- cycle;
                \draw[plotIII, facet]
                        (p16)-- (p27)-- (p28)-- (p26)-- (p7)-- (p6)
                        -- cycle;
                \draw[plotIII, facet]
                        (p9)-- (p8)-- (p14)-- (p19)
                        -- cycle;
                \draw[plotIII, facet]
                        (p16)-- (p27)-- (p23)-- (p11)
                        -- cycle;
                \draw[plotIII, facet]
                        (p12)-- (p15)-- (p21)-- (p29)
                        -- cycle;

                \def\xmin{-3}\def\xmax{5}
                \def\ymin{-7}\def\ymax{10}
                \def\zmin{-3}\def\zmax{10}

                \draw[grey, fill, opacity=.5]
                    (p29) -- (p13) -- (p17) -- (\xmin,\ymax,\zmax) -- (\xmin,\ymax,\zmin) -- (\xmax,\ymax,\zmin) -- cycle;
                \draw[grey, fill, opacity=.5]
                    (p29) -- (p21) -- (p22) -- (\xmax,\ymin,\zmax) -- (\xmax,\ymin,\zmin) -- (\xmax,\ymax,\zmin) -- cycle;
                \draw[grey, fill, opacity=.5]
                    (p22) -- (p20) -- (p26) -- (p27) -- (p23) -- (p18) -- (p17) -- (\xmin,\ymax,\zmax) -- (\xmin,\ymin,\zmax) -- (\xmax,\ymin,\zmax) -- cycle;

                \def\xfid{-0.216} \def\yfid{0.6776} \def\zfid{0.344}
                    
                \draw[axes to fid] (\xfid,\ymax,\zfid) -- (\xfid,\yfid,\zfid);
                \draw[axes to fid] (\xfid,\ymax,\zmin) -- (\xfid,\ymax,\zfid) -- (\xmin,\ymax,\zfid);
                \draw[axes to fid] (\xfid,\yfid,\zmax) -- (\xfid,\yfid,\zfid);
                \draw[axes to fid] (\xfid,\ymin,\zmax) -- (\xfid,\yfid,\zmax) -- (\xmin,\yfid,\zmax);
                    
                \filldraw [black] (\xfid,\yfid,\zfid)circle[radius=1pt];
                
        \end{axis}
    \end{tikzpicture}
    \begin{tikzpicture} 
        \begin{axis}[
                view={170}{50}, xlabel={$10^{3} \Theta_4$},
                colormap/temp, point meta min=0, point meta max = 3500,
                xmin=-3,xmax=5,
                ymin=-7,ymax=10,
                zmin=-3,zmax=10
            ]
                \addplot3 [patch, patch table with point meta={\tresultpath/onlyNNLO/SU2/M0.135/D2/lam8/visualisation/T124/constr_0_table.dat}, opacity=0] 
                            table {\tresultpath/onlyNNLO/SU2/M0.135/D2/lam8/visualisation/T124/constr_0_coords.dat};
                            
                \coordinate (p0) at (-0.205196, -0.341717, 0.723928);
                \coordinate (p1) at (-0.205304, -0.339894, 0.722233);
                \coordinate (p2) at (-0.205191, -0.341709, 0.723911);
                \coordinate (p3) at (-0.205231, 1.08493, -0.702734);
                \coordinate (p4) at (-0.205287, 1.08466, -0.702348);
                \coordinate (p5) at (5.00019, 2.75672, -2.37456);
                \coordinate (p6) at (0.272447, 2.75695, -2.37456);
                \coordinate (p7) at (-0.0273845, -3.45325, 3.83568);
                \coordinate (p8) at (-0.205126, 1.08506, -0.70282);
                \coordinate (p9) at (4.99988, -3.4535, 3.83578);
                \coordinate (p10) at (-2.76477, 9.99959, 8.25719);
                \coordinate (p11) at (-2.89427, 9.16265, 10.0003);
                \coordinate (p12) at (-0.225959, -0.177945, 0.659881);
                \coordinate (p13) at (-3.01487, 10.0004, 10.0004);
                \coordinate (p14) at (-0.225871, 1.11209, -0.630232);
                \coordinate (p15) at (5.0005, 9.99986, -2.3751);
                \coordinate (p16) at (0.272464, 9.99995, -2.37507);
                \coordinate (p17) at (5.00017, 2.75671, -2.37455);
                \coordinate (p18) at (0.272446, 2.75693, -2.37455);
                \coordinate (p19) at (-0.0273848, -3.45329, 3.83573);
                \coordinate (p20) at (-0.203474, -6.53555, 10.0005);
                \coordinate (p21) at (-0.0273846, -3.45327, 3.83571);
                \coordinate (p22) at (-0.217978, -0.22168, 0.655894);
                \coordinate (p23) at (-0.205309, -0.339599, 0.721958);
                \coordinate (p24) at (-0.212417, -0.263607, 0.669902);
                \coordinate (p25) at (-0.205184, -0.341615, 0.723887);
                \coordinate (p26) at (-0.206384, -0.32516, 0.709667);
                \coordinate (p27) at (-0.205344, -0.33974, 0.722082);
                \coordinate (p28) at (-0.208656, -0.299162, 0.690471);
                \coordinate (p29) at (-0.217984, 1.09225, -0.658093);
                \coordinate (p30) at (-0.212381, 1.08393, -0.677571);
                \coordinate (p31) at (-0.206371, 1.08316, -0.698608);
                \coordinate (p32) at (-0.205303, 1.08474, -0.702399);
                \coordinate (p33) at (4.99991, -3.45351, 3.8358);
                \coordinate (p34) at (4.9999, -3.45351, 3.8358);
                \coordinate (p35) at (5.00015, -6.53607, 10.0003);
                
                \draw[color of colormap={3500/3.5}, facet]
                        (p0)-- (p1)-- (p4)-- (p3)-- (p2)
                        -- cycle;
                \draw[color of colormap={3500/3.5}, facet]
                        (p5)-- (p6)-- (p8)-- (p3)-- (p2)-- (p7)-- (p9)
                        -- cycle;
                \draw[color of colormap={3500/3.5}, facet]
                        (p10)-- (p13)-- (p11)-- (p12)-- (p14)
                        -- cycle;
                \draw[color of colormap={367.842/3.5}, facet]
                        (p15)-- (p16)-- (p18)-- (p17)
                        -- cycle;
                \draw[color of colormap={3500/3.5}, facet]
                        (p19)-- (p20)-- (p11)-- (p12)-- (p22)-- (p24)-- (p28)-- (p26)-- (p23)-- (p27)-- (p25)-- (p21)
                        -- cycle;
                \draw[color of colormap={3500/3.5}, facet]
                        (p14)-- (p12)-- (p22)-- (p29)
                        -- cycle;
                \draw[color of colormap={3500/3.5}, facet]
                        (p16)-- (p10)-- (p14)-- (p29)-- (p30)-- (p2)-- (p31)-- (p32)-- (p6)-- (p18)
                        -- cycle;
                \draw[color of colormap={468.756/3.5}, facet]
                        (p33)-- (p34)-- (p19)-- (p20)-- (p35)
                        -- cycle;
                \draw[color of colormap={3500/3.5}, facet]
                        (p24)-- (p28)-- (p2)-- (p30)
                        -- cycle;
                \draw[color of colormap={3500/3.5}, facet]
                        (p26)-- (p28)-- (p2)-- (p31)
                        -- cycle;
                \draw[color of colormap={3500/3.5}, facet]
                        (p23)-- (p26)-- (p31)-- (p32)
                        -- cycle;
                \draw[color of colormap={3500/3.5}, facet]
                        (p22)-- (p24)-- (p30)-- (p29)
                        -- cycle;

                \def\xmin{-3}\def\xmax{5}
                \def\ymin{-7}\def\ymax{10}
                \def\zmin{-3}\def\zmax{10}

                \draw[grey, fill, opacity=.5]
                    (p15) -- (p16) -- (p10) -- (p13) -- (\xmin,\ymax,\zmax) -- (\xmin,\ymax,\zmin) -- (\xmax,\ymax,\zmin) -- cycle;
                \draw[grey, fill, opacity=.5]
                    (p15) -- (p5) -- (p9) -- (p35) -- (\xmax,\ymin,\zmax) -- (\xmax,\ymin,\zmin) -- (\xmax,\ymax,\zmin) -- cycle;
                \draw[grey, fill, opacity=.5]
                    (p35) -- (p20) -- (p11) -- (p13) -- (\xmin,\ymax,\zmax) -- (\xmin,\ymin,\zmax) -- (\xmax,\ymin,\zmax) -- cycle;
                    
                \def\xfid{-0.216} \def\yfid{0.6776} \def\zfid{0.344}
                    
                \draw[axes to fid] (\xfid,\ymax,\zfid) -- (\xfid,\yfid,\zfid);
                \draw[axes to fid] (\xfid,\ymax,\zmin) -- (\xfid,\ymax,\zfid) -- (\xmin,\ymax,\zfid);
                \draw[axes to fid] (\xfid,\yfid,\zmax) -- (\xfid,\yfid,\zfid);
                \draw[axes to fid] (\xfid,\ymin,\zmax) -- (\xfid,\yfid,\zmax) -- (\xmin,\yfid,\zmax);
                    
                \filldraw [black] (\xfid,\yfid,\zfid)circle[radius=1pt];

        \end{axis}
    \end{tikzpicture}
    \caption[NNLO two-derivative bounds on $\Theta_i$ at $\lambda=4$ and $\lambda=8$ for $i=1,2,3,4$.]{
        NNLO two-derivative bounds on $\Theta_i$ at $\lambda=4$ (\textbf{left}) and $\lambda=8$ (\textbf{right}) for $i=1,2,3$ (\textbf{top}) and $1,2,4$ (\textbf{bottom}, note different $\Theta_{1,2}$-axes).
        These specific $i$ choices are used, for if either $\Theta_1$ or $\Theta_2$ is fixed to its reference value, the remaining constraint is not satisfied by any value of the other $\Theta_i$.
        Note the rather different scales on the axes; several facets in the top plots (especially for $\lambda=8$) are in fact nearly perpendicular to the $\Theta_3$ axis, placing upper and lower bounds on $\Theta_3$ that do not depend very strongly on the other parameters.
        
        The \selfcolour[plotIII]{green} facets exclude the reference point.
        The exclusion is by a rather small amount (no orthogonal arrow can be sensibly drawn), and varying the $L_i^r$ within their uncertainties is sufficient to remedy this.
        Note how increasing $\lambda$ to 8 slightly weakens these constraints so that they no longer exclude the reference point.
        }
    \label{fig:SU2-Thetas}
\end{figure}
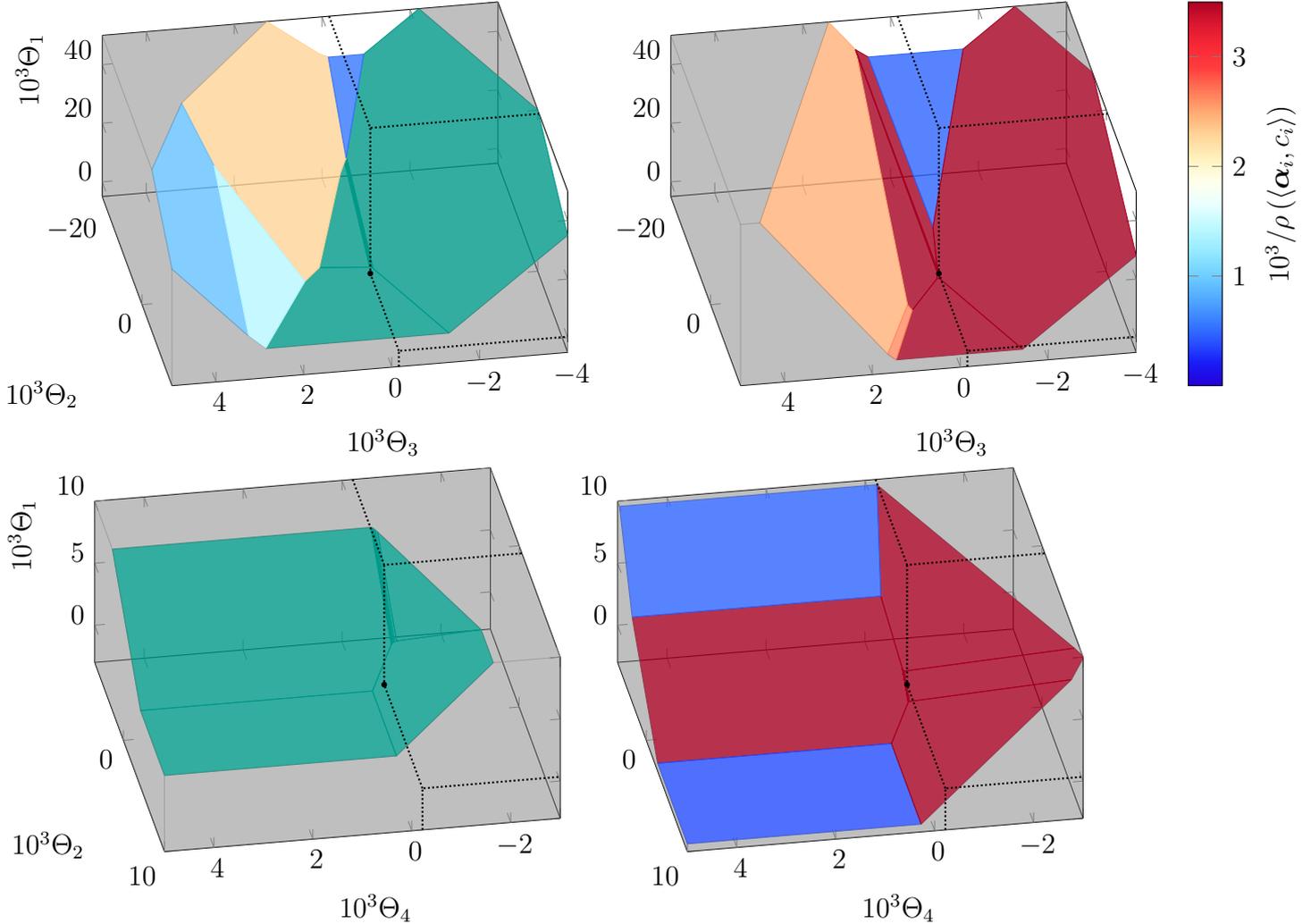
    
Lastly, \cref{fig:SU2-Thetas} shows  bounds on the four NNLO parameters $\Theta_i$, using only two derivatives.
Here, the bounds are not entirely consistent with the reference point, although not too much meaning should be read into this, as the reference values for the NNLO parameters are little more than educated guesses.
It also showcases the phenomenon where some constraints become weaker at larger $\lambda$, at least at the particular values at which we have fixed the NLO LECs.

The NNLO parameter bounds are not particularly strong compared to the magnitude of the reference values, but some are still notable.
$\Theta_4$ has a strict lower bound, with no values of $\Theta_{1,2,3}$ being permitted if $\Theta_4\lesssim -2.5$.
$\Theta_3$ is bounded from both above and below, and the bounds are fairly independent of the values of the other parameters in a large part of parameter space.
Thus, we may write down the tentative single-parameter bounds
\begin{equation}\label{eq:Theta-bound}
    -1.5\lesssim 10^3\Theta_3 \lesssim 1,\qquad -2.5 \lesssim 10^3\Theta_4,
\end{equation}
both of which are satisfied, with a margin of about an order of magnitude, by the reference values in \cref{sec:LEC-details}.

\FloatBarrier

\subsection{Three flavours}\label{sec:results-3flav}
Three-flavour \chpt\ bounds cover 3 parameters at NLO and 12 at NNLO, and three choices for $M_\text{phys}$ (namely $M_\pi,M_K$ and $M_\eta$) present themselves, with no \emph{a priori} indication of which to choose.
This would of course be resolved by working with inequal-mass mesons, but this NLO amplitude is far more complicated (see \cite{GomezNicola:2001as} and references therein) and its NNLO counterpart is so far undetermined in a simple analytic fashion; furthermore, inequal masses have implications for the construction of bounds that we do not address here (see \cite{Mateu:2008gv}).

\begin{figure}[hbtp]
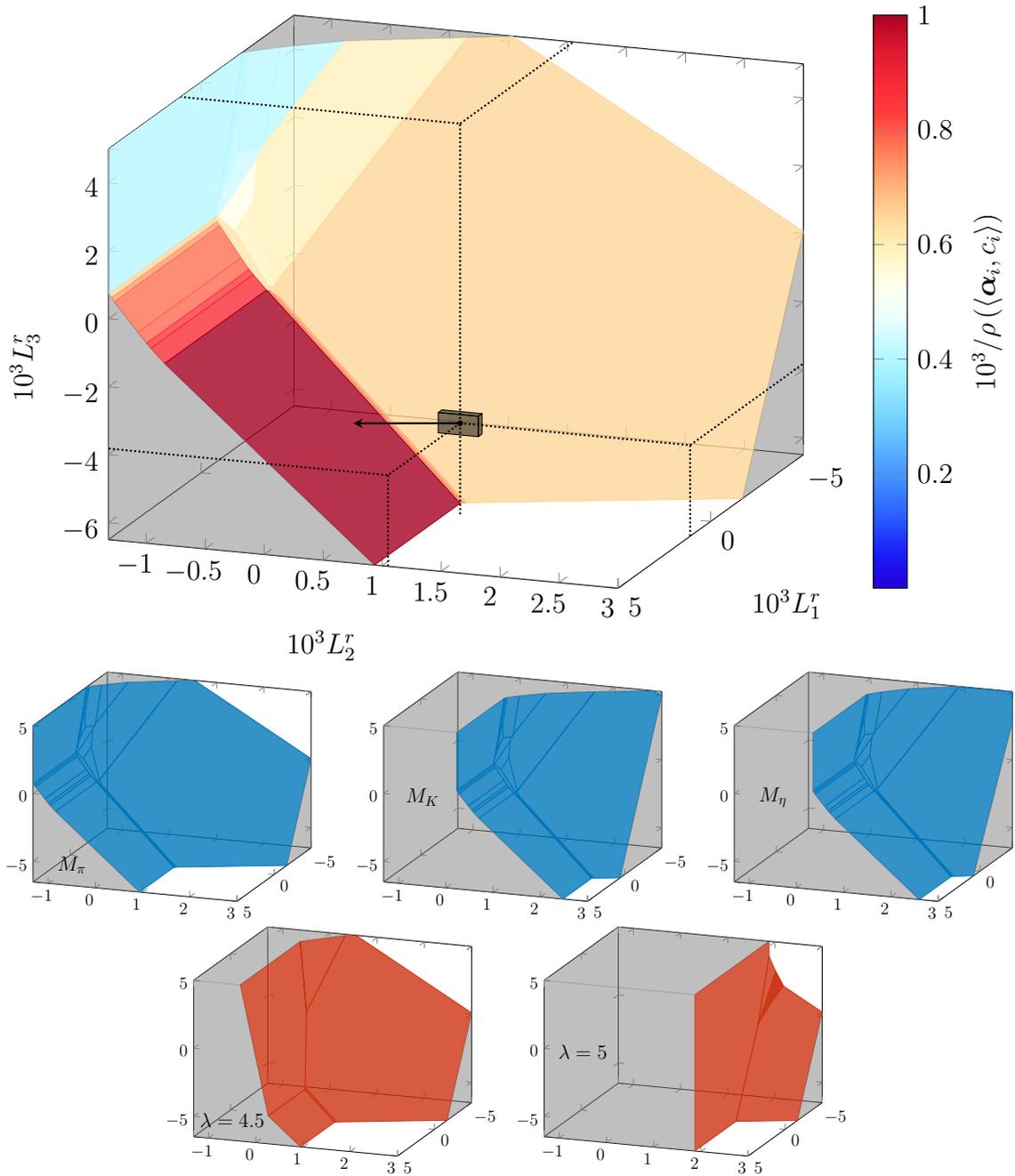

    \centering

    \caption[Two-derivative constraints on the three-flavour NLO LECs $L_i^r$.]{
        Two-derivative constraints on the three-flavour NLO LECs $L_i^r$, visualised similarly to previous figures.
        \textbf{Top:} detailed constraints with $\Mphys=M_\pi, \lambda=4$.
        The visualisation is similar to \cref{fig:SU2-l124}; note the different scales on the axes.
        \textbf{Middle:} Constraints with (from left to right) $\Mphys=M_\pi,M_K$, and $M_\eta$, all with $\lambda=4$.
        Note that the leftmost figure is just a less detailed version of the top figure, and that the $M_K$ and $M_\eta$ surfaces only differ from each other by a very small amount.
        Both these higher-mass constraints exclude the reference point (not drawn), although not by much.
        \textbf{Bottom:} the effect of setting $\lambda=4.5$ (left) or $5$ (right) with $\Mphys=M_\pi$, for comparison with the $\lambda=4$ figure above.
        The $\lambda=5$ constraint excludes the reference point (not drawn) rather severely.}
    \label{fig:SU3-NLO}
\end{figure}

\Cref{fig:SU3-NLO} shows the bounds on $L_1^r,L_2^r$ and $L_3^r$ obtainable at NLO.
With \mbox{$\Mphys=M_\pi$}, the bounds are consistent with the reference point and qualitatively similar to the two-flavour bounds on $\bar l_1,\bar l_2$.
Interestingly, the reference point does not satisfy the bounds at the other choices of $\Mphys$, although only barely --- the smallest distance between the reference point and $\sat(\Omega)$ is $0.5\cdot 10^{-3}$ for $\Mphys=M_K$ and $0.6\cdot 10^{-3}$ for $\Mphys=M_\eta$, which is smaller than the uncertainty in the experimental values (approximately (1-3)$\cdot10^{-3}$).
Therefore, using these $\Mphys$ does not imply any significant inconsistency.

It is worth noting that the three-flavour constraints are much more sensitive to integration than their two-flavour counterparts.
This can be partly understood by noting that integrals scale as roughly the square of the number of flavours (see \cref{fig:integrals}).

\begin{figure}[hbtp]
    \hspace{-2cm}
    \begin{tikzpicture}
        \begin{axis}[
                scale=.9,
                view={-250}{20}, zlabel={$10^{3} L_3^r$}, ylabel={$10^{3} L_2^r$}, xlabel={$10^{3} L_1^r$},
                colormap/temp, point meta min=0, point meta max=1600,
                zmin=-8,zmax=6, ymin=-1.9,ymax=3
            ]

            \def\xfid{1.11} \def\yfid{1.05}  \def\zfid{-3.82}
            \def\xmin{-5}   \def\ymin{-1.9} \def\zmin{-8}
            \def\xmax{5}   \def\ymax{3}   \def\zmax{6}
            
            \addplot3 [
                patch, 
                patch table with point meta={\resultpath/NNLO/SU3/M0.135/D2/lam4/visualisation/L123/constr_0_table.dat}, 
                opacity=0] 
                    table {\resultpath/NNLO/SU3/M0.135/D2/lam4/visualisation/L123/constr_0_coords.dat};
                    
            \coordinate (p0) at (1.57844, -1.7655, 6.00026);
            \coordinate (p1) at (1.2159, -1.72147, 6.00002);
            \coordinate (p2) at (2.25844, -1.72882, 4.24211);
            \coordinate (p3) at (2.36965, -1.75887, 4.47762);
            \coordinate (p4) at (2.3303, -1.77081, 4.72901);
            \coordinate (p5) at (-4.76306, -0.952611, 5.99975);
            \coordinate (p6) at (-4.76307, -0.952615, 5.99977);
            \coordinate (p7) at (-1.18195, -0.953093, -0.39388);
            \coordinate (p8) at (-1.18196, -0.953094, -0.394037);
            \coordinate (p9) at (2.19138, -1.82925, 6.00006);
            \coordinate (p10) at (4.99966, -1.59952, 4.93983);
            \coordinate (p11) at (4.9999, -1.64083, 5.99988);
            \coordinate (p12) at (-5.00006, 1.72385, -2.11692);
            \coordinate (p13) at (-2.56274, 2.20234, -7.99999);
            \coordinate (p14) at (-0.999996, -0.782769, -1.26271);
            \coordinate (p15) at (-0.429337, 1.00935, -7.9997);
            \coordinate (p16) at (-5.00009, -0.819868, 5.9998);
            \coordinate (p17) at (-0.00489809, -0.827492, -1.13497);
            \coordinate (p18) at (2.94048, -0.57118, -2.21579);
            \coordinate (p19) at (4.99984, -0.0208327, -4.36093);
            \coordinate (p20) at (4.99971, 0.950068, -7.99973);
            \coordinate (p21) at (1.62316, -1.54497, 2.82116);
            \coordinate (p22) at (-3.65563, 2.83321, -7.9997);
            \coordinate (p23) at (-5.00008, 2.99988, -6.00814);
            \coordinate (p24) at (-5.00003, 2.56848, -4.75341);
            \coordinate (p25) at (-3.91873, 3.00038, -8.00007);
            \coordinate (p26) at (-0.431624, -1.09548, 0.244735);
            
            \draw[color of colormap={522.938/1.62448}, facet]
                (p0) -- (p4) -- (p3) -- (p2) -- (p1) -- cycle;
            \draw[color of colormap={530.667/1.62448}, facet]
                (p5) -- (p7) -- (p8) -- (p6) -- cycle;
            \draw[color of colormap={494.84/1.62448}, facet]
                (p0) -- (p4) -- (p9) -- cycle;
            \draw[color of colormap={1904.54/1.62448}, facet]
                (p3) -- (p4) -- (p9) -- (p11) -- (p10) -- cycle;
            \draw[color of colormap={556.599/1.62448}, facet]
                (p12) -- (p13) -- (p15) -- (p14) -- (p8) -- (p6) -- (p16) -- cycle;
            \draw[color of colormap={908.402/1.62448}, facet]
                (p17) -- (p14) -- (p15) -- (p20) -- (p19) -- (p18) -- cycle;
            \draw[color of colormap={1624.48/1.62448}, facet]
                (p2) -- (p3) -- (p10) -- (p19) -- (p18) -- (p21) -- cycle;
            \draw[color of colormap={477.304/1.62448}, facet]
                (p22) -- (p24) -- (p23) -- (p25) -- cycle;
            \draw[color of colormap={1250.87/1.62448}, facet]
                (p26) -- (p17) -- (p18) -- (p21) -- cycle;
            \draw[color of colormap={531.15/1.62448}, facet]
                (p1) -- (p2) -- (p21) -- (p26) -- (p7) -- (p5) -- cycle;
            \draw[color of colormap={537.953/1.62448}, facet]
                (p24) -- (p22) -- (p13) -- (p12) -- cycle;
            \draw[color of colormap={897.881/1.62448}, facet]
                (p7) -- (p8) -- (p14) -- (p17) -- (p26) -- cycle;
                
            \coordinate (fid) at (\xfid,\yfid,\zfid);
            
            \draw[grey, fill, opacity=.5]
                (p20) -- (p19) -- (p10) -- (p11) -- (\xmax,\ymin,\zmax) -- (\xmax,\ymin,\zmin) -- cycle;
            \draw[grey, fill, opacity=.5]
                (p25) -- (p23) -- (\xmin,\ymax,\zmin) -- cycle;
            \draw[grey, fill, opacity=.5]
                (p16) -- (p6) -- (p1) -- (p0) -- (p9) -- (p11) -- (\xmax,\ymin,\zmax) -- (\xmin,\ymin,\zmax) -- cycle;
            
            \draw[surface to fid]
                (1.09815, -0.0449969, -4.11219) -- (fid);
                
            \draw[axes to fid]
                    (fid) -- (\xfid,\yfid,\zmax) -- (\xfid,\ymin,\zmax) (\xfid,\yfid,\zmax) -- (\xmin,\yfid,\zmax)
                    (fid) -- (\xfid,\ymax,\zfid) -- (\xfid,\ymax,\zmin) (\xfid,\ymax,\zfid) -- (\xmin,\ymax,\zfid)
                    (fid) -- (\xmax,\yfid,\zfid) -- (\xmax,\ymin,\zfid) (\xmax,\yfid,\zfid) -- (\xmax,\yfid,\zmin)
                    (fid) -- (\xfid,\yfid,\zmin);
                
            \filldraw [black] (\xfid,\yfid,\zfid) circle[radius=1pt] {};

            \draw[black, fill=black, fill opacity=0.5, join=round] 
                (1.21,1.22,-3.52) -- (1.01,1.22,-3.52) -- (1.01,0.88,-3.52) -- (1.21,0.88,-3.52) -- cycle;
            \draw[black, fill=black, fill opacity=0.5, join=round] 
                (1.21,1.22,-3.52) -- (1.21,0.88,-3.52) -- (1.21,0.88,-4.12) -- (1.21,1.22,-4.12) -- cycle;
            \draw[black, fill=black, fill opacity=0.5, join=round] 
                (1.21,1.22,-3.52) -- (1.21,1.22,-4.12) -- (1.01,1.22,-4.12) -- (1.01,1.22,-3.52) -- cycle;
            
        \end{axis}
    \end{tikzpicture}
    \hspace{-.5cm}
    \begin{tikzpicture}
        \begin{axis}[
                scale=.9,
                view={-345}{20}, ylabel={$10^{3} L_2^r$}, xlabel={$10^{3} L_1^r$},
                colormap/temp, colorbar, point meta min=0, point meta max = 1600,
                colorbar style={
                    ymin=0, ytick={500,1000,1500}, yticklabels={0.5,1.0,1.5},ylabel={$10^3/\rho\left(\constr{\v\alpha_i}{c_i}\right)$}
                },
                zmin=-16,zmax=12
            ]
        
            \def\xfid{1.05} \def\yfid{1.11}  \def\zfid{-3.82}
            \def\xmin{-10}   \def\ymin{-6} \def\zmin{-16}
            \def\xmax{10}   \def\ymax{6}   \def\zmax{12}
        
            \addplot3 [
                patch, 
                patch table with point meta={\resultpath/NNLO/SU3/M0.135/D4/lam4/visualisation/L123/constr_0_table.dat}, 
                opacity=.1] 
                    table {\resultpath/NNLO/SU3/M0.135/D4/lam4/visualisation/L123/constr_0_coords.dat};
                    
            \coordinate (p0) at (4.75244, 5.52605, -15.9997);
            \coordinate (p1) at (10.0002, 1.42633, -16.0003);
            \coordinate (p2) at (9.85703, 1.24745, -14.4722);
            \coordinate (p3) at (9.99946, 1.15674, -14.5793);
            \coordinate (p4) at (2.01901, -6.00014, -9.73281);
            \coordinate (p5) at (2.12323, -5.02797, -11.2501);
            \coordinate (p6) at (6.54345, -5.99999, -16);
            \coordinate (p7) at (3.15302, -2.67213, -15.9998);
            \coordinate (p8) at (-2.0227, 6.00009, -1.37287);
            \coordinate (p9) at (9.99965, -6.00053, -4.12405);
            \coordinate (p10) at (-9.95016, 5.99983, 11.9997);
            \coordinate (p11) at (0.441773, -6.00038, 11.9994);
            \coordinate (p12) at (4.11651, 6.00076, -16.0003);
            \coordinate (p13) at (-7.5395, -5.99962, 11.6026);
            \coordinate (p14) at (-7.73757, -4.46538, 10.0166);
            \coordinate (p15) at (-10.0006, -1.57309, 12.0007);
            \coordinate (p16) at (-9.99944, 5.99991, 4.0588);
            \coordinate (p17) at (-8.81202, -4.0849, 12);
            \coordinate (p18) at (-0.950363, 6.00031, -16.0003);
            \coordinate (p19) at (-7.71091, -5.99944, 12.0004);

            \draw[color of colormap={272.137/1.62448}, facet]
                (p0) -- (p2) -- (p3) -- (p1) -- cycle;
            \draw[color of colormap={155.555/1.62448}, facet]
                (p4) -- (p5) -- (p7) -- (p6) -- cycle;
            \draw[color of colormap={562.134/1.62448}, facet]
                (p2) -- (p3) -- (p9) -- (p11) -- (p10) -- (p8) -- cycle;
            \draw[color of colormap={417.512/1.62448}, facet]
                (p2) -- (p0) -- (p12) -- (p8) -- cycle;
            \draw[color of colormap={216.021/1.62448}, facet]
                (p4) -- (p5) -- (p14) -- (p13) -- cycle;
            \draw[color of colormap={233.89/1.62448}, facet]
                (p5) -- (p14) -- (p17) -- (p15) -- (p16) -- (p18) -- (p7) -- cycle;
            \draw[color of colormap={210.352/1.62448}, facet]
                (p13) -- (p14) -- (p17) -- (p19) -- cycle;
                
            \coordinate (fid) at (\xfid,\yfid,\zfid);
            
            \draw[grey, fill, opacity=.5]
                (p6) -- (p4) -- (p13) -- (p19) -- (\xmin,\ymin,\zmax) -- (\xmin,\ymin,\zmin) -- cycle;
            \draw[grey, fill, opacity=.5]
                (p19) -- (p17) -- (p15) -- (\xmin,\ymin,\zmax) -- cycle;
            \draw[grey, fill, opacity=.5]
                (p10) -- (p11) -- (\xmax,\ymin,\zmax) -- (\xmax,\ymax,\zmax) -- cycle;
            \draw[grey, fill, opacity=.5]
                (p11) -- (p9) -- (\xmax,\ymin,\zmax) -- cycle;
            \draw[grey, fill, opacity=.5]
                (p9) -- (p3) -- (p1) -- (\xmax,\ymax,\zmin) -- (\xmax,\ymax,\zmax) -- (\xmax,\ymin,\zmax) -- cycle;
            
            \draw[surface to fid]
                (2.34947, 2.12336, -3.08526) -- (fid);
            \draw[surface to fid]
                (-2.54267, -0.678028, -5.46793) -- (fid);
                
            \draw[axes to fid]
                    (fid) -- (\xfid,\ymin,\zfid) -- (\xmin,\ymin,\zfid) (\xfid,\ymin,\zfid) -- (\xfid,\ymin,\zmin);
                
            \filldraw [black] (\xfid,\yfid,\zfid) circle[radius=1pt] {};

            \draw[black, fill=black, fill opacity=0.5, join=round] 
                (1.21, 1.22, -3.52) -- (1.01, 1.22, -3.52) -- (1.01, 0.88, -3.52) -- (1.21, 0.88, -3.52) -- cycle;
            \draw[black, fill=black, fill opacity=0.5, join=round] 
                (1.21, 1.22, -3.52) -- (1.21, 0.88, -3.52) -- (1.21, 0.88, -4.12) -- (1.21, 1.22, -4.12) -- cycle;
            \draw[black, fill=black, fill opacity=0.5, join=round] 
                (1.21, 0.88, -3.52) -- (1.21, 0.88, -4.12) -- (1.01, 0.88, -4.12) -- (1.01, 0.88, -3.52) -- cycle;
            
        \end{axis}
    \end{tikzpicture}
    \caption[NNLO bounds on $L_1^r,L_2^r$ and $L_3^r$.]{
        NNLO bounds on $L_1^r,L_2^r$ and $L_3^r$ with $\Mphys=M_\pi$ and $\lambda=4$, using two (left) and four (right) derivatives; three derivatives gives very weak bounds. 
        Thus, the left figure is essentially the NNLO version of \cref{fig:SU3-NLO}. 
        Its response to larger $\lambda$ or $\Mphys$ (not shown) is qualitatively similar to that exhibited at NLO.
        
        \qquad In the four-derivative case, $\sat(\Omega)$ is actually a bounded region. It is a lentil-shaped body whose largest dimension is about two orders of magnitude larger than the region shown in the figure. Note that the axes have been rotated relative to the left figure in order to make the inside of $\sat(\Omega)$ reasonably visible.}
    \label{fig:SU3-L123}
\end{figure}
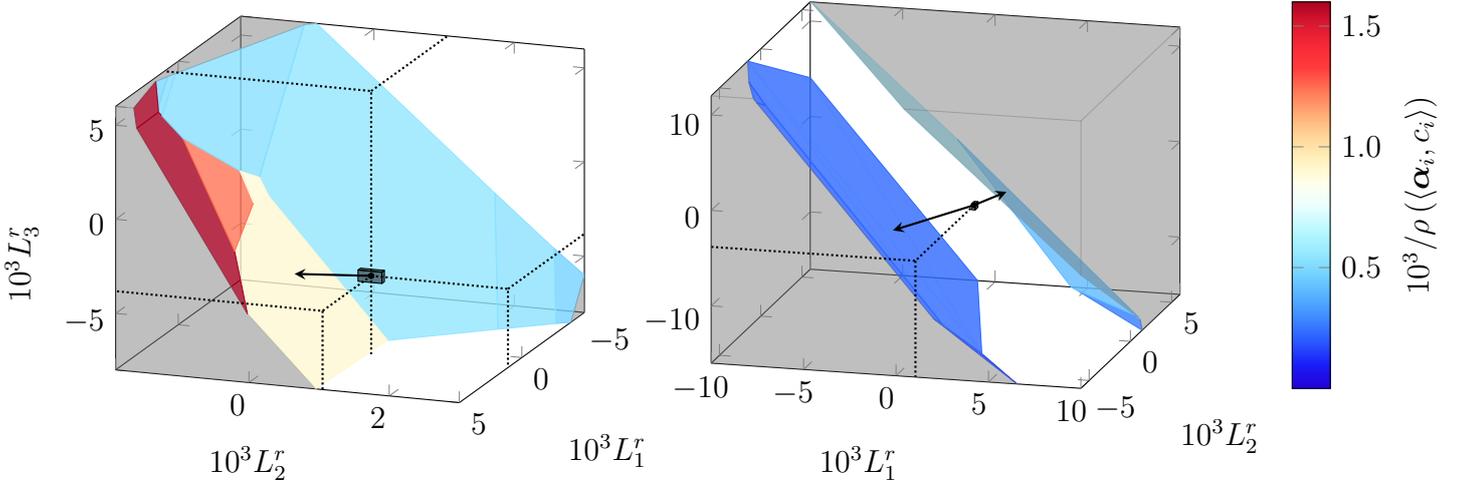

\Cref{fig:SU3-L123} shows NNLO bounds on the same three LECs.
Interestingly, the four-derivative constraints alone confine $L_1^r,L_2^r,L_3^r$ to a bounded region, although it is very large in most directions.
However, the bounds are reasonably strict between the two near-parallel faces shown in the figure, so we may write down another double-ended bound, similar to \cref{eq:Theta-bound}:
\begin{equation}
    \label{eq:L123-bound}
    -27\lesssim 4.9L_1^r + 2.8L_2^r + 2.4L_3^r \lesssim10\qquad\text{for}\quad \Mphys=M_\pi,
\end{equation}
where the linear combination of the LECs is chosen to be roughly orthogonal to the bounding faces.

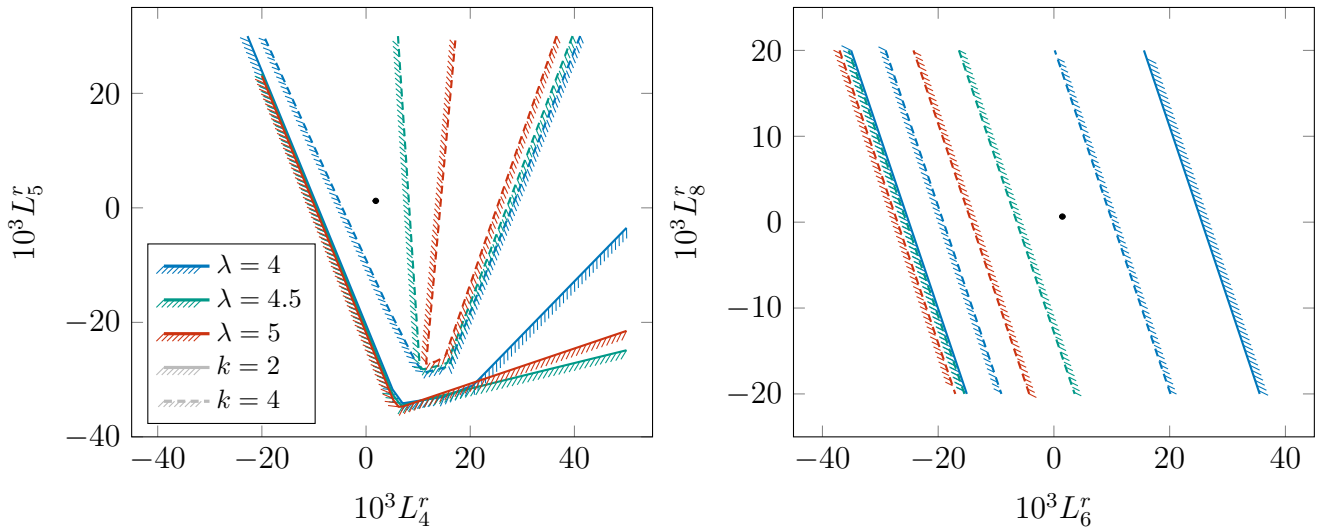
\begin{figure}[hbtp]
    \hspace{-1.7cm}
    \begin{tikzpicture}
        \begin{axis}[ylabel={$10^{3}L_5^r$}, xlabel={$10^{3}L_4^r$},
            xmin=-45,xmax=55, ymin=-40, ymax=35,
            legend pos=south west, legend cell align=left,
            legend image code/.code={
                \draw (0cm,0cm) -- (0.6cm,0cm);
            },
            legend style={font=\footnotesize}]
        
            \addlegendimage{plotI, very thick, hatchborder};
            \addlegendimage{plotIII, very thick, hatchborder};
            \addlegendimage{plotV, very thick, hatchborder};
            \addlegendimage{plotgrey, very thick, hatchborder};
            \addlegendimage{plotgrey, very thick, densely dashed, hatchborder};
            \addlegendimage{plotgrey, very thick, densely dotted, hatchborder};
            
            \draw[plotI, thick, join=round, invhatchborder]
                (50.002,-3.49904)
                -- (20.5444,-30.9933)
                -- (14.9867,-32.9917)
                -- (7.02743,-34.1437)
                -- (5.10546,-31.7419)
                -- (-22.7484,29.9998);
                
            \draw[plotI, thick, join=round, densely dashed, invhatchborder]  
                (41.0973,29.9999)
                -- (15.8446,-27.7145)        
                -- (11.0839,-28.7799)
                -- (9.53156,-26.9266)
                -- (4.32461,-17.6177)
                -- (-11.1988,13.1374)
                -- (-19.6031,29.9995);

            \draw[plotIII, thick, join=round, invhatchborder]
                (49.9984, -24.8736)
                -- (6.7353, -34.5387)
                -- (5.13544, -32.6224)
                -- (-19.9988, 23.4475);

            \draw[plotIII, thick, join=round, densely dashed, invhatchborder]
                (39.6284, 30.0009)
                -- (15.6025, -27.06)
                -- (14.1578, -27.5232)
                -- (10.855, -28.2201)
                -- (10.2857, -27.5647)
                -- (10.2858, -27.5652)
                -- (6.12695, 30.0009);

            \draw[plotV, thick, join=round, invhatchborder]
                (49.9989, -21.4918)
                -- (7.05591, -34.6918)
                -- (6.26907, -34.7571)
                -- (5.20165, -33.5308)
                -- (-19.999, 23.0408);
                
            \draw[plotV, thick, join=round, densely dashed, invhatchborder]                
                (36.5681, 29.998)
                -- (15.1629, -25.9927)
                -- (13.3279, -26.8506)
                -- (11.5337, -27.2243)
                -- (11.5337, -27.2243)
                -- (17.2151, 29.9992);
                
            \filldraw [black] (1.87,1.22)circle[radius=1pt];
            \draw[black, fill=black, fill opacity=0.3] (1.33,1.16) rectangle (2.40,1.28);
                
            \legend{$\lambda=4$,$\lambda=4.5$,$\lambda=5$,$k=2$,$k=4$};
        \end{axis}
    \end{tikzpicture}
    \begin{tikzpicture}
        \begin{axis}[ylabel={$10^{3}L_8^r$}, xlabel={$10^{3}L_6^r$},
            xmin=-45,xmax=45, ymin=-25,ymax=25]

            \draw[plotI, thick, join=round, hatchborder]
                (-35.0465,19.9989)
                -- (-15.0473,-20.0005)
                (35.5571,-20.0019)
                -- (15.5573,19.999);

            \draw[plotI, thick, join=round, densely dashed, hatchborder] 
                (-29.0755,19.9989)
                -- (-9.07505,-19.9994)
                (20.1408,-19.9988)
                -- (0.139763,20.0005);                                  
                
            \draw[plotIII, thick, join=round, densely dashed, invhatchborder]
                (-16.4144, 20.0005)
                -- (3.58624, -19.9997)
                (-15.4681, -19.9988)
                -- (-35.4689, 20.0004);
                
            \draw[plotV, thick, join=round, densely dashed, hatchborder]
                (-4.26803, -20.0006)
                -- (-24.2685, 20.001)
                (-37.0391, 19.9992)
                -- (-17.0392, -20.0003);
                
            \filldraw [black] (1.46,0.65)circle[radius=1pt];
            \draw[black, fill=black, fill opacity=0.3] (1.00,0.58) rectangle (1.92,0.72);
                
        \end{axis}
    \end{tikzpicture}
    \caption{NNLO bounds on $L_4^r,L_5^r$ (left) and $L_6^r,L_8^r$ (right), visualised similarly to \cref{fig:SU2-l12}. 
        The uncertainty region around the reference point is hardly visible at this scale.
        Note how the four-derivative bounds exclude the reference point also for rather small $\lambda>4$.
        However, the integrated two-derivative bounds are weaker and are not visible at all in the right figure.
        The corresponding bounds for $\Mphys=M_K,M_\eta$ are entirely inconsistent with the experimental values.}
    \label{fig:SU3-L4568}
\end{figure}

\Cref{fig:SU3-L4568} shows NNLO bounds on the remaining $L_i^r$. 
These are rather weak, even though $L_6^r$ obtains a double-ended bound.
The weakness is understandable for similar reasons as the weakness of the $\bar l_3$ bounds.
With $\Mphys=M_K,M_\eta$, the bounds are inconsistent, not only with the reference values for $L^r_{4,5,6,8}$, but with \emph{all} values of these parameters, unless $L^r_{1,2,3}$ are removed from their experimental values by an amount roughly one order of magnitude larger than their stated uncertainty (the NNLO parameters have little effect). 
This practically renders this version of the theory self-inconsistent, except at experimentally unreasonable points in parameter space. 
Although the $\Mphys=M_\pi$ version remains consistent, only a small amount of integration excludes the reference point, so its validity is quite dubious.

\begin{figure}[hbtp]
    \centering
    \begin{tikzpicture}
        \begin{axis}[
                view={170}{50}, xlabel={$10^{3} \Xi_1$}, ylabel={$10^{3} \Xi_2$}, zlabel={$10^{3} \Xi_3$},
                colormap/temp, colorbar, point meta min=0, point meta max = 720.041,
                colorbar style={
                    ymin=0, ytick={200,400,600}, yticklabels={0.2,0.4,0.6},ylabel={$10^3/\rho\left(\constr{\v\alpha_i}{c_i}\right)$}
                },
                xmin=-10,xmax=10,
                ymin=-10,ymax=10,
                zmin=-3, zmax=10,
            ]

            \addplot3 [patch, patch table with point meta={\tresultpath/onlyNNLO/SU3/M0.135/D2/lam4/visualisation/X123/constr_0_table.dat}, opacity=0] 
                    table {\tresultpath/onlyNNLO/SU3/M0.135/D2/lam4/visualisation/X123/constr_0_coords.dat};

            \coordinate (p0) at (-10.0007, -2.02727, 10.0007);
            \coordinate (p1) at (0.20276, 2.41665, -1.24568);
            \coordinate (p2) at (0.948376, 1.67097, -0.997299);
            \coordinate (p3) at (-9.99977, 2.90787, 5.06531);
            \coordinate (p4) at (-1.27077, 2.90784, -0.754765);
            \coordinate (p5) at (-2.29861, -7.16112, 10.0003);
            \coordinate (p6) at (7.5731, -7.16176, 3.41908);
            \coordinate (p7) at (0.202796, 9.99971, -1.24581);
            \coordinate (p8) at (0.948354, 10.0004, -0.99721);
            \coordinate (p9) at (0.948375, 1.67097, -0.997298);
            \coordinate (p10) at (0.948374, 1.67097, -0.997297);
            \coordinate (p11) at (0.202759, 2.41664, -1.24567);
            \coordinate (p12) at (10, -10, 7.47722);
            \coordinate (p13) at (6.21639, -9.99967, 9.99967);
            \coordinate (p14) at (9.99994, -8.77911, 5.03696);
            \coordinate (p15) at (-2.29861, -7.16109, 10.0002);
            \coordinate (p16) at (-9.9998, 2.90788, 5.06533);
            \coordinate (p17) at (-10, 10, 5.06535);
            \coordinate (p18) at (-1.27044, 2.90781, -0.75476);
            \coordinate (p19) at (-1.27065, 10.0004, -0.754607);
            \coordinate (p20) at (9.99979, 9.99979, 5.03668);
            \coordinate (p21) at (0.202759, 2.41664, -1.24567);

            \draw[plotIII, facet]
                (p0)-- (p5)-- (p6)-- (p2)-- (p1)-- (p4)-- (p3)
                -- cycle;
            \draw[color of colormap={720.41/0.72041}, facet]
                (p7)-- (p8)-- (p10)-- (p9)-- (p11)
                -- cycle;
            \draw[color of colormap={334.86/0.72041}, facet]
                (p12)-- (p13)-- (p15)-- (p5)-- (p14)
                -- cycle;
            \draw[color of colormap={588.467/0.72041}, facet]
                (p16)-- (p17)-- (p19)-- (p18)
                -- cycle;
            \draw[color of colormap={714.379/0.72041}, facet]
                (p6)-- (p5)-- (p14)
                -- cycle;
            \draw[color of colormap={714.379/0.72041}, facet]
                (p8)-- (p10)-- (p6)-- (p14)-- (p20)
                -- cycle;
            \draw[color of colormap={692.626/0.72041}, facet]
                (p7)-- (p11)-- (p21)-- (p18)-- (p19)
                -- cycle;
            
            \draw[grey, fill, opacity=.5]
                (p20) -- (p8) -- (p7) -- (p19) -- (p17) -- (\xmin,\ymax,\zmin) -- (\xmax,\ymax,\zmin) -- cycle;
            \draw[grey, fill, opacity=.5]
                (p20) -- (p14) -- (p12) -- (\xmax,\ymin,\zmin) -- (\xmax,\ymax,\zmin) -- cycle;
            \draw[grey, fill, opacity=.5]
                (p13) -- (p5) -- (p0) -- (\xmin,\ymin,\zmax) -- cycle;    
                
            \def\xfid{0.2908}   \def\yfid{0.336} \def\zfid{0.2468}
            \def\xmin{-10},\def\xmax{10},
            \def\ymin{-10},\def\ymax{10},
            \def\zmin{-3}, \def\zmax{10},
                        
            \draw[axes to fid] (\xfid,\ymax,\zfid) -- (\xfid,\yfid,\zfid);
            \draw[axes to fid] (\xfid,\ymax,\zmin) -- (\xfid,\ymax,\zfid) -- (\xmin,\ymax,\zfid);
            \draw[axes to fid] (\xfid,\yfid,\zmax) -- (\xfid,\yfid,\zfid);
            \draw[axes to fid] (\xfid,\ymin,\zmax) -- (\xfid,\yfid,\zmax) -- (\xmin,\yfid,\zmax);
            
            \filldraw [black] (\xfid,\yfid,\zfid) circle[radius=1pt];
                    
        \end{axis}
    \end{tikzpicture}
    \caption{
        NNLO two-derivative bounds on $\Xi_1,\Xi_2,\Xi_3$ (those NNLO parameters whose contribution is independent of $t$) with $\lambda=4$ and $\Mphys=M_\pi$. 
        Higher-derivative bounds are extremely weak and are not shown.
        The facets are coloured based on their distance to the reference point in the space of NNLO parameters rather than the full parameter space.
        Like in \cref{fig:SU2-Thetas}, the \selfcolour[plotIII]{green} facet excludes the reference point, although by an extremely small amount, well within the drawn size of the reference point. }
    \label{fig:SU3-X123}
\end{figure}
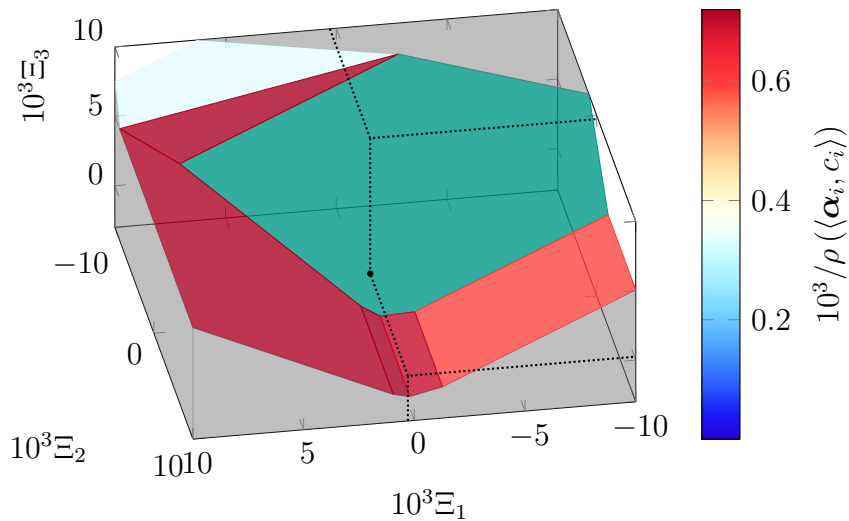

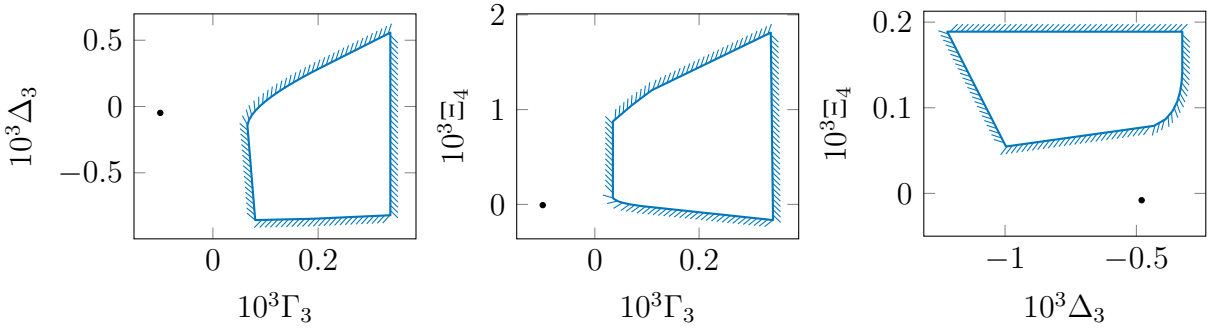
\begin{figure}[hbtp]
    \centering
    \hspace{-1.7cm}
    \begin{tikzpicture}
        \begin{axis}[width=0.35\textwidth,
            xlabel={$10^{3}\Gamma_3$}, ylabel={$10^{3}\Delta_3$},
            xmin=-0.15,]

            \addplot [colormap={cc}{color=(red) color=(red)},
                    name path=lam4onlyNNLO, join=round,
                    patch, patch type=line, very thick, 
                    patch table={\tresultpath/onlyNNLO/SU3/M0.135/D2/lam4/visualisation/GD/constr_0_table.dat},
                    opacity=0
                ] 
                table {\tresultpath/onlyNNLO/SU3/M0.135/D2/lam4/visualisation/GD/constr_0_coords.dat};
                        
            \draw[plotI, thick, join=round, invhatchborder]
                (0.0804473, -0.856614) 
                -- (0.0655725, -0.142755) 
                -- (0.068524, -0.10257) 
                -- (0.0741833, -0.0639481) 
                -- (0.0818041, -0.0269408) 
                -- (0.090816, 0.00846321) 
                -- (0.100794, 0.0422917) 
                -- (0.111461, 0.074652) 
                -- (0.122551, 0.105522) 
                -- (0.133903, 0.135044) 
                -- (0.145355, 0.163142) 
                -- (0.156874, 0.190097) 
                -- (0.168363, 0.215888) 
                -- (0.179637, 0.240376) 
                -- (0.190810, 0.263845) 
                -- (0.201832, 0.286455) 
                -- (0.337266, 0.557333) 
                -- (0.337274, 0.557345) 
                -- (0.337289, -0.819992) 
                -- (0.190725, -0.846547) 
                -- cycle;

            \filldraw [black] (-0.10, -0.048) circle[radius=1pt];
                
        \end{axis}
    \end{tikzpicture}
    \begin{tikzpicture}
        \begin{axis}[width=0.35\textwidth,
            xlabel={$10^{3}\Gamma_3$}, ylabel={$10^{3}\Xi_4$},
            xmin=-0.15,]

            \addplot [colormap={cc}{color=(red) color=(red)},
                    name path=lam4onlyNNLO, join=round,
                    patch, patch type=line, very thick, 
                    patch table={\tresultpath/onlyNNLO/SU3/M0.135/D2/lam4/visualisation/GX/constr_0_table.dat},
                    opacity=0
                ] 
                table {\tresultpath/onlyNNLO/SU3/M0.135/D2/lam4/visualisation/GX/constr_0_coords.dat};
                
            \draw[plotI, thick, join=round, hatchborder]
                (0.0665345, 0.00129436) 
                -- (0.0884575, -0.0173436) 
                -- (0.341953, -0.165808) 
                -- (0.337595, 1.81261) 
                -- (0.109256, 1.2047) 
                -- (0.0713753, 1.04815) 
                -- (0.0346037, 0.875122) 
                -- (0.0346038, 0.0726519) 
                -- (0.0357747, 0.060867) 
                -- (0.0438476, 0.0338243) 
                -- (0.0525874, 0.0183143) 
                -- cycle;

            \filldraw [black] (-0.10, -0.008) circle[radius=1pt];
                
        \end{axis}
    \end{tikzpicture}
    \begin{tikzpicture}
        \begin{axis}[width=0.35\textwidth,
            xlabel={$10^{3}\Delta_3$}, ylabel={$10^{3}\Xi_4$},
            ymin=-0.05,]

            \addplot [colormap={cc}{color=(red) color=(red)},
                    name path=lam4onlyNNLO, join=round,
                    patch, patch type=line, very thick, 
                    patch table={\tresultpath/onlyNNLO/SU3/M0.135/D2/lam4/visualisation/DX/constr_0_table.dat},
                    opacity=0
                ] 
                table {\tresultpath/onlyNNLO/SU3/M0.135/D2/lam4/visualisation/DX/constr_0_coords.dat};
                
            \draw[plotI, thick, join=round, invhatchborder]
                (-0.325217, 0.141974) 
                -- (-0.327536, 0.130739) 
                -- (-0.333217, 0.119272) 
                -- (-0.343692, 0.107733) 
                -- (-0.361158, 0.0966078) 
                -- (-0.389057, 0.0865722) 
                -- (-0.43292, 0.0789008) 
                -- (-0.996461, 0.054738) 
                -- (-1.21992, 0.188876) 
                -- (-0.325218, 0.18882) 
                -- (-0.325219, 0.188821) 
                -- cycle;

            \filldraw [black] (-0.48, -0.008) circle[radius=1pt];
                
        \end{axis}
    \end{tikzpicture}
    \caption{
        NNLO two-derivative bounds on $\Gamma_3,\Delta_3$ and $\Xi_4$ (those NNLO parameters that obtain additional bounds at $t\neq 4$) with $\lambda=4$ and $\Mphys=M_\pi$.
        Note how they are bounded from all directions and confined to a rather small volume;
        this makes three-dimensional representation difficult, so we use two-dimensional slices instead. 
        Note how the bounds exclude the reference point.
        A major source of uncertainty in these bounds is the fixing of the other parameters: as the NLO LECs are varied within their uncertainties, the boundary lines shift in various directions, with the typical amount displacement being roughly $0.2$.
        However, at no values are the bounds consistent with the reference values of $\Gamma_3,\Delta_3$ and $\Xi_4$.}
    \label{fig:SU3-GDX}
\end{figure}

\FloatBarrier
Lastly, \cref{fig:SU3-X123,fig:SU3-GDX} show bounds on the NNLO parameters; for the same reason as above, we keep $\Mphys=M_\pi$.
The former shows some features reminiscent of its $\SU(2)$ analogue, \cref{fig:SU2-Thetas}, although there is not a clear double-ended bound on either parameter. 
The latter is more interesting, since all three parameters are confined to a small bounded region and quite significantly excludes the reference point. 
The bounds on $\Gamma_3,\Delta_3$ in particular are fairly independent of each other and $\Xi_4$, leading to the single-parameter bounds
\begin{equation}
    \label{eq:GD-bound}
    0.08     \lesssim 10^3\Gamma_3   \lesssim 0.34,\qquad 
    -0.8    \lesssim 10^3\Delta_3   \lesssim 0.25 \qquad
    \text{for}\quad \Mphys=M_\pi.
\end{equation}

\FloatBarrier

\subsection{Higher number of flavours}
\chpt\ with more than three flavours are not of direct interest as low-energy QCD, since the large mass of the charm quark makes it entirely invalid as a model of mesons.
An arbitrary number of flavours is useful when developing the methods, though, and is interesting in its own right in the context of EFT studies.
Furthermore, \chpt\ has many uses other than QCD (for a review of some of these, see \cite{Cacciapaglia:2020kgq}).
Besides various numbers of flavours, these commonly use different symmetry breaking patterns than $\SU(n)\times\SU(n)\to\SU(n)$; some of these were treated in \cite{Bijnens:2011fm}, so their amplitudes would be a drop-in replacement into our methods.
However, that is beyond the scope of this paper, in which we are content to show some basic high-flavour results in ``QCD-style'' \chpt.

\begin{figure}[hbtp]
    \begin{tikzpicture}[scale=.8]
        \begin{axis}[
                view={-250}{20}, zlabel={$10^{3} \hat L_3^r$}, ylabel={$10^{3} \hat L_2^r$}, xlabel={$10^{3} \hat L_1^r$},
                colormap/temp, point meta min=400, point meta max=1000, 
                zmin=-6.5,zmax=5,ymin=-1.95,ymax=3
            ]

            \addplot3 [
                patch, 
                patch table with point meta={\resultpath/NLO/SU4/M0.135/D2/lam4/visualisation/L1L2L3/constr_0_table.dat}, 
                opacity=0] 
                    table {\resultpath/NLO/SU4/M0.135/D2/lam4/visualisation/L1L2L3/constr_0_coords.dat};
            
            \coordinate (p0) at (-0.508858, -0.738119, -1.91551);
            \coordinate (p1) at (-0.50187, -0.737594, -1.93132);
            \coordinate (p2) at (-0.503139, -0.747492, -1.88808);
            \coordinate (p3) at (-0.496878, -0.747419, -1.9005);
            \coordinate (p4) at (-0.508559, -0.744952, -1.88759);
            \coordinate (p5) at (-0.509107, -0.743756, -1.89124);
            \coordinate (p6) at (-0.547957, -0.689925, -1.95157);
            \coordinate (p7) at (-0.547961, -0.68993, -1.95159);
            \coordinate (p8) at (-4.23784, 2.99977, -1.9514);
            \coordinate (p9) at (-4.23784, 2.99977, -1.9514);
            \coordinate (p10) at (-0.509569, -0.744915, -1.88528);
            \coordinate (p11) at (-0.527342, -0.750679, -1.78469);
            \coordinate (p12) at (-0.510069, -0.744955, -1.88264);
            \coordinate (p13) at (-0.497749, -0.734099, -1.94839);
            \coordinate (p14) at (-0.465982, -0.742231, -1.95387);
            \coordinate (p15) at (-0.488162, -0.74815, -1.90819);
            \coordinate (p16) at (-0.461643, -0.745123, -1.94804);
            \coordinate (p17) at (-0.461444, -0.742426, -1.95701);
            \coordinate (p18) at (-0.457427, -0.745057, -1.9516);
            \coordinate (p19) at (-0.782654, -0.925695, 1.70021);
            \coordinate (p20) at (-0.315286, -0.863345, -0.294248);
            \coordinate (p21) at (-1.14356, -1.02899, 5.00006);
            \coordinate (p22) at (5.00018, -0.863519, -0.294227);
            \coordinate (p23) at (4.99972, -1.02895, 4.99972);
            \coordinate (p24) at (-0.843907, -0.803471, -0.25526);
            \coordinate (p25) at (-2.15782, -0.803677, 4.99986);
            \coordinate (p26) at (4.99998, -0.745092, -1.95187);
            \coordinate (p27) at (-2.23065, -0.74484, 5.00004);
            \coordinate (p28) at (-4.99995, 2.99985, 1.09706);
            \coordinate (p29) at (-5, 2.02413, 5);
            \coordinate (p30) at (-0.50581, -0.712299, -1.98528);
            \coordinate (p31) at (-0.418264, -0.310112, -2.82113);
            \coordinate (p32) at (-0.111596, 1.52935, -6.50009);
            \coordinate (p33) at (5.00018, 1.52908, -6.50007);
            \coordinate (p34) at (-0.430159, -0.350324, -2.73588);
            \coordinate (p35) at (-0.116483, 1.53218, -6.50009);
            \coordinate (p36) at (-1.58449, 3.00034, -6.5);
            
            \draw[color of colormap={151.067/1.0228}, facet]
                    (p0) -- (p1) -- (p3) -- (p2) -- (p4) -- (p5)
                -- cycle;
            \draw[color of colormap={126.536/1.0228}, facet]
                    (p6) -- (p7) -- (p8) -- (p9)
                -- cycle;
            \draw[color of colormap={110.654/1.0228}, facet]
                    (p4) -- (p5) -- (p10)
                -- cycle;
            \draw[color of colormap={91.713/1.0228}, facet]
                    (p4) -- (p2) -- (p11) -- (p12) -- (p10)
                -- cycle;
            \draw[color of colormap={232.507/1.0228}, facet]
                    (p1) -- (p13) -- (p14) -- (p16) -- (p15) -- (p3)
                -- cycle;
            \draw[color of colormap={341.632/1.0228}, facet]
                    (p17) -- (p14) -- (p16) -- (p18)
                -- cycle;
            \draw[color of colormap={159.164/1.0228}, facet]
                    (p19) -- (p20) -- (p22) -- (p23) -- (p21)
                -- cycle;
            \draw[color of colormap={93.824/1.0228}, facet]
                    (p19) -- (p24) -- (p25) -- (p21)
                -- cycle;
            \draw[color of colormap={204.926/1.0228}, facet]
                    (p20) -- (p15) -- (p16) -- (p18) -- (p26) -- (p22)
                -- cycle;
            \draw[color of colormap={85.092/1.0228}, facet]
                    (p11) -- (p12) -- (p27) -- (p25) -- (p24)
                -- cycle;
            \draw[color of colormap={97.395/1.0228}, facet]
                    (p5) -- (p0) -- (p6) -- (p9) -- (p28) -- (p29) -- (p27) -- (p12) -- (p10)
                -- cycle;
            \draw[color of colormap={139.235/1.0228}, facet]
                    (p2) -- (p3) -- (p15) -- (p20) -- (p19) -- (p24) -- (p11)
                -- cycle;
            \draw[color of colormap={200.054/1.0228}, facet]
                    (p0) -- (p1) -- (p13) -- (p30) -- (p7) -- (p6)
                -- cycle;
            \draw[color of colormap={1022.8/1.0228}, facet]
                    (p17) -- (p31) -- (p32) -- (p33) -- (p26) -- (p18)
                -- cycle;
            \draw[color of colormap={434.29/1.0228}, facet]
                    (p34) -- (p31) -- (p32) -- (p35)
                -- cycle;
            \draw[color of colormap={538.371/1.0228}, facet]
                    (p13) -- (p30) -- (p34) -- (p31) -- (p17) -- (p14)
                -- cycle;
            \draw[color of colormap={309.386/1.0228}, facet]
                    (p7) -- (p30) -- (p34) -- (p35) -- (p36) -- (p8)
                -- cycle;

            \def\xmin{-5}   \def\ymin{-1.95} \def\zmin{-6.5}
            \def\xmax{5}   \def\ymax{3}   \def\zmax{5}
            
            \draw[grey, fill, opacity=.5]
                (p33) -- (p26) -- (p22) -- (p23) -- (\xmax,\ymin,\zmax) -- (\xmax,\ymin,\zmin) -- cycle;
            \draw[grey, fill, opacity=.5]
                (p23) -- (p21) -- (p25) -- (p27) -- (p29) -- (\xmin,\ymin,\zmax) -- (\xmax,\ymin,\zmax) -- cycle;
            \draw[grey, fill, opacity=.5]
                (p36) -- (p8) -- (p9) -- (p28) -- (\xmin,\ymax,\zmin) -- cycle;
        \end{axis}
    \end{tikzpicture}
    \begin{tikzpicture}[scale=.8]
        \begin{axis}[
                view={-250}{20}, zlabel={$10^{3} \hat L_3^r$}, ylabel={$10^{3} \hat L_2^r$}, xlabel={$10^{3} \hat L_0^r$},
                colormap/temp, point meta min=400, point meta max=1000, 
                zmin=-6.5,zmax=5,ymin=-1.95,ymax=3,xmin=-5,xmax=10
            ]

            \addplot3 [
                patch, 
                patch table with point meta={\resultpath/NLO/SU4/M0.135/D2/lam4/visualisation/L0L2L3/constr_0_table.dat}, 
                opacity=0] 
                    table {\resultpath/NLO/SU4/M0.135/D2/lam4/visualisation/L0L2L3/constr_0_coords.dat};
            
            \coordinate (p0) at (-4.50525, 2.99984, -4.936);
            \coordinate (p1) at (-4.50525, 2.99984, -4.936);
            \coordinate (p2) at (-1.38911, -0.116401, -1.81983);
            \coordinate (p3) at (-1.38911, -0.116401, -1.81982);
            \coordinate (p4) at (-0.760473, -0.744918, -0.625045);
            \coordinate (p5) at (-0.767248, -0.738189, -0.649392);
            \coordinate (p6) at (-0.723359, -0.763486, -0.595366);
            \coordinate (p7) at (-0.495356, -0.780846, -0.954693);
            \coordinate (p8) at (-0.244782, -0.738262, -1.69427);
            \coordinate (p9) at (-0.206478, -0.746499, -1.725);
            \coordinate (p10) at (-1.28469, -0.211785, -1.71235);
            \coordinate (p11) at (-1.2703, -0.225523, -1.6943);
            \coordinate (p12) at (-1.37994, -0.125559, -1.80743);
            \coordinate (p13) at (-1.21188, -0.293514, -1.53867);
            \coordinate (p14) at (-0.166161, -0.744879, -1.7855);
            \coordinate (p15) at (3.97085, -1.52493, 5.00016);
            \coordinate (p16) at (8.00788, -1.81307, 4.99988);
            \coordinate (p17) at (10, -1.81347, 3.00796);
            \coordinate (p18) at (4.54826, -0.744949, -6.49967);
            \coordinate (p19) at (10.0003, -1.13433, -6.49952);
            \coordinate (p20) at (-4.50525, 2.99984, -4.936);
            \coordinate (p21) at (-0.760677, -0.74483, 5.00012);
            \coordinate (p22) at (-1.38911, -0.116401, -1.81982);
            \coordinate (p23) at (-4.50528, 2.99984, 4.99973);
            \coordinate (p24) at (-1.3291, -0.163465, -1.78552);
            \coordinate (p25) at (-2.94186, 2.99991, -6.49981);
            \coordinate (p26) at (10.0002, -1.93106, 5.0001);
            \coordinate (p27) at (-0.25708, -0.996465, 5.00019);
            
            \draw[color of colormap={563.389/1.0228}, facet]
                    (p0) -- (p1) -- (p3) -- (p2)
                -- cycle;
            \draw[color of colormap={319.743/1.0228}, facet]
                    (p4) -- (p5) -- (p8) -- (p9) -- (p7) -- (p6)
                -- cycle;
            \draw[color of colormap={264.402/1.0228}, facet]
                    (p10) -- (p11) -- (p13) -- (p12)
                -- cycle;
            \draw[color of colormap={204.926/1.0228}, facet]
                    (p14) -- (p9) -- (p7) -- (p15) -- (p16) -- (p17) -- (p19) -- (p18)
                -- cycle;
            \draw[color of colormap={153.409/1.0228}, facet]
                    (p5) -- (p4) -- (p21) -- (p23) -- (p20) -- (p22) -- (p12) -- (p13)
                -- cycle;
            \draw[color of colormap={430.665/1.0228}, facet]
                    (p5) -- (p8) -- (p11) -- (p13)
                -- cycle;
                -- cycle;
            \draw[color of colormap={139.235/1.0228}, facet]
                    (p16) -- (p26) -- (p17)
                -- cycle;
            \draw[color of colormap={113.989/1.0228}, facet]
                    (p4) -- (p6) -- (p27) -- (p21)
                -- cycle;
            \draw[color of colormap={305.059/1.0228}, facet]
                    (p0) -- (p2) -- (p22) -- (p20)
                -- cycle;
            \draw[color of colormap={159.164/1.0228}, facet]
                    (p6) -- (p7) -- (p15) -- (p27)
                -- cycle;
            \draw[color of colormap={619.33/1.0228}, facet]
                    (p8) -- (p9) -- (p14) -- (p24) -- (p10) -- (p11)
                -- cycle;
            \draw[color of colormap={423.596/1.0228}, facet]
                    (p2) -- (p3) -- (p24) -- (p10) -- (p12) -- (p22)
                -- cycle;
            \draw[color of colormap={1022.8/1.0228}, facet]
                    (p1) -- (p3) -- (p24) -- (p14) -- (p18) -- (p25)
                -- cycle;
            
            \def\xmin{-5} \def\ymin{-1.95} \def\zmin{-6.5}
            \def\xmax{10}   \def\ymax{3}   \def\zmax{5}
            
            \draw[grey, fill, opacity=.5]
                (p19) -- (p17) -- (p26) -- (\xmax,\ymin,\zmax) -- (\xmax,\ymin,\zmin) -- cycle;
            \draw[grey, fill, opacity=.5]
                (p26) -- (p16) -- (p15) -- (p27) -- (p21) -- (p23) -- (\xmin,\ymax,\zmax) -- (\xmin,\ymin,\zmax) -- (\xmax,\ymin,\zmax) -- cycle;
            \draw[grey, fill, opacity=.5]
                (p25) -- (p20) -- (p23) -- (\xmin,\ymax,\zmax) -- (\xmin,\ymax,\zmin) -- cycle;
            
        \end{axis}
    \end{tikzpicture}
    \begin{tikzpicture}[scale=.8]
        \begin{axis}[
                view={-250}{20}, zlabel={$10^{3} \hat L_3^r$}, ylabel={$10^{3} \hat L_0^r$}, xlabel={$10^{3} \hat L_1^r$},
                colormap/temp, point meta min=400, point meta max=1000, 
                zmin=-6.5,zmax=5,ymin=-5
            ]

            \addplot3 [
                patch, 
                patch table with point meta={\resultpath/NLO/SU4/M0.135/D2/lam4/visualisation/L1L0L3/constr_0_table.dat}, 
                opacity=0] 
                    table {\resultpath/NLO/SU4/M0.135/D2/lam4/visualisation/L1L0L3/constr_0_coords.dat};
            
            \coordinate (p0) at (-0.621343, -2.33911, -2.46947);
            \coordinate (p1) at (-0.744668, -2.30704, -2.29003);
            \coordinate (p2) at (-0.487597, -2.38369, -2.64762);
            \coordinate (p3) at (-0.643686, -2.34818, -2.41228);
            \coordinate (p4) at (-2.47121, -1.71909, 0.663345);
            \coordinate (p5) at (-2.45981, -1.68562, 0.496074);
            \coordinate (p6) at (-2.02221, -1.83827, -0.369529);
            \coordinate (p7) at (-2.38684, -1.80327, 0.71024);
            \coordinate (p8) at (-2.04853, -1.91722, 0.0253197);
            \coordinate (p9) at (-0.311023, -2.42725, -2.89964);
            \coordinate (p10) at (-0.269951, -2.27225, -3.13843);
            \coordinate (p11) at (-0.958656, -2.22774, -2.00286);
            \coordinate (p12) at (-0.755659, -2.43064, -1.79999);
            \coordinate (p13) at (-1.92395, -1.85796, -0.585644);
            \coordinate (p14) at (-1.37217, -2.05036, -1.47099);
            \coordinate (p15) at (-0.759204, -2.44883, -1.71554);
            \coordinate (p16) at (-1.98856, -1.95703, 0.00539062);
            \coordinate (p17) at (-2.76298, -1.81648, 2.18841);
            \coordinate (p18) at (-0.730318, -2.47563, -1.67607);
            \coordinate (p19) at (-0.475887, -2.55557, -1.97042);
            \coordinate (p20) at (-0.454703, -2.55548, -2.02182);
            \coordinate (p21) at (-0.210948, -2.47124, -2.96402);
            \coordinate (p22) at (-0.196945, -2.4818, -2.95514);
            \coordinate (p23) at (-3.15546, -1.88371, 4.00383);
            \coordinate (p24) at (-3.30735, -1.96662, 5.00015);
            \coordinate (p25) at (-3.40468, -1.88353, 4.99995);
            \coordinate (p26) at (-0.941465, -2.5554, -0.293917);
            \coordinate (p27) at (-1.10702, -2.46086, -0.435427);
            \coordinate (p28) at (-2.1596, -2.29909, 2.39855);
            \coordinate (p29) at (-2.93134, -2.24217, 5.00013);
            \coordinate (p30) at (-2.2651, -2.55573, 5.00006);
            \coordinate (p31) at (0.0273438, -2.55554, -2.98592);
            \coordinate (p32) at (5.00009, -2.55537, -2.98593);
            \coordinate (p33) at (4.99991, -2.55559, 4.99991);
            \coordinate (p34) at (-1.82932, -1.74243, -0.995169);
            \coordinate (p35) at (-0.121298, -2.47012, -3.07148);
            \coordinate (p36) at (-0.0664736, -2.51529, -3.02614);
            \coordinate (p37) at (-2.49981, -1.6579, 0.55426);
            \coordinate (p38) at (-3.61104, -1.65799, 5.00015);
            \coordinate (p39) at (-2.28777, -1.02201, -0.929539);
            \coordinate (p40) at (-2.28791, 4.54857, -6.4996);
            \coordinate (p41) at (-3.65087, 9.99968, -6.49973);
            \coordinate (p42) at (-5.00053, 9.99993, -1.10301);
            \coordinate (p43) at (-5.00013, 3.89724, 5.00013);
            \coordinate (p44) at (-0.196788, -2.10093, -3.43542);
            \coordinate (p45) at (-0.19152, -2.10523, -3.43664);
            \coordinate (p46) at (-0.19676, 0.963749, -6.49972);
            \coordinate (p47) at (-0.19154, 0.958355, -6.50012);
            \coordinate (p48) at (5.00009, 0.958572, -6.50029);
            
            \draw[color of colormap={220.832/1.0228}, facet]
                    (p0) -- (p1) -- (p3) -- (p2)
                -- cycle;
            \draw[color of colormap={100.424/1.0228}, facet]
                    (p4) -- (p7) -- (p8) -- (p6) -- (p5)
                -- cycle;
            \draw[color of colormap={299.187/1.0228}, facet]
                    (p0) -- (p2) -- (p9) -- (p10)
                -- cycle;
            \draw[color of colormap={125.322/1.0228}, facet]
                    (p1) -- (p3) -- (p12) -- (p11)
                -- cycle;
            \draw[color of colormap={111.369/1.0228}, facet]
                    (p6) -- (p8) -- (p16) -- (p15) -- (p12) -- (p11) -- (p14) -- (p13)
                -- cycle;
            \draw[color of colormap={83.839/1.0228}, facet]
                    (p4) -- (p7) -- (p17)
                -- cycle;
            \draw[color of colormap={145.295/1.0228}, facet]
                    (p2) -- (p3) -- (p12) -- (p15) -- (p18) -- (p19) -- (p20) -- (p22) -- (p21) -- (p9)
                -- cycle;
            \draw[color of colormap={79.669/1.0228}, facet]
                    (p23) -- (p24) -- (p25)
                -- cycle;
            \draw[color of colormap={100.926/1.0228}, facet]
                    (p18) -- (p19) -- (p26) -- (p27)
                -- cycle;
            \draw[color of colormap={96.972/1.0228}, facet]
                    (p15) -- (p16) -- (p28) -- (p27) -- (p18)
                -- cycle;
            \draw[color of colormap={88.578/1.0228}, facet]
                    (p17) -- (p7) -- (p8) -- (p16) -- (p28) -- (p29) -- (p24) -- (p23)
                -- cycle;
            \draw[color of colormap={153.409/1.0228}, facet]
                    (p19) -- (p20) -- (p31) -- (p32) -- (p33) -- (p30) -- (p26)
                -- cycle;
            \draw[color of colormap={188.718/1.0228}, facet]
                    (p13) -- (p14) -- (p34)
                -- cycle;
            \draw[color of colormap={250.662/1.0228}, facet]
                    (p21) -- (p22) -- (p36) -- (p35)
                -- cycle;
            \draw[color of colormap={97.395/1.0228}, facet]
                    (p37) -- (p38) -- (p43) -- (p42) -- (p41) -- (p40) -- (p39)
                -- cycle;
            \draw[color of colormap={360.732/1.0228}, facet]
                    (p9) -- (p10) -- (p44) -- (p45) -- (p35) -- (p21)
                -- cycle;
            \draw[color of colormap={75.334/1.0228}, facet]
                    (p5) -- (p4) -- (p17) -- (p23) -- (p25) -- (p38) -- (p37)
                -- cycle;
            \draw[color of colormap={178.748/1.0228}, facet]
                    (p5) -- (p6) -- (p13) -- (p34) -- (p39) -- (p37)
                -- cycle;
            \draw[color of colormap={155.744/1.0228}, facet]
                    (p20) -- (p22) -- (p36) -- (p31)
                -- cycle;
            \draw[color of colormap={434.29/1.0228}, facet]
                    (p46) -- (p47) -- (p45) -- (p44)
                -- cycle;
            \draw[color of colormap={1022.8/1.0228}, facet]
                    (p35) -- (p36) -- (p31) -- (p32) -- (p48) -- (p47) -- (p45)
                -- cycle;
            \draw[color of colormap={94.214/1.0228}, facet]
                    (p26) -- (p27) -- (p28) -- (p29) -- (p30)
                -- cycle;
            \draw[color of colormap={309.386/1.0228}, facet]
                    (p0) -- (p1) -- (p11) -- (p14) -- (p34) -- (p39) -- (p40) -- (p46) -- (p44) -- (p10)
                -- cycle;
                
            \def\xmin{-5} \def\ymin{-5} \def\zmin{-6.5}
            \def\xmax{5}   \def\ymax{10}   \def\zmax{5}
            
            \draw[grey, fill, opacity=.5]
                (p48) -- (p32) -- (p33) -- (\xmax,\ymin,\zmax) -- (\xmax,\ymin,\zmin) -- cycle;
            \draw[grey, fill, opacity=.5]
                (p33) -- (p30) -- (p29) -- (p24) -- (p25) -- (p38) -- (p43) -- (\xmin,\ymin,\zmax) -- (\xmax,\ymin,\zmax) -- cycle;
            \draw[grey, fill, opacity=.5]
                (p41) -- (p42) -- (\xmin,\ymax,\zmin) -- cycle;

        \end{axis}
    \end{tikzpicture}
    \begin{tikzpicture}[scale=.8]
        \begin{axis}[
                view={-250}{20}, zlabel={$10^{3} \hat L_0^r$}, ylabel={$10^{3} \hat L_2^r$}, xlabel={$10^{3} \hat L_1^r$},
                colormap/temp, point meta min=400, point meta max=1000, 
                zmin=-5.5,zmax=10,ymin=-1.95,ymax=3
            ]

            \addplot3 [
                patch, 
                patch table with point meta={\resultpath/NLO/SU4/M0.135/D2/lam4/visualisation/L1L2L0/constr_0_table.dat}, 
                opacity=0] 
                    table {\resultpath/NLO/SU4/M0.135/D2/lam4/visualisation/L1L2L0/constr_0_coords.dat};

            \coordinate (p0) at (-0.430134, -0.350024, 1.0838);
            \coordinate (p1) at (-0.418246, -0.310221, 0.998811);
            \coordinate (p2) at (-0.123002, 1.46075, -2.5434);
            \coordinate (p3) at (-0.128117, 1.46189, -2.54034);
            \coordinate (p4) at (-0.430165, -0.350049, 1.08387);
            \coordinate (p5) at (4.99992, 1.88367, -3.38959);
            \coordinate (p6) at (0.166194, 1.88395, -3.38948);
            \coordinate (p7) at (0.0240534, 1.82698, -3.27551);
            \coordinate (p8) at (-0.461355, -0.742243, 1.86291);
            \coordinate (p9) at (-0.457514, -0.745059, 1.86842);
            \coordinate (p10) at (-0.461379, -0.742281, 1.86301);
            \coordinate (p11) at (-0.12301, 1.46086, -2.54358);
            \coordinate (p12) at (-0.00326212, 1.80689, -3.23544);
            \coordinate (p13) at (4.99973, -0.744812, 1.8685);
            \coordinate (p14) at (-0.504395, -0.732441, 1.87267);
            \coordinate (p15) at (-0.506819, -0.731013, 1.87146);
            \coordinate (p16) at (-0.5075, -0.720052, 1.85065);
            \coordinate (p17) at (-0.466173, -0.742248, 1.86613);
            \coordinate (p18) at (-0.165237, 2.87866, -4.3841);
            \coordinate (p19) at (-0.178722, 2.99983, -4.50542);
            \coordinate (p20) at (-0.242085, 2.75308, -4.21658);
            \coordinate (p21) at (-0.329058, 2.99978, -4.43609);
            \coordinate (p22) at (-0.508422, -0.745145, 1.93239);
            \coordinate (p23) at (-0.509602, -0.744869, 1.93487);
            \coordinate (p24) at (-0.509033, -0.743856, 1.92862);
            \coordinate (p25) at (-0.510158, -0.74504, 1.93719);
            \coordinate (p26) at (-0.527242, -0.750683, 2.03486);
            \coordinate (p27) at (-0.503093, -0.747391, 1.932);
            \coordinate (p28) at (-0.501101, 3.00007, -4.30992);
            \coordinate (p29) at (-0.736882, 3.00016, -4.13335);
            \coordinate (p30) at (-0.655691, 2.86733, -4.04518);
            \coordinate (p31) at (-0.504397, -0.732443, 1.87268);
            \coordinate (p32) at (-0.508445, -0.730825, 1.87363);
            \coordinate (p33) at (-0.50683, -0.731029, 1.8715);
            \coordinate (p34) at (-0.496898, -0.747571, 1.91936);
            \coordinate (p35) at (-0.152829, 1.9773, -3.37523);
            \coordinate (p36) at (-0.240613, 2.14452, -3.51739);
            \coordinate (p37) at (-0.152831, 1.85174, -3.16639);
            \coordinate (p38) at (-0.22147, 2.48079, -3.91949);
            \coordinate (p39) at (-0.256553, 2.17664, -3.54548);
            \coordinate (p40) at (-2.1753, -1.02532, 10.0002);
            \coordinate (p41) at (0.378345, -1.32542, 9.99971);
            \coordinate (p42) at (-0.488107, -0.748092, 1.91191);
            \coordinate (p43) at (-0.124987, 1.95685, -3.36807);
            \coordinate (p44) at (-0.504396, -0.732441, 1.87267);
            \coordinate (p45) at (-0.465943, -0.742416, 1.86604);
            \coordinate (p46) at (-0.465944, -0.742212, 1.86604);
            \coordinate (p47) at (-0.461558, -0.745148, 1.87186);
            \coordinate (p48) at (4.99974, -0.744815, 1.8685);
            \coordinate (p49) at (4.99998, -1.32627, 9.99995);
            \coordinate (p50) at (-0.457514, -0.745059, 1.86842);
            \coordinate (p51) at (-0.461379, -0.742282, 1.86301);
            \coordinate (p52) at (-0.457511, -0.745055, 1.86841);
            \coordinate (p53) at (0.0243924, 1.82709, -3.27571);
            \coordinate (p54) at (-0.042675, 1.86528, -3.2951);
            \coordinate (p55) at (-0.00331066, 1.80669, -3.23507);
            \coordinate (p56) at (-2.52629, -0.74496, 9.99967);
            \coordinate (p57) at (0.0244037, 1.82709, -3.27571);
            \coordinate (p58) at (0.166312, 1.88396, -3.3895);
            \coordinate (p59) at (0.16635, 1.88396, -3.3895);
            \coordinate (p60) at (4.99973, 2.99984, -4.50528);
            \coordinate (p61) at (4.99994, 1.88368, -3.38961);
            \coordinate (p62) at (4.9999, 1.88366, -3.38958);
            \coordinate (p63) at (-0.506798, -0.730982, 1.87138);
            \coordinate (p64) at (-0.507621, -0.720223, 1.85109);
            \coordinate (p65) at (-0.520484, -0.717345, 1.86832);
            \coordinate (p66) at (-0.00331101, 1.80668, -3.23505);
            \coordinate (p67) at (-0.123003, 1.46077, -2.54344);
            \coordinate (p68) at (-0.128148, 1.46191, -2.54037);
            \coordinate (p69) at (-4.23793, 3.00021, 1.86844);
            \coordinate (p70) at (-0.430234, -0.350227, 1.08405);
            \coordinate (p71) at (-0.12821, 1.46157, -2.53927);
            \coordinate (p72) at (-4.99976, 2.99985, 4.91738);
            \coordinate (p73) at (-5.00047, 1.7298, 9.99981);
            
            \draw[color of colormap={434.29/1.0228}, facet]
                    (p0) -- (p1) -- (p2) -- (p3) -- (p4)
                -- cycle;
            \draw[color of colormap={1022.8/1.0228}, facet]
                    (p5) -- (p6) -- (p7) -- (p12) -- (p11) -- (p2) -- (p1) -- (p8) -- (p10) -- (p9) -- (p13)
                -- cycle;
            \draw[color of colormap={538.371/1.0228}, facet]
                    (p14) -- (p15) -- (p16) -- (p0) -- (p1) -- (p8) -- (p17)
                -- cycle;
            \draw[color of colormap={100.926/1.0228}, facet]
                    (p18) -- (p20) -- (p21) -- (p19)
                -- cycle;
            \draw[color of colormap={110.654/1.0228}, facet]
                    (p22) -- (p23) -- (p24)
                -- cycle;
            \draw[color of colormap={91.713/1.0228}, facet]
                    (p22) -- (p23) -- (p25) -- (p26) -- (p27)
                -- cycle;
            \draw[color of colormap={111.369/1.0228}, facet]
                    (p28) -- (p29) -- (p30)
                -- cycle;
            \draw[color of colormap={151.067/1.0228}, facet]
                    (p22) -- (p24) -- (p32) -- (p33) -- (p31) -- (p34) -- (p27)
                -- cycle;
            \draw[color of colormap={299.187/1.0228}, facet]
                    (p35) -- (p36) -- (p37)
                -- cycle;
            \draw[color of colormap={125.322/1.0228}, facet]
                    (p20) -- (p21) -- (p28) -- (p30) -- (p39) -- (p38)
                -- cycle;
            \draw[color of colormap={139.235/1.0228}, facet]
                    (p26) -- (p40) -- (p41) -- (p42) -- (p34) -- (p27)
                -- cycle;
            \draw[color of colormap={145.295/1.0228}, facet]
                    (p38) -- (p43) -- (p35) -- (p36) -- (p39)
                -- cycle;
            \draw[color of colormap={232.507/1.0228}, facet]
                    (p31) -- (p44) -- (p46) -- (p45) -- (p47) -- (p42) -- (p34)
                -- cycle;
            \draw[color of colormap={204.926/1.0228}, facet]
                    (p48) -- (p49) -- (p41) -- (p42) -- (p47) -- (p50)
                -- cycle;
            \draw[color of colormap={341.632/1.0228}, facet]
                    (p45) -- (p47) -- (p50) -- (p52) -- (p51)
                -- cycle;
            \draw[color of colormap={250.662/1.0228}, facet]
                    (p53) -- (p54) -- (p55)
                -- cycle;
            \draw[color of colormap={85.092/1.0228}, facet]
                    (p25) -- (p26) -- (p40) -- (p56)
                -- cycle;
            \draw[color of colormap={155.744/1.0228}, facet]
                    (p18) -- (p20) -- (p38) -- (p43) -- (p54) -- (p53) -- (p57) -- (p58)
                -- cycle;
            \draw[color of colormap={153.409/1.0228}, facet]
                    (p58) -- (p59) -- (p62) -- (p61) -- (p60) -- (p19) -- (p18)
                -- cycle;
            \draw[color of colormap={263.009/1.0228}, facet]
                    (p33) -- (p63) -- (p64)
                -- (p65) -- (p32)
                -- cycle;
            \draw[color of colormap={360.732/1.0228}, facet]
                    (p54) -- (p55) -- (p66) -- (p67) -- (p68) -- (p37) -- (p35) -- (p43)
                -- cycle;
            \draw[color of colormap={309.386/1.0228}, facet]
                    (p29) -- (p30) -- (p39) -- (p36) -- (p37) -- (p68) -- (p71) -- (p70)
                -- (p65) -- (p69)
                -- cycle;
            \draw[color of colormap={97.395/1.0228}, facet]
                    (p24) -- (p23) -- (p25) -- (p56) -- (p73) -- (p72) -- (p69) -- (p65) -- (p32)
                -- cycle;

            \def\xmin{-5}   \def\ymin{-1.95} \def\zmin{-5.5}
            \def\xmax{5}   \def\ymax{3}   \def\zmax{10}
            
            \draw[grey, fill, opacity=.5]
                (p60) -- (p62) -- (p48) -- (p13) -- (p49) -- (\xmax,\ymin,\zmax) -- (\xmax,\ymin,\zmin) -- (\xmax,\ymax,\zmin) -- cycle;
            \draw[grey, fill, opacity=.5]
                (p49) -- (p41) -- (p40) -- (p56) -- (p73) -- (\xmin,\ymin,\zmax) -- (\xmax,\ymin,\zmax) -- cycle;
            \draw[grey, fill, opacity=.5]
                (p72) -- (p69) -- (p29) -- (p28) -- (p21) -- (p19) -- (p60) -- (\xmax,\ymax,\zmin) -- (\xmin,\ymax,\zmin) -- cycle;
            
        \end{axis}
    \end{tikzpicture}
    \caption{Two-derivative constraints on the 4-flavour NLO LECs $L_0^r,\ldots,L_3^r$ with $\Mphys=M_\pi$ and $\lambda=4$.
        The four possible selections of three LECs are shown, with the top left figure ($L_1^r,L_2^r,L_3^r$) being almost exactly the 4-flavour equivalent of \cref{fig:SU3-NLO}
        (the LEC and colour ranges differ slightly).
        No reference point is shown, and the colour information should not be considered very meaningful, as discussed in the text; therefore, the colour bar has been omitted to save space.}
    \label{fig:SU4-NLO}
\end{figure}
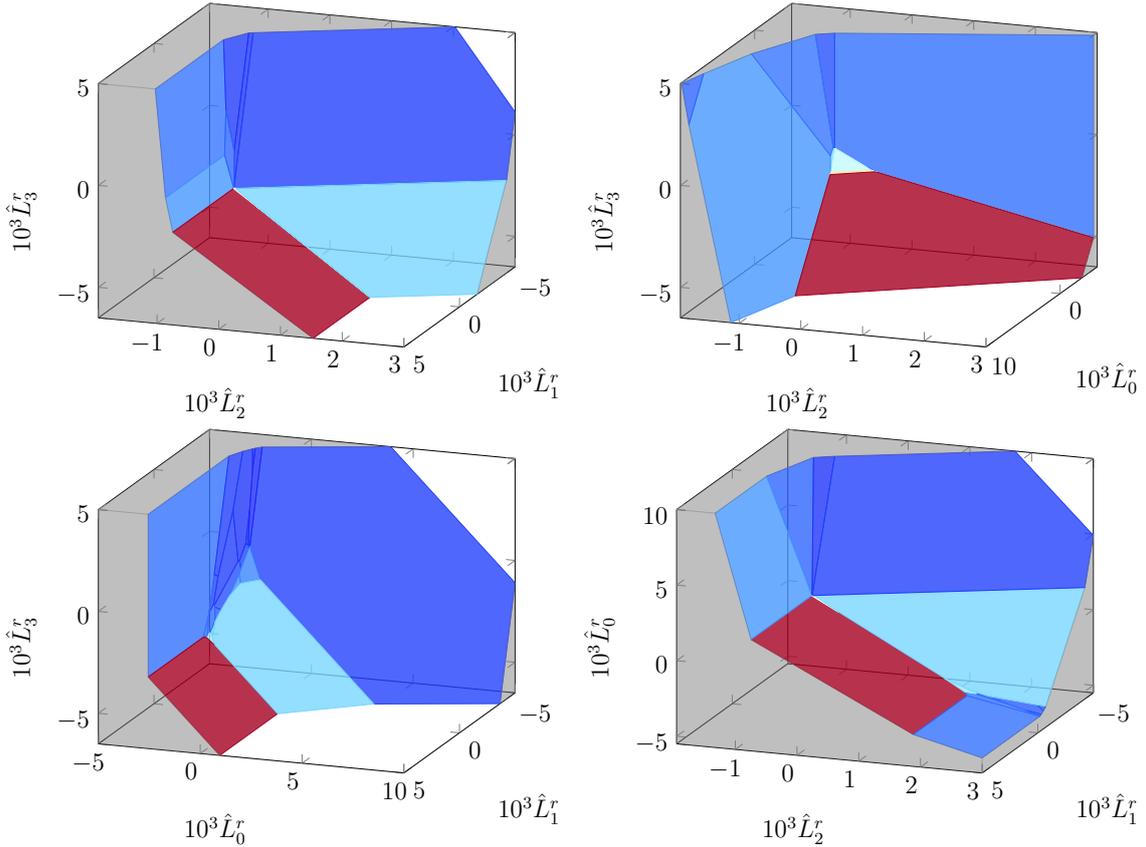

It turns out that the bounds change quite gradually between different $n\geq4$, to the extent that the difference between e.g.\ $n=4$ and $n=5$ is hardly visible upon first inspection of figures like those used here.
Therefore, we have chosen to only display $n=4$ (the ``leading'' high-flavour example) and $n=8$ (a reasonable ``very high flavour'' example, also of historical technicolour relevance \cite{Farhi:1979zx}).
Understandably, there are no experimental reference values for the high-flavour LECs, so to perform two- or three-dimensional slices of parameter space, we provisionally set $\hat L_i^r=L_i^r$ and use the 3-flavour reference values when possible, and use 0 as the reference value for e.g.\ $\hat L_0^r$.
We also retain the distance-to-reference-point colouring of the constraint surfaces to make them more visually distinguishable.
Of course, not too much meaning should be read into these distances.

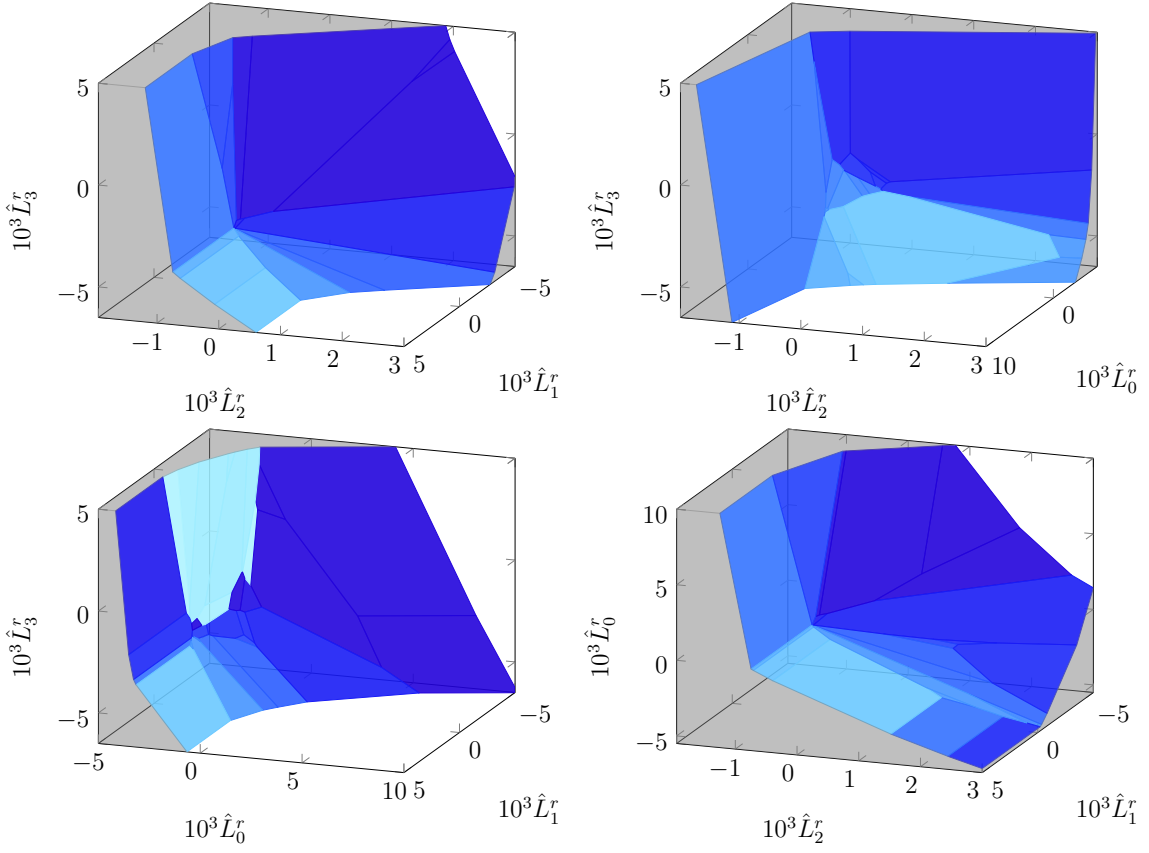
\begin{figure}[hbtp]
    \begin{tikzpicture}[scale=.8]
        \begin{axis}[
                view={-250}{20}, zlabel={$10^{3} \hat L_3^r$}, ylabel={$10^{3} \hat L_2^r$}, xlabel={$10^{3} \hat L_1^r$},
                colormap/temp, point meta min=400, point meta max=1000, 
                zmin=-6.5,zmax=5,ymin=-1.95,ymax=3
            ]

            \addplot3 [
                patch, 
                patch table with point meta={\resultpath/NLO/SU8/M0.135/D2/lam4/visualisation/L1L2L3/constr_0_table.dat}, 
                opacity=0] 
                    table {\resultpath/NLO/SU8/M0.135/D2/lam4/visualisation/L1L2L3/constr_0_coords.dat};

            \coordinate (p0) at (-1.01091, 1.60405, -6.50015);
            \coordinate (p1) at (-0.489418, -0.690414, -3.99664);
            \coordinate (p2) at (-0.391774, 1.14788, -6.49982);
            \coordinate (p3) at (-5.00021, 2.02062, 3.77336);
            \coordinate (p4) at (-4.17897, 1.47323, 2.45944);
            \coordinate (p5) at (-5.00003, 1.94235, 4.33649);
            \coordinate (p6) at (-1.42736, -0.93238, 4.99999);
            \coordinate (p7) at (-1.80322, -0.681633, 5.00025);
            \coordinate (p8) at (-1.21423, -0.89358, 2.83352);
            \coordinate (p9) at (-0.650577, -0.73246, -3.60846);
            \coordinate (p10) at (-0.734173, -0.681759, -3.55275);
            \coordinate (p11) at (5.00015, 0.600577, -6.49995);
            \coordinate (p12) at (4.99975, -0.0673759, -5.31274);
            \coordinate (p13) at (0.976855, 0.600568, -6.50005);
            \coordinate (p14) at (0.420181, -0.0673653, -5.31248);
            \coordinate (p15) at (-0.50601, -0.727189, -3.93004);
            \coordinate (p16) at (-0.489497, -0.743504, -3.92564);
            \coordinate (p17) at (-3.16838, 3.00036, -6.05972);
            \coordinate (p18) at (-2.75541, 3.00011, -6.50038);
            \coordinate (p19) at (5.0003, -0.625218, -4.19697);
            \coordinate (p20) at (-0.277116, -0.625073, -4.19692);
            \coordinate (p21) at (-1.30901, -0.249021, -3.2805);
            \coordinate (p22) at (-0.803508, -0.628097, -3.53323);
            \coordinate (p23) at (-5.00025, 1.87573, 5.00025);
            \coordinate (p24) at (5.00011, -0.744833, -3.92317);
            \coordinate (p25) at (-0.489209, -0.745079, -3.92333);
            \coordinate (p26) at (-0.491461, -0.745027, -3.91802);
            \coordinate (p27) at (0.77237, -1.1913, 5.00014);
            \coordinate (p28) at (-0.268641, -0.893651, -0.949429);
            \coordinate (p29) at (5.00012, -1.19105, 5.00012);
            \coordinate (p30) at (-0.489044, -0.74383, -3.92523);
            \coordinate (p31) at (-4.91909, 3.00012, -2.5582);
            \coordinate (p32) at (-5.00017, 2.99973, -2.10458);

            \draw[color of colormap={138.843/1.0228}, facet=.9]
                    (p0)-- (p1)-- (p2)
                -- cycle;
            \draw[color of colormap={03.255/1.0228}, facet=.9]
                    (p3)-- (p5)-- (p4)
                -- cycle;
            \draw[color of colormap={13.853/1.0228}, facet=.9]
                    (p6)-- (p8)-- (p9)-- (p10)-- (p7)
                -- cycle;
            \draw[color of colormap={283.502/1.0228}, facet=.9]
                    (p11)-- (p13)-- (p14)-- (p12)
                -- cycle;
            \draw[color of colormap={127.461/1.0228}, facet=.9]
                    (p0)-- (p1)-- (p16)-- (p15)-- (p17)-- (p18)
                -- cycle;
            \draw[color of colormap={271.569/1.0228}, facet=.9]
                    (p14)-- (p12)-- (p19)-- (p20)
                -- cycle;
            \draw[color of colormap={08.782/1.0228}, facet=.9]
                    (p21)-- (p22)-- (p10)-- (p7)-- (p23)-- (p5)-- (p4)
                -- cycle;
            \draw[color of colormap={156.906/1.0228}, facet=.9]
                    (p24)-- (p25)-- (p26)-- (p28)-- (p27)-- (p29)
                -- cycle;
            \draw[color of colormap={99.178/1.0228}, facet=.9]
                    (p6)-- (p8)-- (p28)-- (p27)
                -- cycle;
            \draw[color of colormap={201.596/1.0228}, facet=.9]
                    (p1)-- (p2)-- (p13)-- (p14)-- (p20)-- (p30)-- (p16)
                -- cycle;
            \draw[color of colormap={58.546/1.0228}, facet=.9]
                    (p17)-- (p15)-- (p22)-- (p21)-- (p31)
                -- cycle;
            \draw[color of colormap={95.385/1.0228}, facet=.9]
                    (p8)-- (p9)-- (p26)-- (p28)
                -- cycle;
            \draw[color of colormap={48.534/1.0228}, facet=.9]
                    (p15)-- (p16)-- (p30)-- (p25)-- (p26)-- (p9)-- (p10)-- (p22)
                -- cycle;
            \draw[color of colormap={06.804/1.0228}, facet=.9]
                    (p31)-- (p21)-- (p4)-- (p3)-- (p32)
                -- cycle;
            \draw[color of colormap={239.065/1.0228}, facet=.9]
                    (p19)-- (p20)-- (p30)-- (p25)-- (p24)
                -- cycle;
                
            \def\xmin{-5}   \def\ymin{-1.95} \def\zmin{-6.5}
            \def\xmax{5}   \def\ymax{3}   \def\zmax{5}

            \draw[grey, fill, opacity=.5]
                (p11) -- (p12) -- (p19) -- (p24) -- (p29) -- (\xmax,\ymin,\zmax) -- (\xmax,\ymin,\zmin) -- cycle;
            \draw[grey, fill, opacity=.5]
                (p29) -- (p27) -- (p6) -- (p7) -- (p23) -- (\xmin,\ymin,\zmax) -- (\xmax,\ymin,\zmax) -- cycle;
            \draw[grey, fill, opacity=.5]
                (p32) -- (p31) -- (p17) -- (p18) -- (\xmin,\ymax,\zmin) -- cycle;
            
        \end{axis}
    \end{tikzpicture}
    \begin{tikzpicture}[scale=.8]
        \begin{axis}[
                view={-250}{20}, zlabel={$10^{3} \hat L_3^r$}, ylabel={$10^{3} \hat L_2^r$}, xlabel={$10^{3} \hat L_0^r$},
                colormap/temp, point meta min=400, point meta max=1000, 
                zmin=-6.5,zmax=5,ymin=-1.95,ymax=3,xmin=-6.5,xmax=10
            ]

            \addplot3 [
                patch, 
                patch table with point meta={\resultpath/NLO/SU8/M0.135/D2/lam4/visualisation/L0L2L3/constr_0_table.dat}, 
                opacity=0] 
                    table {\resultpath/NLO/SU8/M0.135/D2/lam4/visualisation/L0L2L3/constr_0_coords.dat};
            
            \coordinate (p0) at (-2.16397, -0.203631, -1.75447);
            \coordinate (p1) at (-2.37714, 0.0362262, -2.28772);
            \coordinate (p2) at (-2.16395, -0.119793, -2.42546);
            \coordinate (p3) at (-2.23913, -0.0413583, -2.56353);
            \coordinate (p4) at (-1.70592, -0.570691, -1.68105);
            \coordinate (p5) at (-1.69427, -0.583131, -1.65452);
            \coordinate (p6) at (-1.70603, -0.277409, -2.85417);
            \coordinate (p7) at (-1.02479, -0.515682, -3.26287);
            \coordinate (p8) at (-0.597751, -0.702558, -3.36955);
            \coordinate (p9) at (-0.597769, -0.720204, -3.29922);
            \coordinate (p10) at (-4.22227, 3.00029, -5.54747);
            \coordinate (p11) at (-3.83744, 2.6792, -5.50508);
            \coordinate (p12) at (-4.83059, 3.00019, -4.77047);
            \coordinate (p13) at (-4.28015, 2.45791, -4.57007);
            \coordinate (p14) at (-1.90617, -0.372161, -2.01734);
            \coordinate (p15) at (-1.90647, -0.181499, -2.7797);
            \coordinate (p16) at (-2.25667, -0.00622722, -2.66297);
            \coordinate (p17) at (0.899645, 0.0204799, -6.50028);
            \coordinate (p18) at (0.593764, 0.192371, -6.49986);
            \coordinate (p19) at (0.899143, -0.231763, -5.9955);
            \coordinate (p20) at (-0.43726, -0.74602, -3.4631);
            \coordinate (p21) at (-0.273719, -0.745082, -3.64959);
            \coordinate (p22) at (-3.27019, 3.00023, -6.50023);
            \coordinate (p23) at (-0.852085, 1.18652, -6.50023);
            \coordinate (p24) at (1.23564, -0.158178, -6.49984);
            \coordinate (p25) at (-2.0222, -0.38962, -1.15234);
            \coordinate (p26) at (-1.92904, -0.500242, -0.849697);
            \coordinate (p27) at (-1.65222, -0.647268, -1.40336);
            \coordinate (p28) at (-0.797291, -0.771626, -2.59296);
            \coordinate (p29) at (-1.52233, -1.04341, 3.56897);
            \coordinate (p30) at (2.57713, -0.74466, -6.50013);
            \coordinate (p31) at (10, -1.11608, -6.49979);
            \coordinate (p32) at (-1.55164, -1.11329, 5.00016);
            \coordinate (p33) at (-0.519469, -0.749657, -3.3123);
            \coordinate (p34) at (10.0004, -1.6916, 5.00018);
            \coordinate (p35) at (-5.18286, 2.99969, -4.06666);
            \coordinate (p36) at (-2.60454, 0.363111, -2.89472);
            \coordinate (p37) at (-2.44807, 0.0893662, -2.21681);
            \coordinate (p38) at (-5.68178, 3.0003, -1.5696);
            \coordinate (p39) at (-1.52211, -0.748884, -1.14307);
            \coordinate (p40) at (-2.39695, -0.558971, 5.00009);
            \coordinate (p41) at (-6.27946, 3.00005, 5.00009);

            \draw[color of colormap={70.107/1.0228}, facet=.9]
                    (p0)-- (p2)-- (p3)-- (p1)
                -- cycle;
            \draw[color of colormap={190.877/1.0228}, facet=.9]
                    (p4)-- (p5)-- (p9)-- (p8)-- (p7)-- (p6)
                -- cycle;
            \draw[color of colormap={203.291/1.0228}, facet=.9]
                    (p10)-- (p11)-- (p13)-- (p12)
                -- cycle;
            \draw[color of colormap={171.917/1.0228}, facet=.9]
                    (p4)-- (p6)-- (p15)-- (p14)
                -- cycle;
            \draw[color of colormap={168.48/1.0228}, facet=.9]
                    (p2)-- (p3)-- (p16)-- (p15)-- (p14)
                -- cycle;
            \draw[color of colormap={271.569/1.0228}, facet=.9]
                    (p7)-- (p8)-- (p20)-- (p21)-- (p19)-- (p17)-- (p18)
                -- cycle;
            \draw[color of colormap={201.596/1.0228}, facet=.9]
                    (p11)-- (p10)-- (p22)-- (p23)
                -- cycle;
            \draw[color of colormap={254.331/1.0228}, facet=.9]
                    (p19)-- (p17)-- (p24)
                -- cycle;
            \draw[color of colormap={74.726/1.0228}, facet=.9]
                    (p0)-- (p2)-- (p14)-- (p4)-- (p5)-- (p27)-- (p26)-- (p25)
                -- cycle;
            \draw[color of colormap={156.906/1.0228}, facet=.9]
                    (p20)-- (p21)-- (p30)-- (p31)-- (p34)-- (p32)-- (p29)-- (p28)-- (p33)
                -- cycle;
            \draw[color of colormap={68.181/1.0228}, facet=.9]
                    (p1)-- (p3)-- (p16)-- (p36)-- (p35)-- (p38)-- (p37)
                -- cycle;
            \draw[color of colormap={152.971/1.0228}, facet=.9]
                    (p12)-- (p13)-- (p36)-- (p35)
                -- cycle;
            \draw[color of colormap={55.317/1.0228}, facet=.9]
                    (p26)-- (p27)-- (p39)-- (p29)-- (p32)-- (p40)
                -- cycle;
            \draw[color of colormap={283.502/1.0228}, facet=.9]
                    (p6)-- (p7)-- (p18)-- (p23)-- (p11)-- (p13)-- (p36)-- (p16)-- (p15)
                -- cycle;
            \draw[color of colormap={239.065/1.0228}, facet=.9]
                    (p19)-- (p21)-- (p30)-- (p24)
                -- cycle;
            \draw[color of colormap={42.56/1.0228}, facet=.9]
                    (p26)-- (p25)-- (p37)-- (p38)-- (p41)-- (p40)
                -- cycle;
            \draw[color of colormap={57.691/1.0228}, facet=.9]
                    (p0)-- (p1)-- (p37)-- (p25)
                -- cycle;
            \draw[color of colormap={171.963/1.0228}, facet=.9]
                    (p5)-- (p9)-- (p33)-- (p28)-- (p39)-- (p27)
                -- cycle;
            \draw[color of colormap={142.429/1.0228}, facet=.9]
                    (p39)-- (p28)-- (p29)
                -- cycle;
            \draw[color of colormap={188.542/1.0228}, facet=.9]
                    (p8)-- (p9)-- (p33)-- (p20)
                -- cycle;
                
            \def\xmin{-6.5} \def\ymin{-1.95} \def\zmin{-6.5}
            \def\xmax{10}   \def\ymax{3}   \def\zmax{5}

            \draw[grey, fill, opacity=.5]
                (p31) -- (p34) -- (\xmax,\ymin,\zmax) -- (\xmax,\ymin,\zmin) -- cycle;
            \draw[grey, fill, opacity=.5]
                (p34) -- (p32) -- (p40) -- (p41) -- (\xmin,\ymax,\zmax) -- (\xmin,\ymin,\zmax) -- (\xmax,\ymin,\zmax) -- cycle;
            \draw[grey, fill, opacity=.5]
                (p22) -- (p10) -- (p12) -- (p35) -- (p38) -- (p41) -- (\xmin,\ymax,\zmax) -- (\xmin,\ymax,\zmin) -- cycle;
            
        \end{axis}
    \end{tikzpicture}
    \begin{tikzpicture}[scale=.8]
        \begin{axis}[
                view={-250}{20}, zlabel={$10^{3} \hat L_3^r$}, ylabel={$10^{3} \hat L_0^r$}, xlabel={$10^{3} \hat L_1^r$},
                colormap/temp, point meta min=400, point meta max=1000, 
                zmin=-6.5,zmax=5,ymin=-5
            ]

            \addplot3 [
                patch, 
                patch table with point meta={\resultpath/NLO/SU8/M0.135/D2/lam4/visualisation/L1L0L3/constr_0_table.dat}, 
                opacity=0] 
                    table {\resultpath/NLO/SU8/M0.135/D2/lam4/visualisation/L1L0L3/constr_0_coords.dat};
            
            \coordinate (p0) at (-1.79742, -1.03775, -3.86416);
            \coordinate (p1) at (-1.77525, -0.905869, -4.0397);
            \coordinate (p2) at (-1.69805, -1.51477, -3.46659);
            \coordinate (p3) at (-1.41183, -1.77881, -3.55467);
            \coordinate (p4) at (-0.884784, -2.59342, -3.30185);
            \coordinate (p5) at (-0.802296, -2.75326, -3.22213);
            \coordinate (p6) at (-0.408895, -3.06265, -3.29247);
            \coordinate (p7) at (-0.396994, -2.94437, -3.42203);
            \coordinate (p8) at (-0.884739, 0.604626, -6.50016);
            \coordinate (p9) at (-0.396997, 0.133461, -6.49981);
            \coordinate (p10) at (-2.0528, -1.89119, -2.43905);
            \coordinate (p11) at (-1.95654, -2.03178, -2.43255);
            \coordinate (p12) at (-2.12771, -1.69859, -2.5423);
            \coordinate (p13) at (-2.5191, -1.62312, -1.37678);
            \coordinate (p14) at (-2.47557, -1.89121, -1.0296);
            \coordinate (p15) at (-2.51104, -1.08519, -2.38901);
            \coordinate (p16) at (-1.77509, 1.5545, -6.50015);
            \coordinate (p17) at (-1.22403, -2.73046, -2.62889);
            \coordinate (p18) at (-0.93305, -2.93857, -2.7954);
            \coordinate (p19) at (-0.909902, -2.96615, -2.786);
            \coordinate (p20) at (-0.0821756, -3.2177, -3.3166);
            \coordinate (p21) at (0.0918911, -3.02257, -3.60459);
            \coordinate (p22) at (-0.191775, -3.27586, -3.19977);
            \coordinate (p23) at (-0.317285, -3.22055, -3.18829);
            \coordinate (p24) at (-0.281247, -3.28805, -3.14046);
            \coordinate (p25) at (1.07927, -0.653654, -6.50012);
            \coordinate (p26) at (-3.17104, -1.75252, 2.55892);
            \coordinate (p27) at (-2.48229, -2.00595, -0.739085);
            \coordinate (p28) at (-3.1387, -1.62315, 2.09342);
            \coordinate (p29) at (-0.411047, -3.4766, -1.78487);
            \coordinate (p30) at (-0.185064, -3.92903, 4.99971);
            \coordinate (p31) at (-1.25931, -3.47695, 4.99994);
            \coordinate (p32) at (-0.408479, -3.33337, -2.91408);
            \coordinate (p33) at (-0.783268, -3.07429, -2.79684);
            \coordinate (p34) at (-3.72126, 2.9673, -2.90652);
            \coordinate (p35) at (-3.93401, 5.87173, -6.49971);
            \coordinate (p36) at (-3.67807, 4.1676, -5.3073);
            \coordinate (p37) at (-4.99999, 8.08188, -2.90598);
            \coordinate (p38) at (-5.00031, 10.0006, -6.26256);
            \coordinate (p39) at (-4.96623, 9.99964, -6.49977);
            \coordinate (p40) at (-3.4386, -0.423088, 1.61362);
            \coordinate (p41) at (-3.43879, -1.712, 4.19217);
            \coordinate (p42) at (-3.53395, -1.73611, 5.00022);
            \coordinate (p43) at (-4.99982, 4.1294, 4.99982);
            \coordinate (p44) at (-0.383895, -3.45167, -2.32242);
            \coordinate (p45) at (-0.192447, -3.51556, -2.00305);
            \coordinate (p46) at (0.76235, -4.15203, 5.0001);
            \coordinate (p47) at (-2.31519, -2.7048, 3.66964);
            \coordinate (p48) at (-2.48172, -2.70502, 5.00003);
            \coordinate (p49) at (-0.679146, -3.25644, -2.43264);
            \coordinate (p50) at (-1.0393, -3.06931, -1.92039);
            \coordinate (p51) at (-0.438976, -3.39432, -2.60933);
            \coordinate (p52) at (0.124735, -3.15389, -3.444);
            \coordinate (p53) at (4.99977, -3.27623, -3.19961);
            \coordinate (p54) at (4.99995, -3.15405, -3.44414);
            \coordinate (p55) at (-1.79921, -2.30453, -2.31945);
            \coordinate (p56) at (5.00024, -3.51509, -2.00293);
            \coordinate (p57) at (5.00034, -4.15247, 5.00034);
            \coordinate (p58) at (5.00014, -0.653688, -6.49986);
            \coordinate (p59) at (-2.45351, -2.03481, -0.767889);
            \coordinate (p60) at (-3.17459, -2.15484, 4.9998);
            \coordinate (p61) at (-2.03496, -2.28093, -1.58849);
            \coordinate (p62) at (-2.74304, -2.51492, 4.99992);

            \draw[color of colormap={99.891/1.0228}, facet=.9]
                    (p0)-- (p2)-- (p3)-- (p1)
                -- cycle;
            \draw[color of colormap={138.843/1.0228}, facet=.9]
                    (p4)-- (p8)-- (p9)-- (p7)-- (p6)-- (p5)
                -- cycle;
            \draw[color of colormap={73.374/1.0228}, facet=.9]
                    (p0)-- (p2)-- (p11)-- (p10)-- (p12)
                -- cycle;
            \draw[color of colormap={25.805/1.0228}, facet=.9]
                    (p10)-- (p12)-- (p15)-- (p13)-- (p14)
                -- cycle;
            \draw[color of colormap={127.461/1.0228}, facet=.9]
                    (p1)-- (p3)-- (p4)-- (p8)-- (p16)
                -- cycle;
            \draw[color of colormap={31.828/1.0228}, facet=.9]
                    (p17)-- (p18)-- (p19)
                -- cycle;
            \draw[color of colormap={201.596/1.0228}, facet=.9]
                    (p9)-- (p7)-- (p23)-- (p24)-- (p22)-- (p20)-- (p21)-- (p25)
                -- cycle;
            \draw[color of colormap={398.874/1.0228}, facet=.9]
                    (p13)-- (p14)-- (p27)-- (p26)-- (p28)
                -- cycle;
            \draw[color of colormap={377.201/1.0228}, facet=.9]
                    (p29)-- (p30)-- (p31)
                -- cycle;
            \draw[color of colormap={86.333/1.0228}, facet=.9]
                    (p5)-- (p6)-- (p23)-- (p24)-- (p32)-- (p33)-- (p19)-- (p18)
                -- cycle;
            \draw[color of colormap={03.945/1.0228}, facet=.9]
                    (p34)-- (p37)-- (p38)-- (p39)-- (p35)-- (p36)
                -- cycle;
            \draw[color of colormap={08.782/1.0228}, facet=.9]
                    (p40)-- (p41)-- (p42)-- (p43)-- (p37)-- (p34)
                -- cycle;
            \draw[color of colormap={03.255/1.0228}, facet=.9]
                    (p40)-- (p41)-- (p26)-- (p28)
                -- cycle;
            \draw[color of colormap={397.433/1.0228}, facet=.9]
                    (p44)-- (p45)-- (p46)-- (p30)-- (p29)
                -- cycle;
            \draw[color of colormap={370.667/1.0228}, facet=.9]
                    (p47)-- (p48)-- (p31)-- (p29)-- (p44)-- (p51)-- (p49)-- (p50)
                -- cycle;
            \draw[color of colormap={152.971/1.0228}, facet=.9]
                    (p52)-- (p20)-- (p22)-- (p53)-- (p54)
                -- cycle;
            \draw[color of colormap={72.901/1.0228}, facet=.9]
                    (p2)-- (p3)-- (p4)-- (p5)-- (p18)-- (p17)-- (p55)-- (p11)
                -- cycle;
            \draw[color of colormap={42.56/1.0228}, facet=.9]
                    (p45)-- (p56)-- (p57)-- (p46)
                -- cycle;
            \draw[color of colormap={283.502/1.0228}, facet=.9]
                    (p52)-- (p21)-- (p25)-- (p58)-- (p54)
                -- cycle;
            \draw[color of colormap={06.804/1.0228}, facet=.9]
                    (p13)-- (p15)-- (p36)-- (p34)-- (p40)-- (p28)
                -- cycle;
            \draw[color of colormap={368.885/1.0228}, facet=.9]
                    (p27)-- (p59)-- (p60)-- (p42)-- (p41)-- (p26)
                -- cycle;
            \draw[color of colormap={58.546/1.0228}, facet=.9]
                    (p0)-- (p1)-- (p16)-- (p35)-- (p36)-- (p15)-- (p12)
                -- cycle;
            \draw[color of colormap={68.181/1.0228}, facet=.9]
                    (p32)-- (p51)-- (p44)-- (p45)-- (p56)-- (p53)-- (p22)-- (p24)
                -- cycle;
            \draw[color of colormap={384.26/1.0228}, facet=.9]
                    (p17)-- (p19)-- (p33)-- (p49)-- (p50)-- (p61)-- (p55)
                -- cycle;
            \draw[color of colormap={174.979/1.0228}, facet=.9]
                    (p6)-- (p7)-- (p23)
                -- cycle;
            \draw[color of colormap={367.325/1.0228}, facet=.9]
                    (p59)-- (p61)-- (p50)-- (p47)-- (p62)-- (p60)
                -- cycle;
            \draw[color of colormap={09.251/1.0228}, facet=.9]
                    (p49)-- (p33)-- (p32)-- (p51)
                -- cycle;
            \draw[color of colormap={203.291/1.0228}, facet=.9]
                    (p52)-- (p20)-- (p21)
                -- cycle;
            \draw[color of colormap={365.764/1.0228}, facet=.9]
                    (p47)-- (p62)-- (p48)
                -- cycle;
            \draw[color of colormap={10.163/1.0228}, facet=.9]
                    (p10)-- (p11)-- (p55)-- (p61)-- (p59)-- (p27)-- (p14)
                -- cycle;
 
            \def\xmin{-5} \def\ymin{-5} \def\zmin{-6.5}
            \def\xmax{5}   \def\ymax{10}   \def\zmax{5}
            
            \draw[grey, fill, opacity=.5]
                (p58) -- (p54) -- (p53) -- (p56) -- (p57) -- (\xmax,\ymin,\zmax) -- (\xmax,\ymin,\zmin) -- cycle;
            \draw[grey, fill, opacity=.5]
                (p57) -- (p46) -- (p30) -- (p31) -- (p48) -- (p62) -- (p60) -- (p42) -- (p43) -- (\xmin,\ymin,\zmax) -- (\xmax,\ymin,\zmax) -- cycle;
            
        \end{axis}
    \end{tikzpicture}
    \begin{tikzpicture}[scale=.8]
        \begin{axis}[
                view={-250}{20}, zlabel={$10^{3} \hat L_0^r$}, ylabel={$10^{3} \hat L_2^r$}, xlabel={$10^{3} \hat L_1^r$},
                colormap/temp, point meta min=400, point meta max=1000, 
                zmin=-5.5,zmax=10,ymin=-1.95,ymax=3
            ]

            \addplot3 [
                patch, 
                patch table with point meta={\resultpath/NLO/SU8/M0.135/D2/lam4/visualisation/L1L2L0/constr_0_table.dat}, 
                opacity=0] 
                    table {\resultpath/NLO/SU8/M0.135/D2/lam4/visualisation/L1L2L0/constr_0_coords.dat};
            
            \coordinate (p0) at (-1.94116, 1.35933, -1.3381);
            \coordinate (p1) at (-1.75343, 0.984084, -1.03771);
            \coordinate (p2) at (-2.50391, 1.88475, -1.33817);
            \coordinate (p3) at (-0.805739, -0.681578, 0.420193);
            \coordinate (p4) at (-0.947055, -0.58018, 0.478123);
            \coordinate (p5) at (-0.71765, -0.727347, 0.341748);
            \coordinate (p6) at (-0.50588, -0.727174, -0.110156);
            \coordinate (p7) at (-0.491264, -0.745203, -0.0981451);
            \coordinate (p8) at (-0.489226, -0.744919, -0.103174);
            \coordinate (p9) at (-0.489216, -0.744046, -0.105268);
            \coordinate (p10) at (-0.489569, -0.743542, -0.105785);
            \coordinate (p11) at (0.14252, 2.03832, -4.15159);
            \coordinate (p12) at (0.453536, 1.51966, -3.53);
            \coordinate (p13) at (0.207314, 1.23482, -3.11526);
            \coordinate (p14) at (-1.74313, 1.31532, -1.51382);
            \coordinate (p15) at (-1.22562, 0.280444, -0.685928);
            \coordinate (p16) at (-0.215025, 2.99997, -5.2323);
            \coordinate (p17) at (-0.135351, 2.79722, -5.01592);
            \coordinate (p18) at (-0.348186, 3.00014, -5.14907);
            \coordinate (p19) at (-0.35557, 1.8288, -3.60685);
            \coordinate (p20) at (-3.13237, -0.727315, 10.0001);
            \coordinate (p21) at (-3.20075, -0.681606, 9.99997);
            \coordinate (p22) at (5.00005, 2.44495, -4.6399);
            \coordinate (p23) at (0.0407556, 2.44534, -4.64047);
            \coordinate (p24) at (4.99977, 2.99971, -5.23166);
            \coordinate (p25) at (-0.48957, -0.690517, -0.176477);
            \coordinate (p26) at (-1.11657, 2.06729, -3.18505);
            \coordinate (p27) at (-0.836712, 3.00014, -4.6775);
            \coordinate (p28) at (4.99978, -0.707468, -0.188439);
            \coordinate (p29) at (-0.424272, -0.707627, -0.188396);
            \coordinate (p30) at (-0.162769, -0.39386, -0.746298);
            \coordinate (p31) at (5.00022, -0.393842, -0.746408);
            \coordinate (p32) at (5.00012, -0.744785, -0.103147);
            \coordinate (p33) at (0.266148, -1.25009, 9.99969);
            \coordinate (p34) at (4.99951, -1.24988, 10.0001);
            \coordinate (p35) at (5.00026, 1.51982, -3.52963);
            \coordinate (p36) at (-3.50765, 2.99994, -1.56112);
            \coordinate (p37) at (-4.99979, 1.80538, 4.93451);
            \coordinate (p38) at (-2.92199, 0.593285, 2.51074);
            \coordinate (p39) at (-4.99982, 2.64711, 2.12965);
            \coordinate (p40) at (-1.84847, 0.020745, 1.07899);
            \coordinate (p41) at (-4.99987, 2.99992, 1.42373);
            \coordinate (p42) at (-4.79439, 0.593384, 9.99969);
            \coordinate (p43) at (-5.00026, 0.762702, 10.0005);
            
            \draw[color of colormap={73.374/1.0228}, facet=.9]
                    (p0)-- (p2)-- (p1)
                -- cycle;
            \draw[color of colormap={48.534/1.0228}, facet=.9]
                    (p3)-- (p5)-- (p7)-- (p8)-- (p9)-- (p10)-- (p6)-- (p4)
                -- cycle;
            \draw[color of colormap={203.291/1.0228}, facet=.9]
                    (p11)-- (p12)-- (p13)
                -- cycle;
            \draw[color of colormap={99.891/1.0228}, facet=.9]
                    (p0)-- (p1)-- (p15)-- (p14)
                -- cycle;
            \draw[color of colormap={174.979/1.0228}, facet=.9]
                    (p16)-- (p18)-- (p19)-- (p17)
                -- cycle;
            \draw[color of colormap={13.853/1.0228}, facet=.9]
                    (p3)-- (p5)-- (p20)-- (p21)
                -- cycle;
            \draw[color of colormap={68.181/1.0228}, facet=.9]
                    (p16)-- (p17)-- (p23)-- (p22)-- (p24)
                -- cycle;
            \draw[color of colormap={138.843/1.0228}, facet=.9]
                    (p19)-- (p18)-- (p27)-- (p26)-- (p25)
                -- cycle;
            \draw[color of colormap={127.461/1.0228}, facet=.9]
                    (p6)-- (p10)-- (p25)-- (p26)-- (p14)-- (p15)
                -- cycle;
            \draw[color of colormap={271.569/1.0228}, facet=.9]
                    (p28)-- (p29)-- (p30)-- (p31)
                -- cycle;
            \draw[color of colormap={156.906/1.0228}, facet=.9]
                    (p8)-- (p7)-- (p33)-- (p34)-- (p32)
                -- cycle;
            \draw[color of colormap={152.971/1.0228}, facet=.9]
                    (p11)-- (p12)-- (p35)-- (p22)-- (p23)
                -- cycle;
            \draw[color of colormap={72.901/1.0228}, facet=.9]
                    (p0)-- (p2)-- (p36)-- (p27)-- (p26)-- (p14)
                -- cycle;
            \draw[color of colormap={239.065/1.0228}, facet=.9]
                    (p8)-- (p9)-- (p29)-- (p28)-- (p32)
                -- cycle;
            \draw[color of colormap={06.804/1.0228}, facet=.9]
                    (p37)-- (p38)-- (p40)-- (p39)
                -- cycle;
            \draw[color of colormap={283.502/1.0228}, facet=.9]
                    (p13)-- (p12)-- (p35)-- (p31)-- (p30)
                -- cycle;
            \draw[color of colormap={201.596/1.0228}, facet=.9]
                    (p9)-- (p10)-- (p25)-- (p19)-- (p17)-- (p23)-- (p11)-- (p13)-- (p30)-- (p29)
                -- cycle;
            \draw[color of colormap={95.385/1.0228}, facet=.9]
                    (p5)-- (p7)-- (p33)-- (p20)
                -- cycle;
            \draw[color of colormap={58.546/1.0228}, facet=.9]
                    (p6)-- (p4)-- (p40)-- (p39)-- (p41)-- (p36)-- (p2)-- (p1)-- (p15)
                -- cycle;
            \draw[color of colormap={08.782/1.0228}, facet=.9]
                    (p3)-- (p4)-- (p40)-- (p38)-- (p42)-- (p21)
                -- cycle;
            \draw[color of colormap={03.945/1.0228}, facet=.9]
                    (p37)-- (p38)-- (p42)-- (p43)
                -- cycle;
                
            \def\xmin{-5}   \def\ymin{-1.95} \def\zmin{-5.5}
            \def\xmax{5}   \def\ymax{3}   \def\zmax{10}

            \draw[grey, fill, opacity=.5]
                (p24) -- (p22) -- (p35) -- (p31) -- (p28) -- (p32) -- (p34) -- (\xmax,\ymin,\zmax) -- (\xmax,\ymin,\zmin) -- (\xmax,\ymax,\zmin) -- cycle;
            \draw[grey, fill, opacity=.5]
                (p34) -- (p33) -- (p20) -- (p21) -- (p42) -- (p43) -- (\xmin,\ymin,\zmax) -- (\xmax,\ymin,\zmax) -- cycle;
            \draw[grey, fill, opacity=.5]
                (p24) -- (p16) -- (p18) -- (p27) -- (p36) -- (p41) -- (\xmin,\ymax,\zmin) -- (\xmax,\ymax,\zmin) -- cycle;
            
        \end{axis}
    \end{tikzpicture}
    \caption{The exact 8-flavour equivalent of \cref{fig:SU4-NLO}.}
    \label{fig:SU8-NLO}
\end{figure}

Figures~\ref{fig:SU4-NLO} ($n=4$) and \ref{fig:SU8-NLO} ($n=8$) show the basic NLO bounds, similarly to \cref{fig:SU3-NLO}, to which the bounds are qualitatively similar.
There is a trend towards weaker constraints as $n$ increases, as can be seen from the amplitude: many of the most important terms go as powers of $1/n$ (cf.\ \cref{eq:rescale}; all constraints have $c=-1$).
Note how we wholly abandon the $\Mphys$ debate and use $M_\pi$ throughout.

\begin{figure}[hbtp]
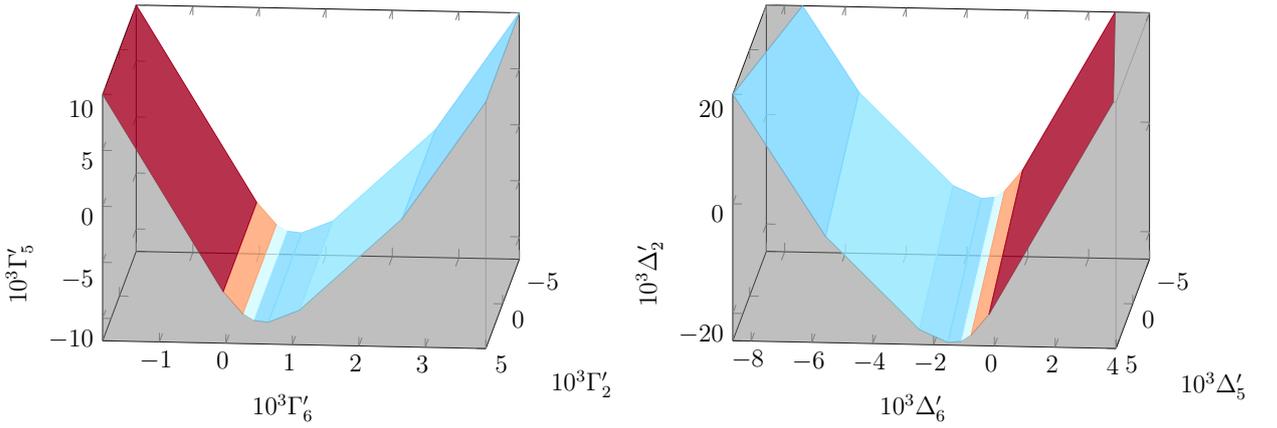


    \caption{
        NNLO 4-flavour bounds on $\Gamma'_i$ (left) and $\Delta'_i$ (right), analogous to \cref{fig:SU2-Thetas}.
        Note the different scales on the axes; $\Gamma'_6$ and $\Delta'_6$ are bounded by almost parallel faces.}
    \label{fig:SU4-GD}
\end{figure}

\begin{figure}[hbtp]
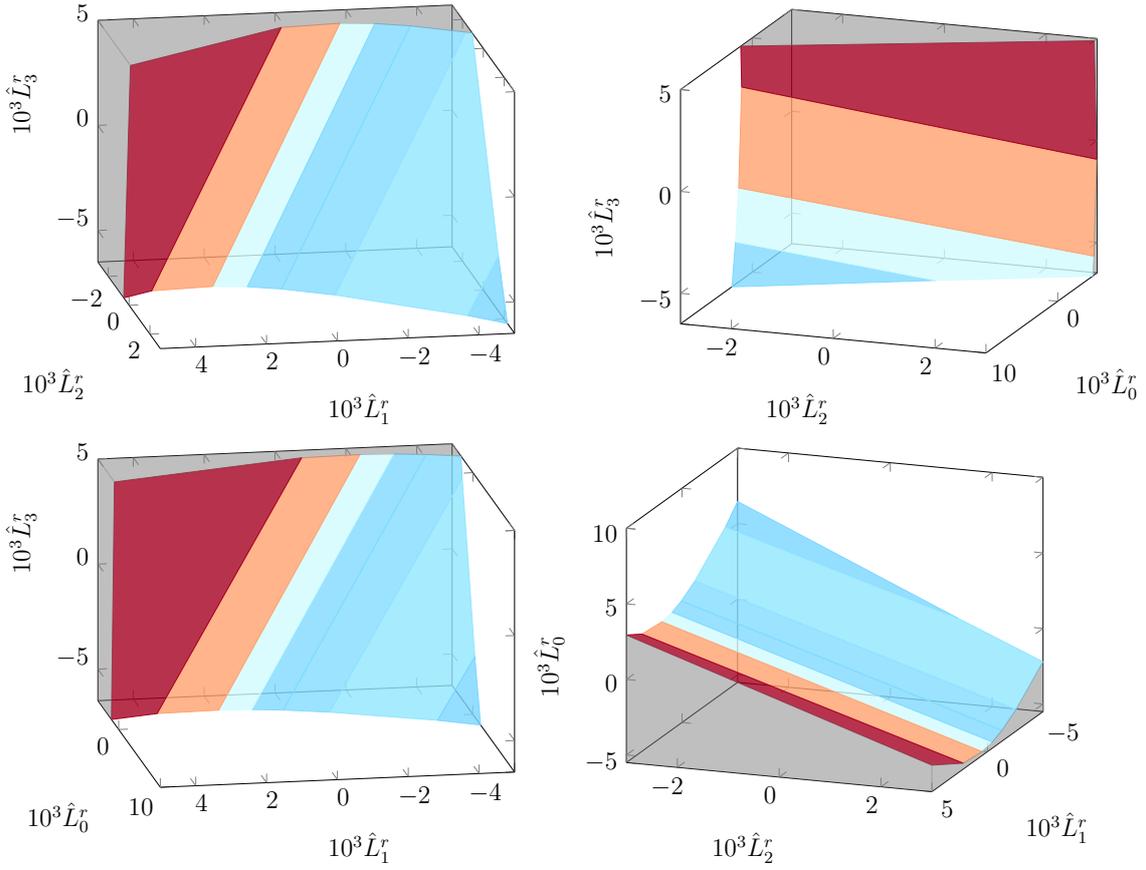

    %
    \caption{
        NNLO 4-flavour bounds on $L_0^r,\ldots,L_3^r$ presented in the same way as \cref{fig:SU4-NLO}, but with some axes rotated to give a better view.
        See \cref{fig:SU3-L123} for the 3-flavour counterpart 
        (no analogous 4-derivative version has been produced due to issues with high-dimensional parameter space).}
    \label{fig:SU4-L0123}
\end{figure}

\begin{figure}[hbtp]
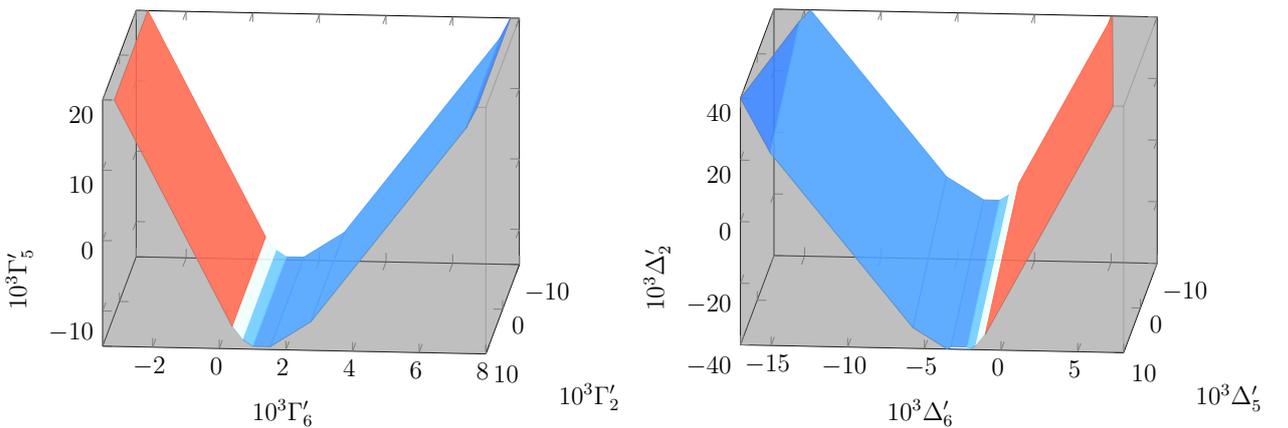

    %
    \caption{The 8-flavour equivalent of \cref{fig:SU4-GD}.
        The axes have been zoomed out to adequately view the weaker constraints.}
    \label{fig:SU8-GD}
\end{figure}

\begin{figure}[hbtp]
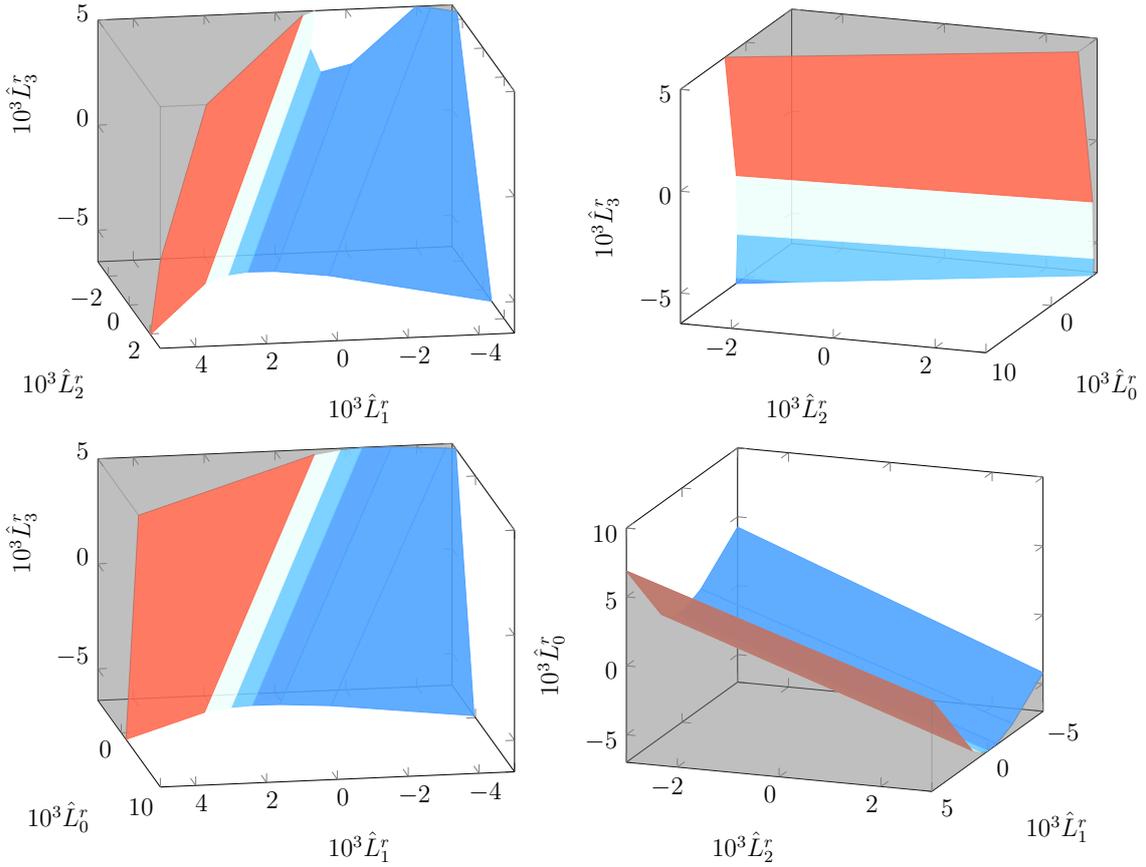

    %
    \caption{The exact 8-flavour equivalent of \cref{fig:SU4-L0123}.}
    \label{fig:SU8-L0123}
\end{figure}

Figures~\ref{fig:SU4-GD},~\ref{fig:SU4-L0123} ($n=4$) and \ref{fig:SU8-GD},~\ref{fig:SU8-L0123} ($n=8$) show the most interesting NNLO bounds, which are again similar to their lower-flavour analogues.
The bounds on $\hat L_4^r,\hat L_5^r,\hat L_6^r$ and $\hat L_8^r$, which are even weaker than the corresponding ones for $n=3$, are not shown.
Likewise, the bounds on $\Gamma_4,\Delta_4$ are very weak and have been omitted.
Note how the prominent lower bound on $\hat L_1^r$ in \cref{fig:SU8-L0123} breaks the trend of weaker bounds at larger $n$.


\FloatBarrier

Finally, one may ask what happens in the limit $n\to\infty$.
Looking at the amplitudes in \cite{Bijnens:2011fm}, we see that they are independent of the LECs at leading order in $n$: at NLO, the amplitude is $\O(n)$ whereas the LEC parts are $\O(1)$, while at NNLO they are $\O(n^2)$ and $\O(n)$, respectively (this of course indicates convergence problems at high $n$, in agreement with \cite{Chivukula:1992gi}).
Thus, the bounds, expressed in the schematic form of \cref{eq:constr-LEC}, will eventually be dominated by $\gamma$, and will therefore asymptotically tend towards either $\constr{\v 0}{1}$ or the trivial $\constr{\v 0}{-1}$; the gradual weakening in \cref{fig:SU4-NLO,fig:SU8-NLO,fig:SU4-GD,fig:SU4-L0123,fig:SU8-GD,fig:SU8-L0123} suggests the latter.

\FloatBarrier

\subsection{Considerations about the integrals}
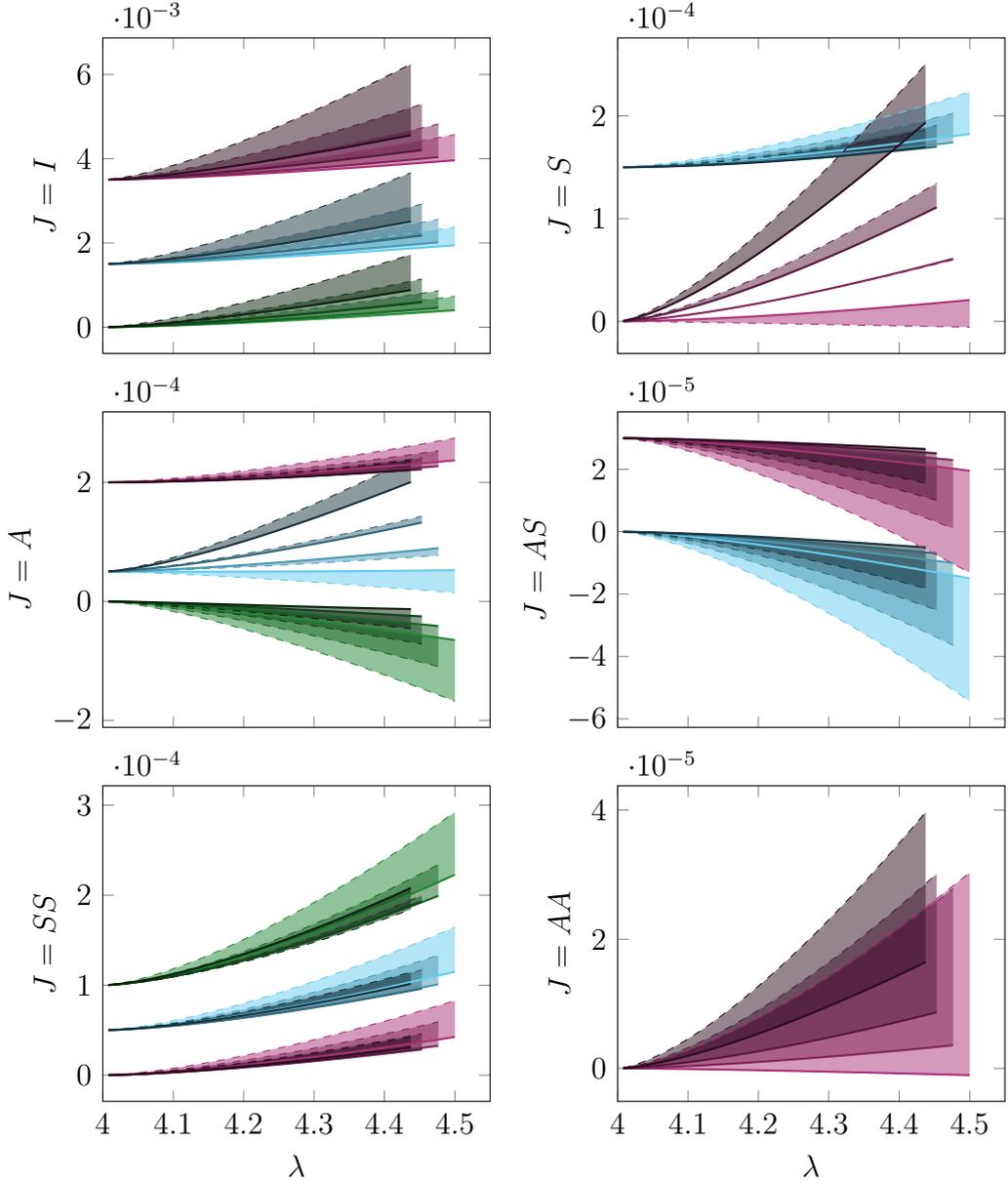
\begin{figure}[p]
    \centering
    \vspace{-1cm}
    \begin{subfigure}[b]{0.45\textwidth}
        \begin{flushright}
            \begin{tikzpicture}
                \begin{axis}[width=\linewidth,
                        xmin=4,xmax=4.55,xtick={4,4.1,...,4.6},
                        xticklabels={},
                        xlabel={},ylabel={$J=I$}]
                        
                    \plotseries{SU2}{DI}{0}

                    \plotseries{SU3}{DI}{0.0015}

                    \plotseries{SU4}{DI}{0.0035}
                \end{axis}
            \end{tikzpicture}
        \end{flushright}
    \end{subfigure}
    \begin{subfigure}[b]{0.45\textwidth}
        \begin{flushright}
            \begin{tikzpicture}
                \begin{axis}[width=\linewidth,
                        xmin=4,xmax=4.55,
                        xtick={4,4.1,...,4.6},xticklabels={},
                        xlabel={},ylabel={$J=S$}]     
                        
                    \plotseries{SU3}{DS}{0.00015}

                    \plotseries{SU4}{DS}{0}
                \end{axis}
            \end{tikzpicture}
        \end{flushright}
    \end{subfigure}\\
    \begin{subfigure}[b]{0.45\textwidth}
        \begin{flushright}
                \begin{tikzpicture}
                    \begin{axis}[width=\linewidth,
                            xmin=4,xmax=4.55,xtick={4,4.1,...,4.6},
                            xticklabels={},
                            xlabel={},ylabel={$J=A$}]
                            
                        \plotseries{SU2}{DA}{0}

                        \plotseries{SU3}{DA}{0.00005}

                        \plotseries{SU4}{DA}{0.0002}
                    \end{axis}
                \end{tikzpicture}
        \end{flushright}
    \end{subfigure}
    \begin{subfigure}[b]{0.45\textwidth}
        \begin{flushright}
            \begin{tikzpicture}
                    \begin{axis}[width=\linewidth,
                            xmin=4,xmax=4.55,xtick={4,4.1,...,4.6},
                            xticklabels={},
                            xlabel={},ylabel={$J=AS$}]
                            
                        \plotseries{SU3}{DAS}{0}

                        \plotseries{SU4}{DAS}{0.00003}
                    \end{axis}
                \end{tikzpicture}
        \end{flushright}
    \end{subfigure}\\
    \begin{subfigure}[b]{0.45\textwidth}
        \begin{flushright}
            \begin{tikzpicture}
                    \begin{axis}[width=\linewidth,
                            xmin=4,xmax=4.55,xtick={4,4.1,...,4.6},
                            xticklabels={4,4.1,4.2,4.3,4.4,4.5},
                            xlabel={$\lambda$},ylabel={$J=SS$}]
                            
                        \plotseries{SU2}{DSS}{0.00010}

                        \plotseries{SU3}{DSS}{0.00005}

                        \plotseries{SU4}{DSS}{0.00000}
                    \end{axis}
                \end{tikzpicture}
        \end{flushright}
    \end{subfigure}
    \begin{subfigure}[b]{0.45\textwidth}
        \begin{flushright}
            \begin{tikzpicture}
                    \begin{axis}[width=\linewidth,
                            xmin=4,xmax=4.55,xtick={4,4.1,...,4.6},
                            xticklabels={4,4.1,4.2,4.3,4.4,4.5},
                            xlabel={$\lambda$},ylabel={$J=AA$}]
                            
                        \plotseries{SU4}{DAA}{0}
                    \end{axis}
                \end{tikzpicture}
        \end{flushright}
    \end{subfigure}
    \caption[The integral as a function of $\lambda$.]{
        The integrals $\frac{1}{n^2}\big[D^J_k(s,t;\lambda)+(-1)^kC^{JI}D^I_k(u,t;\lambda)\big]$, defined in \cref{eq:mod-dispersion}, as functions of $\lambda$ between 4 and 4.5 at $k=2$, $t=4$, for the six possible $J$.
        The factor $1/n^2$ roughly cancels the $n$-dependence, making it possible to show all $n$ at the same vertical scale.
        Each function is represented for $n=2$ (\SUIIcolcol), $n=3$ (\SUIIIcolcol) and $n=4$ (\SUIVcolcol) \chpt, with the latter representing the general $n$-flavour case reasonably well.
        
        \qquad Different values of $s$ between 0 and 1.5 (the range of most relevant constraints) in increments of 0.5 are shown with the colour saturation (brighter colour = smaller $s$).
        The NLO result is drawn with solid lines, and the NNLO result (with the NLO LECs fixed to their reference values) is drawn with dashed lines.
        The difference between them is shaded, and the $\lambda$ ranges are staggered slightly to improve readability.
        
        \qquad The graphs have been vertically shifted to enhance readability; all integrals are equal to 0 at $\lambda=4$. 
        Note that the vertical scale differs greatly between different $J$.}
    \label{fig:integrals}
\end{figure}

\Cref{fig:integrals} demonstrates the integrals of the components of the amplitude over relevant $s,\lambda$ ranges.
We may note that despite the great complexity seen in \cref{sec:integrals}, the graphs are typically quite simple and qualitatively similar.
There is, however, a very wide range of magnitudes; typically, ``lower'' $J$ components have larger integrals.

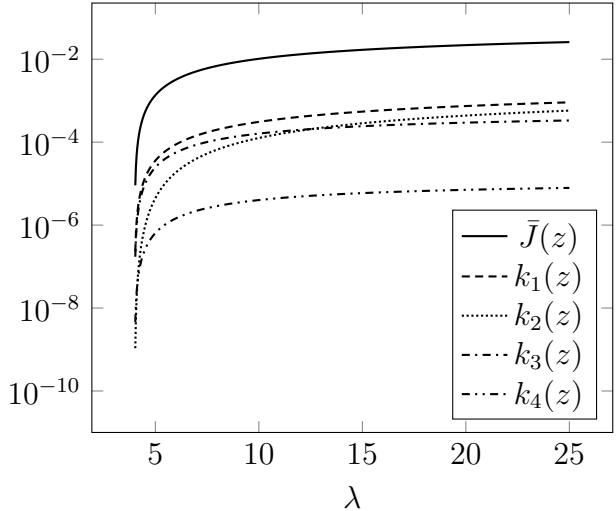
\begin{figure}[hbtp]
    \begin{minipage}[t]{0.6\textwidth}
        \raisebox{-\height}{
            \begin{tikzpicture}
                \begin{semilogyaxis}[xlabel={$\lambda$}, legend pos=south east, ymin=1e-11]
                    \addplot[no marks, thick]                       table[x=lambda, y=Jz0]  {sampled_integrals.dat};
                    \addplot[no marks, thick, densely dashed]       table[x=lambda, y=k1z0] {sampled_integrals.dat};
                    \addplot[no marks, thick, densely dotted]       table[x=lambda, y=k2z0] {sampled_integrals.dat};
                    \addplot[no marks, thick, dashdotted]           table[x=lambda, y=k3z0] {sampled_integrals.dat};
                    \addplot[no marks, thick, dashdotdotted]        table[x=lambda, y=k4z0] {sampled_integrals.dat};
                    
                    \legend{{$\bar J(z)$},{$ k_1(z)$},{$k_2(z)$},{$k_3(z)$},{$k_4(z)$}}
                \end{semilogyaxis}
            \end{tikzpicture}
        }
    \end{minipage}
    \begin{minipage}[t]{0.4\textwidth}
        \caption[Logarithmic plot of the relative size of the different loop integral functions.]{Logarithmic plot of the relative size of the different loop integral functions, covering a rather wide range of $\lambda$.
        Each line represents $\left|\int_4^\lambda X(z)\d z\right|$ for the given function $X$.
        Note how $\bar J\gg k_1\approx k_3 \gg k_2\approx k_4$ for small $\lambda$, which changes to $\bar J\gg k_1\approx k_2 \approx k_3\gg k_4$ for large $\lambda$.
        In either case, ``$\gg$'' is by about two orders of magnitude, and ``$\approx$'' is within one order of magnitude.}
        \label{fig:int-size}
    \end{minipage}
\end{figure}

It is interesting to note that the ratio between NLO and NNLO integrals is approximately constant in $\lambda$, varying only with $s$ and $J$.
This is perhaps unexpected since the NNLO integrand contains terms like $z^3X(z)$, where $X$ is one of $\bar J$ or $k_i$, whereas the NLO amplitude only contains $z^2X(z)$.
Therefore, we would expect the ratio to grow approximately linearly with $\lambda$.
However, the $z^3\bar J(z)$ terms are typically suppressed by small numerical coefficients or NLO LECs, and $\bar J(z)$ dominates $k_i(z)$, as shown in \cref{fig:int-size}.
Therefore, this effect does not manifest until $\lambda$ is much larger than the values relevant to this application, which in practice limits the importance of higher-order corrections to the bounds.

\FloatBarrier

\vspace{-.7cm} 
\subsection{Considerations about $a_J$}\label{sec:aJ-stats}
We have made two innovations in the treatment of $a_J$: 
\begin{enumerate}[label=(\roman*)]
    \item\label[improvement]{imp:aJ}
        not restricting $a_J$ to the physical eigenstates (this was done already in \cite{AlvarezThesis}),\\[-.8cm]
    \item\label[improvement]{imp:s}
        employing the fixed-$s$ (as opposed to all-$s$) constraints \cref{eq:main:cond} on $a_J$.
\end{enumerate}
This section investigates whether these changes actually give any improvements at all --- it would be conceivable that the physical eigenstates, which are allowed for all $s$, were special in a way that guarantees that they generate the strongest bounds.

To measure how significantly \cref{imp:aJ} is used, we consider $a_J$ as a point on the unit sphere and find the angle $\theta(a_J)$ between it and the closest eigenstate point.
Thus, larger $\theta$ indicates, in a sense, more use of \cref{imp:aJ}.
Similarly, we may measure \cref{imp:s} via the fraction of points on $s\in[-4,4]$ for which \cref{eq:main:cond} permits $a_J$.
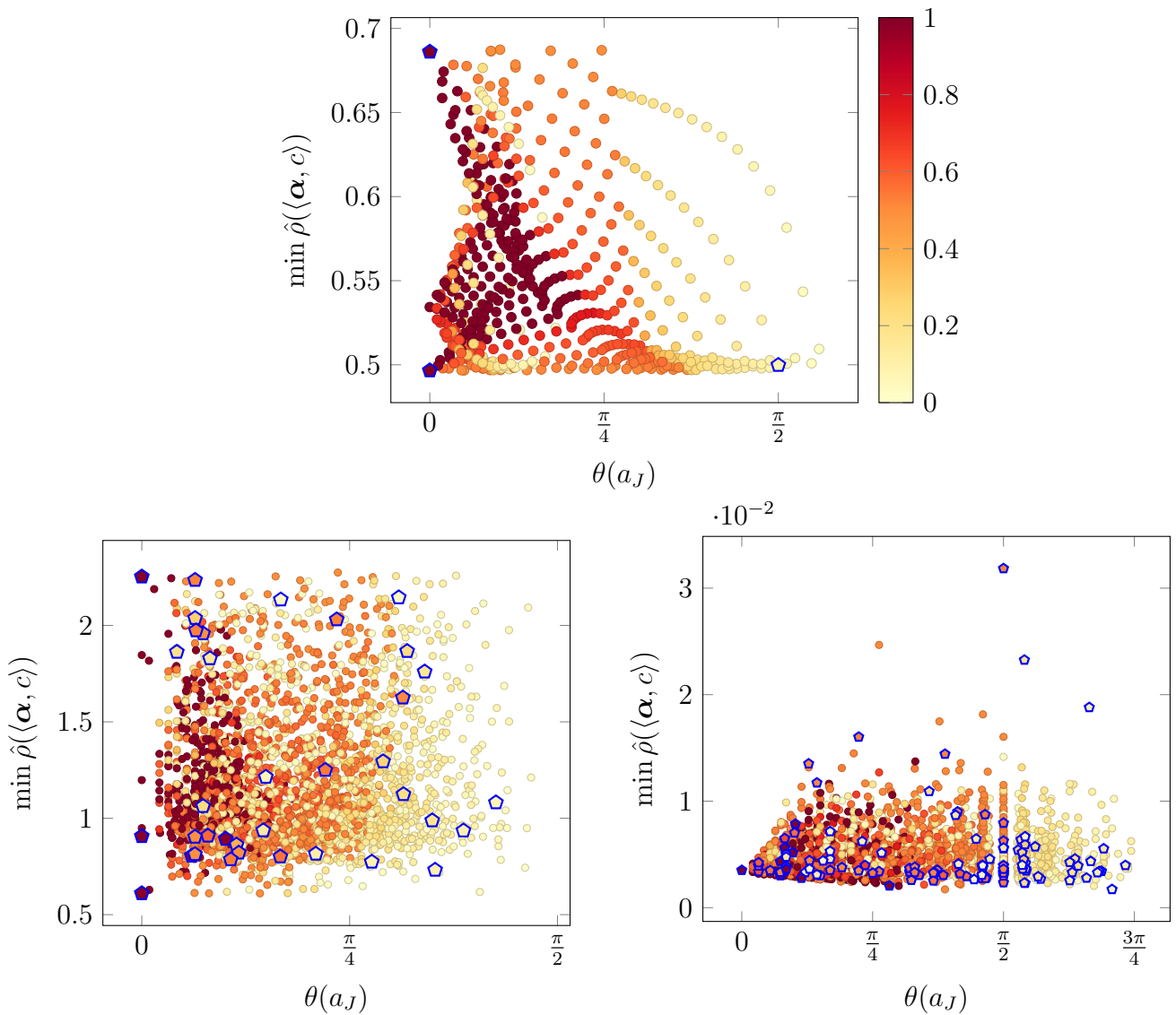
\begin{figure}[hbtp]
    \centering
            \begin{subfigure}[b]{\textwidth}
                \centering
                \begin{tikzpicture}
                    \begin{axis}[
                            xlabel={$\theta(a_J)$},
                            ylabel={$\min\hat\rho(\constr{\v\alpha}{c})$},
                            xtick={0,0.7854, 1.5708, 2.3562, 3.1415},
                            xticklabels={0,$\frac\pi4$, $\frac\pi2$, $\frac{3\pi}{4}$, $\pi$},
                            colormap/YlOrRd, colorbar
                        ]
                        
                        \addplot+[
                                scatter, only marks, mark=*,
                                point meta={\thisrow{count}/246}, point meta max=1, point meta min=0
                            ]
                            table[x=rdist, y=fdist] {\resultpath/NLO/SU2/M0.135/D2/lam4/constr_0.qhc_stats.dat};
                            
                        \addplot+[
                                scatter, thick, only marks, mark=\relmark, mark size=3pt,
                                point meta={\thisrow{count}/246}, point meta max=1, point meta min=0,
                                scatter/use mapped color={
                                    fill=mapped color,
                                    draw=blue,
                                },
                                discard if={relevant}{no}
                            ]
                            table[x=rdist, y=fdist] {\resultpath/NLO/SU2/M0.135/D2/lam4/constr_0.qhc_stats.dat};
                            
                    \end{axis}
                \end{tikzpicture}
            \end{subfigure}\\ 
            \hspace{-1.7cm}
            \begin{subfigure}[b]{0.5\textwidth}
                \begin{tikzpicture}
                    \begin{axis}[
                            xlabel={$\theta(a_J)$},
                            ylabel={$\min\hat\rho(\constr{\v\alpha}{c})$},
                            xtick={0,0.7854, 1.5708, 2.3562, 3.1415},
                            xticklabels={0,$\frac\pi4$, $\frac\pi2$, $\frac{3\pi}{4}$, $\pi$},
                            colormap/YlOrRd, 
                        ]

                        \addplot+[
                                scatter, only marks, mark=*, mark size=1.5pt,
                                point meta={\thisrow{count}/124}, point meta max=1, point meta min=0
                            ]
                            table[x=rdist, y=fdist] {/home/mssjo/Documents/papers/LECbounds/code/constrscan/scans/NLO/SU3/M0.135/D2/lam4/constr_0.qhc_stats.dat};
                            
                        \addplot+[
                                scatter, thick, only marks, mark=\relmark, mark size=3pt,
                                point meta={\thisrow{count}/124}, point meta max=1, point meta min=0,
                                scatter/use mapped color={
                                    fill=mapped color,
                                    draw=blue,
                                },
                                discard if={relevant}{no}
                            ]
                            table[x=rdist, y=fdist] {/home/mssjo/Documents/papers/LECbounds/code/constrscan/scans/NLO/SU3/M0.135/D2/lam4/constr_0.qhc_stats.dat};
                            
                    \end{axis}
                \end{tikzpicture}
            \end{subfigure}\hfill
            \begin{subfigure}[b]{0.5\textwidth}
                \begin{tikzpicture}
                    \begin{axis}[
                            xlabel={$\theta(a_J)$},
                            ylabel={$\min\hat\rho(\constr{\v\alpha}{c})$},
                            xtick={0,0.7854, 1.5708, 2.3562, 3.1415},
                            xticklabels={0,$\frac\pi4$, $\frac\pi2$, $\frac{3\pi}{4}$, $\pi$},
                            colormap/YlOrRd,
                        ]

                        \addplot+[
                                scatter, only marks, mark=*, mark size=1.5pt,
                                point meta={\thisrow{count}/124}, point meta max=1, point meta min=0
                            ]
                            table[x=rdist, y=fdist] {\resultpath/NLO/SU4/M0.135/D2/lam4/constr_0.qhc_stats.dat};
                            
                        \addplot+[
                                scatter, thick, only marks, mark=\relmark, mark size=2pt,
                                point meta={\thisrow{count}/124}, point meta max=1, point meta min=0,
                                scatter/use mapped color={
                                    fill=mapped color,
                                    draw=blue,
                                },
                                discard if={relevant}{no}
                            ]
                            table[x=rdist, y=fdist] {\resultpath/NLO/SU4/M0.135/D2/lam4/constr_0.qhc_stats.dat};
                            
                    \end{axis}
                \end{tikzpicture}
            \end{subfigure}
        \caption[The full set of $a_J$ permitted by \cref{eq:main:cond}.]{
            A discrete sample of the full set of $a_J$ permitted by \cref{eq:main:cond}, plotted over $\theta$ as defined in \cref{sec:aJ-stats}.
            For each $a_J$, the closest distance $\hat\rho$ between the reference point and any constraint generated at that $a_J$ is indicated with a dot.
            Relevant constraints are outlined with a blue pentagon.
            The colour of the dots indicate the fraction of the full range $s\in[-4,4]$ in which \cref{eq:main:cond} permits that $a_J$.
            \textbf{Top:}
            Two flavours.
            The eigenstates, i.e.\ the points at $\theta=0$, are, from top to bottom, $\pi^0\pi^+$, $\pi^0\pi^0$ and $\pi^+\pi^+$.
            \textbf{Lower left:}
            Three flavours.
            The eigenstates are, from top to bottom, $\pi^0\pi^+$, $K\eta$, $\pi^+\pi^+$, $K\pi^0$, $K^\pm\pi^\pm/K^0\pi^\pm/\pi^0\pi^0$  (overlapping) and $\eta\pi$.
            \textbf{Lower right:}
            Four flavours (representing $n>3$ flavours).
            Since high-flavour \chpt\ is not directly applicable to meson physics, we do not use the full set of ``physical'' eigenstates, but for $\theta$ to be defined, we retain a single eigenstate: $\pi^+\pi^+$, whose decomposition (consisting of the $SS$ component only) is uniquely flavour-independent.
            }
        \label{fig:aJ-stats}
\end{figure}

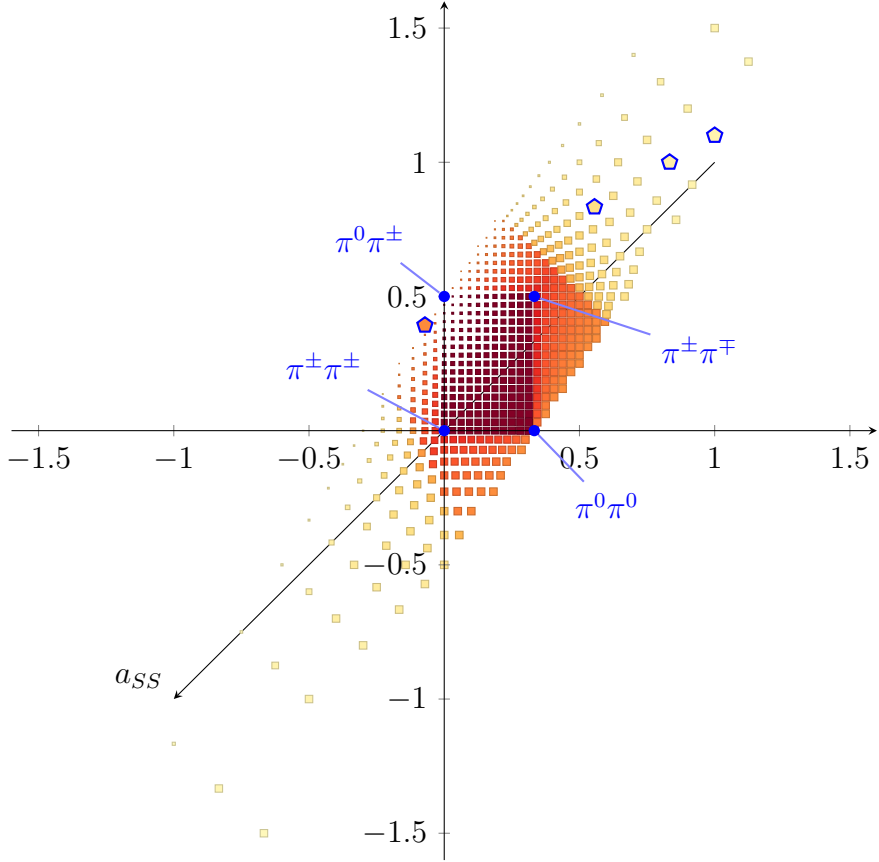
\begin{figure}[hbtp]
    \centering
    \begin{tikzpicture}
        \begin{axis}[
                scale=2, axis equal image,
                axis x line = middle, axis y line = middle,
                xmin=-1.6,xmax=1.6, ymin=-1.6,ymax=1.6,
                point meta min = 0,
                colormap/YlOrRd,
                xlabel={$a_{I}$}, ylabel={$a_A$},
                axis lines=none
            ]
            
            \draw[thin,-stealth] (1,1) -- (-1,-1) node[above left] {$a_{SS}$};
            
            \addplot[scatter, only marks, mark=square*,
                    point meta={\thisrow{count}/246}, point meta max=1, point meta min=0,
                    visualization depends on=\thisrow{fdist}\as\fdist,
                    scatter/@pre marker code/.append style={/tikz/mark size={(.7 - \fdist)*7}}
                ] 
                table [
                        x expr=\thisrow{aI}/(\thisrow{aI}+\thisrow{aA}+\thisrow{aSS}),
                        y expr=\thisrow{aA}/(\thisrow{aI}+\thisrow{aA}+\thisrow{aSS})
                    ] 
                    {../code/results/scans/NLO/SU2/M0.135/D2/lam4/constr_0.qhc_stats.dat};

            \addplot[scatter, thick, only marks, mark=\relmark, mark size=3pt,
                    point meta={\thisrow{count}/246}, point meta max=1, point meta min=0,
                    scatter/use mapped color={
                        fill=mapped color,
                        draw=blue,
                    }]
                table [
                        x expr=\thisrow{aI}/(\thisrow{aI}+\thisrow{aA}+\thisrow{aSS}),
                        y expr=\thisrow{aA}/(\thisrow{aI}+\thisrow{aA}+\thisrow{aSS})
                    ]
                    {../code/results/scans/NLO/SU2/M0.135/D2/constr_0.qhc_stats_rel1.dat};
                        
        \end{axis}
        \begin{axis}[
                scale=2, axis equal image,
                axis x line = middle, axis y line = middle,
                xmin=-1.6,xmax=1.6, ymin=-1.6,ymax=1.6,
            ]            
            \draw[blue] (0,0)    
                node[circle,inner sep=1.5pt, fill, pin={[pin edge={thick,blue!50}, pin distance= 1cm]150:$\pi^\pm\pi^\pm$}] {};
            \draw[blue] (.333,0) 
                node[circle,inner sep=1.5pt, fill, pin={[pin edge={thick,blue!50}, pin distance=.7cm]-60:$\pi^0\pi^0$}] {};
            \draw[blue] (0,.5)   
                node[circle,inner sep=1.5pt, fill, pin={[pin edge={thick,blue!50}, pin distance=.5cm]130:$\pi^0\pi^\pm$}] {};
            \draw[blue] (.333,.5)   
                node[circle,inner sep=1.5pt, fill, pin={[pin edge={thick,blue!50}, pin distance=1.5cm]-15:$\pi^\pm\pi^\mp$}] {};
            
        \end{axis}
    \end{tikzpicture}
    \caption[The same 2-flavour data as in \cref{fig:aJ-stats}, but plotted like in \cref{fig:aJ}]{
        The same 2-flavour data as in \cref{fig:aJ-stats}, but plotted over $a_I,a_A$ plane with $a_I+a_A+a_{SS}=1$, like in \cref{fig:aJ}, rather than as a function of $\theta$.
        The colours convey the same meaning as in \cref{fig:aJ-stats}, whereas the $\hat\rho$ values are indicated by the size of the points: smaller $\hat\rho$ (higher proximity to the reference point) corresponds to larger points.
        Note how the $a_J$ that are permitted at all $s$ can be clearly seen as a uniformly coloured patch, which is also featured in \cref{fig:aJ}. 
        It is interesting to note that most relevant constraints come from the $a_{SS}<0$ region (upper right), which is permitted for significantly fewer $s$ than those points where all $a_J$ are positive.}
    \label{fig:aJ-stats-on-sphere}
\end{figure}

In terms of the \emph{ad hoc} measures together with $\hat\rho$ defined above, \cref{fig:aJ-stats} shows the distribution of NLO bounds for 2, 3 and 4 flavours, and \cref{fig:aJ-stats-on-sphere} shows a geometrically more intuitive version in the 2-flavour case.%
\footnote{
    The points were sampled uniformly over the unit octahedron (i.e.\ the unit sphere under the 1-norm $\sum_J |a_J| = 1$) and its higher-dimensional analogues in $a_J$ space.
    This shape was used rather than the unit sphere to preserve the linearity of bounds on $a_J$; compare the discussion in \cref{sec:aJ-cond}.}
We see that relevant constraints tend to have low $\hat\rho$, albeit with many exceptions --- the orientation of the constraint is another important factor.
The relevant constraints are rather evenly distributed over the $\theta$ range permitted by \cref{eq:main:cond}, indicating that there is, in this regard, nothing special about the eigenstates, validating \cref{imp:aJ}.
Most relevant constraints also occur at $a_J$ that are permitted for very few $s$ (i.e.\ coloured very pale in the plots), validating \cref{imp:s}.

There are, of course, severe limitations to the analysis in this section.
Apart from the roughness of the chosen measures discussed above, it is difficult to assess just how great the benefits of \cref{imp:aJ,imp:s} are.
Stronger constraints are obtained, but not necessarily much stronger: the improvement in \cref{fig:SU2-l12} over \cite{Manohar:2008tc} is very slight, although it seems that the use of \cref{imp:aJ} is limited at $n=2$ but more extensive at e.g.\ $n=3$ by comparing the subfigures of \cref{fig:aJ-stats}.

\FloatBarrier
\pagebreak

\section{Conclusions and outlook}\label{sec:final}
To recapitulate, our method has been as follows: 
We scan over the $s,t$ range depicted in \cref{fig:mandelstam} (or in many cases just the $s$ range, with $t$ fixed to $4$, as discussed in \cref{sec:t=4}). 
At each $(s,t)$-value, we scan over those $a_J$ that are permitted by \cref{eq:main:cond}, and compute the $k$th derivative of the $n$-flavour amplitude to either NLO or NNLO at that point, possibly with above-threshold integration up to $\lambda$, as described in \cref{sec:int}. 
Through \cref{eq:main:constr}, this yields positivity bounds on the LECs (or at NNLO, the parameters derived in \cref{sec:param}), which can be handled as linear constraints using the language and methods of \cref{sec:constr} (implemented as discussed in \cref{sec:constraints}). 
In the end, this yields a manageable set of relevant constraints, which can be visualised and interpreted.

Our results at NLO consist of stronger bounds than in \cite{Manohar:2008tc,Mateu:2008gv}, whereas the comparison to the more recent works \cite{Wang:2020jxr,Tolley:2020gtv} is less clear-cut.
As in previous works, most bounds consist of irregularly shaped and usually infinite regions, although some of the nicer cases allow for more clear-cut bounds such as \cref{eq:Theta-bound,eq:L123-bound,eq:GD-bound}.
Although the bounds themselves are highly uncertain, basic compatibility suggests that one may assign uncertainties of roughly 200-1000\%\ to $\Theta_i,\Gamma_i,\Delta_i$ and therefore also to the NNLO LECs, which are not given any error estimates in \cite{Bijnens:2014lea}.
Obtaining better error estimates is a possible direction for future work.

The employment of above-threshold integration allows for very strong bounds, but carries the risk of going too far beyond the low-energy limit; the difference between our NLO and NNLO bounds, such as in \cref{fig:SU2-l12}, does not inspire much confidence in integrated bounds for $\lambda$ significantly larger than 4. 
Our improved handling of $a_J$, which is evaluated in \cref{sec:aJ-stats}, improves bounds without additional assumptions (\cite{Wang:2020jxr,Tolley:2020gtv} also uses assumption-less improvements). 
Of course, the choice of fixed-order \chpt\ is itself something that relies on the low-energy limit, although it is easier to motivate than a particular choice of $\lambda>4$ is. 

At NNLO, our methods suffer some practical problems due to the very high dimension of the parameter space, so we have only performed rather coarse scans of the available $s,t$ and $a_J$ ranges. 
This is remedied by fixing some parameters and focusing on the lower-dimensional space that remains. 
However, it is important to keep in mind that, even though we may fail to obtain many constraints due to limited scans and technical issues with determining convex hulls (see \cref{sec:practical}), it is guaranteed that the constraints we do find are true --- the method automatically errs on the side of caution, so to speak. 
In particular, this means that if the bounds are inconsistent with the experimental values, then the error must lie either with the values, or with the theory itself.
In general, the main problem with our method is not its ability to produce bounds, but our ability to rely on the assumptions behind them.

The most prominent example of this is the problems encountered at three flavours, as discussed in \cref{sec:results-3flav}.
The obvious remedy is to replace equal-mass \chpt\ with the more realistic unequal-mass version, towards which the main hurdle is some so-far-unsolved two-loop integrals. 
The emergence of distinct mass eigenstates, as well as some other features discussed in \cite{Mateu:2008gv}, could possibly interfere with some of our method innovations, although we are confident that they can be remedied.

Beyond practical improvements and the use of unequal-mass \chpt, a possible step forward is to either go beyond $2\to2$ scattering, or to study NNNLO. 
The former would allow for bounds on LECs that do not appear in the $2\to2$ amplitude, as well as possibly new bounds on those covered here.
While it would be infeasible to manage all the NNNLO LECs in the $2\to2$ amplitude, or even all the NNLO LECs in higher-multiplicity amplitudes, it is not unthinkable that useful results could be obtained by fixing most LECs and studying the rest. 
In any case, further explorations in these directions are mainly hindered by the lack of available amplitudes; 
the NLO 2-flavour $2\to4$ amplitude was recently calculated \cite{Bijnens:2021hpq}, but no higher-order or higher-multiplicity amplitudes are currently known.
Furthermore, \cref{sec:bounds} would need to be generalised to handle the 9-dimensional kinematic space of 6-particle amplitudes.
Lastly, some parts of the NNNLO amplitude would be nonlinear in the NLO LECs, necessitating the development of \cref{proposition 4.2}-like technology for \emph{nonlinear} constraints, i.e.\ $\alpha_i b_i+\alpha_{ij} b_ib_j+ \ldots\geq c$.
We believe it possible that at least some of the tools in \cref{sec:constraints} can be generalised to handle this, but have not investigated it much.

Lastly, these methods could lend themselves to application on EFTs other than \chpt, e.g.\ for beyond-the-Standard-Model (BSM) applications.
This could be particularly promising if there are no experimentally measured values for the LECs, or if experiments have only yielded bounds.
An experimental upper bound coupled with an analytic lower bound could confine the coupling of an unobserved process to a range, or exclude a BSM EFT altogether.

\subsection{Acknowledgements}
Mattias Sjö thanks Torbjörn Lundberg for rewarding discussions resulting in the new treatment of \cref{proposition 4.1,proposition 4.2} as compared to \cite{AlvarezThesis}.
This work is supported in part by the Swedish Research Council grants contract numbers 2016-05996 and 2019-03779.
Colour schemes for the figures, appropriate for colourblind people and monochrome printing, were chosen based on \cite{colourschemes,colorbrewer}.

\pagebreak
\appendix

\crefname{appendix}{section}{sections}
\Crefname{appendix}{Section}{Sections}

\section{LEC details}\label{sec:LEC-details}
This section contains further details about the LECs that are introduced in \cref{sec:lagr}, and the NNLO parameters $\Gamma_i,\Delta_i,\Theta_i,\Xi_i$ that are defined in \cref{sec:param}.
\Cref{tab:fiducial} contains the values determined in \cite{Bijnens:2014lea} that are used as references in \cref{sec:results}. 
Naturally, these do not include estimates of parameters that only appear above 3 flavours; similarly, the $\Gamma_i,\Delta_i$ that only appear in the $n\geq3$ amplitude are just given provisional values based on the 3-flavour data.
\begin{table}[hbtp]
    \centering
    \begin{tabular}{LR@{.}L@{\qquad}cLR@{.}L}
        \hline\hline
        \text{NLO LEC} &\multicolumn{2}{c}{}     &\phantom{OPTLN}& \multicolumn{2}{l}{NNLO parameter}& \\
        \multicolumn{3}{l}{}            &&   \multicolumn{3}{l}{\phantom{$\Theta'_2$}$\cdot 10^3$}     \\
        \hline
        \rule{0pt}{2.5ex}
        \bar l_1    &  -0&4(6)          &&   \Theta_1       &  0&34     \\
        \bar l_2    &   4&3(1)          &&   \Theta_2       &  0&68     \\
        \bar l_3    &   2&9(24)         &&   \Theta_3       & -0&16     \\
        \bar l_4    &   4&4(2)          &&   \Theta_4       & -0&22     \\
        \multicolumn{3}{l}{}            &&   \Xi_1          &  0&29     \\
        \multicolumn{3}{l}{\phantom{$L_1^r$}$\cdot 10^3$}
                                        &&   \Xi_2          &  0&34     \\ 
        \cline{1-3}
        L_1^r       &   1&11(10)        &&   \Xi_3          &  0&25     \\
        L_2^r       &   1&05(17)        &&   \Xi_4          & -0&008    \\
        L_3^r       &  -3&82(30)        &&   \Gamma_1       &  0&008    \\
        L_4^r       &   1&87(53)        &&   \Gamma_2       & -0&71     \\
        L_5^r       &   1&22(06)        &&   \Gamma_3       & -0&10     \\
        L_6^r       &   1&46(46)        &&   \Gamma_4       & -0&22     \\
        L_8^r       &   0&65(07)        &&   \Delta_1       & -0&032    \\
        \multicolumn{3}{l}{}            &&   \Delta_2       & -0&83     \\
        \multicolumn{3}{l}{}            &&   \Delta_3       & -0&048    \\
        \multicolumn{3}{l}{}            &&   \Delta_4       & -0&14
    \end{tabular}
    \caption{
        All LECs and parameters covered by the bounds derived in this paper, along with their experimental reference values, taken from the most general fits in \cite{Bijnens:2014lea}.
        The values in parentheses indicate the uncertainties in the last decimal places.
        No uncertainties are provided for the NNLO parameters, since none are given for the NNLO LEC values in \cite{Bijnens:2014lea}; these values are little more than educated guesses.
        The values of $\Gamma_i$, $\Delta_i$ also depend directly on $n$ for $i<3$ (see \cref{eq:Gi-Di}); the listed values use $n=4$.}
    \label{tab:fiducial}
\end{table}

\pagebreak
\Cref{eq:Gi-Di} shows how the independent NNLO parameters depend on the LECs $K_i^r$, for a general number of flavours $n$.\footnote{These are generated by \form\ with minimal post-processing.
Common factors have been extracted to make the expressions shorter.}
\begin{subequations}\label{eq:Gi-Di}
    \begin{align}
        \tfrac{1}{8}\Gamma_1 &=
              6 K_{1}^r
            - 4 K_{5}^r
            + K_{7}^r
            + K_{11}^r
            + 2 K_{31}^r
            + n K_{8}^r
            + 2 n K_{18}^r\\
        \tfrac{1}{2}\Gamma_2 &=
              12 K_{1}^r
            - 48 K_{3}^r
            - 12 K_{5}^r
            + K_{7}^r
            + K_{11}^r
            - 16 K_{13}^r
            + 32 K_{17}^r
            - 8 K_{19}^r    
            - 8 K_{23}^r    
            - 16 K_{28}^r   \notag\\&\qquad
            + 2 K_{31}^r
            - 8 K_{33}^r
            + 16 K_{37}^r
            + n K_{8}^r
            - 16 n K_{14}^r
            + 34 n K_{18}^r
            - 4 n K_{20}^r\\
        \tfrac{1}{4}\Gamma_3 &=
              K_{1}^r
            - 2 K_{3}^r
            - 2 K_{5}^r\\
        \tfrac{1}{8}\Gamma_4 &=
            - 3 K_{3}^r
            - K_{5}^r\\
        \tfrac{1}{16}\Delta_1 &=
              6 K_{4}^r
            - 2 K_{6}^r
            + K_{15}^r
            + 2 K_{29}^r
            + n K_{16}^r\\
        \tfrac{1}{4}\Delta_2 &=
            - 48 K_{2}^r
            + 12 K_{4}^r
            - 4 K_{6}^r
            - 16 K_{9}^r
            + K_{15}^r
            + 32 K_{18}^r
            + 8 K_{20}^r
            + 16 K_{21}^r   \notag\\&\qquad
            + 2 K_{29}^r
            - 8 K_{32}^r
            + 16 K_{35}^r
            - 8 K_{38}^r
            - 16 n K_{10}^r
            + n K_{16}^r
            + 16 n K_{22}^r\\
        \tfrac{1}{8}\Delta_3 &=
            - 2 K_{2}^r
            + K_{4}^r\\
        \tfrac{1}{8}\Delta_4 &=
            - 6 K_{2}^r
            + K_{6}^r
    \end{align}
\end{subequations}
\Cref{eq:Gi-Di-Xi} shows the same for 3 flavours.
The application of the $n=3$ Cayley-Hamilton identity and the numbering of the $C_i^r$ follows \cite{Bijnens:1999sh}.
\begin{subequations}\label{eq:Gi-Di-Xi}
    \begin{align}
        \tfrac{1}{4} \Xi_1 &=
            - 6 C_{1}^r
            + 12 C_{3}^r
            + 4 C_{4}^r
            - C_{5}^r
            - 3 C_{6}^r
            - C_{8}^r
            + 2 C_{10}^r
            + 6 C_{11}^r
            - 4 C_{12}^r     \notag\\&\qquad
            - 18 C_{13}^r
            + 4 C_{22}^r
            - 4 C_{25}^r\\
        \tfrac{3}{8} \Xi_2 &=
              50 C_{1}^r
            + 27 C_{2}^r
            + 22 C_{3}^r
            - 18 C_{4}^r
            - 48 C_{5}^r
            + 54 C_{6}^r
            + 45 C_{7}^r
            - 21 C_{8}^r
            + 24 C_{9}^r   \notag\\&\qquad
            + 3 C_{10}^r
            + 9 C_{11}^r
            + 6 C_{13}^r
            + 6 C_{16}^r
            - 4 C_{24}^r
            + 12 C_{25}^r
            - 4 C_{26}^r
            + 4 C_{29}^r\\
        \tfrac{3}{8} \Xi_3 &=
              70 C_{1}^r
            - 18 C_{2}^r
            + 68 C_{3}^r
            - 16 C_{4}^r
            + 43 C_{5}^r
            + 21 C_{6}^r
            - 30 C_{7}^r
            + 25 C_{8}^r    \notag\\&\qquad
            - 24 C_{9}^r
            + 10 C_{10}^r
            + 30 C_{11}^r
            - 8 C_{12}^r
            - 18 C_{13}^r
            + 12 C_{16}^r
            - 4 C_{22}^r    \notag\\&\qquad
            + 18 C_{23}^r
            + 4 C_{24}^r
            - 2 C_{25}^r
            - 8 C_{26}^r
            + 8 C_{29}^r\\
        \tfrac{3}{4} \Xi_4 &=
            - 38 C_{1}^r
            + 9 C_{2}^r
            - 28 C_{3}^r
            + 12 C_{4}^r
            - 6 C_{16}^r
            + 6 C_{22}^r
            - 3 C_{23}^r
            - 2 C_{24}^r     \notag\\&\qquad
            + 4 C_{26}^r
            - 4 C_{29}^r\\
        \tfrac{1}{4} \Gamma_3 &=
              C_{1}^r
            - 2 C_{3}^r
            - 2 C_{4}^r\hspace{8cm}\\
        \tfrac{1}{4} \Delta_3 &=
              4 C_{1}^r
            - 5 C_{2}^r
            + 2 C_{3}^r
    \end{align}
\end{subequations}
Lastly, \cref{eq:Ti} shows the same for 2 flavours.
Again, we have followed \cite{Bijnens:1999sh}.
\begin{subequations}\label{eq:Ti}
    \begin{align}
        \tfrac{3}{16} \Theta_1 &=
              14 c_{1}^r
            + 40 c_{2}^r
            - 6 c_{3}^r
            + 15 c_{4}^r
            + 30 c_{5}^r
            - 30 c_{6}^r
            + 12 c_{7}^r
            + 9 c_{12}^r
            - 3 c_{13}^r    \notag\\&\qquad
            - 4 c_{14}^r
            + 4 c_{16}^r\\
        \tfrac{3}{8} \Theta_2 &=
              104 c_{1}^r
            + 22 c_{2}^r
            - 18 c_{3}^r
            + 93 c_{4}^r
            + 15 c_{5}^r
            + 12 c_{7}^r
            + 12 c_{13}^r
            - 4 c_{14}^r
            + 4 c_{16}^r\\
        \tfrac{3}{16} \Theta_3 &=
              6 c_{1}^r
            - 9/2 c_{2}^r
            - 3 c_{3}^r\\
        \tfrac{3}{4} \Theta_4 &=
            - 14 c_{1}^r
            - 40 c_{2}^r
            - 12 c_{7}^r
            + 3 c_{12}^r
            + 4 c_{14}^r
            - 4 c_{16}^r   
    \end{align}
\end{subequations}

\pagebreak
\section{Details and proofs regarding linear constraints}
\label{sec:constraints}
In this appendix, we prove the propositions stated in \cref{sec:constr} and provide some more details on how they may be applied.%
\footnote{
    Like in \cref{sec:constr}, we make use of potentially unfamiliar mathematical notation in this section, so the glossary (\cref{sec:glossary}) may be helpful.}

\subsection{Proof of propositions \ref{proposition 4.1} and \ref{proposition 4.2}}\label{sec:proof}
As has already been mentioned, \cref{proposition 4.1} is a direct consequence of \cref{proposition 4.2}, obtained by leaving all but one $I_c$ empty.
However, directly proving \cref{proposition 4.2} is much less straightforward than the following chain of implications,
\begin{equation}\label{eq:roadmap}
    \text{\Cref{proposition 4.1} ($c=0$ only) $\imp$ \Cref{proposition 4.2} $\imp$ \Cref{proposition 4.1}},
\end{equation}
which we will demonstrate in this section.
First, however, we will show an easily accessible partial result (\cref{sec:trivial}), and then prove some properties that are necessary for the main proof (\cref{sec:closed}).

\subsubsection{The trivial half of the proof}\label{sec:trivial}
One side of \cref{proposition 4.2} is easy to prove, namely that $\constr{\v\beta}{c}\leq\Omega$ if $\v\beta\in\reg_c(\Omega)$. 

Assuming that $\Omega\neq\Omega_\infty$, take any point $\v b$ that satisfies $\Omega$.
Then by \cref{eq:def-gen-reg},
\begin{equation}
    \v\beta\cdot\v b = \sumI{1}\lambda_i\v\alpha_i\cdot\v b + \sumI{0}\lambda_i\v\alpha_i\cdot\v b + \sumI{-1}\lambda_i\v\alpha_i\cdot\v b,
\end{equation}
and since $\v\alpha_i\cdot\v b \geq c$ for $i\in I_c$, we have
\begin{equation}
    \v\beta\cdot\v b \geq \sumI{1}\lambda_i - \sumI{-1}\lambda_i \geq c,
\end{equation}
which uses (and motivates) \cref{eq:gen-condition}.
The corresponding result for \cref{proposition 4.1} follows immediately.

\subsubsection{Proof that $\reg(\omega_0)$ is closed and convex}\label{sec:closed}
As \cref{sec:proof-0} will show, these properties of $\reg(\omega_0)$ are crucial for the main proof.
Convexity is easy to show for $\reg(\omega_0)$: given any points $\v\beta_1,\v\beta_2$ satisfying \cref{eq:def-reg}, their convex combination
\begin{equation}
    \mu\v\beta_1 + (1-\mu)\v\beta_2 = \sum_{i\in I} \big(\mu\lambda_{1i} + (1-\mu)\lambda_{2i}\big)\v\alpha_i,
    \qquad \lambda_{1i}\geq0,\quad \lambda_{2i}\geq0
\end{equation}
satisfies \cref{eq:def-reg} as well.

For the proof of closedness, we only need three basic facts: (i) the intersection of closed sets is closed, (ii) the union of a finite number of closed sets is closed, and (iii) for any $\v\alpha$ and $c$ the set $\sat\left(\constr{\v\alpha}{c}\right)$ is closed.%
\footnote{This last fact is easy to prove: take a point $\v b\not\in\sat\left(\constr{\v\alpha}{c}\right)$, i.e.\ $\v\alpha\cdot\v b < c$.
Then for any vector $\v d$ such that $|\v d|<\e$, 
\begin{equation}
    \v\alpha\cdot(\v b + \v d) \leq \v\alpha\cdot\v b + |\v\alpha\cdot\v d| < \v\alpha\cdot\v b + \e|\v\alpha|,
\end{equation}
where we used the Cauchy-Schwarz inequality in the last step.
For $\e>0$ sufficiently small, this is still less than $c$, so $\v b + \v d\not\in\sat\left(\constr{\v\alpha}{c}\right)$, proving that it is closed. (Note that this also works for $\sat\left(\constr{\v 0}{ 1}\right)=\emptyset$ and $\sat\left(\constr{\v 0}{ -1}\right)=\R^D$.)}

Now, we employ \emph{Carathéodory's theorem for convex cones},%
\footnote{In common mathematical nomenclature, $\reg(\omega_0)$ is a convex cone, $\reg(\omega_1)$ is an affine cone, and $\reg(\omega_\inv)$ is a convex hull.
We have chosen not to use these classifications, partly because neither applies to the general $\reg_c(\Omega)$.} 
which for our purposes can be formulated as follows:
\begin{quotation}
    \noindent\textit{Let $I\subset\N$ be finite, and let $\{\v\alpha_i\}_{i\in I}$ be vectors in $\R^D$. For any point $\v\beta\in\R^D$ fulfilling
    \begin{equation}\label{eq:cara-1}
        \v\beta = \sum_{i\in I}\lambda_i\v\alpha_i,\qquad\lambda_i\geq 0,
    \end{equation}
    there exists a set $J\subseteq I$ with at most $D$ elements such that
    \begin{equation}\label{eq:cara-2}
        \v\beta = \sum_{j\in J}\lambda_j\v\alpha_j,\qquad \lambda_j\geq 0,
    \end{equation}
    where and all $\v\alpha_j$ are linearly independent.}
\end{quotation}
\Cref{eq:cara-1} is clearly equivalent to \cref{eq:def-reg} for $c=0$. 

For any such $J$, there exists a set of vectors $\{\v\gamma_k\}$ such that $\{\v\alpha_j\}_{j\in J} \cup \{\v\gamma_k\}$ is a basis of $\R^D$.
Let $\m A_{J}$ be the invertible matrix whose columns are these basis vectors, and let $\v\lambda$ be the vector whose components are $\lambda_j$, where $\lambda_j=0$ if $j\not\in J$.
Then \cref{eq:cara-2} can be rewritten as $\v\beta = \m A_{J}\v\lambda$, or equivalently $\m A_{J}^\inv\v\beta = \v\lambda$.
Since $\lambda_j\geq 0$, we therefore obtain the inequalities $\v a_j\cdot\v\beta\geq 0$, where $\v a_j$ are the column vectors of $\m A_{J}^\inv$, and $\v a_j\cdot\v\beta= 0$ if $j\not\in J$.
In other words,%
\footnote{
    Note that $\v\beta$, which normally is part of a linear constraint, is itself constrained here.
    This is not a problem; in fact, the expression $\v\alpha\cdot\v b \geq c$ and be interpreted both as $\v b\in\sat\left(\constr{\v\alpha}{c}\right)$ and as $\v\alpha\in\sat\left(\constr{\v b}{c}\right)$.
    We will return to this symmetric interpretation many times below.}  
\begin{equation}
    \v\beta\in\sat\left(\sum_{j}\constr{\v a_j}{0} + \sum_{j\not\in J}\constr{-\v a_j}{0}\right).
\end{equation} 
By facts (i) and (iii), this set is always closed.
Therefore, the set of all $\v\beta\in\sat(\omega_0)$ associated with the same $J$ is closed.
$\reg(\omega_0)$ must then be the union of all such sets, but since $I$ is finite, there are finitely many different subsets $J$, so $\reg(\omega_0)$ is the union of a finite number of closed sets. 
By fact (ii), it is therefore closed. 
\qed

Let us remark that this proof extends to the other cases, so that $\reg_c(\Omega)$ is closed and convex for any $\Omega$ and $c$. 

\subsubsection{Proof of \cref{proposition 4.1} in the $c=0$ case}\label{sec:proof-0}
We will now turn our attention to the statement that $\constr{\v\beta}{0}\not\leq\omega_0$ if $\v\beta\not\in\reg(\omega_0)$, which will complete the proof of \cref{proposition 4.1} for $c=0$.%
\footnote{
    The same method is easy to apply to the $c=-1$ case and, with some slight complications, the $c=1$ case.
    With considerable effort, it can also be extended to \cref{proposition 4.2}.
    However, we will follow the outline \cref{eq:roadmap} and only prove what is necessary.} 
We employ the \emph{separating hyperplane theorem}, which can be formulated as follows:
\begin{quotation}
    \noindent\textit{Let $\s X$ and $\s Y$ be disjoint convex sets, with $\s X$ closed and $\s Y$ compact.
    Then there exists a nonzero vector $\v h$ and a real number $d$ such that}
    \begin{equation}
        \v\chi\cdot\v h > d \mathand \v\psi\cdot\v h < d
    \end{equation}
    \textit{for all $\v\chi\in\s X,\v\psi\in\s Y$.
    (The set $\setbuild{\v\pi}{\v\pi\cdot\v h=d}$ is a hyperplane that separates $\s X$ from $\s Y$, hence the name.)}
\end{quotation}
Since $\reg(\omega_0)$ is closed and convex, and because the set consisting of the single point $\v\beta \notin\reg(\omega_0)$ is compact and convex, the separating hyperplane theorem implies that there exists $\constr{\v h}{d}$ such that 
\begin{equation}
    \forall \v \alpha \in \reg(\omega_0),\quad \v h \cdot \v\alpha > d\mathand \v h \cdot \v\beta <d.
\end{equation}
Since $\v0\in\reg(\omega_0)$, we know that $d<0$. However, we claim that for any $\v \alpha \in \reg(\omega_0)$, we in fact have $ \v h \cdot \v\alpha \geq 0$. Indeed, if we assume that there exists  $\v \alpha \in \reg(\omega_0)$ such that  $ \v h \cdot \v\alpha < 0$, then for any $\lambda \geq \frac{|d|}{|\v\alpha\cdot\v h|}$, \cref{eq:def-reg} implies that $\lambda\v\alpha\in\reg(\omega_0)$. But $\lambda\v\alpha\cdot\v h \leq -|d|$, which  contradicts the fact that $\v\alpha\cdot\v h > d$. Therefore, we have $\v\alpha\cdot\v h\geq 0$ and $\v\beta\cdot\v h < 0$, implying that $\constr{\v \beta}{0} \not\leq\omega_0$. This proves \cref{proposition 4.1}.
\QED

\subsubsection{Proof of \cref{proposition 4.2}}\label{sec:proof-gen}
In order to reduce the general $\Omega$ defined in \cref{eq:def-Omega} to one that can be handled by \cref{proposition 4.1} for $c=0$, we define the ``lifted'' vector
\begin{equation}\label{eq:lift}
    \lift[x]{\v v} = \left( v_1 , v_2 , \ldots , v_D , x \right),
\end{equation}
where $D$ is the dimension of the original vector $\v v$.
Then we note that 
\begin{equation}
    \v\alpha\cdot\v b \geq c \equ \lift[-c]{\v\alpha}\cdot\lift[1]{\v b} = \v\alpha\cdot\v b - c \geq 0.
\end{equation}
Thus, any $D$-dimensional linear constraint $\constr{\v\alpha}{c}$ can be lifted into a $(D+1)$-dimensional linear constraint $\constr{\lift[-c]{\v\alpha}}{0}$.
We can now show that \cref{proposition 4.1} in the lifted space, where we only ever have $c=0$, is equivalent to \cref{proposition 4.2} in the original space.
Define
\begin{equation}
    \lOmega = \sumI{1}\constr{\lift[-1]{\v\alpha_i}}{0} + \sumI{0}\constr{\lift[0]{\v\alpha_i}}{0} + \sumI{-1}\constr{\lift[1]{\v\alpha_i}}{0} + \constr{\lift[1]{\v 0}}{0},
\end{equation}
where the extra constraint $\constr{\lift[1]{\v 0}}{0}$ imposes that $\lift[x]{\v b}$ only satisfies $\lOmega$ if $x\geq 0$. 
$\lOmega$ can be thought of as a lifted version of $\Omega$, and as indicated by the notation, it fulfils the definition of $\omega_0$ so that \cref{proposition 4.1} for $c=0$ applies to it.

Now, assume that $\lift[-c]{\v\beta}\in\reg(\lOmega)$.
Looking at \cref{eq:def-reg}, we find that
\begin{equation}
    \begin{aligned}
        \v\beta &= \sumI{1}\lambda_i\v\alpha_i + \sumI{0}\lambda_i\v\alpha_i + \sumI{-1}\lambda_i\v\alpha_i, + \lambda'\v 0,\\
        -c      &= \sumI{-1}\lambda_i - \sumI{1}\lambda_i + \lambda',\qquad \lambda_i \geq 0,\lambda'\geq 0.
    \end{aligned}
\end{equation}
These exactly reproduce \cref{eq:def-gen-reg,eq:gen-condition}, so we have shown that $\lift[-c]{\v\beta}\in\reg(\lOmega)$ implies $\v\beta\in\reg_c(\Omega)$.

Conversely, let us now assume, $\lift[-c]{\v\beta}\not\in\reg(\lOmega)$. 
\Cref{proposition 4.1} then implies the existence of some $\lift[x]{\v b}\in\sat(\lOmega)$ with $x\geq 0$ such that $\lift[-c]{\v\beta}\cdot\lift[x]{\v b} < 0$, and for any $i\in I$, $\lift[-c_i]{\v\alpha_i}\cdot\lift[x]{\v b} \geq 0$.
We may moreover assume that $x>0$, for if $x=0$, we may choose any $\v a\in\sat(\Omega)$ and consider $\lift[\e]{\v b'} = (1-\e)\lift[0]{\v b} + \e\lift[1]{\v a}$ for $0<\e<1$.
By convexity, $\lift[\e]{\v b'}\in\sat(\lOmega)$, and
\begin{equation}
   \lift[-c]{\v\beta} \cdot \lift[\e]{\v b'} = \v\beta\cdot\v b + \e\left[\v\beta\cdot(\v a-\v b) - c\right],
\end{equation}
so for $\e$ small enough, $\lift[-c]{\v\beta}\cdot\lift[\e]{\v b'} < 0.$
Consequently,
\begin{equation}
    \begin{cases}
        \v\alpha_i\cdot\v b \geq x, & i\in I_1,\\
        \v\alpha_i\cdot\v b \geq 0, & i\in I_0,\\
        \v\alpha_i\cdot\v b \geq -x, & i\in I_\inv,
    \end{cases}
    \qquad\qquad
    \v\beta\cdot\v b < xc.
\end{equation}
Thus, $\frac1x \v b$ is a point that satisfies $\Omega$ but not $\constr{\v\beta}{c}$, which means that $\lift[-c]{\v\beta}\not\in\reg(\lOmega)$ implies $\v\beta\not\in\reg_c(\Omega)$.

We have now shown that $\lift[-c]{\v\beta}\in\reg(\lOmega)$ is equivalent to $\v\beta\in\reg_c(\Omega)$, and since $\constr{\lift[-c]{\v\beta}}{0}\leq\lOmega$ is equivalent to $\constr{\v\beta}{c}\leq\Omega$, we have therefore proven \cref{proposition 4.2} as a consequence of the $c=0$ version of \cref{proposition 4.1}.
As mentioned before, \cref{proposition 4.1} for $c=\pm1$ follows easily.\QED

\subsubsection{An important corollary}
The following interesting result is a consequence of \cref{proposition 4.2}:
\begin{corollary}[boundedness of $\sat(\Omega)$]\label{cor:bounded}
    For $\Omega\neq\Omega_\infty$, the region $\sat(\Omega)$ is bounded
    if and only if the origin, $\v 0$, is in the interior of $\reg_\inv(\Omega)$.
    This happens if and only if $\{\v\alpha_i\}_{i\in I}$ spans the full $D$-dimensional space and there are $\lambda_i$ such that
    \begin{equation}\label{eq:zero-interior}
        \v 0 = \sum_{i\in I}\lambda_i\v\alpha_i,\qquad\lambda_i>0
    \end{equation}
\end{corollary}
\proof We will show the converse, namely that $\sat(\Omega)$ being unbounded is equivalent to $\v 0$ not being in the interior of $\reg_\inv(\Omega)$.

First assume that $\v 0$ is not in the interior of $\reg_\inv(\Omega)$.
Then for all $\e>0$, there exists some $\v\beta_\e$ such that $|\v\beta_\e| < \e$ and $\v\beta_\e \not\in\reg_\inv(\Omega)$.
Since $\constr{\v\beta_\e}{\inv}\not\leq \Omega$, there consequently exists a point $\v b_\e\in\sat(\Omega)$ such that $\v\beta_\e\cdot\v b_\e < -1$.
Now, the Cauchy-Schwarz inequality gives
\begin{equation}
    1 < |\v\beta_e\cdot\v b_e| \leq |\v\beta_e|\,|\v b_e| < \e|\v b_\e|.
\end{equation}
Since this holds for arbitrarily small $\e$, there can be no upper bound on $|\v b_\e|$; therefore, $\sat(\Omega)$ is unbounded.

Conversely, assume that $\sat(\Omega)$ is unbounded.
Then for all $M>0$, there must exist some $\v b_M\in\sat(\Omega)$ such that $|\v b_M|>M$.
Now define
\begin{equation}
    \v\beta_M \equiv -\frac{\v b_M}{|\v b_M|^{3/2}}
    \imp |\v\beta_M| = \frac{1}{|\v b_M|^{1/2}} < \frac{1}{\sqrt{M}}, \quad \v\beta_M\cdot\v b_M = -|\v b_M|^{1/2} < -\sqrt{M}.
\end{equation}
For sufficiently large $M$, the last inequality implies that $\constr{\v\beta_M}{\inv}\not\leq \Omega$, so $\v\beta_M\not\in\reg_\inv(\Omega)$.
However, the inequality before that tells us that $\v\beta_M$ may lie arbitrarily close to the origin.
Therefore, $\v 0\in \reg_\inv(\Omega)$ must lie on the boundary, not the interior, of $\reg_\inv(\Omega)$. 

That completes the main proof, but we must also prove condition about the span of $\{\v\alpha_i\}_{i\in I}$.
If the span was lower-dimensional, then there would exist some vector $\v\beta$ linearly independent of all $\v\alpha_i$, and then clearly $\e\v\beta\not\in\reg_\inv(\Omega)$ for all $\e>0$, implying that $\v 0$ is not in the interior of $\reg_\inv(\Omega)$.
Lastly, we must prove \cref{eq:zero-interior}, which is essentially \cref{eq:def-reg-alt} with $\lambda_i>0$ rather than $\lambda_i\geq 0$.
Since $\{\v\alpha_i\}_{i\in I}$ spans the full space, any vector $\v v$ of sufficiently small magnitude satisfies
\begin{equation}
    \v v \in \hull\left(\{\v\alpha_i\}_{i\in I} \cup \left\{-\textstyle\sum_{i\in I}\v\alpha_i\right\}\right).
\end{equation}
Thus, $\v 0$ is in the interior of $\reg_\inv(\Omega)$ if and only if for sufiiciently small $\e>0$%
\footnote{
    Here, we neglect the condition $\sum_{i\in I}\lambda_ic_i \geq -1$ in the second equality, since it is always possible to multiply both sides by a positive factor to rescale the $\lambda_i$ appropriately.}
\begin{equation}
    -\e\sum_{i\in I}\v\alpha_i\in\reg_\inv(\Omega)
    \equ
    -\e\sum_{i\in I}\v\alpha_i = \sum_{i\in I}\lambda_i\v\alpha_i
    \equ
    \v 0 = \sum_{i\in I}(\lambda_i+\e)\v\alpha_i
\end{equation}
for $\lambda_i\geq 0$, which implies $\lambda_i+\e>0$, thereby producing \cref{eq:zero-interior}.
\QED

\subsection{Some mathematical tools}
Before moving on with proving \cref{proposition 4.3} and deriving further results, we need to establish some tools and terminology that range from useful to crucial in subsequent sections.

\subsubsection{The degenerate constraint framework}\label{sec:degenerate}
In this section, we properly define what it means for a constraint to be degenerate, and derive notations and results that are not only useful for the proof of \cref{proposition 4.3} and its generalisation, but also for many other things later in this appendix.

In a $D$-dimensional space, consider an affine subspace $E$ of dimension $d$.
There exists two sets of vectors $\{\v g_j\}_{j=1}^{d},\{\v\delta_k\}_{k=1}^{D-d}$ whose union forms an orthonormal basis for $\R^D$, that, given an arbitrary point $\v e\in E$, allow $E$ to be expressed in two complementary ways:%
\footnote{
    Note that we have written $\v g_j$ as parameter-space vectors, and $\v\delta_k$ as constraint-space vectors. 
    This is consistent with their use in \cref{eq:aff-subspace}, but \cref{eq:param-decomp,eq:constr-decomp} are in a sense breaking our conventions by adding vectors of different types.
    This is of course no problem when both parameter and constraint space are just $\R^D$, but if we considered constraints in more general spaces, we would have to make appropriate adjustments to our formulae.}
\begin{equation}\label{eq:aff-subspace}
    E = \setbuild{\v e + \sum_{j=1}^d x_j\v g_j}{x_j\in\R},\qquad
    E = \sat\left[\sum_{k=1}^{D-d}\big(\constr{\v\delta_k}{\v\delta_k\cdot\v e}+\constr{-\v\delta_k}{-\v\delta_k\cdot\v e}\big)\right].
\end{equation}
Up to the choice of $\v e$ and the basis vectors, any vector $\v b$ in parameter space can be uniquely decomposed as
\begin{equation}\label{eq:param-decomp}
    \v b = \v e + \sum_{j=1}^d x_j\v g_j + \sum_{k=1}^{D-d} z_k\v\delta_k.
\end{equation}
We then define $\ind[E]{\v b}\equiv(x_1,x_2,\ldots x_d)$ and $\cind[E]{\v b}\equiv(z_1,z_2,\ldots,z_{D-d})$.
These are $d$- and $(D-d)$-dimensional vectors, respectively, and live in spaces separate from the $D$-dimensional space in which $E,\v e$, etc.\ live.
Note that if $\v b\in E$, then $\cind[E]{\v b}=\v 0$ and $\v b$ is uniquely determined by $\ind[E]{\v b}$.
For constraint-space vectors, we instead make the decomposition
\begin{equation}\label{eq:constr-decomp}
    \v\alpha = \sum_{j=1}^d \xi_j\v g_j + \sum_{k=1}^{D-d} \zeta_k\v\delta_k
\end{equation}
and analogously define $\ind[E]{\v\alpha}\equiv(\xi_1,\xi_2,\ldots\xi_d)$ and $\cind[E]{\v\alpha}\equiv(\zeta_1,\zeta_2,\ldots \zeta_{D-d})$.%
\footnote{
    Our notation does not make the choice of reference point $\v e$ explicit, and it is arbitrary for all purposes.
    Replacing $\v e\to\v e'$ simply entails translating all $\ind[E]{\v b}\to\ind[E]{\v b} + \ind[E]{\v e' - \v e}$ while leaving $\cind[E]{\v b}$, $\ind[E]{\v\alpha}$ and $\cind[E]{\v\alpha}$ unchanged.
    Likewise, altering $\{\v g_j\}_{j=1}^{d},\{\v\delta_k\}_{k=1}^{D-d}$ just corresponds to a change of basis in the spaces.}
These can form constraints acting on $\ind[E]{\v b}$ and $\cind[E]{\v b}$, respectively.
As an extension, for any set $\s X$ we define
\begin{equation}\label{eq:induced-set}
    \ind[E]{\s X} \equiv \setbuild{\vphantom{\cind[E]{\v x}}\ind[E]{\v x}}{\v x\in\s X},\qquad
    \cind[E]{\s X} \equiv \setbuild{\cind[E]{\v x}}{\v x\in\s X}.
\end{equation}
We reiterate how important it is to view $\s X$, $\ind[E]{\s X}$ and $\cind[E]{\s X}$ as living in three different spaces.
There is of course a straightforward mapping between $\ind[E]{\s X}$ and $\s X\cap E$ --- indeed, the $d$-dimensional space can be seen as the vector space underlying the affine subspace $E$ --- but the notion of separate spaces makes the proofs below clearer.

Before moving on to constraints, let us make the folloing definition:
\begin{quotation}
    \noindent The \emph{dimension} of any nonempty set $\s X$, written $\dim(\s X)$, is the affine dimension of the smallest (i.e.\ lowest-dimensional) affine subspace that contains $\s X$.
    Equivalently, $\dim(\s X)$ is the dimension of the affine span of the points in $\s X$.%
    \footnote{
        Note that this definition agrees with the usual affine/linear dimension when $\s X$ is itself an affine/linear subspace.}
\end{quotation}
This smallest affine subspace is clearly unique, for if it is not, the intersection of all such subspaces is even smaller.
For any nonempty convex set $\s C\in\R^D,D>0$, the following basic fact holds:
\begin{equation}\label{eq:empty-interior}
    \dim(\s C) < D \equ \Int(\s C) = \emptyset,
\end{equation}
where $\Int(\s C)$ is the interior of $\s C$.

Consider now a constraint $\Omega\neq\Omega_\infty$ and define $\deg{\Omega}\equiv\dim\!\big[\sat(\Omega)\big]$.
We formalise the definition of degeneracy stated in \cref{sec:relevant} as follows:
\begin{quotation}
    \noindent A constraint $\Omega\neq\Omega_\infty$ in $D$-dimensional space is \emph{degenerate} if $\deg{\Omega}<D$ and \emph{non-degenerate} otherwise.
    $\Omega_\infty$, for which $\deg{\Omega_\infty}$ is undefined, does not fall into either category.
\end{quotation}
Let $E_\Omega$ be the unique $\deg{\Omega}$-dimensional affine subspace that contains $\sat(\Omega)$.
Given linear constraint $\constr{\v\alpha}{c}$, we define
\begin{equation}\label{eq:linear-induced}
    \ind[\Omega]{\vphantom{\sum}\constr{\v\alpha}{c}} \equiv \constr{\vphantom{\cind[E]{\v\alpha}}\ind[E_\Omega]{\v\alpha}}{\quad c - \v\alpha\cdot\v e},\qquad
    \cind[\Omega]{\vphantom{\sum}\constr{\v\alpha}{c}} \equiv \constr{\cind[E_\Omega]{\v\alpha}}{\quad c - \v\alpha\cdot\v e}.
\end{equation}
For these, the following holds:
\begin{lemma}\label{lem:basic-induced}
    For any point $\v b$,
    \begin{enumerate}[label=(\alph*)]
        \item If $\cind[E_\Omega]{\v b} = \v 0$, then $\ind[E_\Omega]{\v b}$ satisfies $\ind[\Omega]{\constr{\v\alpha}{c}}$ if and only if $\v b$ satisfies $\constr{\v\alpha}{c}$.
        \item If $\ind[E_\Omega]{\v\alpha} = \v 0$, then $\cind[E_\Omega]{\v b}$ satisfies $\cind[\Omega]{\constr{\v\alpha}{c}}$ if and only if $\v b$ satisfies $\constr{\v\alpha}{c}$.
    \end{enumerate}
\end{lemma}
\proof This follows directly from \cref{eq:param-decomp,eq:constr-decomp,eq:linear-induced} and the ortho\-normality of $\{\v g_j\}_{j=1}^{d}\cup\{\v\delta_k\}_{k=1}^{D-d}$.
\qed

Using this, we define for any $\Omega\neq\Omega_\infty$ with representation $\s S$
\begin{equation}\label{eq:induced}
    \ind{\Omega} \equiv \sum_{\constr{\v\alpha}{c}\in\s S} \ind[\Omega]{\constr{\v\alpha}{c}}.
\end{equation}
This constraint, which acts on the $\deg{\Omega}$-dimensional space of vectors $\ind[E_\Omega]{\v b}$, has three important properties.
Firstly, $\ind{\Omega}$ is satisfied by $\ind[E_\Omega]{\v b}$ if $\Omega$ is satisfied by $\v b$, and the converse holds when $\cind[E_\Omega]{\v b}=\v 0$ (this follows from \cref{lem:basic-induced}), so 
\begin{equation}
    \sat(\ind{\Omega})=\ind[E_\Omega]{\sat(\Omega)}.
\end{equation}
Secondly, $\ind{\Omega}$ is independent of $\s S$ as a consequence of this.
Lastly, $\ind{\Omega}$ is, by construction, non-degenerate.%
\footnote{
    If $\deg{\Omega}=0$, so that $\sat(\Omega)$ is a single point, then $\ind{\Omega}$ is a zero-dimensional constraint.
    This is not conceptually a problem for \cref{prop:gen-relevant} below: $\ind{\Omega}$ is satisfied by $\v 0$, which is the only point in zero-dimensional space, and $\rel(\ind{\Omega}) = \emptyset$.}
Thanks to these properties, $\ind{\Omega}$ is key to all further treatment of degenerate constraints.

\subsubsection{$K$-faces}\label{sec:K-faces}
In this section, we introduce \emph{$K$-faces}, which will be highly useful in subsequent sections; especially \cref{cor:D-1,proposition B.2-relevant,prop:dual} rely heavily on them.
Like several other things introduced here, it is partially based on standard concepts and nomenclature, but has been adapted and extended to fit the context of linear constraints.
$K$-faces are defined as follows:
\begin{quotation}
    \noindent Let $\s F$ be a non-empty convex subset of a closed convex set $\s C$, with $K=\dim(\s F)$. Then $\s F$ is called a \emph{$K$-face of $\s C$} if the following holds: 
    For every $\v\ph\in\s F$, if there exists $\v\eta_{1,2}\in\s C$ such that $\v\ph=\mu\v\eta_1+(1-\mu)\v\eta_2$ with $\mu\in(0,1)$, then $\v\eta_{1,2}\in\s F$.\\[1ex]    
    \noindent The single point in a 0-face of $\s C$ is called an \emph{extreme point} or \emph{vertex} of $\s C$; it is a point that cannot be expressed as a convex combination of any two points in $\s C$ distinct from itself.\\[1ex]
    \noindent A 1-face of $\s C$ is called an \emph{edge}.\\[1ex]
    \noindent A $(D-1)$-face of $\s C$ is called a \emph{facet}.
\end{quotation}
\noindent 
For convex polygons, polyhedra, etc., these definitions agree with the usual concepts of vertices, edges and facets.
A number of useful properties of $K$-faces easily follow from the definitions:
\begin{enumerate}[label={(\roman*)}]
    \item\label[property]{pty:no-subset}
        No $K$-face of $\s C$ is a strict subset of another $K$-face of $\s C$ with the same $K$.
    \item\label[property]{pty:unique} 
        There is a unique $\dim(\s C)$-face of $\s C$, namely $\s C$ itself.
    \item\label[property]{pty:boundary} 
        All $K$-faces of $\s C$ (except possibly $\s C$ itself) are contained in the boundary of $\s C$.
    \item\label[property]{pty:subface}
        If $\s F$ is a $K$-face of $\s C$, then for all $K'\leq K$ the $K'$-faces of $\s F$ are also $K'$-faces of $\s C$.
        Specifically, an edge may have up to two vertices, which are its endpoints.
    \item\label[property]{pty:face-of-subset}
        If $\s F$ is a $K$-face of $\s C$ and $\s C'\subseteq\s C$, then if $\s F\cap\s C'$ is non-empty, it is a $K'$-face ($K'\leq K$) of $\s C'$.
\end{enumerate}
\noindent We will now prove some less obvious properties.
In the remainder of this section, let $\s C$ be any convex set such that there exists a finite set of constraints $\{\constr{\v g_\ell}{c_\ell}\}_{\ell\in L}$ fulfilling%
\footnote{
    This class of sets includes all convex hulls of finite sets, all linear and affine subspaces, as well as most other sets we work with, including $\reg_c(\Omega)$ as we will prove in \cref{sec:practical}.
    For convenience, we express $\s C$ as a subset of constraint space, but all results hold equally well if $\s C$ is a subset of parameter space.}
\begin{equation}\label{eq:C}
    \s C = \sat\left(\sum_{\ell\in L}\constr{\v g_\ell}{c_\ell}\right).
\end{equation}
We then begin with the following technical lemma:
\begin{lemma}\label{lem:technical}
    Let $\s C$ be a convex set defined as in \cref{eq:C}, and let $\s F\subseteq\s C$ be non-empty with dimension $K<D$. Then $\s F$ is a $K$-face of $\s C$ if and only if there exists $J\subset L$ with $|J|=D-K$ such that $\{\v g_j\}_{j\in J}$ are linearly independent and
    \begin{equation}\label{eq:technical}
            \forall\v\ph\in\s F,\forall j\in J,\quad \v g_j\cdot\v\ph = c_j.
    \end{equation}
\end{lemma}
\noindent Note that there may be $\ell\in L\setminus J$ such that $\v g_\ell\cdot\v\ph = c_\ell$.
All such $\v g_\ell$ are contained in the span of $\{\v g_j\}_{j\in J}$, though.

\proof 
If such a $J$ exists, then $\s F'\equiv\setbuild{\v\ph\in\R^D}{\forall j\in J,\v g_j\cdot\v\ph = c_j}$ is clearly an affine subspace of dimension $K$.
Let $\v\ph\in\s F'$ and $\v\eta_{1,2}\in\s C$, and assume $\v\ph=\mu\v\eta_1+(1-\mu)\v\eta_2$ with $\mu\in(0,1)$.
Then
\begin{equation}\label{eq:trivial-facet}
    \v g_j\cdot\v\ph = d = \v g_j\cdot\big[\mu\v\eta_1 + (1-\mu)\v\eta_2\big]
\end{equation}
and since $\v g_j\cdot\v\eta_{1,2}\geq c_j$, this implies $\v g_j\cdot\v\eta_{1,2} = c_j$.
Therefore $\v\eta_{1,2}\in\s F'$, so $\s F=\s F'\cap\s C$ is a $K$-face.

For the less straightforward converse, let $\omega = \sum_{\ell\in L}\constr{\v g_\ell}{c_\ell}$, with $\s C=\sat(\omega)$.
If $\omega$ is degenerate, then $\s C$ clearly has no $K$-faces for $K>\deg{\omega}$, and the unique $\deg{\omega}$-face is $\s C$ itself by \cref{pty:unique}.
In that case, the proof follows trivially from \cref{eq:aff-subspace}.
For the remaining cases, we may substitute $\omega\to\ind{\omega},\s C\to\ind[\omega]{\s C}=\sat(\ind{\omega})$ and thus assume without loss of generality that $\omega$ is non-degenerate.

Let us then note that $J$ cannot be empty.  
Indeed, if it were the case, then $\v\ph\in\s F$ would be in the interior of $\s C$ which via \cref{pty:boundary} contradicts the fact that $\s F$ is a $K$-face with $K < D$.
Also, $|J|\leq D-K$, since otherwise the set $\setbuild{\v\chi\in\s C}{\forall j\in J,\v\chi\cdot\v g_j=c_j}$, being the intersection of more than $D-K$ independent hyperplanes, would have dimension less than $K$.
We will then proceed by induction on $D>K$.
The result is trivial in $K+1$ dimensions, since $|J|=1$ is guaranteed by $J$ being nonempty. 

Assume then that the lemma holds in $n$ dimensions, and consider $\s C,\s F$ in $(n+1)$-dimensional space.
Since $J\neq\emptyset$, there is some $i\in L$ such that $\forall\v\ph\in\s F,\v g_i\cdot\v\ph=c_i$.
Then consider
\begin{equation}\label{eq:EJ}
    E_i \equiv \sat\big(\constr{\v g_i}{c_i} + \constr{-\v g_i}{-c_i}\big),\qquad
    \ind[E_i]{\s C} = \sat\left(\sum_{\ell\in L}\ind[E_i]{\constr{\v g_\ell}{c_\ell}}\right).
\end{equation}
By construction, $\ind[E_i]{\s F}$ is a $K$-face of $\ind[E_i]{\s C}$.
Since these are sets in a $n$-dimensional space, we know by the induction hypothesis that we have $J'$ with $|J'|=n-K$ such that $\{\ind[E_i]{\v g_j}\}_{j\in J'}$ are linearly independent and \cref{eq:technical} is satisfied.
Furthermore, $\ind[E_i]{\v g_i}=\v 0$, which cannot be expressed as a linear combination of $\{\ind[E_i]{\v g_j}\}_{j\in J'}$ with nonzero coefficients.
Therefore, $\v g_i$ is linearly independent of $\{\v g_j\}_{j\in J'}$.
Thus, $J=J'\cup\{i\}$ has $|J|=n+1$, $\{\v g_j\}_{j\in J}$ linearly indpendent and satisfies \cref{eq:technical}.
This proves that the lemma holds in $n+1$ dimensions, and completes the induction.
\qed

Based on this, we can prove two more interesting lemmata:
\begin{lemma}\label{lem:K-face}
    Let $\s C$ be a convex set satisfying \cref{eq:C}. 
    Let $\s F$ be a nonempty convex subset of $\s C$, and let $K=\dim(\s F)$. Then $\s F$ is a $K$-face of $\s C$ if and only if there is some constraint $\constr{\v h}{d}$ such that $\s C\subseteq\sat(\constr{\v h}{d})$ and $\setbuild{\v\chi\in\s C}{\v h\cdot\v\chi = d} = \s F$.%
    \footnote{
        Although we do not use it, this holds for $K=D-1$ for \emph{any} convex set $\s C$, not just those satisfying \cref{eq:C} (this is proven in a later footnote).
        This is not the case for smaller $K$: take e.g.\ the $D=2$ example $\s C=\setbuild{(x,y)\in\R^2}{y\geq\max(0,x^3)}$, for which $(0,0)$ is an extreme point but $\constr{\v h}{d}$ does not exist.}
\end{lemma}
\proof First note that this trivially holds when $\setbuild{\v\chi\in\s C}{\v h\cdot\v\chi = d} = \s C$ (compare \cref{pty:unique}), which may happen even when $\s C$ is not contained in any hyperplane if $\constr{\v h}{d}=\constr{\v 0}{0}$. 

Setting aside the trivial cases, assume that such a $\constr{\v h}{d}$ exists.
Then $\s F$ is a $K$-face by essentially the same argument that was made around \cref{eq:trivial-facet}.

Conversely, if $\s F$ is a $K$-face then \cref{lem:technical} holds.
In the notation of that lemma, let
\begin{equation}
    \v h = \sum_{\ell\in J}\v g_\ell,\qquad d = \sum_{\ell\in J} c_\ell,
\end{equation}
so that $\forall\v\ph\in\s F,\v h\cdot\v\ph = d$.
Let us now consider any $\v\eta\in\s C$ such that $\v h\cdot\v\eta = d$.
Let $E_J$ be defined as in \cref{eq:EJ}.
Then by construction, $\ind[E_J]{\s F}$ has the same dimension as its native space, i.e.\ $D-K$, and has nonempty interior by \cref{eq:empty-interior} (recall that $D-K>0$).
Thus, there exists $\v\ph\in\s F$ such that $\ind[E_J]{\v\ph}$ is in the interior of $\ind[E_J]{\s F}$.
For $\e>0$ small enough, we therefore have
\begin{equation}
    \ind[E_J]{\v\eta_\e} \equiv \e\ind[E_J]{\v\eta} + (1-\e)\ind[E_J]{\v\ph} \in \ind[E_J]{\s F}.
\end{equation}
Since by construction $\cind[E_J]{\v\chi}=\v 0$ for all $\v\chi$ such that $\v h\cdot\v\chi = d$, \cref{lem:basic-induced} implies that
\begin{equation}
    \v\eta_\e = \e\v\eta + (1-\e)\v\ph.
\end{equation}
By the definition of a $K$-face, this implies that $\v\eta\in\s F$, which concludes the proof.%
\footnote{
    When $K=D-1$, the following proof, which does not use \cref{lem:technical} and therefore holds for all convex sets $\s C$, works for the converse:
    Let $\s F'$ be the affine span of $\s F$, which is a hyperplane.
    By a variant of the separating hyperplane theorem (for any disjoint convex sets $\s X$ and $\s Y$ (no closedness/compactness needed), there exists $\constr{\v h}{d}$ with $\v h\cdot\v\chi \leq d'$ for all $\v\chi\in\s X$ and $\v h\cdot\v\psi\geq d'$ for all $\v\psi\in\s Y$) applied to the disjoint convex sets $\s F'$ and $\s C\setminus \s F$, we immediately find our desired $\constr{\v h}{d}$.
    This hinges on the properties of hyperplanes: the separating hyperplane must be parallel to $\s F'$, or else they would intersect.
    Therefore, this does not work if $K<D-1$.}
\qed

\begin{lemma}\label{lem:K-face-subset}
    Let $\s C\subseteq\s C'$ be convex sets satisfying \cref{eq:C}, and let $\s F$ be a $K$-face of $\s C$.
    Then there exists a $K$-face $\s F'$ of $\s C'$ with $\s F\subseteq\s F'$ if there is some $\v\ph\in\s F$ and $\e>0$ such that $\setbuild{\v\chi\in\s C'}{\e\geq|\v\chi-\v\ph|} \subseteq \s C$.
\end{lemma}
\noindent Note how this complements \cref{pty:face-of-subset}.
Note also that $\s F'\supset\s F$ may exist even if $\s F$ fails to satisfy the given conditions.

\proof By \cref{lem:K-face}, there exists $\constr{\v h}{d}$ such that  and $\setbuild{\v\chi\in\s C}{\v h\cdot\v\chi = d} = \s F$ and $\s C\subseteq\sat(\constr{\v h}{d})$.
We moreover claim that $\s C'\subseteq \sat(\constr{\v h}{d})$.
Indeed, let $\v\ph\in\s F$ and $\e>0$ such that $\setbuild{\v\chi\in\s C'}{\e\geq|\v\chi-\v\ph|} \subseteq \s C$.
For any $\v\eta\in\s C'$ it follows from the triangle inequality that
\begin{equation}
    \left|\v\ph - \left[\left(1-\tfrac{\e}{|\v\ph|+|\v\eta|}\right)\v\ph + \tfrac{\e}{|\v\ph|+|\v\eta|}\v\eta\right]\right|
    = \e\frac{|\v\ph - \v\eta|}{|\v\ph|+|\v\eta|} \leq \e,
\end{equation}
so $\left(1-\tfrac{\e}{|\v\ph|+|\v\eta|}\right)\v\ph + \tfrac{\e}{|\v\ph|+|\v\eta|}\v\eta\in\s C$.
In particular,
\begin{equation}
    \v h\cdot\left[\left(1-\tfrac{\e}{|\v\ph|+|\v\eta|}\right)\v\ph + \tfrac{\e}{|\v\ph|+|\v\eta|}\v\eta\right]\geq d
    \imp\v h\cdot\v\eta\geq d.
\end{equation}
Therefore, \cref{lem:K-face} guarantees that $\s F'=\setbuild{\v\eta\in\s C'}{\v h\cdot\v\eta = d}$ is a $K'$-face for some $K'$.
Since $\s F\subseteq\s F'$ we have $K\leq K'$, so it remains to prove that $K\geq K'$.
For any $\v\ph'\in\s F'$, we have $\v\ph + \e\frac{\v\ph'}{|\v\ph'|}\in\s F$ since $\left|\v\ph - \left(\v\ph + \e\frac{\v\ph'}{|\v\ph'|}\right)\right| \leq \e$.
However, since $-\v\ph$ belongs to the span of $\s F$, we can deduce that $\e\frac{\v\ph'}{|\v\ph'|}$, and therefore also $\v\ph'$, belongs to the span of $\s F$.
Thus, the span of $\s F'$ is included in the span of $\s F$, implying that $K\geq K'$, which completes the proof.
\qed

\subsection{Proof and generalisation of \cref{proposition 4.3}}\label{sec:proof-relevant}
To simplify this proof, we will introduce the following terminology:
\begin{quotation}
    \noindent A linear constraint $\constr{\v\alpha}{c}$ with $\v\alpha\neq\v 0$ \emph{supports} a point $\v b$ if $\v\alpha\cdot\v b = c$.%
    \footnote{
        This is inspired by the standard concept of a \emph{supporting hyperplane}.}\\ 
    A linear constraint $\constr{\v 0}{c}$ supports no point (this is natural for $c=\pm1$, but we define it to be so also for $c=0$; this simplifies most statements expressed in terms of support). 
    \\[1ex]
    \noindent Given $\Omega$, a linear constraint $\constr{\v\alpha}{c}$ \emph{uniquely supports} a point $\v b\in\sat(\Omega)$ if it supports $\v b$, and there is no other $\constr{\v\beta}{d}\leq\Omega$ with $\constr{\v\beta}{d}\neq\constr{\v\alpha}{c}$ that supports $\v b$.
    \\[1ex]
    \noindent A constraint representation $\s S$ is \emph{non-redundant} if it contains no trivial constraints, and there are no two elements $\constr{\v\alpha}{c}\in\s S$ and $\constr{\v\beta}{d}\in\s S$ such that $\constr{\v\alpha}{c}=\constr{\v\beta}{d}$.
\end{quotation}
Reducing a representation to a non-redundant one is of course trivial. 
Noting that the second paragraph of \cref{proposition 4.3} can be reduced to ``The elements of $\rel(\Omega)$ are exactly those $\constr{\v\alpha}{c}\leq\Omega$ that uniquely support a point $\v b\in\sat(\Omega)$'', we will then begin with the following lemma:
\begin{lemma}\label{lem:unique-support-1}
    Let $\Omega=\sum_{i\in I}\constr{\v\alpha_i}{c_i}$ be a non-degenerate constraint, as defined in \cref{sec:relevant}, and let the representation $\{\constr{\v\alpha_i}{c_i}\}_{i\in I}$ be non-redundant. Then for any $j\in I$ and $\v b\in\sat(\Omega)$, $\constr{\v\alpha_j}{c_j}$ uniquely supports $\v b$ if and only if $\v\alpha_j\cdot\v b = c_j$ and $\v\alpha_i\cdot\v b > c_i$ for all $i\neq j$.
\end{lemma}
\noindent Note that compared to the definition of unique support, this only concerns the elements of a non-redundant representation rather than all $\constr{\v\beta}{c}\leq\Omega$.

\proof Assume that $\v\alpha_j\cdot\v b = c_j$ and $\v\alpha_i\cdot\v b > c_i$ for all $i\neq j$, and assume there is some $\constr{\v\beta}{c}<\Omega$ that supports $\v b$, i.e.\ that $\constr{\v\alpha_j}{c_j}$ does not support it uniquely. 
By \cref{proposition 4.2}, there exist some positive numbers $\{\lambda_i\}_{i\in I}$ such that
\begin{equation}
    c = \v\beta\cdot\v b = \sum_{i\in I} \lambda_i\v\alpha_i\cdot\v b \imp
    \begin{cases}
        c = \lambda_j c_j &   \text{if $\lambda_i=0$ for all $i\neq j$},\\
        c > \sum_{i\in I}\lambda_i c_i \geq c &   \text{otherwise}.
    \end{cases}
\end{equation}
The second case is a contradiction, so $\v\beta=\lambda_j\v\alpha_j$ and $c=\lambda_j c_j$ with $\lambda_j\geq0$. 
If $c_j=\pm1$, then either $\v\beta=-\v\alpha_j$ and $c=-c_j$ (contradicting non-degeneracy), or $\v\beta=\v 0$ (impossible since $\constr{\v 0}\pml$ supports no point), or $\v\beta=\v\alpha_j$ and $c=c_j$ (contradicting non-redundancy).
If $c_j=0$, then $c=0$ and $\v\beta$ is proportional to $\v\alpha_j$ (again contradicts non-rendundancy, via \cref{eq:rescale}).  
With no non-contradictory cases left, we have proven that $\constr{\v\alpha_j}{c_j}$ supports $\v b$ uniquely if $\v\alpha_j\cdot\v b = c_j$ and $\v\alpha_i\cdot\v b > c_i$ for all $i\neq j$. 
The converse is trivial. 
\qed

Let us then prove the following more significant lemma:
\begin{lemma}\label{lem:unique-support-2}
    Let $\constr{\v\alpha}{c}$ be a linear constraint, and let $\Omega$ be a constraint such that $\Omega+\constr{\v\alpha}{c}$ is non-degenerate. 
    Assume that $\Omega$ uses a non-redundant representation $\s S$ and that $\s S \cup \{\constr{\v\alpha}{c}\}$ is a non-redundant representation for $\Omega + \constr{\v\alpha}{c}$.
    Then $\Omega + \constr{\v\alpha}{c} > \Omega$ if and only if $\constr{\v\alpha}{c}$ uniquely supports some point $\v b \in\sat[\Omega+\constr{\v\alpha}{c}]$.
\end{lemma}
\proof Assume that $\constr{\v\alpha}{c}$ uniquely supports some point $\v b\in\sat(\Omega+\constr{\v\alpha}{c})$, so that $\v\alpha\cdot\v b = c$ and $\v\beta\cdot\v b > d$ for all $\constr{\v\beta}{d}\in\s S$. 
Then define
\begin{equation}
    \v b_\e \equiv \v b - \e\frac{\v\alpha}{|\v\alpha|^2},\quad\e > 0
\end{equation}
so that
\begin{equation}
    \v\alpha\cdot\v b_\e = c - \e < c,\qquad \v\beta\cdot\v b_\e > d - \e\frac{\v\beta\cdot\v\alpha}{|\v\alpha|^2}.
\end{equation}
Thus, $\v b_\e\not\in\sat(\Omega+\constr{\v\alpha}{c})$, but $\v b_\e \in \sat(\Omega)$ for sufficiently small $\e$. Then by definition, $\Omega+\constr{\v\alpha}{c} > \Omega$.

Conversely, assume that $\Omega+\constr{\v\alpha}{c} > \Omega$. 
Then there exists some $\v b\in \sat(\Omega)$ such that $\v b\not\in\sat(\Omega+\constr{\v\alpha}{c})$. 
Furthermore, since $\Omega+\constr{\v\alpha}{c}$ is non-degenerate, the interior of $\sat(\Omega+\constr{\v\alpha}{c})$ is non-empty;%
\footnote{
    Recall that $\Omega$ being degenerate is equivalent to $\sat(\Omega)$ being contained in a hyperplane, and a convex set has empty interior if and only if it is contained in a hyperplane; see also \cref{eq:empty-interior}.
    In general, the existence of a point in the interior of $\sat(\Omega)$ is the \emph{only} property of non-degenerate constraints used in this proof.
    Circumventing this requirement is key to \cref{prop:gen-relevant} below.}
therefore, it contains some point $\v n$. 
These points have the properties
\begin{equation}
    \begin{alignedat}{4}
        \v\alpha\cdot\v b &< c, \qquad  &   \v\beta\cdot\v b &\geq d,\\
        \v\alpha\cdot\v n &> c,         &   \v\beta\cdot\v n &> d. 
    \end{alignedat}
\end{equation}
where again $\constr{\v\beta}{d}\in\s S$. By the intermediate value theorem, there must therefore exist some $\mu\in(0,1)$ such that%
\footnote{
    Specifically, $\mu = \dfrac{\v\alpha\cdot\v n  - c}{\v\alpha\cdot\v n  - \v\alpha\cdot\v b}$.}
\begin{equation}
    \v\alpha\cdot\big[\mu\v b + (1-\mu)\v n\big] = c,\qquad \v\beta\cdot\big[\mu\v b + (1-\mu)\v n\big] > d,
\end{equation}
which, through \cref{lem:unique-support-1}, proves that $\constr{\v\alpha}{c}$ supports $\mu\v b + (1-\mu)\v n$ uniquely.
\qed

We now move on to proving \cref{proposition 4.3}. Let a non-degenerate constraint $\Omega$ be expressed as
\begin{equation}
    \Omega = \sum_{i\in I}\constr{\v\alpha_i}{c_i} = \sum_{j\in J}\constr{\v\gamma_j}{d_j}
\end{equation}
where $\{\constr{\v\alpha_i}{c_i}\}_{i\in I}$ is any representation, whereas $\{\constr{\v\gamma_j}{d_j}\}_{j\in J}$ is minimal.
A minimal representation must exist, since $\Omega$ can be written as a sum of a finite number of constraints.
For any $k\in J$, minimality implies that
\begin{equation}
    \sum_{j\in J} \constr{\v\gamma_j}{d_j} > \sum_{\substack{j\in J\\j\neq k}}\constr{\v\gamma_j}{d_j},
\end{equation}
but by \cref{lem:unique-support-2}, $\constr{\v\gamma_k}{d_k}$ must then uniquely support some $\v b\in \sat(\Omega)$.
Then, there must also be some $i_k\in I$ such that $\constr{\v\alpha_{i_k}}{c_{i_k}}$ also supports $\v b$.
To see this, assume that no $\constr{\v\alpha_i}{c_i},i\in I$ supports $\v b$, and consider 
\begin{equation}
    \v b_\e \equiv \v b - \e\frac{\v\gamma_k}{|\v\gamma_k|^2}
    \imp \v\alpha_i\cdot\v b_\e > c - \e\frac{\v\alpha\cdot\v\gamma_j}{|\v\gamma_j|^2},
\end{equation}
so that for sufficiently small $\e$, $\v\alpha_i\cdot\v b_\e > c$ for all $i\in I$, and thus $\v b_\e\in\sat(\Omega)$.
On the other hand,
\begin{equation}
    \v\gamma_k\cdot\v b_\e = d_k - \e
\end{equation}
implying $\v b\not\in\sat(\Omega)$, a contradiction.
Therefore, $i_k\in I$ does exist.
By the definition of unique support, we must then have $\constr{\v\gamma_k}{d_k} = \constr{\v\alpha_{i_k}}{c_{i_k}}$; that is, they are identical up to normalisation.
By repeating this argument, we see that a distinct $i_k$ exists for each $k\in J$, so $\{\constr{\v\gamma_j}{d_j}\}_{j\in J}\subseteq \{\constr{\v\alpha_i}{c_i}\}_{i\in I}$ up to normalisation.
Having shown this, we may without loss of generality normalise and re-index $\constr{\v\gamma_j}{d_j}$ so that $J\subseteq I$.
Then
\begin{equation}
    \begin{aligned}
        \Omega 
            &= \sum_{i\in I} \constr{\v\alpha_i}{c_i}    \\
            &= \sum_{i \in J} \constr{\v\alpha_i}{c_i} 
                + \sum_{i\in I\setminus J} \constr{\v\alpha_i}{c_i}\\
            &= \Omega + \sum_{i \in I\setminus J} \constr{\v\alpha_i}{c_i}.
    \end{aligned}
\end{equation}
By \cref{lem:unique-support-2}, the sum in the last line can only contain constraints that do not uniquely support any point.
Thus, the minimal representation $\rel(\Omega)=\{\constr{\v\alpha_i}{c_i}\}_{i\in J}$ consists (up to normalisation) of \emph{exactly} those elements of any representation that uniquely support a point.
From this it follows that $\rel(\Omega)$ consists of exactly all those $\constr{\v\gamma}{d}\leq\Omega$ that uniquely support a point, since $\{\constr{\v\alpha_i}{c_i}\}_{i\in I}$ could be made to include all $\constr{\v\alpha}{c}\leq\Omega$, and from that it follows that $\rel(\Omega)$ is unique up to normalisation.
\QED

\subsubsection{The treatment of degenerate constraints}
The following result generalises \cref{proposition 4.3} to all $\Omega$:  
\begin{proposition}[finding relevant constraints, general case]\label{prop:gen-relevant}
    Let $\Omega$ be any constraint in $D$-dimensional space.
    Then a minimal representation $\rel(\Omega)$ can be determined as follows:
    \begin{enumerate}[label=\textrm{(\roman*)}]
        \item\label[case]{case:non-deg}
            If $\Omega$ is non-degenerate, then \cref{proposition 4.3} applies.
            $\rel(\Omega)$ is therefore unique up to normalisation, and is a subset of any representation of $\Omega$.
        
        \item\label[case]{case:infty}
            If $\Omega=\Omega_\infty$, then trivially $\rel(\Omega) = \{\constr{\v 0}{1}\}$.
            This is unique, but not necessarily a subset of other representations.
            
        \item\label[case]{case:deg}
            If $\Omega$ is degenerate, let $E_\Omega$ be the unique $\deg{\Omega}$-dimensional affine subspace that contains $\sat(\Omega)$.
            Let $\big\{\constr{\v\alpha_i}{c_i}\big\}_{i\in I}$ be any set such that \mbox{$\big\{\ind[\Omega]{\constr{\v\alpha_i}{c_i}}\big\}_{i\in I}=\rel(\ind{\Omega})$}, with $\ind[\Omega]{\constr{\v\alpha_i}{c_i}}\neq\ind[\Omega]{\constr{\v\alpha_j}{c_j}}$ for all $i\neq j$.
            Let $\left\{\v\sigma_k\right\}_{k=0}^{\deg{\Omega}}$ be any set of \mbox{$(D-\deg{\Omega}+1)$} vectors with the following properties:
            \begin{itemise}
                \item $\ind[E_\Omega]{\v\sigma_k} = \v 0$;
                \item The dimension of $\Span\!\left(\{\v\sigma_k\}_{k=0}^{D-\deg{\Omega}}\right)$ is $D-\deg{\Omega}$;
                \item There exists a solution to
                \begin{equation}\label{eq:sigma-zero}
                     \v 0 = \sum_{k=0}^{D-\deg{\Omega}} \lambda_k\v\sigma_k,\qquad \lambda_k>0.
                \end{equation}
            \end{itemise}
            Then for arbitrary $\v e\in E_\Omega$,
            \begin{equation}\label{eq:deg-relevant}
                \rel(\Omega) = \big\{\constr{\v\alpha_i}{c_i}\big\}_{i\in I} \cup \big\{\constr{\v\sigma_k}{\v\sigma_k\cdot\v e}\big\}_{k=0}^{D-\deg{\Omega}}.
            \end{equation}
            This $\rel(\Omega)$ is generally not unique, and is not neccesarily a subset of any given representation of $\Omega$.
            However, there is no minimal representation of $\Omega$ that is not of this form.
    \end{enumerate}
\end{proposition}
\noindent The non-unqiueness in \cref{case:deg} comes about in two ways. 
Firstly, $\cind[E_\Omega]{\v\alpha}$ is arbitrary for $\ind[\Omega]{\constr{\v\alpha}{c}}\in\rel(\ind{\Omega})$. 
Secondly, there is clearly freedom in the choice of $\left\{\v\sigma_k\right\}_{k=0}^{D-\deg{\Omega}}$.
Given $\{\v\delta_k\}_{k=1}^{D-\deg{\Omega}}$ as defined above \cref{eq:aff-subspace}, a straightforward choice is
\begin{equation}
    \{\v\sigma_k\}_{k=1}^{D-\deg{\Omega}} = \{\v\delta_k\}_{k=1}^{D-\deg{\Omega}},\qquad -\v\sigma_0 = \sum_{k=1}^{D-\deg{\Omega}}\v\sigma_k.
\end{equation}

\proof We only need to prove \cref{case:deg}.
For brevity, we will omit some sub/superscripts: $\ind{\constr{\v\alpha}{c}}$ should be read as $\ind[\Omega]{\constr{\v\alpha}{c}}$, $\cind{\v b}$ as $\cind[E_\Omega]{\v b}$, and so on.

Let $\big\{\constr{\v\alpha_i}{c_i}\big\}_{i\in J}$ be a minimal representation of $\Omega$, and subdivide it as
\begin{equation}\label{eq:categ}
    J_1 \equiv \setbuild{\vphantom{\sum}i\in J}{\ind{\constr{\v\alpha_i}{c_i}} \neq \constr{\v 0}{0}},\qquad
    J_2 \equiv \setbuild{\vphantom{\sum}i\in J}{\ind{\constr{\v\alpha_i}{c_i}} =    \constr{\v 0}{0}},
\end{equation}
for which the following holds:
\begin{lemma}\label{lem:induced}
    $\{\ind{\constr{\v\alpha_i}{c_i}}\}_{i\in J_1} = \rel(\ind{\Omega})$ (up to normalisation).
\end{lemma}
\proof By \cref{proposition 4.3}, it is clear that $\{\ind{\constr{\v\alpha_i}{c_i}}\}_{i\in J} \supseteq \rel(\ind{\Omega})$, since it it a representation of $\ind{\Omega}$.
Then, for some $i\in J$, consider
\begin{equation}
    \Omega'\equiv\sum_{\substack{j\in J\\j\neq i}} \constr{\v\alpha_j}{c_j}.
\end{equation}
Let us assume that $\ind{\constr{\v\alpha_i}{c_i}} \not\in \rel(\ind{\Omega})$, so that $\ind{\Omega'} = \ind{\Omega}$.
However, $\Omega'\neq\Omega$, since $\{\constr{\v\alpha_i}{c_i}\}_{i\in J}$ is minimal.
Thus, by \cref{lem:basic-induced}, $\Omega'$ must be satisfied by some point $\v a\not\in E_\Omega$.
For each $\v b\in\sat(\Omega)$, consider then $x\v a + (1-x)\v b$ for $x\in (0,1]$.
This point satisfies $\Omega'$ but not $\Omega$, since it lies outside $E_\Omega$.
From this, we conclude that it does not satisfy $\constr{\v\alpha_i}{c_i}$.
Thus, the continuous function 
\begin{equation}
    f_{\v b}(x) = \v\alpha_i\cdot[x\v a + (1-x)\v b]
\end{equation}
has $f_{\v b}(x)<c_i$ for $x>0$.
However, $f_{\v b}(0)\geq c_i$ since $\v b$ satisfies $\constr{\v\alpha_i}{c_i}$, and this is only consistent with continuity if $f_{\v b}(0)=c_i$, i.e.\ that $\v\alpha_i\cdot\v b = c_i$, for all $\v b\in\sat(\Omega)$.
Then $\sat(\Omega)\subseteq \sat(\constr{\v\alpha_i}{c_i}+\constr{-\v\alpha_i}{-c_i})$, so it follows from \cref{eq:aff-subspace} that $\v\alpha_i$ is a linear combination of $\{\v\delta_k\}_{k=1}^{D-\deg{\Omega}}$, i.e.\ that $\ind{\v\alpha_i}=\v 0$.
\Cref{eq:linear-induced} then shows that $i\in J_2$, completing the proof.
\qed

Now, let
\begin{equation}\label{eq:min-rep}
    \Omega = \sum_{i\in J_1} \constr{\v\alpha_i}{c_i} + \sum_{i\in J_2}\constr{\v\alpha_i}{c_i}.
\end{equation}
\Cref{lem:induced} connects the $J_1$ part with $\rel(\ind{\Omega})$, so it remains to study the $J_2$ part.
We claim that \cref{eq:min-rep} holds true if and only if
\begin{equation}\label{eq:sat-0}
    \sat\left(\sum_{i\in J_2} \cind{\constr{\v\alpha_i}{c_i}} \right) = \{\v 0\}.
\end{equation}
It follows immediately from \cref{lem:induced} that
\begin{equation}
    \v 0 \in \sat\left(\sum_{i\in J_2} \cind{\constr{\v\alpha_i}{c_i}} \right).
\end{equation}
Assume then that this set also contains some $\cind{\v v}\neq\v 0$.
Since $\ind{\Omega}$ is non-degenerate, there exists $\ind{\v b}\in\Int\big[\sat(\ind{\Omega})\big]$; choosing $\cind{\v b}=\v 0$, it follows that \mbox{$\forall i\in J_1,\; \v\alpha_i\cdot\v b > c_i$}.
Therefore, for $\e>0$ small enough we have $\forall i\in J_1,\; \v\alpha_i\cdot(\v b+\e\v v) \geq c_i$.
Since $\cind{\v b + \e\v v} = \e\cind{\v v}$, we also have $\forall i\in J_2,\; \v\alpha_i\cdot(\v b+\e\v v) \geq c_i$.
Thus, $\v b + \e\v v\in\sat(\Omega)$, but since $\cind{\v v}\neq \v 0$, we have $\v b+\e\v v\not\in E_\Omega$, a contradiction.
Along with its trivial converse, this proves the equivalence between \cref{eq:min-rep,eq:sat-0}.

What is then the minimal set $\{\constr{\v\alpha_i}{c_i}\}_{i\in J_2}$ that produces \cref{eq:sat-0}?
Since $\{\v 0\}$ is a bounded set, \cref{cor:bounded} states that $\{\cind{\v\alpha_i}\}_{i\in J_2}$ spans the full $(D-\deg{\Omega})$-dimensional space, and that there are $\lambda_i>0,i\in J_2$ such that
\begin{equation}
    \v 0 = \sum_{i\in J_2}\lambda_i\v\alpha_i.
\end{equation}
The span condition requires $|J_2|\geq D-\deg{\Omega}$, but in order for there to be a nontrivial linear combination equal to zero, we must in fact have $|J_2|\geq D-\deg{\Omega}+1$.
It is easy to see that this bound is sufficient (for details, see the proof of \cref{cor:bounded}), so the minimal set must have $|J_2| = D-\deg{\Omega}+1$.
Identifying $\{\v\alpha_i\}_{i\in J_2}$ with $\{\v\sigma_k\}_{k=0}^{D-\deg{\Omega}}$, we see that we have just derived all conditions stated in the proposition, so the proof is complete.
\QED

\subsubsection{An important corollary}
The following interesting result, which is also our first use of the $K$-faces defined in \cref{sec:K-faces}, is a consequence of \cref{proposition 4.3}:
\begin{corollary}[facet supported by relevant element]\label{cor:D-1}
    Let $\Omega\neq\Omega_\infty$ be a non-degenerate constraint, and let $\constr{\v\alpha}{c}\leq\Omega$.
    Then $\constr{\v\alpha}{c}\in\rel(\Omega)$ (up to normalisation) if and only if the set $\s F\equiv\setbuild{\v b\in\sat(\Omega)}{\v\alpha\cdot\v b = c}$ is a facet of $\sat(\Omega)$.%
    \footnote{
        It may seem obvious that this condition is equivalent to uniquely supporting a point, but is in fact rather subtle, and crucially depends on $\Omega$ having a finite representation.
        For instance, if $\sat(\Omega)$ were a closed unit ball, then every point $\v b$ on its surface would be uniquely supported by the constraint $\constr{-\v b}{-1}$.
        For such a constraint, $\s F=\{\v b\}$, which is not a facet in $D>1$ dimensions.}
\end{corollary}
\proof It is guaranteed via \cref{lem:K-face} that $\s F$ is a $K$-face; we only need to show that $K=D-1$ so that it is a facet.

Assume $\constr{\v\alpha}{c}\in\rel(\Omega)$.
Let $\constr{\v\beta}{d}$ be any constraint that supports all of $\s F$, and construct the constraints
\begin{equation}
    \Omega' \equiv \sum_{\constr{\v\alpha'}{c'}\in\rel(\Omega)}\constr{\v\alpha'}{c'} + \constr{\v\beta}{d},\qquad
    \Omega'' \equiv \sum_{\constr{\v\alpha'}{c'}\in\rel(\Omega)}\constr{\v\alpha'}{c'} + \constr{-\v\beta}{d}.
\end{equation}
Then it is clear that $\sat(\Omega) = \sat(\Omega') \cup \sat(\Omega'')$, so at least one of $\Omega',\Omega''$ must be non-degenerate.
Without loss of generality, assume $\Omega'$ is non-degenerate.
Since $\constr{\v\alpha}{c}$ does not uniquely support any point in $\sat(\Omega')$ (all points supported by it are also supported by $\constr{\v\beta}{d}$), \cref{proposition 4.3} (or rather \cref{lem:unique-support-2}) gives
\begin{equation}\label{eq:Omega-prime}
    \Omega' = \sum_{\substack{\constr{\v\alpha'}{c'}\in\rel(\Omega)\\\constr{\v\alpha'}{c'}\neq\constr{\v\alpha}{c}}}\constr{\v\alpha'}{c'} + \constr{\v\beta}{d}.
\end{equation}
Recall that $\constr{\v\alpha}{c}$, viewed as an element of $\Omega$, uniquely supports some point $\v b\in\s F$.
Therefore, $\v\alpha'\cdot\v b > c'$ for all $\constr{\v\alpha'}{c'}\in\rel(\Omega)\setminus\{\constr{\v\alpha}{c}\}$.
By \cref{eq:Omega-prime}, $\constr{\v\beta}{d}$ then uniquely supports $\v b\in\sat(\Omega')$, but since $\constr{\v \alpha}{c}\leq\Omega'$ and $\v\alpha\cdot\v b = c$, the definition of unique support gives $\constr{\v\alpha}{c}=\constr{\v\beta}{d}$.
Thus, the only constraints that support all of $\s F$ are, up to normalisation, $\constr{\v\alpha}{c}$ and $\constr{-\v\alpha}{-c}$, confirming via \cref{lem:technical} that $\dim(\s F) = D-1$ (compare also \cref{eq:aff-subspace}).
The converse is trivial.
\qed

\subsection{Practical construction of $\reg_c(\Omega)$ and $\rel(\Omega)$}\label{sec:practical}
This section describes how to leverage \cref{proposition 4.2,proposition 4.3} for the practical management of linear constraints.
The results stated here, namely \cref{proposition B.2,proposition B.2-relevant} along with their corollaries, double as the algorithms which we used in practice to obtain the results presented in \cref{sec:results}.%
\footnote{
    The implementation code is available from Mattias Sjö upon request.}


\subsubsection{Construction of $\reg_c(\Omega)$}
When using \cref{proposition 4.2} to determine if $\constr{\v\beta}{c}\leq\Omega$ for some $\v\beta$, $c$ and $\Omega$, the rather indirect definition in \cref{eq:def-gen-reg} is of little practical use.
Instead, we will take the approach of finding a constraint that is satisfied by $\v\beta$ if and only if $\v\beta\in\reg_c(\Omega)$.

To understand why such a constraint exists and has a finite representation, note that \cref{eq:gen-condition} along with $\lambda_i\geq 0$ are nothing more than an obfuscated set of linear constraints on the set $\reg(\Omega)$.
Provided a finite set of $\v\alpha_i$, a considerable amount of linear algebra will determine a finite representation this way.
Here, however, we present a simpler method in which the only complicated operation is the determination of the convex hull of a set of points.
Highly efficient algorithms for determining convex hulls exist; we use the QuickHull algorithm \cite{barber1996quickhull} and the associated \texttt{qhull} implementation.\footnote{For up-to-date information about \texttt{qhull}, see \url{http://www.qhull.org/}.}

The key to the construction is that, given a finite set of points $\{\v\beta_j\}_{j\in J}$, a side-effect of the QuickHull algorithm is the creation of a set of constraints $\{\constr{\v n_\ell}{ r_\ell}\}_{\ell\in L}$ such that
\begin{equation}
    \sat\left(\textstyle\sum_{\ell\in L} \constr{\v n_\ell}{ r_\ell}\right) = \hull\left(\{\v\beta_j\}_{j\in J}\right),
\end{equation}
since $\v n_\ell,r_\ell$ are the normals and offsets of the facets of the hull.%
\footnote{\texttt{qhull} uses a different sign convention, but the conversion to the format given above is trivial.} 
This is, by construction, a minimal representation.
If we choose $\v\beta_j$ such that $\hull\left(\{\v\beta_j\}_{j\in J}\right)$ is a suitable subset of $\reg_c(\Omega)$, we will see that it is possible to write a simple rule that selects a subset $M\subseteq L$ such that
\begin{equation}
    \sat\left(\textstyle\sum_{\ell\in M}\constr{\v n_\ell}{ r_\ell}\right) = \reg_c(\Omega).
\end{equation}
In order to do this, let $\Omega$ be given as in \cref{eq:def-Omega}.
Importantly, assume without loss of generality that $\constr{\v\alpha_i}{-1}=\constr{\v 0}{-1}$ for some $i\in I_\inv$, but that $\v\alpha_i\neq\v 0$ for all $i\in I_0$.%
\footnote{
    For maximum efficiency, the representation used for $\Omega$ should otherwise have as few elements as possible. 
    The best easily accessible one is
    \begin{equation*}
        \rel\left(\sum_{i\in I_1}\constr{\v\alpha_i}{c_i}\right)\cup\rel\left(\sum_{i\in I_0}\constr{\v\alpha_i}{c_i}\right)\cup\rel\left(\sum_{i\in I_\inv}\constr{\v\alpha_i}{c_i}\right)\cup\constr{\v 0}{-1},
    \end{equation*}
    where the minimal representations are determined with \cref{proposition B.2-relevant} applied to \cref{cor:construct-simple}.}
Then, define the following sets of points:
\begin{equation}\label{eq:point-sets}
    \begin{aligned}
        \psP{\pm}(\Omega) &= \setbuild{\v\alpha_i}{i\in I_\pml},\\
        \psZ{p}{\pm}(\Omega) &= \setbuild{\v\alpha_i + (p-1)\v\alpha_j}{i\in I_\pml, j\in I_0},\\
        \psN{p}{\pm}(\Omega) &= \setbuild{p\v\alpha_i + (p-1)\v\alpha_j}{i\in I_\pml, j\in I_\mpl}
    \end{aligned}
\end{equation}
for integer $p>1$.
Also define 
\begin{equation}\label{eq:point-sets-hull}
    \psH{p}{\pm}(\Omega) = \hull\big[\psP{\pm}(\Omega)\cup\psZ{p}{\pm}(\Omega)\cup\psN{p}{\pm}(\Omega)\big],
\end{equation}
where $\hull$ denotes the convex hull; see \cref{eq:hull}.
From now on, we will often drop the ``($\Omega$)'' for brevity.
In terms of these, we have the following result:
\pagebreak
\begin{proposition}[constructing $\reg_c(\Omega)$]\label{proposition B.2}
    Let $\Omega$ be a constraint, and arbitrarily select an integer $p>1$.
    Construct a minimal representation $\{\constr{\v n_\ell}{ r_\ell}\}_{\ell\in\isL{p}{\pm}}$ such that\,%
    \footnote{
        To avoid clutter, we do not indicate any $p$-dependence on $\constr{\v n_\ell}{r_\ell}$, but one should bear in mind that they may be entirely different constraints for different $p$ (and different $\pm$).
        To remember this, it can be useful to think of $\isL{p}{\pm}$ as disjoint sets for different $p,\pm$.}
    \begin{equation}\label{eq:def-n}
        \sat\Bigg[\sum_{\ell\in\isL{p}{\pm}} \constr{\v n_\ell}{ r_\ell}\Bigg] = \psH{p}{\pm}(\Omega).
    \end{equation}
    with $\psH{p}{\pm}(\Omega)$ defined as in \cref{eq:point-sets-hull}.
    Let $\isM{p}{\pm}\subseteq\isL{p}{\pm}$ be the set of all $\ell$ for which $\constr{\v n_\ell}{r_\ell}$ supports at least one point $\v\pi\in\psP{\pm}$.
    Then
    \begin{equation}\label{eq:construct}
        \sat\Bigg[\sum_{\ell\in\isM{p}{\pm}} \constr{\v n_\ell}{ r_\ell}\Bigg] = \reg_\pml(\Omega),\qquad
        \sat\Bigg[\sum_{\ell\in\isM{p}{\pm}} \constr{\v n_\ell}{0}\Bigg] = \reg_0(\Omega).
    \end{equation}
    Note that $\reg_0(\Omega)$ can be constructed from either $\isM{p}{+}$ or $\isM{p}{-}$. 
    The exception is when $I_1=\emptyset$, in which case the construction of $\reg_1(\Omega)$ fails; $\reg_1(\Omega)=\emptyset$ trivially, and $\reg_0(\Omega)$ can only be constructed from $\isM{p}{-}$.    
\end{proposition}
\noindent 
This result (along with \cref{cor:construct-0,cor:construct-simple,proposition B.2-relevant} below) outlines the procedure we use in practice to obtain minimal representations.
An example of this construction for $c=-1$ is given in \cref{fig:construct}.

\renewcommand{\nangle}{60}
\renewcommand{\zangle}{305}
\renewcommand{\pangle}{0}
\renewcommand{\regioncoords}[1][]{%
    \coordinate (orig) at (0,0 #1);
    \coordinate (r1)  at (   0,3   #1);%
    \coordinate (r2)  at (  1,2.4  #1);%
    \coordinate (r3)  at ( 1  ,1.7 #1);%
    \coordinate (r4)  at (  .2,1.3 #1);%
    \coordinate (r5)  at (- .3,1.4 #1);%
    \coordinate (r6)  at (- .7,2   #1);%
    \coordinate (r7)  at (- .7,2.7 #1);%
    \coordinate (r8)  at (  .4,1.5 #1);%
    \coordinate (r9)  at (  .3,2   #1);%
    \coordinate (r10) at (  .4,2.3 #1);%
    \coordinate (r11) at (- .2,2.5 #1);%
    \coordinate (r12) at (- .2,1.7 #1);%
    \rotatecoordinate{1};%
    \rotatecoordinate{2};%
    \rotatecoordinate{3};%
    \rotatecoordinate{4};%
    \rotatecoordinate{5};%
    \rotatecoordinate{6};%
    \rotatecoordinate{7};%
    \rotatecoordinate{8};%
    \rotatecoordinate{9};%
    \rotatecoordinate{10};%
    \rotatecoordinate{11};%
    \rotatecoordinate{12};%
}
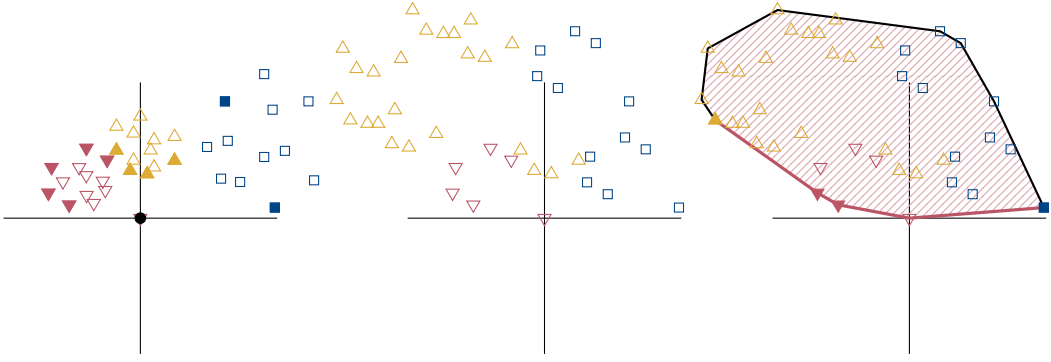
\begin{figure}[hbtp]
        \centering
        \begin{tikzpicture}[scale=.45]
            \regioncoords
            \draw[thin] (-4,0) -- (4,0);
            \draw[thin] (0,-4) -- (0,4);
            
            \foreach \i in {1,...,12}{
                \path[pcolour] (p\i) node[pdot] {};
                \path[ncolour] (n\i) node[ndot] {};
                \path[zcolour] ($ 2*(z\i) $) node[zdot] {};
            }
            \path[ncolour] (orig) node[ndot] {};

            \path[pcolour, every node/.style={pdot=pcolour}]
                (p3) node {}  (p4) node {}  (p5) node {} (p6) node {};
            \path[ncolour, every node/.style={ndot=ncolour}]
                (n1) node {}  (n2) node {}  (n3) node {}  (n6) node {}  (n7) node{};
            \path[zcolour, every node/.style={zdot=zcolour}]
                ($ 2*(z3) $) node {} ($ 2*(z6) $) node {};
                
            \draw[black] (orig) node[dot=black] {};
            
        \end{tikzpicture}
        \begin{tikzpicture}[scale=.45]
            \regioncoords
            \draw[thin] (-4,0) -- (4,0);
            \draw[thin] (0,-4) -- (0,4);
            
            \path[ncolour, every node/.style={ndot}]
                (n1) node {} -- (n2) node {} -- (n3) node {} -- (orig)node{} -- (n6) node {} -- (n7) node{} -- cycle;
                           
            \foreach \z in {(z3),(z6)}{
                \foreach \n in {(n1),(n2),(n3),(n6),(n7),(orig)}{
                    \draw[zcolour] \n ++ \z ++ \z node[zdot] {};
                }
            }
            \foreach \p in {(p3),(p4),(p5),(p6)}{
                \foreach \n in {(orig),(n1),(n2),(n3),(n6),(n7)}{
                    \draw[pcolour] \n ++ \n ++ \p node[pdot] {};
                }
            }
            
        \end{tikzpicture}
        \begin{tikzpicture}[scale=.45]
            \regioncoords
            \draw[thin] (-4,0) -- (4,0);
            \draw[thin] (0,-4) -- (0,4);
            
            \path[pattern=north east lines, pattern color=ncolour!50]
                ($(p5) + 2*(n7)$) -- ($(p6) + 2*(n7)$) -- ($(p6) + 2*(n1)$) -- ($(p6) + 2*(n2)$) 
                -- ($2*(z6) + (n2)$) -- ($2*(z6) + (n3)$) -- ($2*(z6)$) -- ($2*(z3)$)
                -- (orig) -- (n6) -- (n7) -- cycle;
                
            \path[ncolour, every node/.style={ndot}]
                (n1) node {} -- (n2) node {} -- (n3) node {} -- (orig) node{} -- (n6) node[ndot=ncolour] {} -- (n7) node[ndot=ncolour]{} -- cycle;
            \draw[very thick, ncolour] ($2*(z3)$) -- (orig) -- (n6) -- (n7) -- ($(p5) + 2*(n7)$);
            \draw[thick, black]
                ($(p5) + 2*(n7)$) -- ($(p6) + 2*(n7)$) -- ($(p6) + 2*(n1)$) -- ($(p6) + 2*(n2)$) 
                -- ($2*(z6) + (n2)$) -- ($2*(z6) + (n3)$) -- ($2*(z6)$) -- ($2*(z3)$);

            \foreach \z in {(z3),(z6)}{
                \foreach \n in {(n1),(n2),(n3),(n6),(n7),(orig)}{
                    \draw[zcolour] \n ++ \z ++ \z node[zdot] {};
                }
            }
            \draw[zcolour] ($ 2*(z3) $) node[zdot=zcolour] {};
            \foreach \p in {(p3),(p4),(p5),(p6)}{
                \foreach \n in {(orig),(n1),(n2),(n3),(n6),(n7)}{
                    \draw[pcolour] \n ++ \n ++ \p node[pdot] {};
                }
            }
            \draw[pcolour] ($(p5) + 2*(n7)$) node[pdot=pcolour] {};
            
        \end{tikzpicture}
    \caption[Example of the construction of $\reg_\inv(\Omega)$ as described in \cref{proposition B.2}]%
    {
        Example of the construction of $\reg_\inv(\Omega)$ as described in \cref{proposition B.2}. 
        \textbf{Left:} Three sets of points $\v\alpha_i$ for $i\in I_\inv$ (\nMark), $i\in I_0$ (\zMark) and $i\in I_1$ (\pMark).
        The sets are the same up to rescaling and rotation about the origin, and the addition of $\constr{\v 0}{-1}$.
        The relevant elements of the respective $\omega_c$ are marked as filled points, and the rest are left empty; compare to the similar sets in \cref{fig:alvarez}. 
        \textbf{Middle:} The sets $\psP{-}$ (\nmark), $\psZ{2}{-}$ (\zmark) and $\psN{2}{-}$ (\pmark) constructed using only the relevant elements, as remarked above \cref{eq:point-sets}.
        \textbf{Right:} The convex hull thereof (\ncolcol), i.e.\ $\psH{2}{-}$.
        The segments of its boundary correspond to $\constr{\v n_\ell}{r_\ell},\ell\in\isL{2}{-}$.
        The ones for which $\ell\in\isM{2}{-}$ are highlighted in \ncolcol.
        Removing the other segments leaves behind the unbounded region  $\reg_\inv(\Omega)$, which can be seen in \cref{fig:alvarez-n} below. 
        The relevant elements of $\Omega$ have been marked as filled points.
        They follow from \cref{proposition B.2-relevant}, although it can be intuitively seen that they alone influence the shape of $\reg_\inv(\Omega)$.
        }
    \label{fig:construct}
\end{figure}

\proof For brevity, we will write 
\begin{equation}\label{eq:mho}
    \mho^{(p)}_\pm\equiv\sum_{\ell\in \isM{p}{\pm}}\constr{\v n_\ell}{r_\ell}.
\end{equation}
The goal is then to show that $\sat(\mho^{(p)}_\pm) = \reg_\pml(\Omega)$, which we will do by showing that $\sat(\mho^{(p)}_\pm) \supseteq \reg_\pml(\Omega)$ followed by $\sat(\mho^{(p)}_\pm) \subseteq \reg_\pml(\Omega)$.
We will then show the $c=0$ case as a consequence of the others.
First, however, we will establish some lemmata.
\begin{lemma}\label{lem:sign}
    $r_\ell \geq 0$ for all $\ell\in\isM{p}{+}$, and $r_\ell\leq 0$ for all $\ell\in\isM{p}{-}$. 
\end{lemma}
\proof The latter inequality is trivial, since $\v 0\in \psH{p}{-}$.
In the $\isM{p}{+}$ case, recall that by definition, $\v n_\ell\cdot\v\alpha_{i_\ell} = r_\ell$ for some $i_\ell\in I_1$.
We must also have $r_\ell \leq \v n_\ell\cdot (p\v\alpha_{i_\ell} + \v0) = pr_\ell$ since $(p\v\alpha_{i_\ell} + \v 0)\in\psN{p}{+}$, but $r_\ell\leq pr_\ell$ for $p>1$ implies that $r_\ell\geq 0$.
\qed

\begin{lemma}\label{lem:support}
    $\v n_\ell\cdot\v\alpha_i \geq c_i |r_\ell|$ for all $i\in I, \ell\in \isM{p}{\pm}$.
\end{lemma}
\proof For $\ell\in \isM{p}{\pm}$, let $i_\ell\in I_\pml$ be such that $\v n_\ell\cdot\v\alpha_{i_\ell} = r_\ell$.
Then consider two specific points in $\psZ{p}{\pm}$ and $\psN{p}{\pm}$:
\begin{alignat}{4}
    \forall j\in I_0,\quad&&
        \v n_\ell\cdot(\v\alpha_{i_\ell} + (p-1)\v\alpha_j) \geq r_\ell &\imp \v n_\ell\cdot\v\alpha_j \geq 0 
        \\
    \forall {j'}\in I_\mpl,\quad&&
        \v n_\ell\cdot(p\v\alpha_{i_\ell} + (p-1)\v\alpha_{j'}) \geq r_\ell &\imp \v n_\ell\cdot\v\alpha_{j'} \geq \frac{1-p}{p-1}r_\ell=-r_\ell.
\end{alignat}
This, together with \cref{lem:sign}, implies \cref{lem:support}.\qed

Now for the main proof.
Let $\v\beta\in \reg_\pml(\Omega)$ (note that this excludes the exceptional case $\reg_1(\Omega)=\emptyset$).
Using \cref{eq:def-reg-alt}, it can therefore be written
\begin{equation}
    \v\beta = \sum_{i\in I}\lambda_i\v\alpha_i,\qquad \lambda_i\geq 0, \quad \sum_{i\in I}\lambda_ic_i\geq c,
\end{equation}
so for all $\ell\in \isM{p}{\pm}$, \cref{lem:support} gives
\begin{equation}
    \v n_\ell\cdot\v\beta \geq |r_\ell|\sum_{i\in I}\lambda_i c_i \geq |r_\ell| c.
\end{equation}
Since $|r_\ell|c = r_\ell$ for $c=\pm1$ by \cref{lem:sign}, this means that all $\v\beta\in\reg_\pml(\Omega)$ satisfy $\constr{\v n_\ell}{r_\ell}$ for all $\ell\in \isM{p}{\pm}$, thereby proving that $\reg_\pml(\Omega) \subseteq \sat(\mho^{(p)}_\pm)$.

For the converse, 
we first note a direct consequence of \cref{eq:def-gen-reg,eq:point-sets},
\begin{equation}\label{eq:H-subset}
    \psH{p}{\pm}(\Omega)\subseteq\reg_\pml(\Omega),\qquad
    \psH{p}{\pm}\subseteq\psH{q}{\pm}\quad\text{if }q\geq p.
\end{equation}
Then, the proof hinges on the following deceptively simple result:
\begin{lemma}\label{lemma B.10}
    For any $\v\beta\in\sat(\mho_\pm^{(p)})$, there is some $q$ such that $\v\beta\in\psH{q}{\pm}$.
\end{lemma}
\noindent We will postpone its lengthy proof until after the main proof is complete.
\Cref{lemma B.10}, along with \cref{eq:H-subset}, shows that
\begin{equation}\label{eq:keystone}
    \sat\big(\mho_\pm^{(p)}\big) \subseteq \bigcup_{q=p}^\infty\psH{q}{\pm} \subseteq \reg_\pml(\Omega).
\end{equation}
This proves that $\sat(\mho_\pm^{(p)})=\reg_\pml(\Omega)$.

For $\reg_0$, let us turn to \cref{eq:def-reg-alt}, which lets us straightforwardly generalise $\reg_c(\Omega)$ to non-integer $c$.
One easily finds the following generalisation of \cref{eq:change-c}:
\begin{equation}\label{eq:non-integer}
    \constr{\v\alpha}{0}\leq\Omega \equ
    \exists\e > 0,\; \constr{\v\alpha}{+\e} \leq \Omega \equ
    \forall\e > 0,\; \constr{\v\alpha}{-\e} \leq \Omega,
\end{equation}
from which it follows that
\begin{equation}\label{eq:zoom-out-1}
    \bigcup_{n=1}^\infty \reg_{+\frac1n}(\Omega) = \reg_0(\Omega) = \bigcap_{n=1}^\infty \reg_{-\frac1n}(\Omega).
\end{equation}
Using that we have proven \cref{proposition B.2} for $c=\pm1$, which generalises to all $c\neq 0$ by rescaling, we have for all positive integers $n$
\begin{equation}
    \reg_{\pm\frac1n}(\Omega) = \sat\left(\textstyle\sum_{\ell\in \isM{p}{\pm}}\constr{\v n_l}{\tfrac1n r_l}\right),
\end{equation}
so by extension,
\begin{equation}\label{eq:zoom-out-2}
    \begin{aligned}
        \bigcup_{n=1}^\infty \reg_{\pm\frac1n}(\Omega) = \bigcup_{n=1}^\infty \sat\left(\textstyle\sum_{\ell\in \isM{p}{\pm}}\constr{\v n_l}{\tfrac1n r_l}\right),\\
        \bigcap_{n=1}^\infty \reg_{\pm\frac1n}(\Omega) = \bigcap_{n=1}^\infty \sat\left(\textstyle\sum_{\ell\in \isM{p}{\pm}}\constr{\v n_l}{\tfrac1n r_l}\right).
    \end{aligned}
\end{equation}
With the sign of $r_\ell$ given by \cref{lem:sign}, we have
\begin{equation}
    \begin{aligned}
        \bigcup_{n=1}^\infty \sat\left(\textstyle\sum_{\ell\in L'_+}\constr{\v n_l}{\tfrac1n r_l}\right) 
            = \sat\left(\textstyle\sum_{\ell\in L'_+}\constr{\v n_l}{0}\right),\\
        \bigcap_{n=1}^\infty \sat\left(\textstyle\sum_{\ell\in L'_-}\constr{\v n_l}{\tfrac1n r_l}\right) 
            = \sat\left(\textstyle\sum_{\ell\in L'_-}\constr{\v n_l}{0}\right),
    \end{aligned}
\end{equation}
so by \cref{eq:zoom-out-1,eq:zoom-out-2}, both of these sets are equal to $\reg_0(\Omega)$, which completes the proof.
\QED

\subsubsection{Proof of \cref{lemma B.10}}
This lemma is the key to proving \cref{proposition B.2}, and relies on several other lemmata that we will now establish.
They also serve to elucidate some aspects of the proposition and its proof; for instance, \cref{lem:q-independent} explains why the choice of $p$ is arbitrary.

\begin{lemma}\label{lem:support-related}
    Let $\constr{\v m}{s}$ be a constraint such that $\psH{p}{\pm}\in\sat(\constr{\v m}{s})$, and which supports a point \mbox{$\v\alpha_i+(p-1)\v\alpha_j\in\psZ{p}{\pm}$} or a point \mbox{$p\v\alpha_i+(p-1)\v\alpha_{j'}\in\psN{p}{\pm}$}.
    Then $\constr{\v m}{s}$ either supports $\v\alpha_i\in\psP{\pm}$ for that same $i\in I_\pml$, or supports no point in $\psP{\pm}$ at all.
\end{lemma}
\proof Assume $\constr{\v m}{s}$ supports some $\v\pi\in\psP{\pm}$, but that it does not support $\v\alpha_i$. 
Then
\begin{equation}
    \v m\cdot(p\v\alpha_i+(p-1)\v\alpha_j) = s  \imp   \v m\cdot(p\v\pi + (p-1)\v\alpha_j) < s,
\end{equation}
since $\v m\cdot\v\alpha_i > \v m\cdot\v\pi = s$.
This is a contradiction, since 
\begin{equation}
    p\v\pi + (p-1)\v\alpha_j\in\psN{p}{\pm} \subseteq \psH{p}{\pm} \subseteq \sat(\constr{\v m}{s}).
\end{equation}
The argument for $\v\alpha_i+(p-1)\v\alpha_j\in\psZ{p}{\pm}$ is the same.
\qed

\begin{lemma}\label{lem:support-all-p}
    Let $\constr{\v m}{s}$ be a constraint with $\psH{p}{\pm}\subseteq\sat(\constr{\v m}{s})$, and let it support at least one point in $\psP{\pm}$.
    Then if it supports $\v\alpha_i+(p-1)\v\alpha_j\in\psZ{p}{\pm}$, it also supports $\v\alpha_i+(q-1)\v\alpha_j\in\psZ{q}{\pm}$ for all $q>1$.
    Likewise, if it supports $p\v\alpha_i+(p-1)\v\alpha_j\in\psN{p}{\pm}$, it also supports $q\v\alpha_i+(q-1)\v\alpha_j\in\psN{q}{\pm}$ for all $q>1$.%
    \footnote{
        \Cref{lem:support-all-p,lem:q-independent-dimension} actually work for all $q\neq1$. 
        The only places where \cref{proposition B.2} actually requires $p>1$ rather than $p<1$ are in \cref{lem:sign,lem:support} and in \cref{eq:H-subset}.}
\end{lemma}
\proof Assume $\v m\cdot [p\v\alpha_i+(p-1)\v\alpha_j] = s$. 
Then by \cref{lem:support-related}, $\constr{\v m}{s}$ also supports $\v\alpha_i$, and thus $\v m\cdot(\v\alpha_i+\v\alpha_j) = 0$. 
Adding $(q-p)(\v\alpha_i+\v\alpha_j)$ therefore gives $\v m\cdot [q\v\alpha_i+(q-1)\v\alpha_j] = s$. 
The argument for $\v\alpha_i+(p-1)\v\alpha_j\in\psZ{p}{\pm}$ is the same.
\qed

\begin{lemma}\label{lem:q-independent-dimension}
    $\dim(\psH{q}{\pm})$, as defined in \cref{sec:degenerate}, is independent of $q$ for $q>1$. Furthermore, for any $J_c\subseteq I_c$, $\dim\left[\psJ{q}{J_\pml,J_0,J_\mpl}\right]$ is independent of $q$ for $q>1$, where
    \begin{equation}\label{eq:J}
        \psJ{q}{J_\pml,J_0,J_\mpl} \equiv \big\{\v\alpha_i\big\}_{i\in J_\pml} \cup \big\{\v\alpha_i+(q-1)\v\alpha_j\big\}_{\substack{i\in J_\pml\\j\in J_0\phantom{\pm}}} \cup \big\{q\v\alpha_i+(q-1)\v\alpha_j\big\}_{\substack{i\in J_\pml\\j\in J_\mpl}}.
    \end{equation}
\end{lemma}
\proof The main statement is actually a special case of the ``furthermore'' statement, since $\psP{\pm}\cup\psZ{q}{\pm}\cup\psN{q}{\pm} = \psJ{q}{I_\pml,I_0,I_\mpl}$,
and since for any set $\s X$, $\dim[\hull(\s X)]=\dim(\s X)$ because convex combinations are a special case of affine combinations.

By definition, the affine span of $\psJ{q}{J_\pml,J_0,J_\mpl}$ has the same dimension as the linear span of $\setbuild{\v\alpha-\v\beta}{\v\alpha,\v\beta\in\psJ{q}{J_\pml,J_0,J_\mpl}}$; according to \cref{eq:J}, this set consists of
\begin{equation}\label{eq:span-set}
    \begin{mcases}
        \v\alpha_i - \v\alpha_j         &   \text{for }i,j\in J_\pml,\\
        (q-1)\v\alpha_j                 &   \text{for }j\in J_0,\\
        (q-1)(\v\alpha_i-\v\alpha_j)    &   \text{for }i\in J_\pml\cup J_0\cup J_\mpl\text{ and }j\in J_0\cup J_\mpl,\\
        \multicolumn{2}{l}{\text{various linear combinations of the above}.}
    \end{mcases}
\end{equation}
The linear span is unaffected by the inclusion of extra linear combinations or nonzero scale factors, so as long as $(q-1)\neq 0$ we can drop these and be left with the span of 
\begin{equation}
    \setbuild{\v\alpha_i-\v\alpha_j}{i,j\in J_\pml\cup J_0\cup J_\mpl}\cup\setbuild{\v\alpha_j}{j\in J_0},
\end{equation}
which is clearly $q$-independent.
\qed

\begin{lemma}\label{lem:q-independent}
    If $\dim(\psH{q}{\pm})=D$ for some $q>1$, then $\mho^{(q)}_\pm$ is non-degenerate for all $q>1$ and $\mho^{(p)}_\pm=\mho^{(q)}_\pm$ for all $p,q>1$.
\end{lemma}
\noindent This makes it quite clear why $p$ is arbitrary in \cref{proposition B.2}. 

\proof By \cref{lem:q-independent-dimension}, $\dim(\psH{q}{\pm})$ equals $D$ for all $q>1$ if it does for some $q>1$.
Then $\mho^{(q)}_\pm$ is non-degenerate, since $\psH{p}{\pm}\subseteq\sat(\mho^{(p)}_\pm)$ has dimension $D$.

Now for the converse.
Given $\ell\in\isM{p}{\pm},p>1$, define 
\begin{equation}\label{eq:G}
    \psG{p}{\ell} = \setbuild{\v\pi \in \psP{\pm}\cup\psZ{p}{\pm}\cup\psN{p}{\pm} }{\v n_\ell\cdot\v\pi = r_\ell}.
\end{equation}
By the definition of $\isM{p}{\pm}$ along with \cref{lem:support-related}, there are some $\isJ{\ell}{c}\subseteq I_c$ such that $\psG{p}{\ell}=\psJ{p}{\isJ{\ell}\pml,\isJ{\ell}{0},\isJ{\ell}\mpl}$ as defined in \cref{eq:J}.
Then by \cref{lem:support-all-p}, that same $\constr{\v n_\ell}{r_\ell}$ also supports $\psG{q}{\ell}\equiv\psJ{q}{\isJ{\ell}\pml,\isJ{\ell}{0},\isJ{\ell}\mpl}$ for all $q>1$.
From this and \cref{eq:H-subset}, it follows that $\constr{\v n_\ell}{r_\ell}\leq\mho^{(q)}_\pm$.

Recall now that $\constr{\v n_\ell}{r_\ell}\in\rel\big(\mho^{(p)}_\pm\big)$, so \cref{cor:D-1} implies that $\dim\big[\psG{p}{\ell}\big]=D-1$.%
\footnote{
    What \cref{cor:D-1} calls ``$\s F$'' is here $\hull\big(\psG{p}{\ell}\big)$, but $\dim[\s X] = \dim\big[\hull(\s X)\big]$ for any set $\s X$ since convex combinations are a special case of affine combinations.}
Then by \cref{lem:q-independent-dimension}, $\dim[\psG{q}{\ell}]=D-1$ for all $q>1$,
so again by \cref{cor:D-1}, $\constr{\v n_\ell}{r_\ell}\in\rel(\mho^{(q)}_\pm)$.
By doing this for all $\ell\in\isM{p}{\ell}$ and repeating with $p$ and $q$ exchanged, we see that $\rel(\mho^{(q)}_\ell)=\rel(\mho^{(p)}_\ell)$ up to normalisation.
This implies  $\mho^{(q)}_\ell=\mho^{(p)}_\ell$.
\qed

Thanks to \cref{lem:q-independent}, we will drop the ``$(p)$'' superscript on $\mho_\pm$ from now on.
However, we face the problem that $\mho_\pm$ may be degenerate, which would make \cref{lem:q-independent} inapplicable.
It can be circumvented by using the notion of induced constraints developed in \cref{sec:degenerate}:
just substitute
\begin{equation}\label{eq:construct-induced}
   \mho_\pm \to \ind{\mho_\pm},\qquad
   \reg_\pml(\Omega) \to \ind[E_{\mho_\pm}]{\reg_\pml(\Omega)}, 
\end{equation}
since it follows from \cref{lem:basic-induced} (along with $\reg_\pml(\Omega)\subseteq E_{\mho_\pm}$ which we proved earlier) that%
\footnote{
    For practical applications, there is the additional problem that \texttt{qhull} does not function properly when its output would be degenerate.
    This has not been a problem for us, and is of course no issue for the present proof, but if needed, one could identify the affine subspace $E$ containing all points defined in \cref{eq:point-sets}, apply $\ind[E]{\cdots}$ to them, and work entirely in the lower-dimensional space where there are no degeneracies.}
\begin{equation}
    \ind[E_{\mho_\pm}]{\reg_\pml(\Omega)}\supseteq \sat\!\left(\ind{\mho_\pm}\right)
    \equ
    \reg_\pml(\Omega)\supseteq\sat\!\left(\mho_\pm\right).
\end{equation}
Thus, we may for the remainder assume that $\mho_\pm$ is non-degenerate.

We are now, at long last, ready to prove \cref{lemma B.10} itself. 
Consider $\psH{q}{\pm}$ for arbitrary $q$, and presume $\v\beta\not\in\psH{q}{\pm}$.
Select some $\v\alpha\in\psH{q}{\pm}$, and draw the line segment joining $\v\alpha$ and $\v\beta$.
It must intersect the boundary of $\psH{q}{\pm}$ in some point $\v\gamma$, which is supported by one or more $\constr{\v n_\ell}{r_\ell},\ell\in\isL{q}{\pm}$, at least one of which is not satisfied by $\v\beta$.
We therefore have some $\ell\in\isL{q}{\pm}$ such that
\begin{equation}
    \v n_\ell\cdot\v\alpha\geq r_\ell,\qquad \v n_\ell\cdot\v\beta < r_\ell,\qquad \v n_\ell\cdot\v\gamma = r_\ell.
\end{equation}
If $\ell\in\isM{q}{\pm}$, then $\v\beta\not\in\sat(\mho_\pm)$; this is where the $q$-independence of $\mho_\pm$ proven in \cref{lem:q-independent} is crucial.
Thus, we can assume that $\ell\not\in\isM{q}{\pm}$.

The magnitude of $\v\gamma$ is bounded from above by $|\v\gamma|\leq\max(|\v\alpha|,|\v\beta|)$.
We will now attempt to prove that $|\v\gamma| > \max(|\v\alpha|,|\v\beta|)$, which leads to a contradiction, proving that $\v\gamma$ does not exist and consequently that $\v\beta\in\psH{q}{\pm}$.

For any $\ell\in \isL{q}{\pm}$, let $\psG{q}{\ell}\subseteq\big(\psP{\pm}\cup\psZ{q}{\pm}\cup\psN{q}{\pm}\big)$ be defined as in \cref{eq:G}, and let us think about the structure of $\psG{q}{\ell}$.
Each of its elements is of the form $\v\alpha_i + (q-1)\v\alpha'$, where $i\in I_\pml$ and $\v\alpha'$ may take the form $\v 0,\v\alpha_j$ or $\v\alpha_i+\v\alpha_j$ depending on whether the element is part of $\psP{\pm},\psZ{q}{\pm}$ or $\psN{q}{\pm}$.
This observation allows us to write the Minkowski sum
\begin{equation}\label{eq:minkowski}
    \hull\big(\psG{q}{\ell}\big) = \s U^{(q)}_\ell + (q-1)\s V^{(q)}_\ell,
\end{equation}
where $\s U^{(q)}_\ell\subseteq \hull\big(\psP{\pm}\big)$.
$\s V^{(q)}_\ell$ is the convex hull of a subset of $\{\v\alpha_i\}_{i\in I}$; the details are messy and unimportant, so we will not write it explicitly, but the important thing is that it only depends on $q$ and $\ell$ through the specific choice of subset.
This means that when counted over the infinitely many choices of $q$ and $\ell$, there is only a finite number of distinct $\s V^{(q)}_\ell$ that appear: at most as many as there are subsets of $I$.

Now, focus on the case $\ell\in\isL{q}{\pm}\setminus\isM{q}{\pm}$, where we find that $\v 0\not\in\s V^{(q)}_\ell$:
otherwise, $\hull\big(\psG{q}{\ell}\big)\cap\hull\big(\psP{\pm}\big)\neq\emptyset$, which straightforwardly leads to a contradiction of the definition of $\isM{q}{\pm}$.
We may also observe that $\s V^{(q)}_\ell$ (and $\s U^{(q)}_\ell$) are closed sets, being the convex hulls of finite sets of points.

Consider then the finite set
\begin{equation}
    \setbuild{\s V}{\s V^{(q)}_\ell = \s V \quad\text{for some }q\geq 2,\ell\in\isL{q}{\pm}\setminus\isM{q}{\pm}},
\end{equation}
and let $\s W_\pm$ be the union of all elements of this set.
By the observations we have made about $\s V^{(q)}_\ell$, this is a closed set (being the finite union of closed sets) that does not contain $\v 0$.
Consequently, there is some $m>0$ such that $\s W_\pm$ contains no vector of magnitude less than $m$.

Recall now that $\v\gamma\in\psG{q}{\ell}$ for some $\ell\in\isL{q}{\pm}\setminus\isM{q}{\pm}$.
Thus, \cref{eq:minkowski} gives
\begin{equation}
    \v\gamma = \v\pi + (q-1)\v\eta,\qquad \v\pi\in\hull\big(\psP{\pm}\big),\v\eta\in\s W_\pm,
\end{equation}
and this is true no matter the value of $q$ and no matter which $\v\alpha,\v\beta$ are used to obtain $\v\gamma$.
As derived above, $|\v\eta|\geq m$, and since $\hull\big(\psP{\pm}\big)$ is a bounded set, $|\v\pi|\leq M$ for sufficiently large $M$.
For $q$ sufficiently large that $(q-1)m>M$, the triangle inequality then gives
\begin{equation}
    |\v\gamma| \geq (q-1)m - M,
\end{equation}
which can be made arbitrarily large by further increasing $q$, thereby providing the desired contradiction and completing the proof.
\qed

\subsubsection{Some important corollaries of \cref{proposition B.2}}\label{sec:practical-cor}
We can refine the treatment of $\reg_0(\Omega)$ with the following:%
\footnote{
    An intuitive understanding of the construction of $\reg_0(\Omega)$ can be gained by noting that in a sense, one can make $c\to 0$ by ``zooming out'' on constraint space.
    This is the principle that is formalised in the end of the proof of \cref{proposition B.2}.
    Each facet of the body $\reg_c(\Omega)$, consisting of the points supported by one $\constr{\v n_\ell}{r_\ell}$, is thereby shifted so that it passes through the origin (hence $r_\ell\to0$), and if it was bounded, it shrinks down to a point.
    \Cref{cor:construct-0} identifies those bounded facets and removes them.
    It is easy to see why it works: all facets of $\psH{p}{\pm}$ are bounded, and only by being adjacent to a facet that is removed in the restriction $\isL{p}{\pm}\to \isM{p}{\pm}$ can a facet become unbounded.
    The removed facets are those that only support points in $\psZ{p}{\pm}\cup\psN{p}{\pm}$, so their neighbours are the ones that support at least one point in these sets.}
\begin{corollary}[construction of $\reg_0(\Omega)$]\label{cor:construct-0}
    Let $\Omega$, $\isM{p}{\pm}$, etc.\ be as in \cref{proposition B.2}.
    Let $\isN{p}{\pm}\subseteq\isM{p}{\pm}$ be the set of those $\ell$ for which $\constr{\v n_\ell}{r_\ell}$ also supports at least one point in $\big(\psZ{p}{\pm}\cup\psN{p}{\pm}\big)$.
    If $\sum_{\ell\in\isM{p}{\pm}} \constr{\v n_\ell}{ 0}$ is a non-degenerate constraint,%
    \footnote{
        This is not equivalent to $\sum_{\ell\in\isM{p}{\pm}} \constr{\v n_\ell}{r_\ell}$ being non-degenerate.
        Consider as a counterexample $\Omega=\constr{\binom{1}{1}}{-1}+\constr{\binom{1}{-1}}{-1}+\constr{\binom{1}{0}}{0}$, for which $\reg_0(\Omega)$ is contained in a hyperplane (i.e.\ a line) whereas $\reg_\inv(\Omega)$ is not.} 
    then
    \begin{equation}\label{eq:construct-0}
        \reg_0(\Omega) = \sat\bigg(\sum_{\ell\in\isN{p}{\pm}} \constr{\v n_\ell}{ 0}\bigg).
    \end{equation}
    As in \cref{proposition B.2}, the $\isM{p}{+}$ construction does not work when $I_1=\emptyset$.
\end{corollary}
\proof For $\ell$ containted in $\isM{p}{\pm}$ but not in $\isN{p}{\pm}$, \cref{lem:support} more specifically gives
\begin{equation}
    \forall i\in I_\pml,\;  \v n_\ell\cdot\v\alpha_i\geq r_\ell,\qquad 
    \forall i\in I_{0},\;   \v n_\ell\cdot\v\alpha_i > 0,\qquad 
    \forall i\in I_\mpl,\;  \v n_\ell\cdot\v\alpha_i > -r_\ell
\end{equation}
with \cref{lem:sign} dictating the sign of $r_\ell$.
Then if $\v n_\ell\cdot\v\beta=0$ for $\v\beta\in\reg_0(\Omega)$, a look at \cref{eq:def-gen-reg} tells us that all $\lambda_i,i\in I_0\cup I_1$ must be zero due to the above inequalities, and \cref{eq:gen-condition} then implies that also $\lambda_i,i\in I_\inv$ must be zero.
Thus, the only point $\v\beta\in\reg_0(\Omega)$ supported by $\constr{\v n_\ell}{0}$ is the trivial $\v\beta=\v 0$, which is supported by $\constr{\v\alpha}{0}$ for all $\v\alpha\neq\v 0$.
Therefore, $\constr{\v n_\ell}{0}$ does not uniquely support any point, so by \cref{proposition 4.3} (which requires non-degeneracy), it can be omitted.
\QED

One also easily finds the following simplification:
\begin{corollary}[construction of $\reg_c(\omega_c)$]\label{cor:construct-simple}
    For those $\omega_c$ covered by \cref{proposition 4.1}, \cref{proposition B.2} reduces down to the following:
    \begin{description}
        \item[$c=+1:$]
            $\psH{p}{+} = \hull\left(\{\v\alpha_i, p\v\alpha_i\}_{i\in I}\right)$. $\isM{p}{+}$ consists of those $\ell$ for which $r_\ell\geq 0$.       
        \item[$c=\phantom{+}0:$]
            $\psH{p}{-} = \hull\left(\{\v 0\}\cup\{p\v\alpha_i\}_{i\in I}\right)$. $\isN{p}{-}=\isM{p}{-}$ consists of those $\ell$ where $r_\ell=0$.
        \item[$c=-1:$]
            $\psH{p}{-} = \hull\left(\{\v 0\}\cup\{\v\alpha_i\}_{i\in I}\right)$. All $\constr{\v n_\ell}{r_\ell}$ are kept, since $\isM{p}{-} = \isL{p}{-}$.
        \QED
    \end{description}
\end{corollary}
\noindent An alternative to using \cref{proposition B.2} is to apply \cref{cor:construct-simple} to $\lOmega$ as defined in \cref{sec:proof-gen}, and then ``unlifting'' the result.
This requires the treatment of a much smaller number of points, which makes the QuickHull algorithm run faster; on the other hand, lifting increases the dimension, which makes the QuickHull algorithm run slower and be less numerically stable.
The time complexity of the \texttt{qhull} implementation suggests that asymptotically, lifting should be the faster method, but since \texttt{qhull} is vastly more efficient in 2 and 3 dimensions, not lifting should be preferable when the number of dimensions is small.
In practice, we only used \cref{proposition B.2} directly without lifting.

\subsubsection{Construction of $\rel(\Omega)$}\label{sec:practical-relevant}
The marriage of \cref{proposition 4.3,proposition B.2} makes for a practical way of determining the minimal representation $\rel(\Omega)$ of any non-degenerate constraint $\Omega$.
As remarked before, degenerate constraints are of little practical relevance, although if needed, the degenerate case can be covered by adapting \cref{prop:gen-relevant}.

When forming a convex hull, \texttt{qhull} produces a list of its vertices and readily checks if two vertices form the endpoints of an edge.
Based on that, we devise the following:
\begin{proposition}[constructing $\rel(\Omega)$]\label{proposition B.2-relevant}
    Let $\Omega$ be a non-degenerate constraint, and let $\psH{p}{\pm}$, $\isM{p}{\pm}$, etc.\ be defined as in \cref{proposition B.2}. Then the unique minimal representation $\rel(\Omega)$ is determined as follows:
    \begin{itemise}
        \item For $i\in I_\pml$, $\constr{\v\alpha_i}\pml\in\rel(\Omega)$ if and only if $\v\alpha_i$ is an extreme point of $\reg_\pml(\Omega)$. 
        (The exception is $\constr{\v 0}{-1}$, which is of course not in $\rel(\Omega)$.)
        
        Equivalently, $\constr{\v\alpha_i}{\pml}\in\rel(\Omega)$ if and only if $\v\alpha_i\in\psP{\pm}$ is an extreme point of $\psH{p}{\pm}$ for some $p>1$, with the same exception.
        
        \item For $j\in I_0$, $\constr{\v\alpha_j}{0}\in\rel(\Omega)$ if and only if there is some $i\in I_\pml$ such that the ray $\setbuild{\v\alpha_i + \lambda\v\alpha_j}{\lambda\geq0}$ is an edge of $\reg_\pml(\Omega)$ that contains no point in $\psN{p}{\pm}(\Omega)$. 
        (This is up to normalisation; several equivalent $\constr{\v\alpha_i}{0}$ may satisfy this condition.)
        
        Equivalently, $\constr{\v\alpha_j}{0}\in\rel(\Omega)$ if and only if there is some $i\in I_\pml$ such that the line segment between  $\v\alpha_i$ and $\v\alpha_i+(p-1)\v\alpha_j$ is an edge of $\psH{p}{\pm}(\Omega)$  that contains no point in $\psN{p}{\pm}(\Omega)$. 
        (This breaks the normalisation ambiguity: if several $\constr{\v\alpha_j}{0}$ are equivalent under \cref{eq:rescale}, then only the one with the largest $|\v\alpha_j|$ will form the endpoint of their edge and be included in $\rel(\Omega)$.)
    \end{itemise}
\end{proposition}
\noindent 
Note how this can be applied to \cref{cor:construct-simple} without modification.
Illustrations can be found in \cref{fig:construct,fig:alvarez-n,fig:alvarez-p,fig:alvarez-z}.

\proof We will prove that the stated conditions are equivalent to $\constr{\v\alpha_i}{c_i}$ uniquely supporting a point; the rest follows from \cref{proposition 4.3}. 
We will focus on proving the conditions based on $\psH{p}{\pm}$ rather than $\reg_\pml(\Omega)$; that they are equivalent follows easily from \cref{pty:face-of-subset} of $K$-faces along with \cref{lem:K-face-subset} ($\v\alpha_i\in\psP{\pm}$ serves as the point $\v\ph$).%
\footnote{
    There is one subtlety for $\constr{\v\alpha_j}{0}\in\rel(\Omega)$: the line segment between $\v\alpha_i$ and $\v\alpha_i+(p-1)\v\alpha_j$ may fail to be an edge of $\psH{p}{\pm}$ even though $\setbuild{\v\alpha_i + \lambda\v\alpha_j}{\lambda\geq0}$ is an edge of $\reg_\pml(\Omega)$, if it is contained in the line segment between $\v\alpha_i$ and some $\v\alpha_{i'},i'\in I_\pml$.
    This can be remedied by using sufficiently large $p$.
    For $\constr{\v\alpha_i}{\pm1}\in\rel(\Omega)$, all $p$ work equally well, as should be apparent from the proof.}

Consider first $i\in I_\pml$.
If $\constr{\v\alpha_i}{\pm1}\in\rel(\Omega)$, then there is some $\v h$ uniquely supported by $\constr{\v\alpha_i}{\pm1}$; in particular, for any $\v\beta\in\reg_\pml(\Omega),\v\beta\neq\v\alpha$ we have $\v\beta\cdot\v h>\v\alpha_i\cdot\v h = \pm1$. 
Since by \cref{eq:keystone}
\begin{equation}\label{eq:union-H}
    \reg_\pml(\Omega) = \bigcup_{p=2}^\infty \psH{p}{\pm}
\end{equation}
we have that for any $p>1$
\begin{equation}\label{eq:extreme}
    \setbuild{\v\eta\in\psH{p}{\pm}}{\v\eta\cdot\v h = \pml} = \{\v\alpha_i\}.
\end{equation}
From \cref{lem:K-face}, we conclude that $\v\alpha_i$ is an extreme point of $\psH{p}{\pm}$.

Conversely, if $\v\alpha_i$ is an extreme point of $\psH{p}{\pm}$, then \cref{eq:extreme} holds.
In particular, if $j\in I_\mpl$ then $\v h\cdot\big(p\v\alpha_i + (p-1)\v\alpha_j\big)>\pml$ so $\v\alpha_j\cdot\v h > \mpl$.
Similarly, for any $j\in I_0$ we have $\v\alpha_j\cdot\v h > 0$.
It then follows from \cref{lem:unique-support-1,proposition 4.3} that $\constr{\v\alpha_i}{\pm1}\in\rel(\Omega)$.

Consider then $j\in I_0$.
If the line segment between $\v\alpha_i$ and $\v\alpha_i+(p-1)\v\alpha_j$ is an edge of $\psH{p}{\pm}$, then by \cref{lem:K-face} there exists $\constr{\v h}{d}$ such that
\begin{equation}
    \setbuild{\v\eta\in\psH{p}{\pm}}{\v\eta\cdot\v h = d} = \setbuild{\v\alpha_i + \lambda\v\alpha_j}{\lambda \in [0,(p-1)]}
\end{equation}
and $\psH{p}{\pm}\subseteq\sat(\constr{\v h}{d})$.
We then immediately find that $\v h\cdot\v\alpha_i = d,\v h\cdot\v\alpha_j = 0$.

What about the other $\v\alpha_k,k\in I$?
If $k\in I_\mp$ then $p\v\alpha_i+(p-1)\v\alpha_k\in\psN{p}{\pm}$ which is by assumption not supported by $\constr{\v h}{d}$, so it follows that $\v h\cdot\v\alpha_k>-d$.
In the same way, for any $k\in I_0$ we have $\v\alpha_i + (p-1)\v\alpha_k\in\psZ{p}{\pm}$, implying that $\v h\cdot\v\alpha_k\geq 0$. 
If $\v h\cdot\v\alpha_k=0$ then it follows that $\v\alpha_k$ is collinear with $\v\alpha_j$, so either $\constr{\v\alpha_i}{0}=\constr{\v\alpha_k}{0}$ or $\constr{\v\alpha_i}{0}=\constr{-\v\alpha_k}{0}$ under \cref{eq:rescale}; the latter is excluded by non-degeneracy.
Therefore, we can without loss of generality assume $\v h\cdot\v\alpha_k>0$ for all $k\in I_0\setminus\{j\}$ by omitting equivalent constraints.
Lastly, if $k\in I_\pm$ then $\v h\cdot\v\alpha_k\geq d$; here, the non-strict inequality is unavoidable but does not pose a problem. 

If $i\in I_+$, then the fact that $p\v\alpha_i\in\psN{p}{+}$ implies that $d>0$.
If $i\in I_-$, we can use the fact that $\v 0\in\psP{-}$ to prove that $d\leq 0$, with equality only if $\v\alpha_i=\v 0$.
Combining this with the previous paragraph, we conclude that
\begin{equation}
    \forall k\in I\setminus\{i,j\},\; \v h\cdot\v\alpha_k  \geq c_k |d|
\end{equation}
(equality only possible if $k\in I_\pm$). For sufficiently small $\e>0$, we may replace it by $|d|\to|d|\mp\e$ without invalidating the above inequality; this also guards against problems when $d=0$.
Then
\begin{equation}
    \tfrac{1}{|d|\mp\e}\v h\cdot\v\alpha_i > \pm1,\qquad
    \tfrac{1}{|d|\mp\e}\v h\cdot\v\alpha_j = 0,\qquad
    \forall k\in I\setminus\{i,j\},\;\tfrac{1}{|d|\mp\e}\v h\cdot\v\alpha_k  > c_k.
\end{equation}
Thus, $\constr{\v\alpha_j}{0}$ uniquely supports $\frac{1}{|d|\mp\e}\v h$ by \cref{lem:unique-support-1}.

Conversely, assume $\v\alpha_j$ uniquely supports a point $\v b$.
Then
\begin{equation}\label{eq:alphaj}
    \v b\cdot\v\alpha_j=0,\qquad \forall i\in I\setminus{j},\;\v b\cdot\v\alpha_i > c_i.
\end{equation}
In particular,
\begin{equation}
    \s F = \setbuild{\v\beta\in\reg_0(\Omega)}{\v b\cdot\v\beta = 0} = \setbuild{\lambda\v\alpha_j}{\lambda\geq 0}.
\end{equation}
(Non-degeneracy ensures $\lambda\geq 0$).
By \cref{lem:K-face,proposition B.2}, $\s F$ is an edge of $\reg_0(\Omega) = \sat\big(\sum_{\ell\in\isM{\pm}{p}}\constr{\v n_\ell}{0}\big)$.
Therefore, \cref{lem:technical} guarantees that there exists $Q_\pm\subseteq\isM{p}{\pm}$ such that
\begin{equation}
    \forall\ell\in Q_\pm,\; \v n_\ell\cdot\v\alpha_j = 0,\qquad
    \forall\ell\in \isM{p}{\pm}\setminus Q_\pm,\; \v n_\ell\cdot\v\alpha_j > 0
\end{equation}
where $\{\v n_\ell\}_{\ell\in Q_\pm}$ contains a subset of $D-1$ linearly independent vectors.
Now, consider the set
\begin{equation}
    \s C \equiv \sat\bigg(\sum_{\ell\in Q_\pm}\constr{\v n_\ell}{r_\ell}\bigg)\supseteq\reg_\pml(\Omega).
\end{equation}
This set must have at least one edge, since the intersection of $D-1$ independent hyperplanes is a line.
By \cref{lem:technical}, there is then a subset $J_\pm\subseteq Q_\pm$ with $|J_\pm|=D-1$ and $\{\v n_\ell\}_{\ell\in J_\pm}$ linearly independent, that describes that edge according to \cref{eq:technical} (if there are several, we choose one arbitrarily).
By construction, $\s C$ has no extreme points: $\{\v n_\ell\}_{\ell\in Q_\pm}$ has no subset of $D$ linearly independent vectors.
Therefore, the chosen edge can be written like
\begin{equation}
    \s F_\pm \equiv \setbuild{\v\beta\in\R^D}{\forall\ell\in J_\pm,\;\v n_\ell\cdot\v\beta=r_\ell}.
\end{equation}
We now claim that
\begin{equation}\label{eq:F-in-A}
    \s F_\pm\cap\reg_\pml(\Omega) = \s F_\pm\cap\sat\bigg(\sum_{\ell\in\isM{p}{\pm}}\constr{\v n_\ell}{r_\ell}\bigg) \neq\emptyset.
\end{equation}
All points in $\s F_\pm$ are satisfied by $\constr{\v n_\ell}{r_\ell}$ for $\ell\in Q_\pm$ by the arguments made above about $\s C$.
For $\ell\not\in Q_\pm$, we note that for any $\v\beta\in\s F$, we also have $\v\beta+\lambda\v\alpha_j\in\s F$ for all $\lambda\in\R$, so since $\v n_\ell\cdot\v\alpha_j > 0$, $\v n_\ell\cdot(\v\beta+\lambda\v\alpha_j)\geq r_\ell$ for sufficiently large $\lambda$ no matter what $\v n_\ell\cdot\v\beta$ is.
Therefore, at least some subset of $\s F_\pm$ is satisfied by all $\constr{\v n_\ell}{r_\ell},\ell\in\isM{p}{\pm}\setminus Q_\pm$, which proves \cref{eq:F-in-A}.
By then considering the constraint 
\begin{equation}
    \constr{\v h}{d} \equiv \constr{\sum_{\ell\in J_\pm}\v n_\ell}{\sum_{\ell\in J_\pm}r_\ell},
\end{equation}
it follows from \cref{lem:K-face} that $\s F_\pm\cap\reg_\pml(\Omega)$ is a $K$-facet of $\reg_\pml(\Omega)$.

Let us now note that $\s F_\pm\cap\reg_\pml(\Omega)$ is an edge, not an extreme point, since for any $\v\beta\in\s F_\pm$ and $\lambda\geq 0$,%
\footnote{
    It also follows from this that $\s F_\pm\cap\reg_\pml(\Omega)$ is not a line segment.
    That it is a ray rather than a full line is not difficult to prove from non-degeneracy, and it is then possible to show that $\s F_\pm\cap\reg_\pml(\Omega)=\setbuild{\v\alpha_i+\lambda\v\alpha_j}{\lambda\geq 0}$ which completes the proof, but we choose an easier path.}
\begin{equation}
    \v\beta+\lambda\v\alpha_j\in\s F_\pm\cap\reg_\pml(\Omega).
\end{equation}
From \cref{eq:F-in-A,eq:union-H,eq:H-subset}, there is some $q>1$ such that $\s F_\pm\cap\psH{p}{\pm}\neq\emptyset$ for all $p\geq q$, and by \cref{pty:face-of-subset} of $K$-facets, $\s F_\pm\cap\psH{p}{\pm}$ is an edge of $\psH{p}{\pm}$.
Unlike $\reg_\pml(\Omega)$, $\psH{p}{\pm}$ is always a compact set, so all its edges must be line segments with exactly two endpoints, which are extreme points of $\psH{p}{\pm}$ by \cref{pty:subface}.
Thus, $\s F_\pm\cap\psH{p}{\pm}$ is the line segment between two extreme points of $\psH{p}{\pm}$, and these must clearly be elements of $\psP{\pm}\cup\psZ{p}{\pm}\cup\psN{p}{\pm}$.%
\footnote{
    This is made rigorous by comparing \cref{eq:point-sets-hull} with the \emph{Krein-Milman theorem}:
    \begin{quotation}
       \noindent\textit{Let $\s C$ be a compact convex set and $\s E$ be the set of its extreme points.
        Then $\s C=\hull(\s E)$, and if $\s C=\hull(\s E')$ then $\s E'\supseteq\s E$.}
    \end{quotation}
}

Then, let us prove that at least one of these endpoints of $\s F_\pm\cap\psH{p}{\pm}$ is in $\psP{\pm}$ by considering the alternatives.
For $\psN{p}{\pm}$, i.e.\ if $p\v\alpha_i + (p-1)\v\alpha_k$ is one endpoint for some $i\in I_\pml,k\in I_\mpl$, then it follows from \cref{lem:support-related} that $\v\alpha_i\in\s F_\pm\cap\psH{p}{\pm}$, but also (see the proof of \cref{lem:support-all-p}) that $(p-1)(\v\alpha_i+\v\alpha_k)=\lambda\v\alpha_j$ for some $\lambda$.
This contradicts the fact that $\lambda$ uniquely supports a point, since any point in $\sat(\Omega)$ it supports is also supported by $\v\alpha_i$ and $\v\alpha_k$.
For $\psZ{p}{\pm}$, i.e.\ if $\v\alpha_i + (p-1)\v\alpha_k$ is one endpoint for some $i\in I_\pml,k\in I_0$, then it similarly follows that $\v\alpha_i\in\s F_\pm\cap\psH{p}{\pm}$, and that $\v\alpha_k=\lambda\v\alpha_j$, where $\lambda>0$ by non-degeneracy.

Thus, $\psN{p}{\pm}$ is excluded and $\psZ{p}{\pm}$ can account for at most one endpoint of \mbox{$\s F_\pm\cap\psH{p}{\pm}$}, since the line segment between $\v\alpha_i+\lambda(p-1)\v\alpha_j$ and $\v\alpha_{i'}+\lambda'(p-1)\v\alpha_j$ cannot contain both $\v\alpha_i$ and $\v\alpha_{i'}$, which are clearly contained in $\s F_\pm\cap\psH{p}{\pm}$.
Hence, at least one of the endpoints is $\v\alpha_i$ for some $i\in I_\pml$, and the other may either be $\v\alpha_{i'},i'\in I_\pml$ or $\v\alpha_i+(p-1)\v\alpha_j$ (possibly after exchanging $\constr{\v\alpha_j}{0}$ for an equivalent constraint).
The former case was covered in the first paragraph of this proof, and is removed by considering sufficiently large $p$; the latter completes our proof.
\QED

\subsubsection{Visualisation of $\reg_c(\Omega)$}\label{sec:visual}
We provided \cref{fig:alvarez} for illustration along with the statement of \cref{proposition 4.1}, since the shapes of $\reg(\omega_c)$ are quite simple to interpret, and provide some insight into the result.
The same cannot be said for the general $\reg_c(\Omega)$, however, so we have put off a similar display until now.
\Cref{fig:construct} showed a single example in great detail, and now \cref{fig:alvarez-n,fig:alvarez-p,fig:alvarez-z} illustrate $\reg_c(\Omega)$ in a similar manner to how \cref{fig:alvarez} illustrated $\reg(\omega_c)$.
For legibility, we have omitted the supporting sets $\psP{\pm}$, $\psZ{p}{\pm}$ and $\psN{p}{\pm}$, but if one wishes, it is not difficult to imagine them in the figures like in \cref{fig:construct} to make sense of the shapes.
\begin{figure}[hbtp]
    \begin{minipage}[c]{0.33\textwidth}
        \centering
        \begin{tikzpicture}[scale=.6]
            \regioncoords
            \coordaxes
            
            \coordinate (pn) at ($ -1*(p5) $);
            \draw[thick, ncolour, pattern = north east lines, pattern color=ncolour!50]
                (0,0) -- ($ 10*(z3) $) -- (0,10) -- ($ (pn)!10!(n7) $) -- (n7) -- (n6) -- cycle;
            
            \draw[pcolour, every node/.style={pdot}]
                (p1) node {} -- (p2) node {} -- (p3) node {} -- (p4) node {} -- (p5) node[pdot=pcolour] {} --(p6) node {} -- (p7) node {} -- cycle;
            \draw[ncolour, every node/.style={ndot}]
                (n1) node {} -- (n2) node {} -- (n3) node {} -- (n4) node {} -- (n5) node {} --(n6) node[ndot=ncolour] {} -- (n7) node[ndot=ncolour] {} -- cycle;
            \draw[zcolour, every node/.style={zdot}]
                (z1) node {} -- (z2) node {} -- (z3) node[zdot=zcolour] {} -- (z4) node {} -- (z5) node {} --(z6) node {} -- (z7) node {} -- cycle;
            
            \path[pcolour, every node/.style={pdot}]
                (p8) node{} (p9) node{} (p10) node{} (p11) node{} (p12) node{};
            \path[ncolour, every node/.style={ndot}]
                (n8) node{} (n9) node{} (n10) node{} (n11) node{} (n12) node{};
            \path[zcolour, every node/.style={zdot}]
                (z8) node{} (z9) node{} (z10) node{} (z11) node{} (z12) node{};
                
            \draw (0,0) node[dot=black] {};
        \end{tikzpicture}
    \end{minipage}
    \qquad
    \begin{minipage}{0.67\textwidth}
        \centering
        \begin{tikzpicture}[scale=.4]
            \regioncoords
            \coordaxes
            
            \draw[thick, ncolour, pattern = north east lines, pattern color=ncolour!50]
                ($ (n1) + 10*(z6) $) -- (n1) -- (n7) -- (n6) -- (0,0) -- ($ 10*(z3) $) -- (0,10) -- cycle;
            
            \draw[ncolour, every node/.style={ndot}]
                (n1) node[ndot=ncolour] {} -- (n2) node {} -- (n3) node {} -- (n4) node {} -- (n5) node {} --(n6) node[ndot=ncolour] {} -- (n7) node[ndot=ncolour] {} -- cycle;
            \draw[zcolour, every node/.style={zdot}]
                (z1) node {} -- (z2) node {} -- (z3) node[zdot=zcolour] {} -- (z4) node {} -- (z5) node {} --(z6) node[zdot=zcolour] {} -- (z7) node {} -- cycle;
                                
            \path[ncolour, every node/.style={ndot}]
                (n8) node{} (n9) node{} (n10) node{} (n11) node{} (n12) node{};
            \path[zcolour, every node/.style={zdot}]
                (z8) node{} (z9) node{} (z10) node{} (z11) node{} (z12) node{};
            
            \draw (0,0) node[dot=black] {};
        \end{tikzpicture}
        \begin{tikzpicture}[scale=.4]
            \regioncoords
            \coordaxes
            
            \coordinate (pn) at ($ -1*(p5) $);
            
            \draw[thick, ncolour, pattern = north east lines, pattern color=ncolour!50]
                ($ (n7)!-10!(pn) $) -- (n7) -- (n6) -- (0,0) -- ($ 10*(p3) $) -- cycle;
            
            \draw[pcolour, every node/.style={pdot}]
                (p1) node {} -- (p2) node {} -- (p3) node[pdot=pcolour] {} -- (p4) node {} -- (p5) node[pdot=pcolour] {} --(p6) node {} -- (p7) node {} -- cycle;
            \draw[ncolour, every node/.style={ndot}]
                (n1) node {} -- (n2) node {} -- (n3) node {} -- (n4) node {} -- (n5) node {} --(n6) node[ndot=ncolour] {} -- (n7) node[ndot=ncolour] {} -- cycle;
            
            \path[pcolour, every node/.style={pdot}]
                (p8) node{} (p9) node{} (p10) node{} (p11) node{} (p12) node{};
            \path[ncolour, every node/.style={ndot}]
                (n8) node{} (n9) node{} (n10) node{} (n11) node{} (n12) node{};
                            
            \draw (0,0) node[dot=black] {};
        \end{tikzpicture}
        \begin{tikzpicture}[scale=.4]
            \regioncoords
            \coordaxes
                        
            \draw[thick, ncolour, pattern = north east lines, pattern color=ncolour!50]
                ($ 10*(p6) $) --  (0,0) -- ($ 10*(z3) $) -- cycle;
            
            \draw[pcolour, every node/.style={pdot}]
                (p1) node {} -- (p2) node {} -- (p3) node {} -- (p4) node {} -- (p5) node {} --(p6) node[pdot=pcolour] {} -- (p7) node {} -- cycle;
            \draw[zcolour, every node/.style={zdot}]
                (z1) node {} -- (z2) node {} -- (z3) node[zdot=zcolour] {} -- (z4) node {} -- (z5) node {} --(z6) node {} -- (z7) node {} -- cycle;
            
            \path[pcolour, every node/.style={pdot}]
                (p8) node{} (p9) node{} (p10) node{} (p11) node{} (p12) node{};
            \path[zcolour, every node/.style={zdot}]
                (z8) node{} (z9) node{} (z10) node{} (z11) node{} (z12) node{};
                            
            \draw (0,0) node[dot=black] {};
        \end{tikzpicture}\\
        \begin{tikzpicture}[scale=.4]
            \regioncoords
            \coordaxes
            
            \draw[thick, ncolour, pattern = north east lines, pattern color=ncolour!50]
                (n1) -- (n7) -- (n6) -- (0,0) -- (n3) -- (n2) -- cycle;
            
            \draw[ncolour, every node/.style={ndot}]
                (n1) node[ndot=ncolour] {} -- (n2) node[ndot=ncolour] {} -- (n3) node[ndot=ncolour] {} -- (n4) node {} -- (n5) node {} --(n6) node[ndot=ncolour] {} -- (n7) node[ndot=ncolour] {} -- cycle;
                
            \path[ncolour, every node/.style={ndot}]
                (n8) node{} (n9) node{} (n10) node{} (n11) node{} (n12) node{};
            
            \draw (0,0) node[dot=black] {};
        \end{tikzpicture}
        \begin{tikzpicture}[scale=.4]
            \regioncoords
            \coordaxes
                        
            \draw[thick, ncolour, pattern = north east lines, pattern color=ncolour!50]
                ($ 10*(z6) $) --  (0,0) -- ($ 10*(z3) $) -- cycle;
            
            \draw[zcolour, every node/.style={zdot}]
                (z1) node {} -- (z2) node {} -- (z3) node[zdot=zcolour] {} -- (z4) node {} -- (z5) node {} --(z6) node[zdot=zcolour] {} -- (z7) node {} -- cycle;
            
            \path[zcolour, every node/.style={zdot}]
                (z8) node{} (z9) node{} (z10) node{} (z11) node{} (z12) node{};
                            
            \draw (0,0) node[dot=black] {};
        \end{tikzpicture}
        \begin{tikzpicture}[scale=.4]
            \regioncoords
            \coordaxes
            
            \draw[thick, ncolour, pattern = north east lines, pattern color=ncolour!50]
                ($ 10*(p6) $) --  (0,0) -- ($ 10*(p3) $) -- cycle;
            
            \draw[pcolour, every node/.style={pdot}]
                (p1) node {} -- (p2) node {} -- (p3) node[pdot=pcolour] {} -- (p4) node {} -- (p5) node {} --(p6) node[pdot=pcolour] {} -- (p7) node {} -- cycle;
            
            \path[pcolour, every node/.style={pdot}]
                (p8) node{} (p9) node{} (p10) node{} (p11) node{} (p12) node{};
                
            \draw (0,0) node[dot=black] {};
        \end{tikzpicture}
    \end{minipage}
    \caption[Example of the region $\reg_\inv(\Omega)$.]{%
        Example of the region $\reg_\inv(\Omega)$ (\ncolcol).
        The $\v\alpha_i$, identical (up to $c_i=0$ rescaling) to those in \cref{fig:construct}, are marked with \nmark\ for $i\in I_\inv$, \zmark\ for $I_0$ and \pmark\ for $I_1$, and have their convex hulls outlined.
        The $\v\alpha_i$ that are identified as relevant from the construction of $\reg_\pml(\Omega)$, as per \cref{proposition B.2-relevant}, are filled; the rest are left empty.
        Note how the $c_i=+1$ ones can strictly speaking only be deduced from $\reg_1(\Omega)$ (\cref{fig:alvarez-p} below), although their relevance can be easily seen in ``nice'' cases such as this.
    
        \qquad In the large figure, the full $\reg_\inv(\Omega)$ is drawn.
        The smaller figures demonstrate the effect of omitting $\v\alpha_i$ from the definition of $\Omega$, with $i$ in various combinations of $I_1,I_0$ and $I_\inv$.
        The only-$I_\inv$ figure (bottom left) is thus analogous to $\reg(\omega_\inv)$ in \cref{fig:alvarez}.
        Note how in the only-$I_1$ figure (bottom right), \cref{proposition B.2} still works since $\psP{-1} = \{\v 0\}$ rather than being empty.
    }
    \label{fig:alvarez-n}
\end{figure}
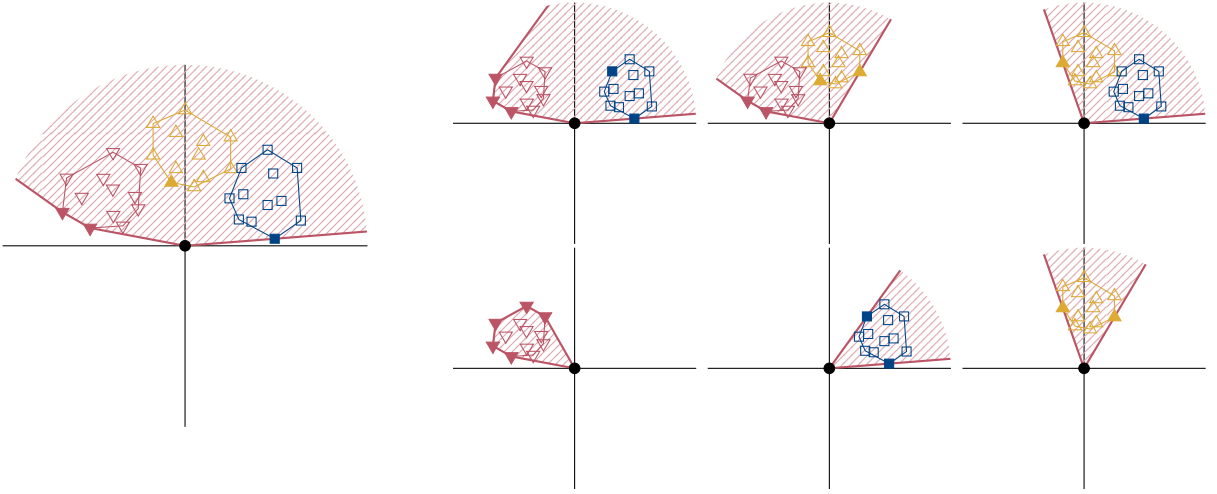 

\begin{figure}[hbtp]
    \begin{minipage}[c]{0.33\textwidth}
        \centering
        \begin{tikzpicture}[scale=.6]
            \regioncoords
            \coordaxes
            
            \coordinate (nn) at ($ -1*(n7) $);
            
            \draw[thick, pcolour, pattern = north east lines, pattern color=pcolour!50]
                ($ (nn)!10!(p5) $) -- (p5) -- (p4) -- +($ 10*(z3) $) -- (0,10) -- cycle;
                
            \draw[pcolour, every node/.style={pdot}]
                (p1) node {} -- (p2) node {} -- (p3) node {} -- (p4) node[pdot=pcolour] {} -- (p5) node[pdot=pcolour] {} --(p6) node {} -- (p7) node {} -- cycle;
            \draw[ncolour, every node/.style={ndot}]
                (n1) node {} -- (n2) node {} -- (n3) node {} -- (n4) node {} -- (n5) node{} --(n6) node{} -- (n7) node[ndot=ncolour] {} -- cycle;
            \draw[zcolour, every node/.style={zdot}]
                (z1) node {} -- (z2) node {} -- (z3) node[zdot=zcolour] {} -- (z4) node {} -- (z5) node {} -- (z6) node {} -- (z7) node {} -- cycle;
            
            \path[pcolour, every node/.style={pdot}]
                (p8) node{} (p9) node{} (p10) node{} (p11) node{} (p12) node{};
            \path[ncolour, every node/.style={ndot}]
                (n8) node{} (n9) node{} (n10) node{} (n11) node{} (n12) node{};
            \path[zcolour, every node/.style={zdot}]
                (z8) node{} (z9) node{} (z10) node{} (z11) node{} (z12) node{};
                
            \draw (0,0) node[dot=black] {};
        \end{tikzpicture}
    \end{minipage}
    \qquad
    \begin{minipage}{0.67\textwidth}
        \centering
        \begin{tikzpicture}[scale=.4]
            \regioncoords
            \coordaxes
            
            \draw[ncolour, every node/.style={ndot}]
                (n1) node {} -- (n2) node {} -- (n3) node {} -- (n4) node {} -- (n5) node{} --(n6) node{} -- (n7) node {} -- cycle;
            \draw[zcolour, every node/.style={zdot}]
                (z1) node {} -- (z2) node {} -- (z3) node {} -- (z4) node {} -- (z5) node {} -- (z6) node {} -- (z7) node {} -- cycle;
            
            \path[ncolour, every node/.style={ndot}]
                (n8) node{} (n9) node{} (n10) node{} (n11) node{} (n12) node{};
            \path[zcolour, every node/.style={zdot}]
                (z8) node{} (z9) node{} (z10) node{} (z11) node{} (z12) node{};
                
            \draw (0,0) node[dot=black] {};
        \end{tikzpicture}
        \begin{tikzpicture}[scale=.4]
            \regioncoords
            \coordaxes
            
            \coordinate (nn) at ($ -1*(n7) $);
            
            \draw[thick, pcolour, pattern = north east lines, pattern color=pcolour!50]
                ($ 5*(p3) $) -- (p3) -- (p4) -- (p5) -- ($ (nn)!10!(p5) $)  -- cycle;

            \draw[pcolour, every node/.style={pdot}]
                (p1) node {} -- (p2) node {} -- (p3) node[pdot=pcolour] {} -- (p4) node[pdot=pcolour] {} -- (p5) node[pdot=pcolour] {} --(p6) node {} -- (p7) node {} -- cycle;
            \draw[ncolour, every node/.style={ndot}]
                (n1) node {} -- (n2) node {} -- (n3) node {} -- (n4) node {} -- (n5) node{} --(n6) node{} -- (n7) node[ndot=ncolour] {} -- cycle;
            
            \path[pcolour, every node/.style={pdot}]
                (p8) node{} (p9) node{} (p10) node{} (p11) node{} (p12) node{};
            \path[ncolour, every node/.style={ndot}]
                (n8) node{} (n9) node{} (n10) node{} (n11) node{} (n12) node{};
                                
            \draw (0,0) node[dot=black] {};
        \end{tikzpicture}
        \begin{tikzpicture}[scale=.4]
            \regioncoords
            \coordaxes
            
            \draw[thick, pcolour, pattern = north east lines, pattern color=pcolour!50]
                ($ 10*(p6) $) -- (p6) -- (p5) -- (p4) -- (p3) -- (p4) -- +($ 10*(z3) $) -- cycle;

            \draw[pcolour, every node/.style={pdot}]
                (p1) node {} -- (p2) node {} -- (p3) node {} -- (p4) node[pdot=pcolour] {} -- (p5) node[pdot=pcolour] {} --(p6) node[pdot=pcolour] {} -- (p7) node {} -- cycle;
            \draw[zcolour, every node/.style={zdot}]
                (z1) node {} -- (z2) node {} -- (z3) node[zdot=zcolour] {} -- (z4) node {} -- (z5) node {} -- (z6) node {} -- (z7) node {} -- cycle;
            
            \path[pcolour, every node/.style={pdot}]
                (p8) node{} (p9) node{} (p10) node{} (p11) node{} (p12) node{};
            \path[zcolour, every node/.style={zdot}]
                (z8) node{} (z9) node{} (z10) node{} (z11) node{} (z12) node{};
            \draw (0,0) node[dot=black] {};
        \end{tikzpicture}\\
        \begin{tikzpicture}[scale=.4]
            \regioncoords
            \coordaxes
            
            \draw[ncolour, every node/.style={ndot}]
                (n1) node {} -- (n2) node {} -- (n3) node {} -- (n4) node {} -- (n5) node{} --(n6) node{} -- (n7) node {} -- cycle;
            
            \path[ncolour, every node/.style={ndot}]
                (n8) node{} (n9) node{} (n10) node{} (n11) node{} (n12) node{};
                
            \draw (0,0) node[dot=black] {};
        \end{tikzpicture}
        \begin{tikzpicture}[scale=.4]
            \regioncoords
            \coordaxes
            
            \draw[zcolour, every node/.style={zdot}]
                (z1) node {} -- (z2) node {} -- (z3) node {} -- (z4) node {} -- (z5) node {} -- (z6) node {} -- (z7) node {} -- cycle;
            
            \path[zcolour, every node/.style={zdot}]
                (z8) node{} (z9) node{} (z10) node{} (z11) node{} (z12) node{};
                
            \draw (0,0) node[dot=black] {};
        \end{tikzpicture}
        \begin{tikzpicture}[scale=.4]
            \regioncoords
            \coordaxes
            
            \draw[thick, pcolour, pattern = north east lines, pattern color=pcolour!50]
                ($ 10*(p6) $) -- (p6) -- (p5) -- (p4) -- (p3) -- ($ 10*(p3) $) -- cycle;
                            
            \draw[pcolour, every node/.style={pdot}]
                (p1) node {} -- (p2) node {} -- (p3) node[pdot=pcolour] {} -- (p4) node[pdot=pcolour] {} -- (p5) node[pdot=pcolour] {} --(p6) node[pdot=pcolour] {} -- (p7) node {} -- cycle;
            
            \path[pcolour, every node/.style={pdot}]
                (p8) node{} (p9) node{} (p10) node{} (p11) node{} (p12) node{};
                
            \draw (0,0) node[dot=black] {};
        \end{tikzpicture}
    \end{minipage}\hfill
    \caption[Example of the region $\reg_1(\Omega)$.]{%
        Example of the region $\reg_1(\Omega)$ (\pcolcol), with the same $\v\alpha_i$ and analogous presentation as in \cref{fig:alvarez-n}.
        Note how some relevant $\v\alpha_i, i\not\in I_1$ lie outside $\reg_1(\Omega)$; this is also possible for $\reg_0(\Omega)$, but not for $\reg_\inv(\Omega)$, since it contains all $\v\alpha_i$.
            
        \qquad As in \cref{fig:alvarez-n}, the smaller figures illustrate the effect of omitting some $\v\alpha_i$.
        The only-$I_1$ figure (bottom right) is thus analogous to $\reg(\omega_1)$ in \cref{fig:alvarez}.
        Note how $\reg_1(\Omega)=\emptyset$ when $i\in I_1$ are omitted, as can be seen from \cref{eq:gen-condition} with $\sumI{1}\lambda_i = 0$ (this is the exception to \cref{proposition B.2}).
    }
    \label{fig:alvarez-p}
\end{figure}
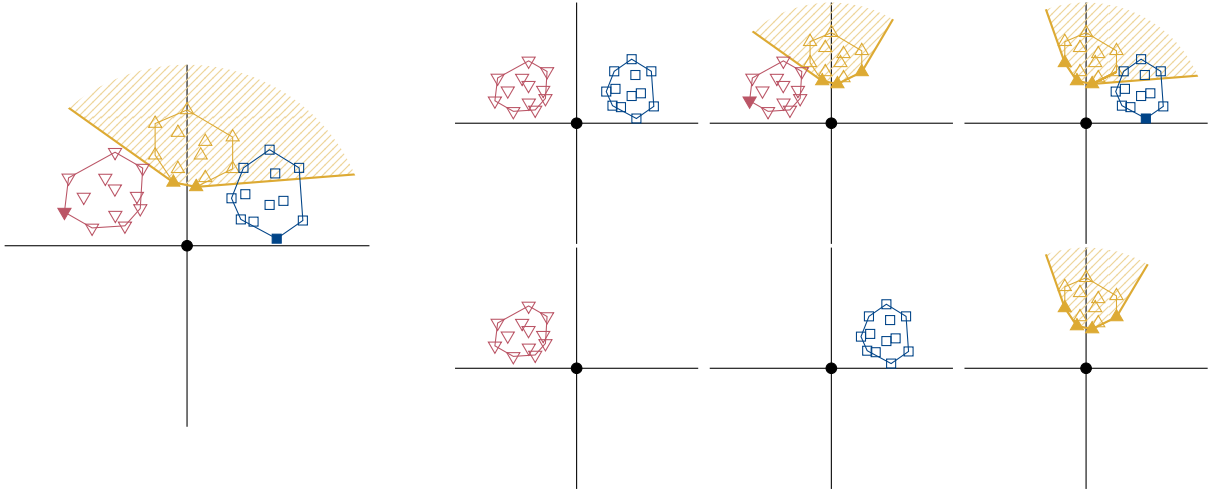 

\begin{figure}[hbtp]
    \begin{minipage}[c]{0.3\textwidth}
        \centering
        \begin{tikzpicture}[scale=.6]
            \regioncoords
            \coordaxes
                            
            \coordinate (pn) at ($ -1*(p5) $);
            \coordinate (nn) at ($ -1*(n7) $);
            
            \draw[thick, ncolour]
                ($ (pn)!10!(n7) $) -- (n7) -- (n6) -- (0,0) -- ($ 10*(z3) $);
            \draw[thick, pcolour]
                ($ (nn)!10!(p5) $) -- (p5) -- (p4) -- +($ 10*(z3) $);
            \draw[thick, zcolour, pattern = north east lines, pattern color=zcolour!50]
                ($ 10*(z3) $) -- (0,0) -- ($ 5*(n7) + -5*(pn) $) -- cycle;
                
            \draw[pcolour, every node/.style={pdot}]
                (p1) node {} -- (p2) node {} -- (p3) node {} -- (p4) node[pdot=pcolour] {} -- (p5) node[pdot=pcolour] {} --(p6) node {} -- (p7) node {} -- cycle;
            \draw[ncolour, every node/.style={ndot}]
                (n1) node {} -- (n2) node {} -- (n3) node {} -- (n4) node {} -- (n5) node {} --(n6) node[ndot=ncolour]{} -- (n7) node[ndot=ncolour] {} -- cycle;
            \draw[zcolour, every node/.style={zdot}]
                (z1) node {} -- (z2) node {} -- (z3) node[zdot=zcolour] {} -- (z4) node {} -- (z5) node {} --(z6) node{} -- (z7) node {} -- cycle;
            
            \path[pcolour, every node/.style={pdot}]
                (p8) node{} (p9) node{} (p10) node{} (p11) node{} (p12) node{};
            \path[ncolour, every node/.style={ndot}]
                (n8) node{} (n9) node{} (n10) node{} (n11) node{} (n12) node{};
            \path[zcolour, every node/.style={zdot}]
                (z8) node{} (z9) node{} (z10) node{} (z11) node{} (z12) node{};

            \draw (0,0) node[dot=black] {};
        \end{tikzpicture}
    \end{minipage}
    \qquad
    \begin{minipage}{0.67\textwidth}
        \centering
        \begin{tikzpicture}[scale=.4]
            \regioncoords
            \coordaxes
            
            \draw[thick, ncolour]
                ($ (n1) + 10*(z6) $) -- (n1) -- (n7) -- (n6) -- (0,0) -- ($ 10*(z3) $) -- (0,10);
            \draw[thick, zcolour, pattern = north east lines, pattern color=zcolour!50]
                (0,0) -- ($ 10*(z6) $) -- ($ 5*(z3) $) -- cycle;
                
            \draw[ncolour, every node/.style={ndot}]
                (n1) node[ndot=ncolour] {} -- (n2) node {} -- (n3) node {} -- (n4) node {} -- (n5) node {} --(n6) node[ndot=ncolour]{} -- (n7) node[ndot=ncolour] {} -- cycle;
            \draw[zcolour, every node/.style={zdot}]
                (z1) node {} -- (z2) node {} -- (z3) node[zdot=zcolour] {} -- (z4) node {} -- (z5) node {} --(z6) node[zdot=zcolour] {} -- (z7) node {} -- cycle;
            
            \path[ncolour, every node/.style={ndot}]
                (n8) node{} (n9) node{} (n10) node{} (n11) node{} (n12) node{};
            \path[zcolour, every node/.style={zdot}]
                (z8) node{} (z9) node{} (z10) node{} (z11) node{} (z12) node{};
                                
            \draw (0,0) node[dot=black] {};
        \end{tikzpicture}
        \begin{tikzpicture}[scale=.4]
            \regioncoords
            \coordaxes
                            
            \coordinate (pn) at ($ -1*(p5) $);
            \coordinate (nn) at ($ -1*(n7) $);
            
            \draw[thick, ncolour]
                ($ (n7)!-10!(pn) $) -- (n7) -- (n6) -- (0,0) -- ($ 10*(p3) $);
            \draw[thick, pcolour]
                ($ 5*(p3) $) -- (p3) -- (p4) -- (p5) -- ($ (nn)!10!(p5) $);
            
            \draw[thick, zcolour, pattern = north east lines, pattern color=zcolour!50]
                ($ 10*(p3) $) -- (0,0) -- ($ 5*(n7) + -5*(pn) $) -- cycle;
                
            \draw[pcolour, every node/.style={pdot}]
                (p1) node {} -- (p2) node {} -- (p3) node[pdot=pcolour] {} -- (p4) node[pdot=pcolour] {} -- (p5) node[pdot=pcolour] {} --(p6) node {} -- (p7) node {} -- cycle;
            \draw[ncolour, every node/.style={ndot}]
                (n1) node {} -- (n2) node {} -- (n3) node {} -- (n4) node {} -- (n5) node {} --(n6) node[ndot=ncolour]{} -- (n7) node[ndot=ncolour] {} -- cycle;
            
            \path[pcolour, every node/.style={pdot}]
                (p8) node{} (p9) node{} (p10) node{} (p11) node{} (p12) node{};
            \path[ncolour, every node/.style={ndot}]
                (n8) node{} (n9) node{} (n10) node{} (n11) node{} (n12) node{};
                                
            \draw (0,0) node[dot=black] {};
        \end{tikzpicture}
        \begin{tikzpicture}[scale=.4]
            \regioncoords
            \coordaxes
            
            \draw[thick, pcolour]
                ($ 10*(p6) $) -- (p6) -- (p5) -- (p4) -- (p3) -- (p4) -- +($ 10*(z3) $);
            \draw[thick, zcolour, pattern = north east lines, pattern color=zcolour!50]
                (0,0) -- ($ 10*(p6) $) -- ($ 5*(z3) $) -- cycle;
                                
            \draw[pcolour, every node/.style={pdot}]
                (p1) node {} -- (p2) node {} -- (p3) node {} -- (p4) node[pdot=pcolour] {} -- (p5) node[pdot=pcolour] {} --(p6) node[pdot=pcolour] {} -- (p7) node {} -- cycle;
            \draw[zcolour, every node/.style={zdot}]
                (z1) node {} -- (z2) node {} -- (z3) node[zdot=zcolour] {} -- (z4) node {} -- (z5) node {} --(z6) node{} -- (z7) node {} -- cycle;
            
            \path[pcolour, every node/.style={pdot}]
                (p8) node{} (p9) node{} (p10) node{} (p11) node{} (p12) node{};
            \path[zcolour, every node/.style={zdot}]
                (z8) node{} (z9) node{} (z10) node{} (z11) node{} (z12) node{};
                                
            \draw (0,0) node[dot=black] {};
        \end{tikzpicture}\\
        \begin{tikzpicture}[scale=.4]
            \regioncoords
            \coordaxes
            
            \draw[thick, ncolour]
                (n1) -- (n7) -- (n6) -- (0,0) -- (n3) -- (n2) -- cycle;
            \draw[ncolour, every node/.style={ndot}]
                (n1) node[ndot=ncolour] {} -- (n2) node[ndot=ncolour] {} -- (n3) node[ndot=ncolour] {} -- (n4) node {} -- (n5) node {} --(n6) node[ndot=ncolour] {} -- (n7) node[ndot=ncolour] {} -- cycle;
            
            \path[ncolour, every node/.style={ndot}]
                (n8) node{} (n9) node{} (n10) node{} (n11) node{} (n12) node{};
                                
            \draw (0,0) node[dot=black] {};
        \end{tikzpicture}
        \begin{tikzpicture}[scale=.4]
            \regioncoords
            \coordaxes
            
            \draw[thick, zcolour, pattern = north east lines, pattern color=zcolour!50]
                (0,0) -- ($ 10*(z6) $) -- ($ 5*(z3) $) -- cycle;
                            
            \draw[zcolour, every node/.style={zdot}]
                (z1) node {} -- (z2) node {} -- (z3) node[zdot=zcolour] {} -- (z4) node {} -- (z5) node {} --(z6) node[zdot=zcolour] {} -- (z7) node {} -- cycle;
            
            \path[zcolour, every node/.style={zdot}]
                (z8) node{} (z9) node{} (z10) node{} (z11) node{} (z12) node{};
                                
            \draw (0,0) node[dot=black] {};
        \end{tikzpicture}
        \begin{tikzpicture}[scale=.4]
            \regioncoords
            \coordaxes
            
            \draw[thick, pcolour]
                ($ 10*(p6) $) -- (p6) -- (p5) -- (p4) -- (p3) -- ($ 10*(p3) $);
            \draw[thick, zcolour, pattern = north east lines, pattern color=zcolour!50]
                (0,0) -- ($ 10*(p6) $) -- ($ 5*(p3) $) -- cycle;
                            
            \draw[pcolour, every node/.style={pdot}]
                (p1) node {} -- (p2) node {} -- (p3) node[pdot=pcolour] {} -- (p4) node[pdot=pcolour] {} -- (p5) node[pdot=pcolour] {} --(p6) node[pdot=pcolour] {} -- (p7) node {} -- cycle;
            
            \path[pcolour, every node/.style={pdot}]
                (p8) node{} (p9) node{} (p10) node{} (p11) node{} (p12) node{};
                                
            \draw (0,0) node[dot=black] {};
        \end{tikzpicture}
    \end{minipage}
    \caption[Example of the region $\reg_0(\Omega)$.]{
        Example of the region $\reg_0(\Omega)$ (\zcolcol), with the same $\v\alpha_i$ as in \cref{fig:alvarez-n,fig:alvarez-p}.
        $\reg_\pml(\Omega)$ are outlined to demonstrate how $\reg_0(\Omega)$ can be constructed by taking their unbounded facets and shifting them so they pass through the origin, as discussed in \cref{sec:practical-cor}.
        Note that unlike in \cref{fig:alvarez-n,fig:alvarez-p}, not all relevant $\v\alpha_i$ influence the shape of $\reg_0(\Omega)$.
            
        \qquad As in \cref{fig:alvarez-n,fig:alvarez-p}, the smaller figures illustrate the effect of omitting some $\v\alpha_i$.
        The only-$I_0$ figure (bottom centre) is thus analogous to $\reg(\omega_0)$ in \cref{fig:alvarez}.
        Note how in the only-$I_\inv$ figure (bottom left), $\reg_0(\Omega)=\{\v0\}$ since $\reg_\inv(\Omega)$ has no unbounded faces (this is one of the exceptions to \cref{cor:construct-0}).}
    \label{fig:alvarez-z}
\end{figure} 


\subsection{The duality between $\reg_c(\Omega)$ and $\sat(\Omega)$}\label{sec:dual}
As has been used many times above, parameter space and constraint space are dual in the sense that points in one correspond to hyperplanes in the other.
This extends to $\sat(\Omega)$ and $\reg_c(\Omega)$, which are similarly related in ways we will explore in this section.

A taste of this duality can be found in the following result, which will also be useful further on: 
\begin{corollary}[duality]\label{cor:dual}
    With $\Omega\neq\Omega_\infty$, let $\mho_\pm$ be defined as in \cref{eq:mho}, so that \mbox{$\sat(\mho_\pm)=\reg_\pml(\Omega)$}. 
    Then \mbox{$\sat(\Omega)=\reg_1(\mho_+)\cap\reg_\inv(\mho_-)$} if $I_1\neq\emptyset$, and \mbox{$\sat(\Omega)=\reg_\inv(\mho_-)$} otherwise.
\end{corollary}
\noindent Loosely, one can think of this as ``\cref{proposition B.2} is its own inverse'': applying it to $\Omega$ gives $\mho_\pm$, and applying it to $\mho_\pm$ gives $\Omega$.

\proof By definition and \cref{proposition B.2},
\begin{equation}
    \begin{alignedat}{4}
        \v b&\in\sat(\Omega)        
            &&\equ   \forall\constr{\v\alpha}{c}\leq\Omega,\quad \v\alpha\cdot\v b   \geq c,\\
        \v b&\in\reg_\pml(\mho_\pml)
            &&\equ  \reg_\pml(\Omega)\subseteq \sat(\constr{\v b}{\pm1}).
    \end{alignedat}
\end{equation}
Looking at \cref{eq:def-reg-alt}, we see that the latter is equivalent to
\begin{equation}\label{eq:duality-cond}
    \forall\{\lambda_i\}_{i\in I} \text{ with }\lambda_i\geq 0\text{ and  }\sum_{i\in I}\lambda_ic_i\geq\pm1,\qquad
    \v b\cdot\sum_{i\in I}\lambda_i\v\alpha_i \geq \pm1.
\end{equation}
This is true if $\forall i\in I,\:\v b\cdot\v\alpha_i\geq c_i$, but it only implies $\forall i\in I\setminus I_\mpl,\:\v b\cdot\v\alpha_i\geq c_i$, as some straightforward algebra shows.
By using $\reg_1(\mho_+)\cap\reg_\inv(\mho_-)$, we avoid this shortcoming and complete the proof.
When $I_1=\emptyset$, we cannot use $\mho_+$ but also do not need it, since $I\setminus I_1=I$.
\qed

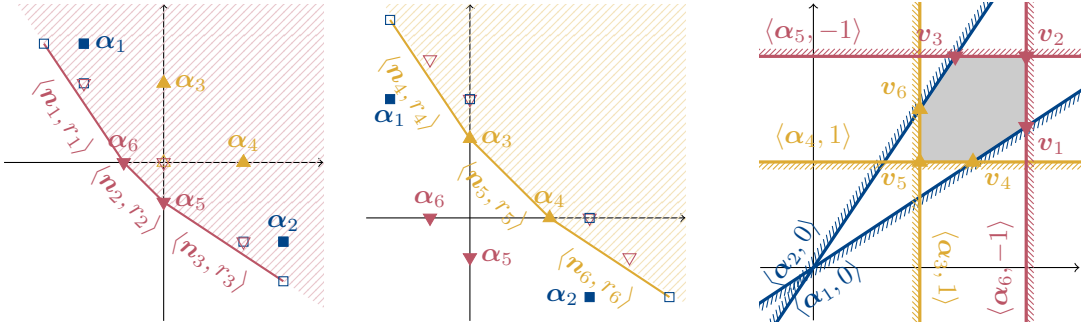
\begin{figure}[hbtp]
    \centering
    \begin{tikzpicture}[scale=1.05]
        \draw[->] (-2,0) -- (2,0);
        \draw[->] (0,-2) -- (0,2);
        \clip (-2,-2) rectangle (2,2);
        
        \path[pattern=north east lines, pattern color=ncolour!33] (-2.5,3) -- (-.5,0) -- (0,-.5) -- (3,-2.5) -- (3,3) -- cycle;
        \draw[thick, ncolour] (-1.5,1.5) 
            -- (-.5,0)      node[midway, sloped, below] {\footnotesize $\constr{\v n_1}{r_1}$} 
            -- (0,-.5)      node[midway, sloped, below] {\footnotesize $\constr{\v n_2}{r_2}$} 
            -- (1.5,-1.5)   node[midway, sloped, below] {\footnotesize $\constr{\v n_3}{r_3}$};
                
        \draw[zcolour] 
            (-1,1.5) node[zdot=zcolour] {} node[right] {\footnotesize $\v\alpha_1$} 
            (1.5,-1) node[zdot=zcolour] {} node[above] {\footnotesize $\v\alpha_2$};
        \draw[pcolour] 
            (1,0)    node[pdot=pcolour] {} node[above] {\footnotesize $\v\alpha_4$} 
            (0,1)    node[pdot=pcolour] {} node[right] {\footnotesize $\v\alpha_3$};
        \draw[ncolour] 
            (-.5,0)  node[ndot=ncolour] {} node[above] {\footnotesize $\v\alpha_6$} 
            (0,-.5)  node[ndot=ncolour] {} node[right] {\footnotesize $\v\alpha_5$};
        
        \draw[pcolour, every node/.style={pdot}] (0,0) node {} (1,0) node {} (0,1) node {};
        \draw[zcolour, every node/.style={zdot}] (-1.5,1.5) node {} (-1,1) node {} (1,-1) node {} (1.5,-1.5) node {} (-1,1.5) node {} (1.5,-1) node {};
        \draw[ncolour, every node/.style={ndot}] (0,0) node {} (1,-1) node{} (-1,1)node {};
    \end{tikzpicture}
    \quad
    \begin{tikzpicture}[scale=1.05]
        \draw[->] (-1.3,0) -- (2.7,0);
        \draw[->] (0,-1.3) -- (0,2.7);
        \clip (-1.3,-1.3) rectangle (2.7,2.7);
        
        \path[pattern=north east lines, pattern color=pcolour!33] (-1.08,2.7) -- (0,1) -- (1,0) -- (2.7,-1.08) -- (3,3) -- cycle;
        \draw[thick, pcolour] (-1,2.5) 
            -- (0,1)      node[midway, sloped, below] {\footnotesize $\constr{\v n_4}{r_4}$} 
            -- (1,0)      node[midway, sloped, below] {\footnotesize $\constr{\v n_5}{r_5}$} 
            -- (2.5,-1)   node[midway, sloped, below] {\footnotesize $\constr{\v n_6}{r_6}$};

        \draw[zcolour] 
            (-1,1.5) node[zdot=zcolour] {} node[below] {\footnotesize $\v\alpha_1$} 
            (1.5,-1) node[zdot=zcolour] {} node[left]  {\footnotesize $\v\alpha_2$};
        \draw[pcolour] 
            (1,0)    node[pdot=pcolour] {} node[above] {\footnotesize $\v\alpha_4$} 
            (0,1)    node[pdot=pcolour] {} node[right] {\footnotesize $\v\alpha_3$};
        \draw[ncolour] 
            (-.5,0)  node[ndot=ncolour] {} node[above] {\footnotesize $\v\alpha_6$} 
            (0,-.5)  node[ndot=ncolour] {} node[right] {\footnotesize $\v\alpha_5$};
                
        \draw[ncolour, every node/.style={ndot}] (1.5,0) node {} (0,1.5) node {} (2,-.5) node {} (-.5,2) node{};
        \draw[zcolour, every node/.style={zdot}] (0,1.5) node {} (1.5,0) node {} (-1,2.5) node {} (2.5,-1) node {};
    \end{tikzpicture}
    \qquad
    \begin{tikzpicture}[scale=1.4]
        \draw[->] (-.5,0) -- (2.5,0);
        \draw[->] (0,-.5) -- (0,2.5);
        \clip (-.51,-.51) rectangle (2.51,2.51);

        \fill[black, opacity=.2] (1,1) -- (1.5,1) -- (2,1.333) -- (2,2) -- (1.333,2) -- (1,1.5) -- cycle;
        
        \constrline[zcolour, relevant]{-1,1.5}{0} node[opacity=1, midway, sloped, below] {\footnotesize $\constr{\v\alpha_1}{ 0}$};
        \constrline[zcolour, relevant]{1.5,-1}{0} node[opacity=1, midway, sloped, above] {\footnotesize $\constr{\v\alpha_2}{ 0}$};
        
        \constrline[pcolour, relevant]{1.0,0}{+1} node[opacity=1, midway, sloped, above] {\footnotesize $\constr{\v\alpha_3}{ 1}$};
        \constrline[pcolour, relevant]{0,1.0}{+1} node[opacity=1, midway, sloped, above] {\footnotesize $\constr{\v\alpha_4}{ 1}$};
        \constrline[ncolour, relevant]{0,-.5}{-1} node[opacity=1, midway, sloped, above] {\footnotesize $\constr{\v\alpha_5}{ -1}$};
        \constrline[ncolour, relevant]{-.5,0}{-1} node[opacity=1, midway, sloped, above] {\footnotesize $\constr{\v\alpha_6}{ -1}$};

        \draw[thick,zcolour] (1.5,1) -- (2,1.333);
        \draw[thick,zcolour] (1,1.5) -- (1.333,2);
        \draw[thick,ncolour,every node/.style={ndot=ncolour}] (1.333,2) node{} -- (2,2) node {} -- (2,1.333) node{};
        \draw[thick,pcolour,every node/.style={pdot=pcolour}] (1.5,1) node {} -- (1,1) node {} -- (1,1.5) node {};
        \draw[ncolour] 
            (1.333,2) node[above left]   {\footnotesize $\v v_3$} 
            (2,2) node[above right] {\footnotesize $\v v_2$} 
            (2,1.333) node[below right]   {\footnotesize $\v v_1$};
        \draw[pcolour] 
            (1.5,1) node[below right]     {\footnotesize $\v v_4$}  
            (1,1) node[below left]  {\footnotesize $\v v_5$}  
            (1,1.5) node[above left]      {\footnotesize $\v v_6$};
            
    \end{tikzpicture}
    \caption[Example of how $\reg_\pml\Omega$ are dual to $\sat(\Omega)$.]{Example of how $\reg_\pml\Omega$ are dual to $\sat(\Omega)$.
    A different, simpler $\Omega$ than in previous figures is used.
    \textbf{Left:} Constraint space, with $\reg_\inv(\Omega)$ constructed in a similar way to \cref{fig:construct}.
    The points $\v\alpha_i$ are marked with filled dots and labelled; points in $\psP{-}(\Omega)$, $\psZ{2}{-}(\Omega)$ and $\psN{2}{-}(\Omega)$ are marked with empty dots.
    The facets corresponding to constraints $\constr{\v n_\ell}{r_\ell}$ are also labelled.
    \textbf{Middle:} Constraint space, with $\reg_1(\Omega)$ constructed similarly.
    \textbf{Right:} Parameter space, drawn similarly to \cref{fig:constr}. $\sat(\Omega)$ is shaded; the six constraints $\constr{\v\alpha_i}{c_i}$ and vertices $\v v_\ell = \v n_\ell\frac\pml{r_\ell}$ are labelled.
    Vertices from $\reg_\inv(\Omega)$ are marked with \nMark, and those from $\reg_1(\Omega)$ with \pMark.
    Note that two vertices sharing an edge of $\sat(\Omega)$ corresponds to two constraints supporting a common point in $\reg_\pml{\Omega}$.
    In the case of $c=0$ edges, it corresponds to supporting $(\v\alpha_i+p\v\alpha_j)\in \psZ{p}{+}(\Omega)$ and $(\v\alpha_{i'}+p\v\alpha_j)\in\psZ{p}{-}(\Omega)$ for a common $j\in I_0$.
    Note also how \cref{cor:dual} applies (compare \cref{fig:alvarez}).}
    \label{fig:dual}
\end{figure}

This relation can be extended into a geometric duality between $\sat(\Omega)$ and $\rel_c(\Omega)$: the vertices of one correspond directly to the facets of the other.%
\footnote{
    This is closely related to the concept of \emph{dual polytopes}, where a polytope is a $D$-dimensional generalisation of a polygon or polyhedron; $\sat(\Omega)$ and $\reg_c(\Omega)$ are polytopes, if the definition is relaxed to permit unbounded polytopes.}
\pagebreak
\begin{proposition}[precise duality relations]\label{prop:dual}
    Let $\isM{p}{\pm}$, etc.\ be defined as in \cref{proposition B.2}, and let $\mho_\pm$ be defined as in \cref{eq:mho}.
    Let $\pm1=1$ if $I_1\neq\emptyset$ and $\pm1=-1$ otherwise, as in \cref{cor:dual}.
    If $\Omega$ and $\mho_\pm$ are non-degenerate, then the following correspondences hold:%
    \begin{enumerate}[label={(\roman*)}]        
        \item\label[correspondence]{corr:vertex-of-A} $\v\alpha_i$ for $i\in I_\pml$ is a vertex of $\reg_\pml(\Omega)$ if and only if $\constr{\v\alpha_i}{\pm1}\in\rel(\Omega)$.
                
        \item\label[correspondence]{corr:vertex-of-B} $\frac{1}{|r_\ell|}\v n_\ell$ is a vertex of $\sat(\Omega)$ if and only if $\ell\in\isM{p}{\pm}$ and $r_\ell\neq 0$.
        
        \item\label[correspondence]{corr:facet-of-A} $\s F\equiv\setbuild{\v\beta\in\reg_\pml(\Omega)}{\v n_\ell\cdot\v\beta=r_\ell}$ is a facet of $\reg_\pml(\Omega)$ if and only if $\ell\in\isM{p}{\pm}$.
        
        \item\label[correspondence]{corr:facet-of-B} $\s F\equiv\setbuild{\v b\in\sat(\Omega)}{\v\alpha\cdot\v b=c}$ is a facet of $\sat(\Omega)$ if and only if $\constr{\v\alpha}{c}\in\rel(\Omega)$.
    \end{enumerate}
    All relations are exhaustive: there is no vertex of $\reg_\pml(\Omega)$ that is not covered by \cref{corr:vertex-of-A}, etc.
    There is the single exception that $\v 0$ may be a vertex without corresponding to a facet.%
    \footnote{
        The origin is a vertex of $\sat(\Omega)$ if $I_1=\emptyset$ and $\{\v\alpha_i\}_{i\in I_0}$ is a basis of $\R^D$.
        It is never a vertex of (or even contained in) $\reg_1(\Omega)$, but is always a vertex of $\reg_\inv(\Omega)$ if $\Omega$ is non-degenerate.}
    Also, the exceptions to \cref{proposition B.2} apply.
\end{proposition}
\noindent This was used to obtain the visualisations of $\sat(\Omega)$ in \cref{sec:results}. 
An illustrative example is given in \cref{fig:dual}.

Note how \cref{prop:dual} is rather negligent of $c=0$ constraints, partly because they are complicated to handle, and partly because an exact zero is an unlikely thing when constraints are generated with numerical inaccuracies.
For similar reasons, we do not consider the degenerate case.

\proof \Cref{corr:vertex-of-A} is just \cref{proposition B.2-relevant}.

\Cref{corr:vertex-of-B}, which is the most useful correspondence for visualisation, follows by applying \cref{proposition B.2-relevant} to $\mho_\pm$ instead.
To do this, we first apply the normalisation
\begin{equation}
    \constr{\v n_\ell}{r_\ell}\quad\to\quad
    \begin{cases}
        \constr{\tfrac{1}{|r_\ell|}\v n_\ell}{\pm1} &\text{if }r_\ell\neq 0,\\
        \constr{\v n_\ell}{0}                       &\text{otherwise},
    \end{cases}
\end{equation}
recalling \cref{lem:sign}. 
Granted that $\mho_\pm$ is non-redundant, \cref{proposition B.2-relevant} therefore states that the extreme points (vertices) of $\reg_\pml(\mho_\pm)$, except $\v 0$, are exactly $\frac{1}{|r_\ell|}\v n_\ell$, since $\constr{\tfrac{1}{|r_\ell|}\v n_\ell}{\pm1}$ are exactly the relevant elements of $\mho_\pm$ with $r_\ell\neq0$.
The correspondence then follows from \cref{cor:dual}, as long as we can ensure that the intersection does not introduce any new vertices that are not vertices of $\reg_\pml(\mho_\pm)$.
Such a vertex would, by \cref{lem:K-face}, be uniquely supported by some constraint that is unaccounted for by \cref{proposition B.2-relevant}, so it does not exist.

\Cref{corr:facet-of-A,corr:facet-of-B} both follow from \cref{proposition 4.3}, since each facet contains all points that are uniquely supported by some relevant constraint.
It follows from \cref{cor:D-1,lem:K-face} that a point $\v b$ is uniquely supported by some constraint if and only if $\v b$ is contained in some $(D-1)$-face, but not in any $K$-face for $K<D-1$.
It also follows from the definition of a $K$-face that all $K$-faces contain at least one point that is not contained in any $K'$-face, $K'<K$.
Thus, every facet gives rise to a relevant constraint, and no constraint can be relevant if it does not give rise to a facet.
\QED

\subsubsection{Duality with a bounding box}
It would be possible to extend \cref{prop:dual} to relate the $\v n_\ell$ with $r_\ell=0$ to unbounded edges of $\sat(\Omega)$ by equating them to the ray $\setbuild{\frac{1}{|r_k|}\v n_k + \lambda\v n_\ell}{\lambda\geq 0}$ for some $k$ (see \cref{proposition B.2-relevant}).
However, it is not obvious how to find $k$, and it is not guaranteed that we exhaustively cover the unbounded edges this way.
In either case, unbounded edges are far less pleasant to deal with than vertices, even though they do provide shape information that the vertices alone cannot provide.

\begin{figure}[hbtp]
    \centering
    \begin{tikzpicture}
        \draw[->] (-3,0) -- (3,0);
        \draw[->] (0,-1) -- (0,3);
        
        \path[pattern=north east lines, pattern color=pcolour!50] (3,2) -- (1,1) -- (-1,1) -- (-3,2) -- (-3,2.9) -- (3,2.9) -- cycle;
        \draw[thick, pcolour] (2,1.5) 
            -- (1,1)    node[midway, sloped, below] {\footnotesize $\constr{\v n_3}{ r_3}$} 
            -- (-1,1)   node[midway, sloped, below,yshift=.5ex] {\footnotesize $\constr{\v n_1}{ r_1}$} 
            -- (-2,1.5) node[midway, sloped, below] {\footnotesize $\constr{\v n_2}{ r_2}$};
        
        \draw[pcolour, every node/.style={pdot=pcolour}] (1,1) node{} (-1,1) node{};
        \draw[pcolour] (1,1) node[above] {\footnotesize $\v\alpha_2$} (-1,1) node[above] {\footnotesize $\v\alpha_1$};
        \foreach \x in {-.5,.5} {
            \draw[ncolour, every node/.style={ndot=ncolour}] (\x,0) node {} (0,\x) node {};
            \foreach \y in {2,-2} {
                \draw[ncolour, every node/.style={ndot}] 
                    (\y,2) + (\x,0) node {} 
                    (\y,2) + (0,\x) node {};
            }
        }
        
    \end{tikzpicture}
    \qquad
    \begin{tikzpicture}
        \draw[->] (-2.5,0) -- (2.5,0);
        \draw[->] (0,-2.5) -- (0,2.5);
        \clip (-2.51,-2.51) rectangle (2.51,2.51);
        
        \fill[black, opacity=.2] (1,2) -- (0,1)  -- (-1,2) -- cycle;
        
        \constrline[ncolour, relevant]{-.5,0}{-1};
        \constrline[ncolour, relevant]{+.5,0}{-1};
        \constrline[ncolour, relevant]{0,-.5}{-1};
        \constrline[ncolour, relevant]{0,+.5}{-1};
        
        \constrline[pcolour, relevant]{-1,1}{1} node[opacity=1, sloped, above, midway] {\footnotesize $\constr{\v\alpha_1}{1}$};
        \constrline[pcolour, relevant]{+1,1}{1} node[opacity=1, sloped, above, midway] {\footnotesize $\constr{\v\alpha_2}{1}$};
        
        \path[pcolour] 
            (1,2) node[ above] {\footnotesize $\v v_2$} 
            (0,1) node[below, yshift=-1ex] {\footnotesize $\v v_1$} 
            (-1,2) node[ above] {\footnotesize $\v v_3$};
        \draw[thick, pcolour, every node/.style={pdot=pcolour}] (1,2) node{} -- (0,1) node{} -- (-1,2) node{};

    \end{tikzpicture}
    \caption[Example of how a $\sat(\Omega)$ is made more manageable by imposing a bounding box.]{Example of how $\sat(\Omega)=\constr{\v\alpha_1}{1}+\constr{\v\alpha_2}{1}$, which would normally have a single vertex $\v v_1$ and two unbounded edges corresponding to $r_k=0$ constraints, is made more manageable by imposing an artifical set of bounding-box constraints (\nMark), giving rise to some points in $\psN{2}{+}(\Omega)$ (\nmark).
    Like in \cref{fig:dual}, $\reg_1(\Omega)$ is shown on the left, and $\sat(\Omega)$ is shown on the right.}
    \label{fig:bbox}
\end{figure}
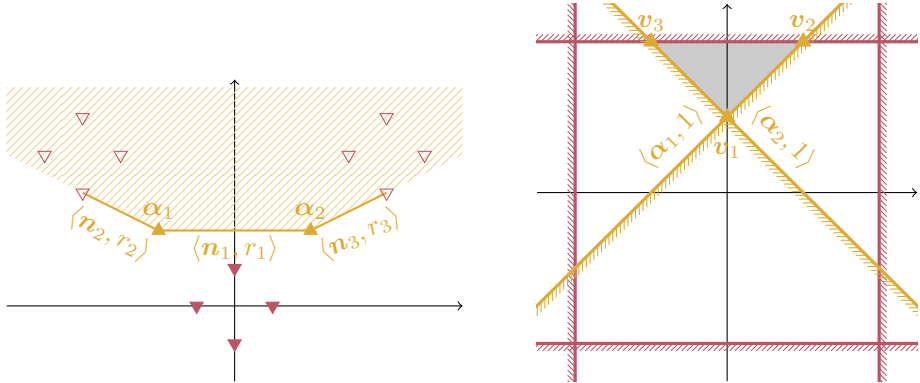

Unbounded edges can be wholly avoided by artificially introducing $2D$ extra constraints that constrain $\sat(\Omega)$ to a $D$-dimensional bounding box.
All unbounded facets of $\sat(\Omega)$ are cropped, and \cref{prop:dual}, \cref{corr:vertex-of-B} seamlessly provides all the vertices that define the intersections between $\sat(\Omega)$ and the walls of the box.
This method, demonstrated in \cref{fig:bbox}, was used extensively in \cref{sec:results}.

Everything mentioned here can of course be applied equally well to $\rel_\pml(\Omega)$.

\subsection{A note on infinite sums of constraints}\label{sec:infinite}
It is extremely important that we only ever consider sums of a finite number of constraints, not only for the validity of our proofs, but also for the validity of the propositions themselves. 
Consider as a counterexample the following countably infinite sum of one-dimensional constraints:
\begin{equation}
    \omega_\inv = \sum_{k=1}^\infty \constr{1 - \tfrac1k}{-1} \imp \reg(\omega_\inv) = [0,1),\quad\sat(\omega_\inv)=[-1,\infty).
\end{equation}
Here, $\reg(\omega_\inv)$ is not a closed set. 
On the other hand, $\sat(\omega_\inv)$ is closed, since for any $\e>0$, the point $-(1+\e)$ fails to satisfy the element $\constr{1-\tfrac1k}{-1}$ for sufficiently large $k$ (specifically, larger than $\frac{1+\e}{\e}$). 
Thus, $\sat(\omega_\inv) = \sat(\constr{1}{-1})$, so $\constr{1}{-1}\leq\omega_\inv$ even though $1\not\in \reg(\omega_\inv)$.
Thus, \cref{proposition 4.1}, and therefore \cref{proposition 4.2}, fails.

There is, however, a straightforward generalisation. 
We let \cref{eq:def-Omega,eq:def-gen-reg} define $\Omega$ and $\reg_c(\Omega)$ as before, but now with infinite sums permitted, so that $\{\v\alpha_i\}_{i\in I_c}$ may be any subset of $\R^D$. 
Then we have the following:
\begin{proposition}[determining if constraint is weaker, generalised]\label{prop:infinite}
    Let $\constr{\v\beta}{c}$ be a linear constraint, and let $\Omega\neq\Omega_\infty$ be a possibly infinite sum of linear constraints. Then $\constr{\v\beta}{c}\leq\Omega$ if and only if $\v\beta\in\cl\left[\reg_c(\Omega)\right]$.
\end{proposition}
\noindent The simple introduction of the closure solves all issues with infinite combinations, such as the counterexample above. 
Note how the only part of \cref{sec:closed,sec:proof-0,sec:proof-gen} that relies on the finiteness of $I$ is the proof that $\reg_c(\Omega)$ is closed. 
The closure is by definition closed, so the remaining arguments in these sections remain valid when applied to $\cl\left[\reg_c(\Omega)\right]$ instead.

The only part of the proof of \cref{proposition 4.2} that does not immediately carry over is \cref{sec:trivial}.
However, by the limit definition of closure, for any $\v\beta\in\cl[\reg_c(\Omega)]$ there exists a sequence $\v\beta_n$ in $\reg_c(\Omega)$ such that $\lim_{n\to\infty}\v\beta_n=\v\beta$.
Since the function $\v\beta\to\v\beta\cdot\v b$ is continuous, it follows that for any $\v b$, $\lim_{n\to\infty}\v\beta_n\cdot\v b = \v\beta\cdot\v b$.
Thus, since the arguments of \cref{sec:trivial} hold for all $\v\beta_n$, they also hold for $\v\beta\in\cl[\reg_c(\Omega)]$.
\QED

\Cref{cor:bounded} does not require any adjustment, since the interior of any convex set in $\R^D$ is equal to the interior of its closure.

The generalisation of \cref{proposition 4.3}, for which counterexamples abound, is less straightforward, partly because there does not necessarily exist a minimal representation if all representations are infinite.
However, the notion of a minimal representation is mainly motivated as being the most practical format of a constraint, so it is not very useful in the infinite case, which is anyway only of theoretical interest.
For the same reason, it is not relevant to our study to attempt to generalise any of the propositions presented in this appendix.

\subsection{Mathematical glossary}\label{sec:glossary}
The table below contains a list of notations and terms that may be unfamiliar to some readers, depending on their background.
(We have chosen to employ such notation, since it makes some things much more brief and expressive, even though it necessitates this table.) 
\newcommand{\glossentry}[3]{#1&#2 \footnotesize #3}
\begin{longtable}{p{0.22\textwidth}p{0.7\textwidth}}
    \hline\hline
    \textsc{Concept}  &   \textsc{Description}   \\
    \hline
    \endhead
    \glossentry
        {$\square$, $\blacksquare$}
        {End of proof.}
        {We use $\blacksquare$ for main proofs and $\square$ for lemmata.}\\
    \glossentry
        {$\forall$}
        {The universal quantifier.}
        {Informally, short for ``for all''.}\\
    \glossentry
        {$\exists$}
        {The existence quantifier.}
        {Informally, short for ``there exists''.}\\
    \glossentry
        {$x\in\s X$}
        {The object $x$ is contained in (is an element of) the set $\s X$.}
        {We typically denote sets using a calligraphic font, but use ordinary italics for sets of indices.}\\
    \glossentry
        {$\emptyset$}
        {The empty set.}
        {The set that contains no elements.}\\
    \glossentry
        {$|\s X|,|\v v|,|s|$}
        {Cardinality or magnitude.}
        {For a set, this indicates its cardinarlity (number of elements).
            For a vector or scalar, the same notation indicates its magnitude $\sqrt{\v v\cdot\v v}$ or $\sqrt{s^2}$.}\\
    \glossentry
        {$\s X \subseteq \s Y$}
        {$\s X$ is a subset of $\s Y$.}
        {All elements of $\s X$ are also contained in $\s Y$.
            Just $\s X\subset\s Y$ means the same, except that $\s X \neq \s Y$; that is, $\s Y$ has at least one element not contained in $\s X$.}\\
    \glossentry
        {$\s X \cup \s Y$}
        {The union of two sets.}
        {The set that contains all elements contained in either $\s X$ or $\s Y$, or both.
            Clearly, $\s X\subseteq \s X\cup\s Y$ and $\s Y\subseteq \s X\cup\s Y$.}\\
    \glossentry
        {$\s X \cap \s Y$}
        {The intersection of two sets.}
        {The set that contains all elements contained in \emph{both} $\s X$ or $\s Y$.
            Clearly, $\s X\supseteq \s X\cup\s Y$ and $\s Y\supseteq \s X\cup\s Y$.}\\
    \glossentry
        {Disjoint sets}
        {Said of two sets $\s X,\s Y$ if $\s X\cap\s Y=\emptyset$.}\\
    \glossentry
        {$\s X\setminus\s Y$}
        {Relative complement.}
        {Consists of all elements of $\s X$ that are not also elements of $\s Y$. 
            Has the properties $(\s X\setminus\s Y)\cap\s Y=\emptyset$, $(\s X\setminus\s Y)\cup\s Y = \s X$.}\\
    \glossentry
        {Complement}
        {}
        {The complement of $\s X$ consists of all elements not contained in $\s X$; equal to $\s U\setminus\s X$ where $\s U$ is the implicit ``universal'' set, e.g.\ $\R^D$ for sets of $D$-dimensional vectors.}\\
    \glossentry
        {$\bigcup, \bigcap$}
        {The union/intersection of many sets.}
        {Used similarly to $\sum$, $\prod$.}\\
    \glossentry
        {$\setbuild{x}{A,B,\ldots}$}
        {Set-builder notation.}
        {Denotes the set of all objects $x$ for which all conditions $A,B,\ldots$ are true.
            $\setbuild{x}{x\in\s X,A,\ldots}$ is more compactly written $\setbuild{x\in\s X}{A,\ldots}$.}\\
    \glossentry
        {$\{a,b,c\}$, $\{a_i\}_{i\in I}$, $\{a_n\}_{n=0}^N$}
        {Various shorthands used to define sets.}
        {The latter two are equivalent to $\setbuild{a_i}{i\in I}$ and $\setbuild{a_n}{0\leq n\leq N}$, respectively.}\\
    \glossentry
        {Open set}
        {}
        {A subset $\s X\subseteq\R^D$ is open if for each $\v\chi\in\s X$, there is some $\e>0$ such that $\v\xi\in\s X$ for all $|\v\xi-\v\chi|<\e$. 
            This is for subsets of $\R^D$; definitions of openness exist for more general sets, but we do not use them.}\\
    \glossentry
        {Closed set}
        {A set whose complement is open.}
        {}\\
    \glossentry
        {Finite set}
        {A set that contains a finite (or zero) number of elements.}
        {Many intuitive properties, such as the existence of a smallest subset with a given property, are only guaranteed for finite sets.}\\
    \glossentry
        {Bounded set}
        {}
        {A set $\s X\subseteq\R^D$ is bounded if there is some $M$ such that $|\v\chi|<M$ for all $\v\chi\in\s X$.}\\
    \glossentry
        {Compact set}
        {A set that is closed and bounded.}
        {This is for subsets of $\R^D$; we do not use the more general versions.
            Many properties of finite sets carry over to compact sets, such as having a (not necessarily unique) element that is the minimum or maximum of some property.}\\
    \glossentry
        {$\cl(\s X)$}
        {Closure.}
        {The smallest closed set that has $\s X$ as a subset.
            Equal to $\s X$ itself if it is closed.}\\
    \glossentry
        {$\Int(\s X)$}
        {Interior.}
        {The largest open set that is a subset of $\s X$.
            Equal to $\s X$ itself if it is open.}\\
    \glossentry
        {Boundary}
        {The boundary of $\s X$ is $\cl(\s X)\setminus\Int(\s X)$.}
        {Note that a non-closed set does not necessarily contain its boundary, and that \emph{boundary} and \emph{boundedness} are unrelated concepts.}\\
    \glossentry
        {$(a,b)$, $[a,b)$, $[a,b]$}
        {Open, half-open and closed intervals.}
        {Denotes the range between $a$ and $b$.
            A square bracket indicates that the endpoint is included in the interval, a parenthesis that it is not.
            Note the imperfect agreement with the concept of open and closed sets: $(a,b)$ is open and $[a,b]$ is closed for finite $a,b$, but $[a,b)$ is neither open nor closed.
            Also, intervals such as $[a,\infty)$ and $(-\infty,\infty)$ are closed.}\\
    \glossentry
        {Linear/affine/convex combination}
        {A sum of vectors of the form $\sum_i a_i\v v_i$.}
        {It is a linear combination for all $a_i$.
            It is an affine combination if $\sum_i a_i = 1$.
            It is a convex combination if $\sum_i a_i = 1$ and $a_i\geq 0$ for all $i$.}\\
    \glossentry
        {Convex set}
        {A set that contains all convex combinations of its elements.}
        {}\\
    \glossentry
        {$\hull(\s X)$}
        {Convex hull.}
        {The set of all convex combinations of the elements of $\s X$.
            Equivalently, the smallest convex set that contains $\s X$.}\\
    \glossentry
        {Linear/affine span}
        {Of a set of vectors: the set of all linear/affine combinations thereof.}
        {The linear span is often just called ``span''.}\\
    \glossentry
        {Linear/affine dimension}
        {}
        {The linear/affine dimension of a set $\s X\subseteq\R^D$ is the smallest number of vectors whose linear/affine span contains $\s X$.
            The dimension is $\leq D$.
            When it is clear from context, just ``dimension'' is often used.}\\
    \glossentry
        {Linear/affine subspace}
        {A subset of $\R^D$ that contains all linear/affine combinations of its elements.} 
        {Equivalently, the linear/affine span of some set of vectors.}\\
    \glossentry
        {Hyperplane}
        {An affine subspace of affine dimension $D-1$.}
        {Generalises the notion of a plane in 3D space and a line in 2D space.
            Given a nonzero vector $\v v$, the set $\setbuild{\v x}{\v v\cdot\v x = u}$ is a distinct hyperplane for each $u$.
            The intersection of $K$ hyperplanes in $\R^D$ with linearly independent $\v v$ vectors is an affine subspace of dimension $D-K$.}\\
    \glossentry
        {$\s A+\s B$}
        {Minkowski sum.}
        {The set $\setbuild{\v a + \v b}{\v a\in\s A,\v b\in\s B}$ (assuming the elements of $\s A,\s B$ support addition).}\\
    \glossentry
        {$x\s A$}
        {}
        {The set $\setbuild{x\v a}{\v a\in\s A}$ (assuming the elements of $\s A$ support scalar multiplication).}\\
\end{longtable}


\section{The loop integral functions}\label{sec:Jbar}
This appendix contains details on the functions appearing in the NLO and NNLO amplitudes, which originate in loop integrals. \Cref{sec:Jbar-s0,sec:Jbar-s4} contain expansions important to their numerical evaluation, and \cref{sec:integrals} contains the derivation of their analytic integrals over the Mandelstam variables.

Using the conventions of \cite{Bijnens:2011fm},\footnote{We use $\beta$ rather than $\sigma$ for consistency with \cref{eq:optical}.} the function $\bar J$ can be defined as
\begin{equation}\label{eq:orig-def}
    \bar J = \pix(\beta^2 h + 2), \qquad h = \frac{1}{\beta}\ln\frac{\beta-1}{\beta+1},\qquad \beta = \sqrt{1-\frac{4}{s}},
\end{equation}
where $\pix\equiv1/16\pi^2$.
To reduce clutter, we define $\hat J \equiv \bar J/\pix$.
Similarly, with $\hat k_i \equiv k_i/\pix^2$ for brevity, the additional functions at NNLO are defined as
\begin{equation}\label{eq:ki-def}
    \begin{alignedat}{4}
        \hat k_1 &= \beta^2 h^2, &
        \hat k_3 &= \frac{\beta^2 h^3}{s} + \frac{\pi^2 h}{s} - \frac{\pi^2}{2}, \\
        \hat k_2 &= \beta^4h^2 - 4,\qquad&
        \hat k_4 &= \frac{1}{s\beta^2}\left[\frac{\hat k_1}{2}+\frac{\hat k_3}{3} + \hat J + \frac{(\pi^2-6)s}{12}\right].
    \end{alignedat}
\end{equation}
$\bar J$ and $k_i$ are real below threshold, and are finite as $s\to 4$ from below.
However, $\beta$ is not real when $0<s<4$, which poses a problem for numerical evaluation.
This can be remedied by defining $\tbe = -i\beta$ and rewriting $h$ as
\begin{equation}\label{eq:th}
    h = \frac{2\tan^\inv(1/\tbe)}{\tbe},
\end{equation}
which can be evaluated for $0<s<4$ using only real numbers. 

The functions have further numerical problems.
$\beta$ diverges at $s=0$, which leaves $\bar J$ and $k_i$ with removable singularities there.
These are rendered harmless with a series expansion, as shown below.
As $s\to 4$ from below, $h$ diverges while $\bar J$ and $k_i$ stay finite.
Reliable evaluation of this limit also requires series expansion.
The derivatives of $\bar J$ and $k_i$ diverge in this limit (starting at the second derivative for $k_2$ and the first derivative for the others), which also necessitates series expansion for reliable handling.

\subsection{Expanding around $s=0$}\label{sec:Jbar-s0}
Near $s=0$, we make an expansion in $s=4\e^2$, where $\e$ may be complex.
To make the NNLO unitarity corrections numerically well-behaved in all cases used by us, an $\O(\e^8)=\O(s^4)$ expansion is needed.
For $\beta$ and $1/\beta$, it is
\begin{alignat}{12}
    \beta = \frac{\sqrt{\e^2-1}}{\e} &= \phantom{-}\frac{i}{\e}\Big(1 \:&-&\: \frac{\e^2}{2} \:&-&\: \frac{\e^4}{8} \:&-&\: \frac{\e^6}{16} \:&-&\: \frac{5\e^8}{128}&+& \O(\e^{10})\Big),\\
    \frac{1}{\beta} = \frac{\e}{\sqrt{\e^2-1}} &= -i\e\Big(1\:&+&\:\frac{\e^2}{2} \:&+&\: \frac{3\e^4}{8} \:&+&\: \frac{5\e^6}{16} \:&+&\: \frac{35\e^8}{128}&+& \O(\e^{10})\Big).
\end{alignat}
By using the expansion
\begin{equation}
    h = -2\sum_{n=1}^\infty \frac{1}{(2n-1)\beta^{2n}},
\end{equation}
it follows that
\begin{subequations}
    \begin{alignat}{10}
        h &= \phantom{-}2\e^2 
            \:&+&\: \tfrac{4}{3}\e^4 
            \:&+&\: \tfrac{16}{15}\e^6 
            \:&+&\: \tfrac{32}{35}\e^8, \\
        \hat J &= \phantom{-}\tfrac{2}{3}\e^2 
            \:&+&\: \tfrac{4}{15}\e^4 
            \:&+&\: \tfrac{16}{105}\e^6 
            \:&+&\: \tfrac{32}{35}\e^8\\
        \hat k_1 &= - 4\e^2 
            \:&-&\: \tfrac{4}{3}\e^4 
            \:&-&\: \tfrac{32}{45}\e^6 
            \:&-&\: \tfrac{16}{35}\e^8\\
        \hat k_2 &= - \tfrac{8}{3}\e^2 
            \:&-&\: \tfrac{28}{45}\e^4 
            \:&-&\: \tfrac{64}{63}\e^6 
            \:&-&\: \tfrac{592}{175}\e^8\\
        \hat k_3 &= \Big(\tfrac{\pi^2}{3}-2\Big)\e^2
                \:&+&\: \Big(\tfrac{4\pi^2}{15}-2\Big)\e^4
                \:&+&\: \Big(\tfrac{8\pi^2}{45}-\tfrac{28}{15}\Big)\e^6
                \:&+&\: \tfrac{328}{189}\e^8\\
        \hat k_4 &= \Big(1 - \tfrac{\pi^2}{9}\Big)\e^2
                \:&+&\: \Big(\tfrac{19}{15}- \tfrac{\pi^2}{15}\Big)\e^4
                \:&+&\: \Big(\tfrac{464}{315} - \tfrac{16\pi^2}{105}\Big)\e^6
                \:&+&\: \Big(\tfrac{820}{567} - \tfrac{16\pi^2}{105}\Big)\e^8,
    \end{alignat}
\end{subequations}
where we have omitted ``$+\O(\e^{10})$'' for brevity.
These functions are real, so only even powers of $\e$ appear.
Note also how all numerical coefficients stay roughly order 1.

\subsection{Expanding around $s=4$}\label{sec:Jbar-s4}
As $s\to4$ from below, we expand $s = 4(1-\de^2)$ with $\de>0$, and again need an eighth-order expansion at NNLO.
The expansion of $\beta$ and $1/\beta$ is
\begin{alignat}{12}
    \beta = \frac{\de}{\sqrt{\de^2-1}} &= -i\de\Big(1\:&+&\:\tfrac{1}{2}\de^2 \:&+&\: \tfrac{3}{8}\de^4 \:&+&\: \tfrac{5}{16}\de^6 \:&+&\: \tfrac{35}{128}\de^8&+& \O(\de^{10})\Big),\\
    \frac{1}{\beta} = \frac{\sqrt{\de^2-1}}{\de} &=  \phantom{-}\frac{i}{\de}\Big(1 \:&-&\: \tfrac{1}{2}\de^2 \:&-&\: \tfrac{1}{8}\de^4 \:&-&\: \tfrac{1}{16}\de^6 \:&-&\: \tfrac{5}{128}\de^8&+& \O(\de^{10})\Big) .
\end{alignat}
By using the expansion
\begin{equation}
    h = -\frac{i\pi}{\beta} + 2i\sum_{n=0}^\infty \frac{\beta^{2n}}{2n+1},
\end{equation}
it follows that (suppressing ``$+\O(\de^{10})$'')
\begin{subequations}
    \begin{alignat}{24}
        h &= \frac{\pi}{\de} 
            \:&-&\: 2&\Big[&\:1 
            \:&+&\: \tfrac{1}{3}\de^2 
            \:&+&\: \tfrac{2}{15}\de^4 
            \:&+&\: \tfrac{8}{105}\de^6 
            \:&+&\: \tfrac{16}{315}\de^8&\Big]& 
            \:&-&\: \frac{\pi\de}{2}&\Big[&\: 1 
            \:&+&\: \tfrac{1}{4}\de^2 
            \:&+&\: \tfrac{1}{8}\de^4 
            \:&+&\: \tfrac{5}{64}\de^6&\Big]&,\\
        \hat J &= 
            \:&&\: 2&\Big[&\:1 
            \:&+&\: \de^2
            \:&+&\: \tfrac{2}{3}\de^4 
            \:&+&\: \tfrac{8}{15}\de^6 
            \:&+&\: \tfrac{16}{35}\de^8&\Big]& 
            \:&-&\: \pi\de&\Big[&\: 1 
            \:&+&\: \tfrac{1}{2}\de^2 
            \:&+&\: \tfrac{3}{8}\de^4 
            \:&+&\: \tfrac{5}{16}\de^6&\Big]&,\\
        \hat k_1 &= 
            \:&-&\: 4&\Big[&\:\frac{\pi^2}{4}
            \:&+&\: \de^2
            \:&+&\: \tfrac{1}{3}\de^4 
            \:&+&\: \tfrac{8}{45}\de^6 
            \:&+&\: \tfrac{4}{35}\de^8&\Big]& 
            \:&+&\: \pi\de&\Big[&\: 4 
            \:&+&\: \tfrac{2}{3}\de^2 
            \:&+&\: \tfrac{3}{10}\de^4 
            \:&+&\: \tfrac{5}{28}\de^6&\Big]&,\\
        \hat k_2 &= 
            \:&-&\:  4&\Big[&\:1
            \:&&\: 
            \:&-&\: \de^4 
            \:&-&\: \tfrac{4}{3}\de^6
            \:&-&\: \tfrac{68}{45}\de^8&\Big]&
            \:&-&\: \pi\de&\Big[&\: 
            \:&&\: 4\de^2
            \:&+&\: \tfrac{14}{3}\de^4 
            \:&+&\: \tfrac{149}{30}\de^6&\Big]&\notag\\
                &
            \:&+&\: \pi^2&\Big[&\:
            \:&&\: \de^2
            \:&+&\: \de^4
            \:&+&\: \de^6
            \:&+&\: \de^8&\Big]&,
    \end{alignat}
\end{subequations}
\addtocounter{equation}{-1}
\begin{subequations}
    \addtocounter{equation}{4}
    \begin{alignat}{24}
        \hat k_3 &= 
            \:&&\:  2&\Big[&\:
            \:&&\:  \de^2
            \:&+&\: \de^4 
            \:&+&\: \tfrac{14}{15}\de^6
            \:&+&\: \tfrac{164}{189}\de^8&\Big]&
            \:&-&\: \pi\de&\Big[&\: 3
            \:&+&\: \tfrac{5}{2}\de^2
            \:&+&\: \tfrac{259}{120}\de^4 
            \:&+&\: \tfrac{3229}{1680}\de^6&\Big]&\notag\\
                &
            \:&+&\: \pi^2&\Big[&\:\tfrac12
            \:&+&\: \tfrac{2}{3}\de^2
            \:&+&\: \tfrac{8}{15}\de^4
            \:&+&\: \tfrac{16}{35}\de^6
            \:&+&\: \tfrac{128}{315}\de^8&\Big]&,\\
        \hat k_4 &= 
            \:&-&\: \frac{1}{3}&\Big[&\:2
            \:&+&\: \de^2
            \:&+&\: \de^4 
            \:&+&\: \tfrac{128}{105}\de^6
            \:&+&\: \tfrac{844}{945}\de^8&\Big]&
            \:&+&\: \frac{\pi\de}{4}&\Big[&\: 1
            \:&+&\: \tfrac{17}{18}\de^2
            \:&+&\: \tfrac{311}{360}\de^4 
            \:&+&\: \tfrac{2227}{2800}\de^6&\Big]&\notag\\
                &
            \:&+&\: \frac{\pi^2}{3}&\Big[&\:\tfrac{1}{12}
            \:&-&\: \tfrac{2}{15}\de^2
            \:&-&\: \tfrac{4}{35}\de^4
            \:&-&\: \tfrac{32}{315}\de^6
            \:&+&\: \tfrac{187}{1260}\de^8&\Big]&
            \:&-&\: \frac{\pi^3\de}{4}&\Big[&\:
            \:&&\:
            \:&&\:
            \:&&\:  \tfrac{121}{1536}\de^6&\Big]&.
    \end{alignat}
\end{subequations}
Since $\de$ is real, odd powers of $\de$ are permitted.
Note again how the coefficients stay roughly order 1.
Due to $\d\de/\d s = -1/8\de$, the derivative of the linear terms diverges as $\de\to0$.
Since $\hat k_2$ lacks a linear term, its derivative remains finite.
Since the derivative changes the powers in steps of two, only odd negative powers appear in derivatives of any order.

\subsection{Integrals above threshold}\label{sec:integrals}
Here, we seek to analytically determine the function $D_k^J(\lambda, v, t)$ defined in \cref{eq:mod-dispersion}.
In the relevant $z$ range, $\beta(z)$ remains real while $h(z)$ obtains an imaginary part.\footnote{We will drop the dependence on $z$ from now on; everything in this section implicitly depends on it unless otherwise specified.} The most convenient form of $h$ is
\begin{equation}
    h(z) = \frac{H(\beta) + i\pi}{\beta},\qquad H(\beta) \equiv \ln\frac{1-\beta}{1+\beta} = -\tfrac12 \tanh^\inv(\beta),
\end{equation}
where we take the branch with positive imaginary part.
We make the following easily verifiable observations about the real function $H(\beta)$:
\begin{subequations}\label{eq:Hrel}
    \setlength{\abovedisplayskip}{0pt}
    \setlength{\belowdisplayskip}{0pt}
    \setlength{\columnsep}{2em}
    \begin{multicols}{2}
        \vspace*{-2\baselineskip}
        \begin{flalign}
            \quad
            H' = \frac{-2}{1-\beta^2} = -\tfrac12 z,\label{eq:Hrel:H'z}
            &&\\
            H'' = -\beta{H'}^2,\label{eq:Hrel:H''H'}
            &&
        \end{flalign}
        \begin{flalign}
            \quad
            \beta^2 = 1 + \frac{2}{H'},\label{eq:Hrel:H'sigma}
            &&\\
            \d z = -2H''\d\beta = 2\beta{H'}^2\d\beta.\label{eq:Hrel:dz}
            &&
        \end{flalign}
    \end{multicols}
\end{subequations}%
\noindent $D_k^J(\lambda, v, t)$ will be a linear combination of integrals of the form $\int_4^\lambda \frac{z^n\:\d z}{(z-v)^{k+1}} \Im X(z)$, where $X$ is one of $\bar J$ and $k_i$.
We will ignore the denominator for now, and show later how to reduce all integrals to the form $\int z^n \Im X(z)\:\d z$.
This will involve a wide range of values for $n$, so it is easiest to treat general $n$ and then read off the special cases.
Reading from \cref{eq:orig-def,eq:ki-def}, we find that above threshold,
\begin{equation}
    \begin{gathered}
            \Im\hat J = \pi\beta,\qquad
            \Im\hat k_1 = 2\pi H,\qquad
            \Im\hat k_2 = 2\pi\beta^2 H,\\
            \Im\hat k_3 = \frac{3\pi H^2}{z\beta},\qquad
            \Im\hat k_4 = \frac{1}{z\beta^2}\left[H + \frac{H^2}{z\beta} + \beta\right].
    \end{gathered}
\end{equation}
Using \cref{eq:Hrel}, we see that everything can be expressed in terms of the functions
\begin{equation}\label{eq:def-Z}
    \Ze[p]{m}{n}(\beta) \equiv \int\ze[p]{m}{n}(\beta)\:\d\beta,\qquad 
    \ze[p]{m}{n}(\beta) \equiv \frac{H^m {H'}^n}{\beta^p}.
\end{equation}
Thanks to \cref{eq:Hrel:H''H'}, the family of functions $\ze[p]{m}{n}$ is closed under derivatives:
\begin{equation}\label{eq:zederiv}
    \frac{\d}{\d\beta}\ze[p]{m}{n} = m\ze[p]{m-1}{n+1} - n\ze[p-1]{m}{n+1} - p\ze[p+1]{m}{n}.
\end{equation}
It is therefore our hope that the highly nontrivial integral $\Ze[p]{m}{n}$ can mostly be expressed as a sum of $\ze[p']{m'}{n'}\!$, plus some special cases (for instance, $\ze[p]{m}{0}$ is not the derivative of another $\ze[p']{m'}{n'}$).
We will therefore attempt to find recurrence relations on $m,n,p$ that allow $Z$ to be reduced to $\zeta$'s and a few special cases. 

\subsubsection{The integral $\Ze[p]{m}{n}$ for $p=0$}
First, we note that $n$ can be assumed non-negative, since 
\begin{equation}\label{eq:incr-n}
    \Ze[ p]{m}{ n} = \tfrac12\big[\Ze[ p-2]{m}{ n+1} - \Ze[ p]{m}{ n+1}\big], 
\end{equation}
according to \cref{eq:Hrel:H'sigma}.\footnote{We will not need to treat negative $m$ when integrating the loop integral functions.
As we will see below, $\Ze[p]{m}{n}$ is tractable for all integer $n$ and $p$, but $m$ has to stay non-negative.} Then, before treating general $p$, we consider $\Ze{m}{n}\equiv \Ze[0]{m}{n}$ for $n,m\geq0$.
The integral of \cref{eq:zederiv} gives
\begin{equation}
    \begin{aligned}
        \ze{m-1}{n-1} &= (m-1)\Ze{m-2}{n} - (n-1)\Ze[-1]{m-1}{n},\\
        \ze[-1]{m}{n-1} &= m\Ze[-1]{m-1}{n} - (n-1)\Ze[-2]{m}{n} + \Ze{m}{n-1},
    \end{aligned}
\end{equation}
where the second line allows for the removal of $\Ze[-1]{m-1}{n}$ in the first.
\Cref{eq:incr-n} can be invoked to turn $\Ze[-2]{m}{n}$ into $2\Ze{m}{n-1} + \Ze{m}{n}$, and after extracting $\Ze{m}{n}$, we get a recurrence relation where both $m$ and $n$ decrease:
\begin{equation}\label{eq:Zmn}
    \Ze{m}{n} = \frac{3-2n}{n-1}\Ze{m}{n-1} - \frac{\ze[-1]{m}{n-1}}{n-1} + \frac{m}{(n-1)^2}\big[(m-1)\Ze{m-2}{n} - \ze{m-1}{n-1}\big].
\end{equation}
This is valid for all $m\geq 0, n > 1$.
The $n=1$ case is covered by the trivial identity\footnote{We suppress the constant of integration here and everywhere else.}
\begin{equation}\label{eq:Zm1}
    \Ze{m}{1} = \frac{\ze{m+1}{0}}{m+1}.
\end{equation}
$\Ze{m}{0}$ requires some more thought and will be treated later (see \cref{eq:Zm0p}).

\subsubsection{Reduction of $\Ze[p]{m}{n}$ to $\Ze[0]{m}{n}$ and $\Ze[1]{m}{n}$}
We can integrate and restructure \cref{eq:zederiv} into the recurrence relation
\begin{equation}\label{eq:Zmnp-decr}
    \Ze[p]{m}{n} = \frac{m}{p-1}\Ze[p-1]{m-1}{n+1} - \frac{n}{p-1} \Ze[p-2]{m}{n+1} - \frac{\ze[p-1]{m}{n}}{p-1}, \qquad p \neq 1.
\end{equation}
This allows any $p>0$ to be reduced to the cases $p=0$ and $p=1$.
When $p$ is negative, \eqref{eq:incr-n} furnishes the simpler relation
\begin{equation}\label{eq:Zmnp-incr}
    \Ze[p]{m}{n} = \Ze[p+2]{m}{n} + 2\Ze[p+2]{m}{n-1}.
\end{equation}
This reduces any $p<0$ to $p=0$ and $p=1$, as long as $n$ stays positive.
In the $n=0$ case, we again turn to \cref{eq:zederiv}:
\begin{equation}
    \Ze[p]{m}{0} = (p+1)\Ze[p+2]{m}{-1} - m\Ze[p+1]{m-1}{0} - \ze[p+1]{m}{-1}.
\end{equation}
Using \cref{eq:incr-n} to treat the $n=-1$ terms, we arrive at
\begin{equation}\label{eq:Zm0p-incr}
    \Ze[p]{m}{0} = \frac{p+1}{p-1}\Ze[p+2]{m}{0} + \frac{2m}{p-1}\Ze[p+1]{m-1}{0} + \frac{\ze[p+1]{m}{0} - \ze[p-1]{m}{0}}{p-1},
\end{equation}
which works for $n=0,p\neq 1$.\footnote{It could also be adapted to other $n\neq \frac{1-p}{2}$, but in those cases \cref{eq:Zmnp-incr} is simpler.}

In the case $p=1$, $n > 0$ we integrate by parts:
\begin{equation}\label{eq:Zmn1}
    \Ze[1]{m}{n} = \frac{\Ze{m}{n}}{\beta} + \int \frac{\Ze{m}{n}}{\beta^2}\d\beta.
\end{equation}
Through \cref{eq:Zmn,eq:Zmnp-decr}, the integral will reduce to $p=0$ terms that are easy to handle, plus various $\Ze[1]{m'}{n'}$ where $m'\leq m,n'\leq n$.
At least one of the inequalities is strict, so we will eventually arrive at $\Ze[1]{0}{n''}$ and $\Ze[1]{m''}{0}$.
The former can be run through the recurrence again, but the latter requires separate consideration.

\subsubsection{Special cases for $n=0$} 
The relations above are capable of reducing almost all $Z$'s to $\zeta$'s, but they are unable to get rid of $\Ze[p]{m}{0}$ for $p=0,1,2$.
If we could define a function $\Ph{p}{\ell}$ with derivative $-H'\Ph{p}{\ell-1}$ and base case $\frac{\d}{\d\beta}\Ph{p}{0}=\beta^{-p}$, then an elegant solution to this would be
\begin{equation}\label{eq:Zm0p}
    \Ze[p]{m}{0} = \sum_{\ell=0}^m \frac{m!}{(m-\ell)!} \Ph{p}{\ell}H^{m-\ell},\qquad m,p\geq 0.
\end{equation}
Such a function can be constructed using polylogarithms, since 
\begin{equation}
    \frac{\d}{\d\beta}\Li_\ell[f(\beta)] = \Li_{\ell-1}[f(\beta)] \frac{f'(\beta)}{f(\beta)}.
\end{equation} The correct recurrence relation is obtained by solving a simple differential relation for $f(\beta)$, which gives
\begin{equation}\label{eq:K}
    f(\beta) = K\frac{1+\beta}{1-\beta},\qquad \Li_0[f(\beta)] = \frac{K(1+\beta)}{(1-K) - \beta(1+K)}.
\end{equation}
We immediately see that $K=-1$ is suitable for $p=0$, and $K=1$ for $p=2$:\footnote{This process can in principle be continued to treat all $p\geq 2$, but it is more practical to rely on the recurrence relations to get rid of larger $p$.}$^,$\footnote{Note that $\Ph{2}{\ell}$ is complex-valued for $\ell>0$, since $\frac{1+\beta}{1-\beta}>1$ for $z$ above threshold.}
\begin{equation}
    \Ph{0}{\ell} = -2\Li_\ell\left(\frac{\beta+1}{\beta-1}\right),\qquad
    \Ph{2}{\ell} = 2\Li_\ell\left(\frac{1+\beta}{1-\beta}\right),
\end{equation}
We cannot handle $p=1$ directly this way, but using the close connection between $\Li_1(x)$ and the natural logarithm, we find that
\begin{equation}
    \Ph{1}{\ell} = -\frac{\Ph{0}{\ell+1} + \Ph{2}{\ell+1}}{2}
\end{equation}
gives the correct result.\footnote{Although $\frac{\d}{\d\beta}\Ph{1}{0} = 1/\beta$, we have $\Ph{1}{0}=\ln(-\beta)$, not $\ln\beta$, so one must be careful to place $\beta$ on the correct side of the branch cut.}

\subsubsection{The treatment of $(z-v)$}
The factors that arise due to the dispersion relations are highly problematic when $v\neq0$ (which corresponds to $s\neq 0$ or $u\neq 0$ in \cref{eq:dispersion}).
We only need to consider $k=0$, since 
\begin{equation}\label{eq:s0}
    \int \frac{\ze[p]{m}{n}}{(z-v)^{k+1}}\d z = \frac{1}{k!}\frac{\d^k}{\d v^k}\int \frac{\ze[p]{m}{n}}{z-v}\d z;
\end{equation}
the derivative can be taken after evaluating the integral.
Trying to get rid of the last power of $(z-v)$ is futile, so we instead change variables to $\beta$ and find
\begin{equation}
    \int\frac{\ze[p]{m}{n}}{z-v}\d z = 2\int\frac{4}{zv}\frac{\ze[p-1]{m}{n+2}}{\tfrac4v - \tfrac4z}\d\beta = \frac{-4}{v}\int \frac{\ze[p-1]{m}{n+1}}{\beta^2 - \beta_v^2}\d \beta,
\end{equation}
where $\beta_v \equiv \beta(v) = \sqrt{1-4/v}$.\footnote{The same relation is useful if one wishes to explicitly evaluate a partial-wave expansion like \cref{eq:part-wave}, since the Legendre polynomials consist of powers of $1/(z-4)$.
It is further simplified by $\beta(4)=0$.} This relation is singular when $v=0$, but then $\frac{z^n}{(z-v)^{k+1}}=z^{n-(k+1)}$ so it is not needed.\footnote{Note that $\beta_v$ is imaginary for $v\in(0,4)$, which forces a more involved detour through the complex plane than above.
For instance, the functions $\Ps[\pm]{\ell}$ below will be complex-valued, although in the end the integral will of course remain real.}

It turns out that $\int H^m/(\beta^2-\beta_v^2)\:\d\beta$ is tractable, so the strategy is to separate that from the rest of the integral by repeatedly applying partial fractions:
\begin{equation}\label{eq:part-frac-n}
    \frac{\ze[p]{m}{n}}{\beta^2 - \beta_v^2} 
    = \frac{v}{4}\left[\ze[p]{m}{n} - \frac{2\ze[p]{m}{n-1}}{\beta^2-\beta_v^2}\right].
\end{equation}
If $n$ is negative, we instead use \cref{eq:incr-n}.
Once $n$ has been reduced to zero in the $(\beta^2-\beta_v^2)$-containing term this way, we remove $p$ similarly:
\begin{equation}\label{eq:part-frac-p}
    \frac{\ze[p]{m}{0}}{\beta^2-\beta_v^2} 
    = \ze[p+2]{m}{0}\left[\frac{\beta_v^2}{\beta^2-\beta_v^2} + 1\right]
    = \frac{1}{\beta_v^2}\left[\frac{\ze[p-2]{m}{0}}{\beta^2-\beta_v^2} - \ze[p]{m}{0}\right].
\end{equation}
This reduces $p$ to 0 or $-1$; the latter can be handled with the expansion
\begin{equation}
    \frac{\beta}{\beta^2-\beta_v^2} = \frac{1-\beta_v}{(1-\beta)(\beta-\beta_v)} - \frac{\beta_v}{\beta^2 - \beta^2_0} + \tfrac12(1+\beta) H'.
\end{equation}
The second term on the right-hand side corresponds to $p=0$, and the third does not involve $\beta_v$ at all.
In the spirit of \cref{eq:Zm0p}, the first is equal to $\frac{\d}{\d\beta}\Ps[-]{0}$ if we identify
\begin{equation}
    \Ps[\pm]{\ell} = -\Li_{\ell+1}\left[\frac{(1+\beta)(1\pm\beta_v)}{(1-\beta)(1\mp\beta_v)}\right]
\end{equation}
guided by \cref{eq:K}.
Since $\frac{\d}{\d\beta}\Ps[\pm]{\ell} = -H'\Ps[\pm]{\ell-1}$, 
\begin{equation}
    \int\frac{\ze[-1]{m}{0}}{\beta^2-\beta_v^2}\d\beta = \frac{\Ze{m}{1}+\Ze[-1]{m}{1}}{2} - \int\frac{\beta_v\ze{m}{0}}{\beta^2-\beta_v^2}\d\beta + \sum_{\ell=0}^m \frac{m!}{(m-\ell)!}\Ps[-]{\ell}H^{m-\ell}.
\end{equation}
This leaves $p=0$, for which a similar solution is
\begin{equation}
    \int\frac{\beta_v\ze{m}{0}}{\beta^2-\beta_v^2}\d\beta = \sum_{\ell=0}^m \frac{m!}{(m-\ell)!}\frac{\Ps[-]{\ell} - \Ps[+]{\ell}}{2}H^{m-\ell}.
\end{equation}
The very last piece in the puzzle of evaluating $D_k^J(\lambda, v, t)$ is the conceptually simple derivative in \cref{eq:s0}:
\begin{equation}
    \frac{\d\beta_v}{\d v} = \frac{2}{\beta_v v^2},\qquad \frac{\d}{\d v}\Ps[\pm]{\ell} = \pm\frac{v}{\beta_v}\Ps[\pm]{\ell-1},\qquad \frac{\d\beta}{\d v} = \frac{\d}{\d v}\Ze[p]{m}{n} = 0.
\end{equation}

\subsubsection{The completed integral}
The above recurrence relations allow for the integration of all terms that appear in $D^J_k(\lambda,v,t)$.
$\Ze[p]{m}{n}$ diverges at $z\to 4$ for some values of $p,m,n$ (in particular those that contain $\Ph{2}{1}$ or negative powers of $\beta$), so obtaining a finite lower limit of the overall integral requires careful (albeit straightforward) extraction and cancellation of those divergences.
The resulting expressions are very lengthy in most cases, so we do not reproduce them here.%
\footnote{
    The \form\ implementation of the relations, and the expressions produced by it, are available from Mattias Sjö upon request.}

\addcontentsline{toc}{section}{References}
\bibliographystyle{myJHEP}
\bibliography{references}

\end{document}